\documentclass[]{aa}
\usepackage[english]{babel}
\usepackage{graphicx}
\usepackage{newlfont}
\usepackage{amssymb}
\usepackage{subfig}
\usepackage{commath}
\usepackage{footnote}
\usepackage{enumitem}
\usepackage{booktabs}
\usepackage{grffile}
\usepackage{threeparttable}
\usepackage{soul}
\usepackage{subcaption}

\usepackage{amsmath}
\usepackage{latexsym}
\usepackage{amsthm}
\usepackage{natbib}
\usepackage{microtype}
\bibpunct{(}{)}{;}{a}{}{,} 
\usepackage{longtable}
\usepackage{rotating}
\usepackage{comment}

\usepackage[usenames,dvipsnames]{color}
\usepackage[bookmarks, colorlinks, breaklinks]{hyperref}  
\hypersetup{linkcolor=blue,citecolor=MidnightBlue,filecolor=black,urlcolor=MidnightBlue} 
\usepackage{stfloats} 
\usepackage[varg]{txfonts}
\usepackage{breakcites}

\def\YSZ {Y_{\textrm SZ}}
\def\Mv {M_{500}}
\def\Lv {L_{500}}
\def\Rv {R_{500}}
\def\Tv {\theta_{500}}
\def\Yv {Y_{500}}
\def\YL {$Y_{500}$--$L_{500}$}
\def\zs{z_{\rm spec}}
\def\zp{z_{\rm phot}}
\def\dzn{\left|\Delta(z)\right|/(1+z)}

\def\Ytot {Y_{5R500}}
\def\ts{\theta_{s}}

\def\xmm{XMM-{\it Newton}}
\def\planck{{\it Planck}}
\def\chandra{{\it Chandra}}
\usepackage{color}

\def\msun{M_{\odot}}

\DeclareOldFontCommand{\rm}{\normalfont\rmfamily}{\mathrm} 

\begin{document}

\title{PSZ: The Meta-catalogue of \planck\ Sunyaev-Zeldovich sources}

\author{ P. Tarr\'io\inst{1} \corrauth{p.tarrio@oan.es}\and 
M. Arnaud\inst{2} \email{monique.arnaud@cea.fr}\and  
J.-B. Melin\inst{3} \email{jean-baptiste.melin@cea.fr}\and 
G.W. Pratt\inst{2} \email{gabriel.pratt@cea.fr}\and 
T. Sadibekova\inst{2} \email{tatyana.sadibekova@cea.fr}}
\institute{Observatorio Astron\'omico Nacional, IGN, Calle Alfonso XII 3, E-28014 Madrid, Spain	
\and Université Paris-Saclay, Université Paris Cité, CEA, CNRS, AIM, 91191 Gif-sur-Yvette, France
\and Université Paris-Saclay, CEA, Département de Physique des Particules, 91191, Gif-sur-Yvette, France
}
\date{}

\abstract{ 
We present the PSZ, a meta-catalogue of 1962 clusters and cluster candidates produced by the \planck\ Collaboration. The PSZ contains fully-updated validation information for all official \planck\ catalogue detections, together with redshift estimates for confirmed clusters, with no duplicate entries. The validation is derived from a synthesis of optical and X-ray follow-up campaigns, supplemented  by cross-matching  with external catalogues with redshift information, and with \xmm\ archive data. The external catalogues considered include the all-sky X-ray catalogue MCXC-II, the eROSITA X-ray cluster catalogue, the  RASS-MCMF and PSZ-MCMF catalogues, and the extended \planck\ catalogues of \citeauthor{bur17}.
A total of 281 clusters are newly-confirmed owing to this process; conversely, 262 \planck\ candidates have been invalidated. Of the 1500 confirmed clusters, 274 have updated redshifts, and 278 have newly-assigned redshifts.   An MCXC-II counterpart is assigned to 631 clusters, updating the MCXC cross-match published in the \planck\ catalogues. Differences with the PSZ2 update of \citeauthor{BH24}  are discussed.
We further introduce a new, homogeneously-derived mass estimate, corrected for selection effects owing to intrinsic scatter and the properties of the underlying mass function. New posterior probability contours in the $Y_{5R500} - \theta_s$ plane are provided for all sources, in addition to the corresponding $\Mv(z)$ degeneracy curves. The PSZ includes both corrected and uncorrected $\Mv$ values for confirmed clusters. A methodology for cross-identification between catalogues is presented.  We show that simple fixed-distance matching is insufficient for this task, and demonstrate the necessity for additional consistency checks. Mass proxies, redshifts, and distance versus angular size are the key quantities to be compared; these enable subsequent detailed investigation at the individual cluster level when necessary. The final PSZ comprises 1500 confirmed clusters,  262 noise-dominated detections, and 200 candidates awaiting validation. The PSZ contains 33 fields per object. Beyond the distinction between confirmed clusters, invalidated detections, and candidates, the new {\tt STATUS}  field identifies complex situations (confusion cases, complex clusters, or problems in counterpart association), along with a {\tt COMMENT} field that provides additional detail.} 
\titlerunning{The Metacatalogue of \planck\ SZ clusters}
\authorrunning{Tarr\'io et al.}

\maketitle
\nolinenumbers

\section{Introduction}\label{sec:intro}

In recent years, the detection of galaxy clusters through the Sunyaev-Zeldovich effect (SZE), produced by interaction of Cosmic Microwave Background photons with the hot ionised gas of the intracluster medium (ICM), has become a competitive way of assembling large cluster samples. Large-scale surveys with \planck\ (hereafter collectively referred to as PSZ), the South Pole Telescope (SPT), and the Atacama Cosmology Telescope (ACT) have now yielded samples of several thousands of clusters and cluster candidates \citep{esz,psz1, psz1rev,psz2,spt,Bocquet2019,sptecs,Huang2020,spt500,act,act_data, Hilton2018, Hilton2021}.

The PSZ catalogues are unique in that they are all-sky (indeed, they are the only all-sky cluster catalogues since the Rosat All-Sky Survey [RASS] in the 1990s), enabling the detection of the rarest, most massive objects, a population that is a particularly sensitive probe of cosmology and structure formation. Such a sample is important in the context of ongoing wide-area cluster surveys in the SZE (see references above), X--ray \citep[][eROSITA]{eRASS}, and optical \citep[][]{Euclid_Q1} and remains competitive with Stage IV surveys in terms of cosmological constraints despite containing fewer objects \citep{aym26}.

Several updates and extensions to the \planck\ cluster catalogues have been published since their original appearance. As the catalogues contained both clusters and cluster candidates, a concerted spectroscopic follow-up effort has been undertaken by the community, resulting in further validation and redshift information becoming available. \citet{BH24} and references therein constitute  perhaps the most comprehensive compilation 
to date of the current state of the spectroscopic follow-up of the second \planck\ cluster catalogue (PSZ2). Another avenue of investigation has resulted in extensions to the \planck\ catalogues, generally to lower S/N detections, in combination with wide-field optical imaging or spectroscopic data \citep[e.g.][]{bur17,PSZMCMF2023}. Alternative \planck\ catalogues have also been produced using e.g. neural networks \citep{Bonjean2020,MILCANN,Meshcheryakov2022} or iterative Multifrequency Matched Filters \citep{SZiFi}. Finally, catalogues have been produced using the \planck\ maps  as part of various multi-wavelength / multi-instrument cluster detection strategies \citep[e.g.][]{Aghanim2019,comprass,ComPACT,PSZSPT}.

In this paper, we revisit the cluster catalogues produced by the \planck\ Collaboration. We describe the construction and properties of the PSZ meta-catalogue, derived from the ESZ, PSZ1, and PSZ2 \planck\ cluster catalogues, with updated validation and redshift information, and a new mass estimate that is corrected for selection effects.  The PSZ includes consolidated cross-identification between the individual \planck\ sub-catalogues and updated validation and redshift estimates obtained from critical compilation of such information from optical follow-up campaigns. A complete cross-identification with the all-sky X--ray meta-catalogue MCXC-II \citep{MCXC2024} was undertaken. Further validation and redshift information was obtained from external catalogues including the extended \planck\ catalogue of \citet{bur17}, the eROSITA All-Sky Survey  \citep[][eRASS]{eRASS}, the RASS and PSZ Multi-Component Matched Filter catalogues \citep[][the RASS-MCMF and PSZ-MCMF, respectively]{RASSMCMF2023,PSZMCMF2023}, and \xmm\ archival data.
Critical comparison with the PSZ2 update by \citet{BH24} was also performed. \citeauthor{BH24}'s study  
concerns uniquely the PSZ2, and could not cross-identify with the more recently-published MCXC-II and eRASS X--ray selected cluster catalogues. In addition, they used a fixed distance criterion for cross-identification, which we show in this paper to be sub-optimal when applied to the \planck\ survey, which has only moderate angular resolution. 

Our introduction of a new mass estimate is motivated by the fact that the published \planck\ cluster masses were not corrected for selection effects. Neglect of such effects can result in biased mass measurements \citep[e.g.][]{aem11}, and can complicate comparison between different SZE-selected cluster samples \citep[e.g.][]{bat16}. The new mass estimate for each object is homogeneously-derived, and includes correction for selection effects due to intrinsic scatter and the properties of the underlying mass function. Such a mass estimate will be useful for future cross-comparison with other SZE-selected cluster samples.

The paper is organised as follows. Section~\ref{sec:method} describes the methodology used for object matching between catalogues, use of ancillary data, redshift updates, and the definition of the new keyword {\tt STATUS}. Particular attention is paid to the  cross-matching procedure, for which we implement an approach that is more sophisticated than a simple fixed-distance matching. Section~\ref{sec:PSZconstruction} describes the construction of the PSZ, including consolidated cross-identification between \planck\ sub-catalogues and the addition of firmly confirmed \planck-detected sources below each individual sub-catalogue S/N threshold. Section~\ref{sec:PSZupfu} details the validation updates and new redshift estimates obtained from various optical follow-up campaigns; Sect.~\ref{sec:mcxc} describes the cross-matching with MCXC-II; and Sect.~\ref{sec:PSZupcat} reports the new validation and redshift information obtained from other external catalogues and \xmm\ information. A critical comparison with the PSZ2 update of \citet{BH24} is presented in Sect.~\ref{sec:BH24}. The new mass estimate is described in Sect.~\ref{sec:mass}, and Sect.~\ref{sec:PSZprop} presents the properties of the PSZ, including figures summarizing the construction steps and differences with the published information. Finally, Sect.~\ref{sec:conclusion} presents our conclusions.
 
We adopt a $\Lambda$CDM cosmology with $H_0 = 70~\rm{km}~\rm{s^{-1}}~\rm{Mpc^{-1}}$, $\Omega_{\rm m} = 0.3$ and $\Omega_{\Lambda} = 0.7$ throughout the paper. The quantity $E(z)$ is the ratio of the Hubble constant at redshift $z$ to its present value, $H_0$, i.e., $E(z)^2 = \Omega_{\rm m}(1 + z)^3 + \Omega_{\Lambda}$. The quantities $\Tv$, $\Mv$ and $\Yv$, $\Lv$ are the angular radius, integrated mass, integrated Compton and X--ray luminosity parameter, respectively, at $\Rv$.

\section{Methodoloy}
\label{sec:method}

\subsection{Object matching}
\label{sec:matching}

Matching objects  between catalogues is a fundamental but challenging task, whether it is for validation of cluster candidates or for scientific identification of counterparts at other wavelengths.  

First, we have  to identify  the  pairs of  potentially  matching objects between  two inputs catalogues  ({\tt cat1} and {\tt cat2}). Unless otherwise stated, we  performed a two-way cross-match, i.e. we determined the closest {\tt cat1} cluster for each {\tt cat2} cluster, the closest {\tt cat2} cluster for each {\tt cat1} cluster, and we retained the {\tt cat1}-{\tt cat2} pairs that were identified in both directions. The two-way crossmatch may give different cluster pairs than those given by a single-way crossmatch. This occurs when a cluster of one catalogue, e.g. {\tt cat1}, is the closest to two or more clusters of the other catalogue, {\tt cat2}. This may happen when {\tt cat1} is much sparser than {\tt cat2};  in that case, the two-way process allow us to correctly discard the un-matching pairs. It may also indicate a {\tt cat1} confusion case, where two clusters, well resolved in {\tt cat2},  cannot be separated by  {\tt cat1} lower resolution.  When appropriate,  we thus also consider the second closest pair (e.g., Sect.~\ref{sec:PSZerass}). 

\begin{figure*}[t]
\centering
\includegraphics[width=0.9\textwidth]{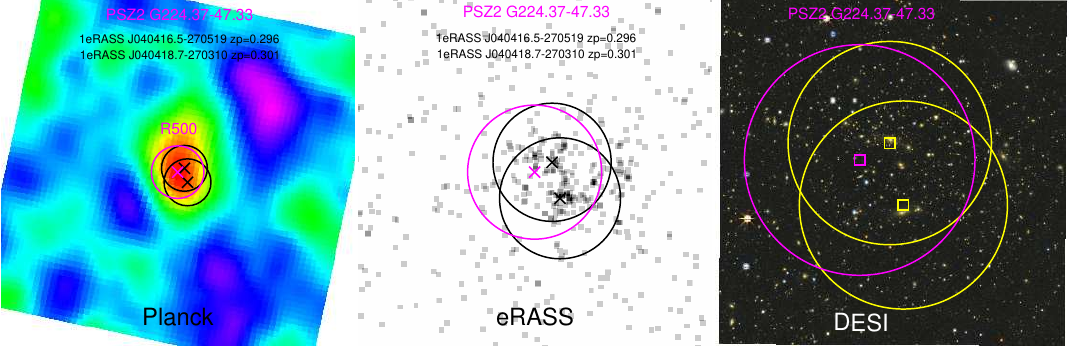}
\caption{\footnotesize Example of a PSZ cluster with  {\tt STATUS=Complex}.
\planck, X--ray and DESI images from left to right. The objects from different surveys are color-coded and listed with redshift on each panel.  The cluster centers are marked with points. Big circles have  a radius of $\Rv$ computed from corresponding X--ray or SZE survey data.
 PSZ2 G224.37-47.33 is a double cluster at $z=0.3$.  The two components, 1eRASS J040416.5-270519  at  $\zp=0.2955\pm0.0055$ and  1eRASS J040418.7-270310 at $\zp=0.3006\pm0.0057$  are  resolved by  eROSITA (middle panel) but not by \planck\ (left panel). }
\label{fig:PSZstatus_complex}
\end{figure*}

\begin{figure*}[t]
\centering
\includegraphics[width=0.9\textwidth]{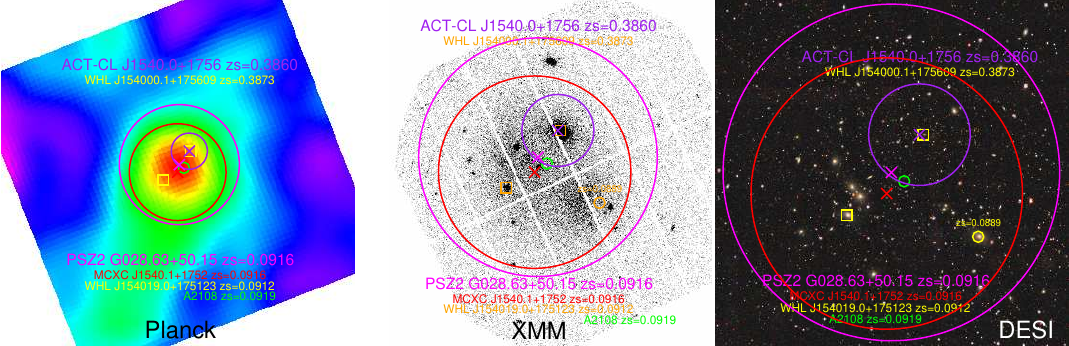}
\caption{\footnotesize Example of a PSZ cluster with {\tt STATUS=Confusion}. \planck, X--ray and DESI images from left to right with layout as in Fig.~\ref{fig:PSZstatus_complex}.    PSZ2 G028.63+50.15 appears as a single diffuse emission in the \planck\ image (left).  It is associated with MCXC J1540.1+1752 (red cross and circle) at  $\zs=0.0916$, coincident with A2108 (green point).   ACT-CL J1540.0+1756   is a background cluster at $\zs=0.386$ (purple) and a separate entry in PSZ. Both clusters are detected in the SDSS data, as WHL J154019.0+175123 ($\zs=0.0912$) and   WHL J154000.1+175609 ($\zs=0.3873$),  clearly visible in the DESI image (yellow squares in right panel).  
 The \xmm\ image (middle panel) reveals the complexity of the region, with 3 diffuse components, not resolved in the \planck\ and  RASS images.
 The North component is the ACT cluster. The two other components correspond to the $z=0.09$ cluster which appears bimodal, with a North-East 
 component around the BCG of WHL J154019.0+175123 and a South-West component centred on a second 
 bright galaxy, SDSS J153946.46+175007.8 at the same redshift ($\zs=0.0889$), marked by yellow/orange circle in middle and right panel. 
 PSZ2 G028.63+50.15 and MCXC J1540.1+1752 centers are located between the 3 X--ray peaks.
 The signal of PSZ2 G028.63+50.15 and MCXC J1540.1+1752 is the projection  
 of the $z=0.09$ and $z=0.39$  clusters.}
\label{fig:PSZstatus_confusion}
\end{figure*}

Second, we must  define which pairs are true associations.  The simple criteria of fixing an upper limit on distance is generally not sufficient:\\
\noindent {\it Distance matching:} The distance, $D$, between the same cluster detected in two surveys  is the result of   1) the spatial resolution of the surveys 2) possible physical offset between emission peaks  (a priori smaller than the cluster size) if surveys were conducted at different wavelengths, e.g. between SZE and X--ray or optical  peak. It thus depends both on the surveys and cluster properties.  We therefore  examine the pair positions in the $D$-$D/\Tv$ plane (e.g.  Fig.~\ref{fig:PSZ_MCXC_crossmatch}).  Ideally  the 'true' and 'bad' associations would be identified by two well separated clouds. This is generally not the case, with a 'grey' zone where each cluster is manually checked. \\
\noindent {\it Consolidation:}  Furthermore, this distance matching needs to be consolidated.  When validating  candidates, the main issue is chance association of a bona-fide cluster with a false candidate (noise dominated source). In cluster matching, the main issue is having two distinct clusters close in projection on the sky  or sub-clustering (e.g. Fig. \ref{fig:PSZstatus_complex} and Fig. \ref{fig:PSZstatus_confusion}). We therefore check both the redshift consistency for independently  confirmed clusters (keeping in mind that inconsistency may be due to redshift error) and mass proxy consistency. The mass proxy is any quantity related to the mass, e.g $\YSZ, \Lv, \Mv$, as published in the catalogue. For cluster candidates, the \planck\ mass and $\Yv$ are recomputed at the redshift of the possibly matching cluster. Inconsistency between mass proxies is defined as significant deviation from the empirical scaling relation between the two catalogue quantities (e.g $M_{500, \rm cat1}$--$M_{500, \rm cat2}$,  mass ratio, $\Lv$--$\YSZ$), taking into account intrinsic scatter and statistical errors (see Sect.~\ref{sec:PSZerass} for an example). Note that the scaling relation does not need to be de-biased for selection effects or 'calibrated'. The goal is simply to identify chance associations as 'outliers', not to measure the underlying physical relation.

Our approach is conservative.
In the case of  line-of-sight structure of two clusters, the prime contributor to the detection may be difficult to assess  and furthermore not the same in SZE and X--ray.  Due to the different $z$ dimming, a more distant and more massive cluster may dominate the SZE signal, while a less massive but closer component dominates the X--ray detection.  In  ambiguous cases we do not associate the SZE object and the X--ray cluster.

\subsection{Manual inspection and ancillary data}
Ambiguous cases for object-matching and/or redshift determination are individually studied. We consider detailed properties of the SZE detection, like the SZE morphology, the position error, and importantly, $Q_{\rm neural}$. $Q_{\rm neural}$ indicates the reliability of the SZE detection as a real galaxy cluster, and was defined for PSZ1 and PSZ2 candidates as $Q_{\rm neural}=1-Q_{\rm bad}$, following the definition in \cite{Aghanim2015}. $Q_{\rm neural}<0.4$ identifies low-reliability detections with a high degree of success.  We  search for close-by known clusters, considering  optical catalogues of \citet{1989ApJS...70....1A}, \citet[][RM]{Rozo2015},  \citet[][]{Rykoff16},  \citet[][WH]{2015ApJ...807..178W}, \citet[][WHY]{2018MNRAS.475..343W}, in addition to  \planck, ACT, SPT\footnote{For these specific cases, we consider  all the published ACT and SPT catalogues, mentioned in Sect.\ref{sec:intro}, but a full cross-match study of \planck, ACT and SPT  clusters will be discussed  in a forthcoming paper presenting MCSZ, a meta-catalogue of SZ clusters.}, PSZ-MCMF, RASS-MCMF, eRASS and MCXC-II catalogues and critically compare the catalogue cluster properties (mass-proxy,  redshift, size... ). 
The study  also includes visual inspections of optical, X--ray and SZE maps, with  the positions of clusters from various catalogues overlaid, as well as  galaxies with spectroscopically-measured redshifts if necessary (from NED and SDSS data base).  Depending on data availability and case complexity, we used  images from  
 XMM-Newton, Chandra, SWIFT,  PanSTARRS (available in M2C data base\footnote{\url{https://www.galaxyclusterdb.eu/m2c/}}),  SDSS-DR18\footnote{\url{https://skyserver.sdss.org/dr18/}}, eRASS\footnote{\url{https://erosita.mpe.mpg.de/dr1/}} and DESI\footnote{\url{https://alasky.cds.unistra.fr/DESI-legacy-surveys/DR10/}} color image from Aladin server.

\subsection{Redshift} 
\label{sec:redshift}
When several redshifts are available, we favour spectroscopic redshift (hereafter $\zs$), which is more accurate  than photometric redshift (hereafter $\zp$), after checking the  consistency between the two values\footnote{We use either the published errors or typical photometric errors \citep{Rozo2015}}. Although our work is not exhaustive, we try to identify cases of prominent  redshift discrepancies (typically $\Delta z/(1+z)>0.05$). Understanding  the origin of the differences allows the identification  of 'incorrect' redshifts (e.g. redshift of  background or foreground galaxy), and also cases of confusion.

\subsection{Object status} 
\label{sec:status}
As the input catalogues  contain both clusters and cluster candidates, PSZ provides  the {\tt STATUS}  of the detection. This fundamental keyword distinguishes between confirmed objects and candidates, and also flags complex situations: 

\noindent  {\bf U:} (Unknown): The detection is still a cluster candidate. There is no identified cluster counterpart from cross-matched catalogues. The search in optical/IR galaxy surveys is inconclusive and there is  no or insufficiently deep  follow-up. In the following text such cases will also be referred to as unvalidated detections.  

\noindent {\bf False:} The candidate is noise dominated. We keep these detections in the PSZ for completeness.  In the following text such cases will also be referred to as invalidated candidates.  

\noindent {\bf C1:} (Confirmed): The candidate is robustly confirmed as a bona fide cluster. 

\noindent {\bf C2:}  There is 
a potential optical or X--ray counterpart but the association is uncertain. For instance, its mass proxy is low for  the estimated SZE mass at the same redshift,  the information is not available, or there is  likely a high Malmquist bias. Such cases are identified in the various step of the PSZ construction, as described in the corresponding sections.

\noindent {\bf Complex:} The counterpart has several components at the same redshift (see Fig.~\ref{fig:PSZstatus_complex}).

\noindent {\bf Confusion:} More than one cluster contributes to the SZE signal with the SZE peak not centered on any of the clusters (see Fig.~\ref{fig:PSZstatus_confusion}). 

\noindent {\bf Secondary:} A secondary detection (sub-structure) in a massive large cluster. This is for  very nearby clusters. 

For all PSZ clusters, the cases of contribution  of  secondary component(s)  or  projection on the sky of a second cluster at different redshift (separation less than cluster size) are  indicated in the {\tt COMMENT}  field.  This information is not exhaustive.

\section{PSZ construction: the union of Planck catalogues} \label{sec:PSZconstruction}

\subsection{Input data}
The ESZ, PSZ1 and PSZ2 catalogues are extracted  from   the \planck\ survey, giving complete sky coverage. They correspond to increasing depth, being based on the  first 10, 15.5 months and 29 months full-mission data, respectively, with a higher detection threshold for ESZ (signal-to-noise ratio  $S/N > 6$) than the nominal $S/N> 4.5$ for PSZ1 and PSZ2. 

The validation was  performed using  previous cluster catalogues and existing optical surveys, and the results of \xmm\ validation campaign~\citep{xmmfu1, xmmfu_pip1, xmmfu_pip4} 
and two optical follow-up campaigns conducted on PSZ1 clusters \citep{rttfu_pip, enofu_pip}.

\subsection{Duplicates and secondary detections in individual \planck\ catalogues}
\label{sec:PSZ_dup}
Each \planck\ catalogue is constructed using three independent extraction algorithms, two  implementations of the matched multi-frequency filter technique (MMF3 and MMF1) and a method based on Bayesian inference (PwS for PowellSnakes). The detections made by the three methods, above the chosen S/N threshold,  are combined into a union catalogue.
A simple distance criteria is used for the cross-match of MMF1, MMF3 and PwS sources, merging detections  separated by less than 5\arcmin, in view of the \planck\ resolution. However, larger separations are possible for nearby large clusters or complex systems, for which the system physical extent can be  larger than $5\arcmin$.   
 
We thus look for such possible residual "duplicate detections" in PSZ1 and PSZ2 catalogues. We determined the  closest source to each source. For angular separations  less than 13\arcmin, we visually inspected the \planck\ SZE maps, compared the distance to $\theta_{500}$ (if $z$ is known for one source), checked the consistency of the S/N,  used ancillary information from other SZE catalogues and/or MCXC or \xmm\ observations when available. One duplicate case was identified in PSZ1, and four in PSZ2. They are detailed  in Appendix~\ref{app:psz12dup}.
The first case corresponds to the complex A2752 region, including \hyperlink{PSZ1 G093.84-38.80}{PSZ1 G093.84-38.80}, \hyperlink{PSZ1 G094.04-38.85}{PSZ1 G094.04-38.85=PSZ2\,G093.94-38.82} and ACT-CL J2318.3+1844.
Three cases correspond to the detection of the same nearby cluster by two different methods:  \hyperlink{PSZ2 G096.78-50.20}{PSZ2\,G096.78-50.20}  and \hyperlink{PSZ2 G096.77-50.29}{PSZ2\,G096.77-50.29};  \hyperlink{PSZ2 G302.49+21.53}{PSZ2\,G302.49+21.53} and \hyperlink{PSZ2 G302.41+21.60}{PSZ2\,G302.41+21.60}; \hyperlink{PSZ2 G332.11-23.63}{PSZ2\,G332.11-23.63} and \hyperlink{PSZ2 G332.29-23.57}{PSZ2\,G332.29-23.57}. The separation exceeds the 5\arcmin\ threshold but is  much smaller than   $\Tv$.  The fifth  case corresponds to the complex field  including, within 6\arcmin\ in diameter,    \hyperlink{PSZ2 G280.76-52.30}{PSZ2\,G280.76-52.30},  \hyperlink{PSZ2 G280.78-52.22}{PSZ2\,G280.78-52.22}, SPT-CLJ0240-5952 and SPT-CLJ0240-5946. 

For PSZ2 and PSZ1 catalogues, we further search for  cases where the nearest  source is  within $\theta_{500}$, without absolute separation limit.  This search is restricted by construction to confirmed clusters with redshift. We identified two cases of secondary detections in the massive nearby clusters, Coma and A3667, primarily detected  by the three methods at their X--ray position as \hyperlink{PSZ2 G057.80+88.00}{PSZ2\,G057.80+88.00} and  \hyperlink{PSZ2 G340.88-33.36}{PSZ2\,G340.88-33.36}, respectively.  These cases are detailed  in Appendix~\ref{app:psz2sec}. The three secondary sources, PSZ2 G056.62+88.42, PSZ2 G061.75+88.11 and PSZ2 G341.09-33.15,  are kept in PSZ and their redshift updated to that of the parent cluster. However their mass is set to -1 (undefined) as the SZE signal cannot be disentangled from the main cluster signal.

\subsection{Cross identification of PSZ2,  PSZ1 and ESZ objects}
\label{sec:PSZ1PSZ2ESZ}
The PSZ2 catalogue includes cross-identification with PSZ1 sources, as detailed by \citet{psz2}.  Each PSZ1 source is matched to its closest PSZ2 object if its distance is less than 5\arcmin. Higher separations are accepted up to 10\arcmin, with  further condition on the consistency between  PSZ2 and PSZ1 S/N.  We essentially relied on this work. However we identified  two cases of missing or wrong PSZ2-PSZ1  associations, after noting inconsistency with the further cross-match with the MCXC catalogue (see Appendix~\ref{app:psz1psz2}). We newly associated \hyperlink{PSZ1 G135.39-61.98}{PSZ1~G135.39-61.98}  and \hyperlink{PSZ2 G135.76-62.03}{PSZ2 G135.76-62.03} (MCXC J0115.2+0019). We associated \hyperlink{PSZ1 G224.09-76.43}{PSZ1~G224.09-76.43} (MCXC J0152.5-2853) with \hyperlink{PSZ2 G224.03-76.42}{PSZ2\,G224.03-76.42} instead of PSZ2\,G224.40-76.54.   
 
\begin{figure}[!t]
\centering
\includegraphics[width=0.6\columnwidth, trim=100 50 90 55, clip]{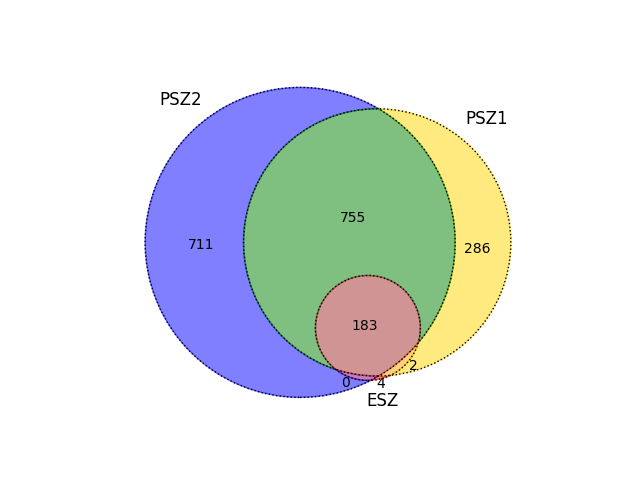}
\caption{\footnotesize Overlap between \planck\ catalogues. }
\label{fig:planck_venn}
\end{figure}

There are 288 PSZ1 clusters not present in PSZ2, either due to noise fluctuations or different PSZ1 and PSZ2 construction criteria (survey mask, point source contamination cuts, PWs cuts). These reasons are discussed in Sect 6.3 of \citet{psz2} and given for each ‘missing’ PSZ1 cluster, listed in Appendix E. There are 711 new \planck\ detected clusters in PSZ2. The overlap between PSZ2 and PSZ1 catalogues is illustrated on Fig.~\ref{fig:planck_venn}

All ESZ clusters are confirmed and have a redshift except PLCKESZ G189.84-37.24 = PSZ2 G189.79-37.25 = PSZ1 G189.82-37.25 which is a false detection \citep{say12}. The cross-match of ESZ clusters with PSZ1 or PSZ2 clusters is not explicitly performed in the PSZ1 and PSZ2 catalogues. The histogram of the distances between each ESZ cluster and its closest  neighbour in the union of PSZ2 and PSZ1 catalogues\footnote{Here, we took  the PSZ2 position for common PSZ2 and PSZ1 clusters} is bimodal with a large gap between  two components,  at $D<4\arcmin$ and $D>70\arcmin$ respectively.  Distances less than $4\arcmin$ unambiguously identify  matching objects, consistent with the  \planck\  position accuracy. The second component  at very large distance includes 4 ESZ clusters without PSZ1 or PSZ2 counterpart: PLCKESZ G228.49+53.12=Zw 3179 ($z=0.1434$),   PLCKESZ G241.85+51.53=A1066 ($z=0.07$), PLCKESZ G269.51+26.42=A1060  ($z=0.0126$) and PLCKESZ G275.21+43.92=A1285 ($z=0.1068$). These  four  clusters are the ‘missing’ ESZ clusters in the PSZ1 catalogue  discussed by \citet[][Sect.~6.1.3]{psz1}. They fall in the mask used for PSZ1 extraction, but  not used for ESZ construction. All other (matching) clusters are PSZ1 objects but two are not present in the PSZ2 catalogue: PLCKESZ G115.71+17.52 = PSZ1 G115.70+17.51 ($z=0.11$) and 
PLCKESZ G282.49+65.17= PSZ1 G282.45+65.18 ($z =0.077$). This is due to the more stringent mask used in the PSZ2 construction \citep[][Table E1]{psz2}.

\subsection{Suplementary \planck\ confirmed clusters}
\label{sec:plckg}
PSZ also includes  21 \planck\ clusters, with firm validation based on \xmm\ observations or optical counterparts with mass estimates consistent with the SZE signal. Those are  clusters from the ESZ or PSZ1 surveys, which did not meet the final catalogue S/N threshold criteria, or candidates from specific search of high z clusters in \planck\ maps at lower S/N \citep[e.g.][]{Zoh19}.

The X--ray validated sample  includes 4 \planck\ sources from the \xmm\ DDT programme of validation of \planck\  sources \citep{xmmfu1, xmmfu_pip1,xmmfu_pip4} and PLCK G260.7-26.3 from an early version of the PSZ1 catalogue with XMM follow-up \citep{bar19}. Two objects, PLCK G214.6+37.0 and PLCK G334.8-38.0, are triplet-cluster systems which are not resolved by \planck\ \citep{xmmfu1}. We give the redshift in PSZ but do not define a SZE mass, as individual clusters are not resolved by \planck\  and total mass  cannot be derived from the SZE signal with standard procedure. Properties of the systems, based on combined \xmm\ and \planck\ analysis  can be found  in and \citet{2013A&A...550A.132P} and  \cite{kol21}, respectively. The \planck\ mass of the other 3 clusters are  extracted from the last \planck\ maps  at the X--ray position. 

The optically validated sample includes PLCK G183.33-36.69 at $z=0.163$  \citep{amo18}, 13 sources from the optical follow-up of \citet{vdB16, Zoh19} targeted at potentially high--z candidates.  We used the  published \planck\ masses.  We also include PLCK G031.41+53.90 at $z=0.693$  and PLCK G087.10+43.77 at $z=0.871$ from MMF3 search down to $S/N=3$ in the latest \planck\ map and identified with ClG-J152741.9+204443  \citep{Bud15} and  SpARCS J161315+564930 \citep{2010ApJ...711.1185D}, respectively. Their masses are extracted at the SZE position.

\subsection{Physical properties}
\label{sec:PSZ_physprop}

The resulting PSZ meta-catalogues includes 1962 objects with 1233 confirmed clusters with redshift and two proved false candidates.  The physical quantities of each object are taken from the original catalogue, with the following hierarchy for objects belonging to several catalogues: PSZ2, then PSZ1, then ESZ (4 objects), following the respective survey depths.  

There is no consistency issue  between PSZ1 and PSZ2. The SZE data are from the same telescope  and extracted  with the same algorithm, apart from a few  PSZ2 improvements \citep[see][App.~C]{psz2}. The PSZ2 redshift of  common PSZ2-PSZ1 objects is  mostly inherited from PSZ1, or the same as in the updated version \citep{psz1rev}, with  the exception of  five  major revisions  \citep[see][App.~B]{psz2}.    $\Mv$ is identically  estimated from SZE data using the SZE mass proxy: the flux-size degeneracy is broken using the $\Mv$--$\Yv$ scaling relation as prior \citep[for details see][Sect.~7.2.2]{psz1}. As expected, the PSZ1 and PSZ2 mass estimates are consistent within  the errors for common objects \citep[][Sect.~5.3]{psz2}. 

The redshift type, {\tt Z\_TYPE}, is provided in the PSZ1 catalogue. It is not in the PSZ2 catalogue but can be inferred from the {\tt validation} field.  For PSZ1 clusters or PSZ2 clusters with $z$ inherited from PSZ1 ({\tt validation}=10, 11, 20, 25),  {\tt Z\_TYPE} is retrieved from the PSZ1 revised catalogue   \citep{psz1rev}.   For other PSZ2 clusters with redshift from a cross-matched catalogue,  {\tt Z\_TYPE} is defined from the catalogue information:  spec for MCXC ({\tt validation}=21), spec or phot for  SPT/ACT ({\tt validation} =22/23), and phot for redMaPPer ({\tt validation}=24). Redshift from counterpart search in PanSTARRS ({\tt validation} =12) or  SDSS  ({\tt validation} =13, 14) at \planck\ position  are photometric, except for 2 high-z clusters \citep[][Table A1]{psz2}.  Finally the type of redshift from NED  at the time of PSZ2 publication cannot be recovered and is left blank. 

At that stage, we updated 15 redshifts from SPT catalogues using the  latest values from  \citet{Bocquet2019}. Similarly we updated 66 photometric redshifts from redMaPPer to spectroscopic redshift using the updated version of  \citet{Rykoff16}. The cross-match of remaining RM clusters (distance of less than 2\arcsec) with the WHL12 catalogue \citep{2012ApJS..199...34W}, updated by \citet{2015ApJ...807..178W} further provides 37 spectroscopic redshifts. We then re-compute  the mass, using the same method as  \planck\ collaboration.  In practice, for PSZ1 clusters,  we used the tabulated flux-size degeneracy curve of each cluster and the  $\Yv$--$\Mv$ relation published by \citet{psz1cosmo}.  For PSZ2 clusters, we directly interpolate the z-$\Mv$ curve, provided for each cluster. 
We found two clusters, PSZ1 G206.52-26.37 and PSZ1 G249.01+73.75 with negative mass  at the catalogue redshift. The SZE detection is of low quality ($Q_{\rm neural}\,=\, 0.21$ and $3\cdot10^{-5}$, respectively) and the extent is very large for the assumed redshift.  We thus set {\tt STATUS=False} for these objects. 
Finally, the 3 secondary detections in Coma and A1367 are assigned the main cluster redshift and their masses are set to -1.

The PSZ1 and PSZ2 validation is consolidated from mass proxy comparison, when available (i.e X-ray luminosity, SZE mass, or richness for optical clusters).  We thus set {\tt STATUS=C1} for the clusters with a redshift in the original \planck\ catalogue, except the objects  with $Q_{\rm neural}<0.4$ for which we set {\tt STATUS=C2}. Those only concern 27 optical confirmations,  without mass proxy consolidation (e.g. early follow-up, PanSTARRS data), as well as a few SSDS clusters. This status may be changed from new information when updating the PSZ.

\begin{table*}[t]
\centering
\caption{\footnotesize \label{tab:PSZval} Inconsistent validation from optical follow-up.}
\resizebox{\textwidth}{!} {
\begin{tabular}{llllcccc}
\toprule
     \multicolumn{1}{l}{{NAME}} &
     \multicolumn{1}{c}{{$Q_{\mathrm neural}$}} &
    \multicolumn{1}{c}{{Invalidation source }} &  
    \multicolumn{1}{c}{{Validation source}}  & 
    \multicolumn{1}{c}{{z}}  & 
    \multicolumn{1}{c}{{z type}} &  
        \multicolumn{1}{c}{{Flag}}& 
    \multicolumn{1}{c}{{  {\tt  STATUS}}}    \\
\midrule
  \multicolumn{3}{c}{} & \multicolumn{4}{c}{Same potential counterpart} \\
 \cline{4-7}
  PSZ1 G081.56$+$31.03 &0.87 &invalidated   in   \citet{vdB16} &     \citet{Bar18} &  0.790 & phot &   2  & False \\
   PSZ2 G157.07$-$33.63 &0.07& invalidated   in   \citet{vdB16} &     \citet{Bar18} &  0.620 & phot &  2 & False \\
PSZ2 G165.41$+$25.93 &0.99& $M_{500c,\lambda}/ M_{500,SZ}<0.25$  in \citet{Zoh19} &  \citet{Agu19} &  0.670 & phot & 2 &False \\
   \multicolumn{3}{c}{} & \multicolumn{4}{c}{Different  potential counterpart} \\
  \cline{4-7}
  PSZ2 G098.38$+$77.22 &0.98&Flag=3 in  \citet{Bar20}  & \citet{boa19} &  0.726   & spec&& Unknown\\  
   PSZ2 G139.00$+$50.92 &0.78&$M_{500c,\lambda}/ M_{500,SZ}<0.25$ in \citet{Zoh19}  &   \citet{Agu19} &  0.784 & spec &   1 & C1\\
PSZ2 G191.82$-$26.64 & $10^{-3}$ &Flag=3 in  \citet{Bar20} &  \citet{boa19} &  0.170& phot  & &False \\
 \bottomrule
\end{tabular}
}
\end {table*}

\section{New validation and redshift update from optical follow-up of PSZ clusters}

\label{sec:PSZupfu}
\subsection{Data compilation}
Since the original \planck\ catalogue publications, a very large effort has been conducted by team originally in \planck\ consortium and the general community to continue  the validation process.  The PSZ includes information from the following recently published works (in chronological order):
\begin{itemize}[noitemsep,topsep=0pt,label=$-$]
\item
A. Validation follow-up targetted at finding new high z clusters, first with CHFT \citep{vdB16}  and WHT \citep{Zoh19}. Both studies provide optical mass estimates of possible counterparts. We also include new redshift estimates from \citet{vdB18}.
\item
B. The systematic and largest by far follow-up (413 PSZ1 or PSZ2 sources in total) conducted at the Canary Island Observatories (CIO) by \citet{Bar18,str19,Agu19, Bar20}. This work extends the ENO follow-up started by the \planck\ consortium \citep{enofu_pip}. We also consider redshift updates of previously confirmed PSZ1 clusters using SDSS DR12 spectroscopic data by  \citet[][Tables  A3]{Str18} and their new validation and z estimate  of PSZ2 clusters not included in CIO follow-up (Table 1, 3 clusters). 
\item
C. Gemini and Keck follow-up of 20 clusters published by \citet{amo18} with velocity dispersion. First results were already included in the updated PSZ1 catalogue \citep{psz1rev} and we adopt the latest published values.
\item
D. Extension of the \planck\ RTT follow-up, initially conducted within the \planck\ consortium \citep{rttfu_pip}, by \citet{Bur18} (targetted at z>0.7 clusters), by \citet{Zaz19}, and by \citet{Zaz20}. 
\item
E. Kitt Peak 4M telescope follow-up of \citet{boa19}.
\item
F. We also consider unpublished NOT spectroscopic results by Dahle et al. (private communication), part of them being included in  \citet{psz1rev}. They are only included when no other data is available. 
\end{itemize}
We first compiled all the above data, carefully examining possible discrepancies and deciding a strategy to adopt the a-priori most secure  and precise data. 

The validation of \planck\ cluster candidates may be ambiguous, specially at low S/N, due to the Eddington bias and the large \planck\ beam ($\sim5\arcmin$). Noise can significantly enhance the signal of a poor system within the search radius, boosting it above the catalogue threshold, with a continuous transition between a noise dominated signal and SZE cluster dominated signal \citep{Zoh19}. \citet{vdB16} show the importance of considering optical mass indicators to consolidate the validation and define quantitative criteria for candidate confirmation, so that the signal is not noise dominated. Such information is available for the A,B,C campaigns above and we use in priority A (direct comparison of SZE and optical mass), B\footnote{There are no  clusters in common among the different B studies, except for two clusters listed in Table 3 of \citet{Agu19}, also validated by \citet{Bar20}: PSZ2 G196.65-45.51 and PSZ2 G249.14+28.98. In these cases we prioritize the most recent data from \citet{Bar20}.} (distance and velocity criteria) and C (velocity) to define the validation status.  We then favoured spectroscopic redshift over photometric values. 
 
Following confirmation criteria of the A and B publications, we set {\tt STATUS = FALSE} for candidates invalidated by \citet{vdB16}, candidates with optical to SZE mass ratio $<1/4$ from \citet{Zoh19}, and labelled as 'ND' (non detected) or Class-3 for B campaigns. Six of these invalidated objects are considered as valid with a $z$ measurement from another follow-up. They are listed in Tab.~\ref{tab:PSZval} together with relevant information, including the detection quality, $Q_{\rm neural}$, and the final assigned {\tt STATUS}.  For 3 objects listed, the two studies considered the same potential counterpart. The validation discrepancy for PSZ1 G081.56$+$31.03, and PSZ2 G165.41$-$25.93 is discussed by \citet{Bar18} and \cite{Agu19} and is mostly due to slight differences in the validation threshold. We keep these objects as {\tt False}, specially since the former is not in PSZ2. PSZ2 G157.07$-$33.63 is also set to {\tt False}, in view of its very low PSZ2 $Q_{\rm neural}$. The counterpart is not the same for the other 3 objects. As discussed by \citet{Agu19}, there is another rich counterpart for PSZ2 G139.00$+$50.92 and we set the {\tt STATUS=C1}. The $Q_{\rm neural}$ value of PSZ2 G191.82-26.64 is extremely low and we kept the {\tt STATUS} as {\tt False}. The status of PSZ2 G098.38$+$77.22 is uncertain: the high $z$ potential counterpart found by \citet{boa19} is  $5.8\arcmin$ away from the SZE position, its richness is unknown,  and there are other lower redshift objects that may contaminate the signal as discussed by the authors. 
We also set {\tt STATUS = False} for candidates without an optical counterpart in \cite{enofu_pip} that were not followed-up with the B campaigns and have very poor quality, $Q_{\rm neural}<0.1$. This includes three candidates: PSZ2 G105.82-38.36, PSZ2 G167.44-38.06, and PSZ1 G127.55+20.84.

\begin{figure*}[!t]
\begin{centering}
\resizebox{0.9\textwidth}{!} {
\includegraphics[width=\columnwidth]{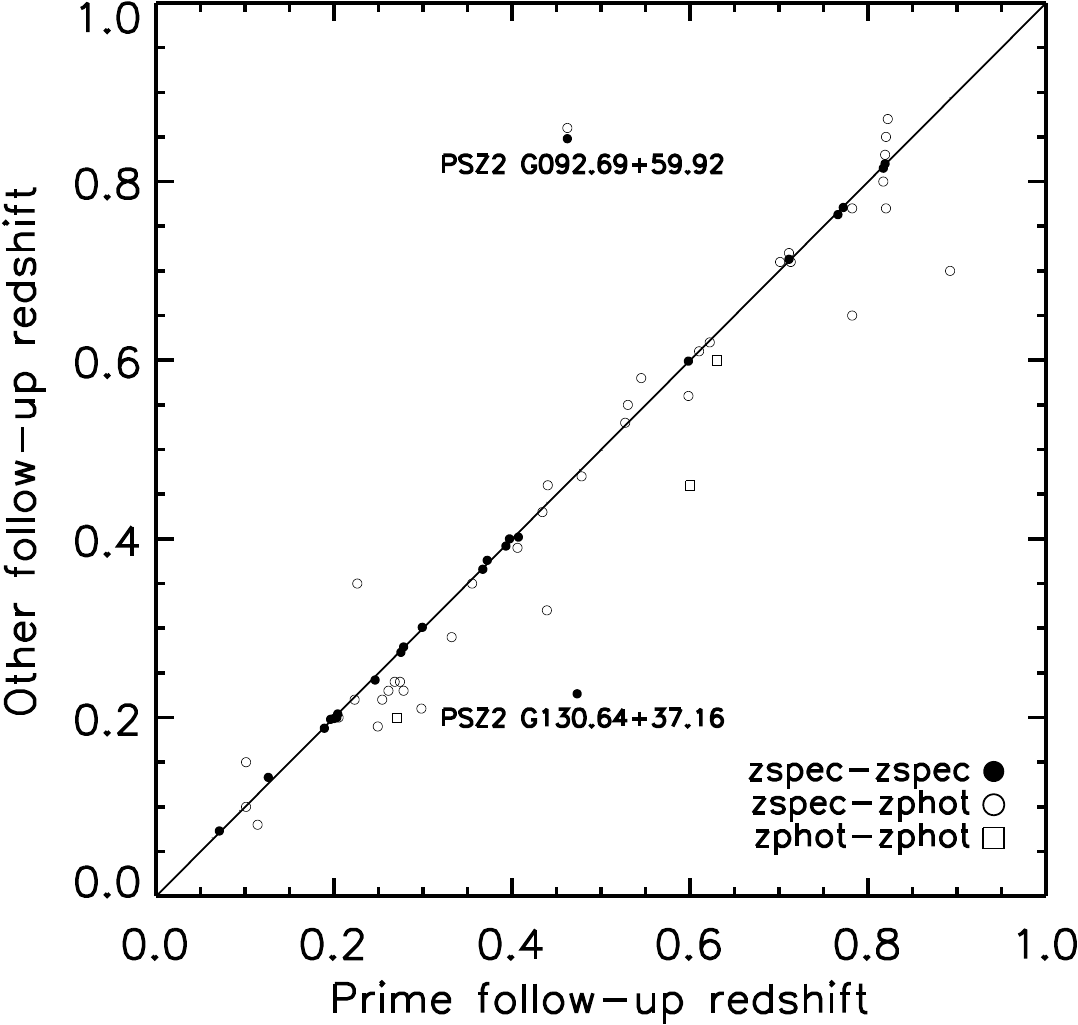}
\hspace{1.5cm}
\includegraphics[width=\columnwidth]{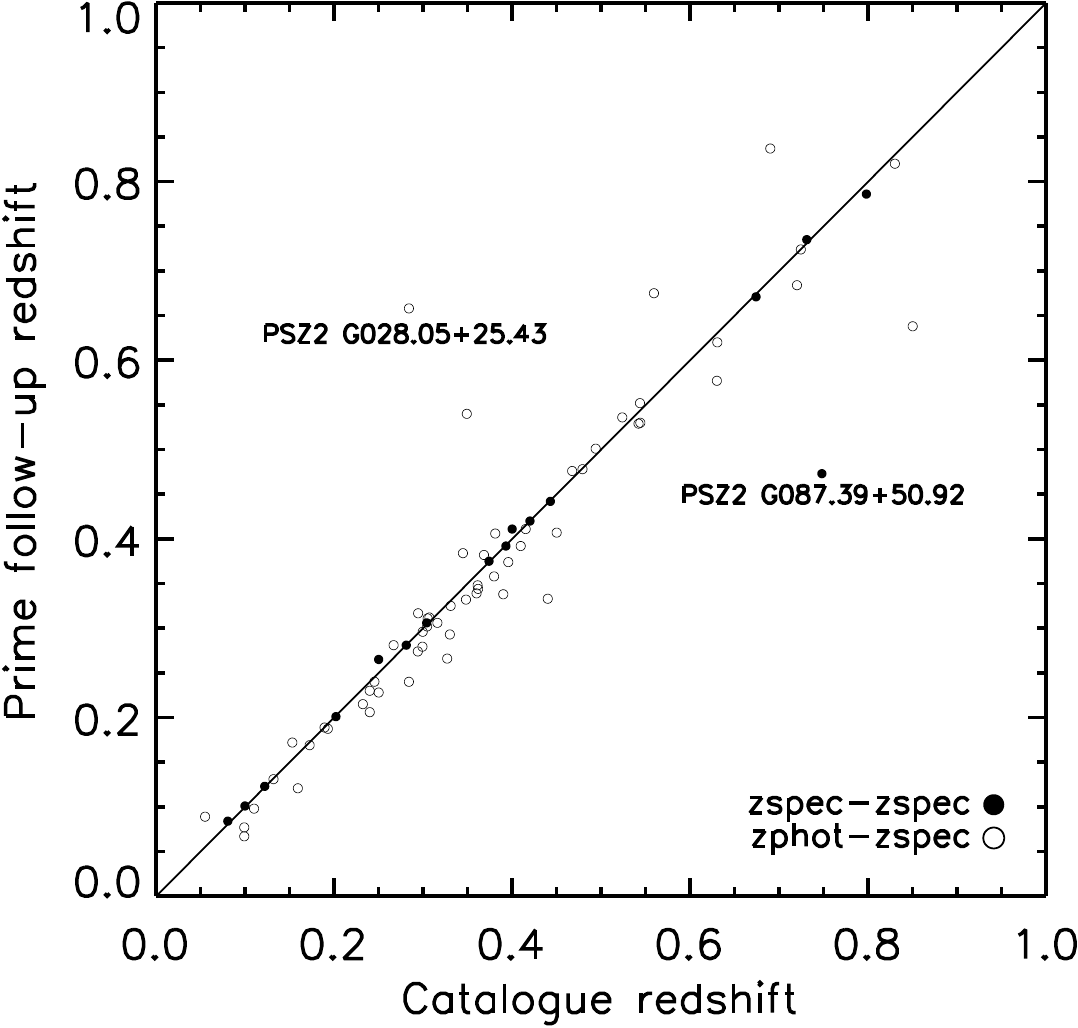}}
\caption{\footnotesize Comparison of redshift from follow-up published after the original PSZ catalogues. Prominent outliers are labelled and discussed in the text. {\it Left panel:} Comparison of the favored redshift to redshift from other follow-up campaigns. {\it Right panel:} Comparison of  spectroscopic redshift with original catalogue value if it was available. Most of those are photometric and can be replaced by the new redshift.}
\label{fig:PSZupfu}
\end{centering}
\end{figure*}

There are 66 validated clusters with multiple redshift measurements. As shown in the left panel of Fig.~\ref{fig:PSZupfu}, there is a good agreement between the favoured redshift (mostly spectroscopic) and the redshift from other follow-ups (mostly photometric), and perfect agreement between spectroscopic redshifts, albeit for two prominent outliers. Those are confusion cases, with published $z$ corresponding to different components.
PSZ2 G092.69+59.92 is a confusion of two components at $z=0.463$ and $z=0.848$, the latter being a low mass system, while PSZ2 G130.64+37.16 is a confusion between z=0.473 (dominant) and a  $z=0.24$ component \citep{Agu19}.

\subsection{PSZ catalogue update}

This follow-up  compilation validates and provides redshift for 175  PSZ candidates and invalidates 214 candidates.  If the validation reference is not same as the (spectroscopic) redshift reference, we provide the two references in the catalogue field {\tt Z\_REF}. 

The compilation  also includes 100 previously validated PSZ clusters. The validation of 8 clusters from PanSTARRS study \citep{Liu15}  and one cluster from NOT are now rejected based on the follow-up by \citet{Bar18} or  \citet{Bar20}  (the potential counterpart is too distant and/or too faint)  and are flagged as {\tt FALSE}.  For the other clusters, there is a good redshift agreement, except for two prominent outliers (Fig.~\ref{fig:PSZupfu}, right panel). Again these discrepancies  are due of the presence of close-by potential counterparts. The case of PSZ2 G087.39+50.92 is discussed by \citet{psz2} and we retain the catalogue value. On the other hand, we adopt the revised $z=0.658$ for PSZ2 G028.05+25.43 (PSZ1 G028.01$+$25.46), as discussed by \citet[][Fig.~6]{Bar18}.  For the other clusters, the catalogue values are updated when more precise/robust values are available: photometric redshifts or X--ray redshifts  from XMM follow-up are updated to spectroscopic redshifts, spectroscopic values from NOT and Gemini campaigns to latest values from these follow-up. In total 65 redshifts are updated, after the initial update described  Sect.~\ref{sec:PSZ_physprop}. 

The {\tt STATUS} of objects with new redshift is set according to the mass proxy information. We set {\tt STATUS=C1} for objects from \citet{vdB16}; {\tt STATUS=C1} and {\tt STATUS=C2} for objects from \citet{Zoh19} with $M_{500c,\lambda}/ M_{500,SZ}>0.5$ and $0.25<M_{500c,\lambda}/ M_{500,SZ}<0.5$, respectively; {\tt STATUS=C1} and {\tt STATUS=C2} for objects in B campaigns with Flag=1 and Flag=2, respectively; and {\tt STATUS=C1} for objects from \citet{amo18}. We do not have mass information from  \citet{boa19} or NOT campaign. However, we verified that they are matching eRASS or RASS-MCMF objects (following the method described in Sect.~\ref{sec:PSZerass}) with consistent SZE and X-ray mass proxies. Their status was set to {\tt STATUS=C1}. For objects with updated redshifts, the {\tt STATUS} from optical data defined as above agrees with that defined in Sect.~\ref{sec:PSZ_physprop}, except for 7 cases from B campaign. Those are updated.

\begin{figure*}[t]
\centering
\includegraphics[width=0.99\columnwidth]{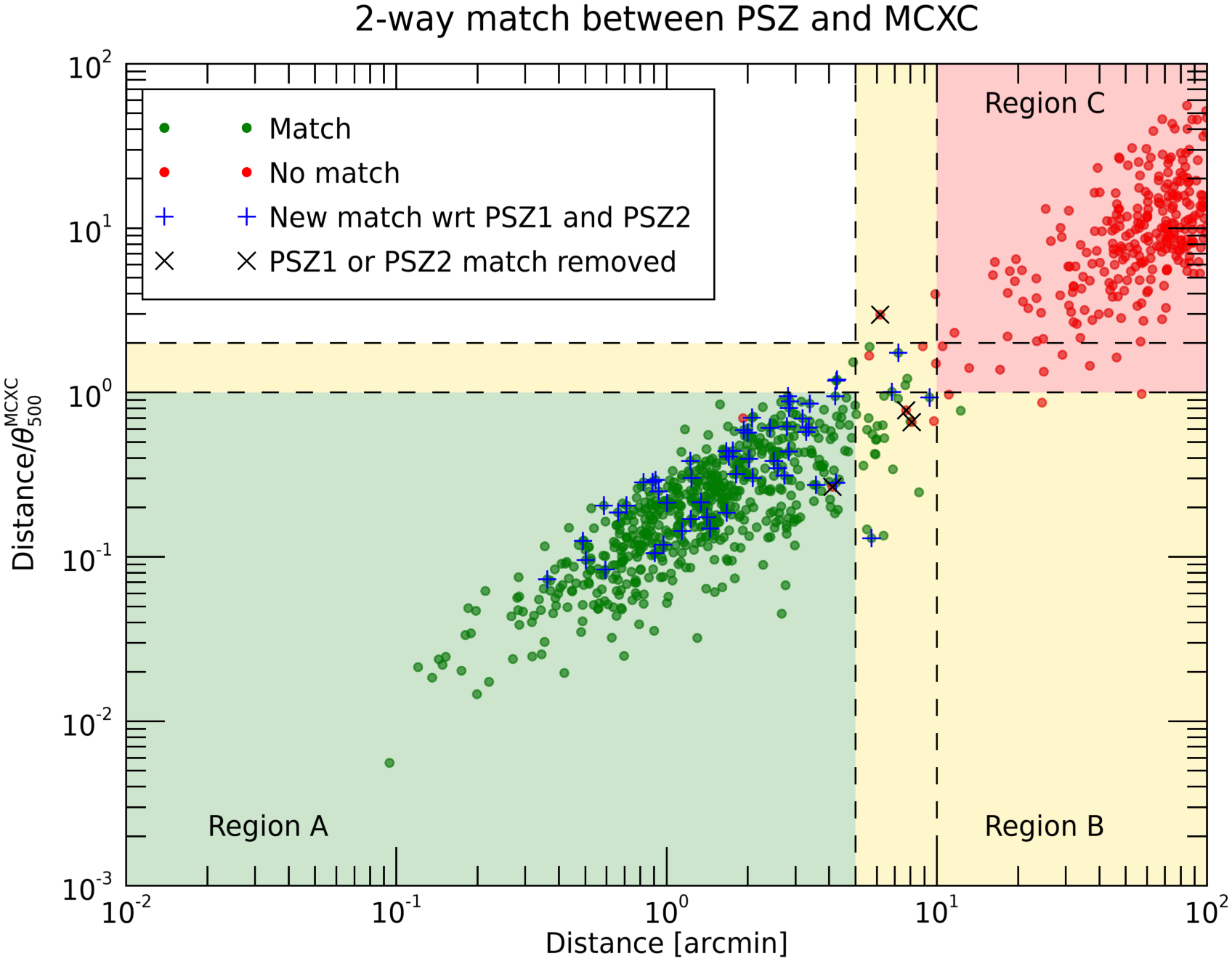}
\includegraphics[width=0.99\columnwidth]{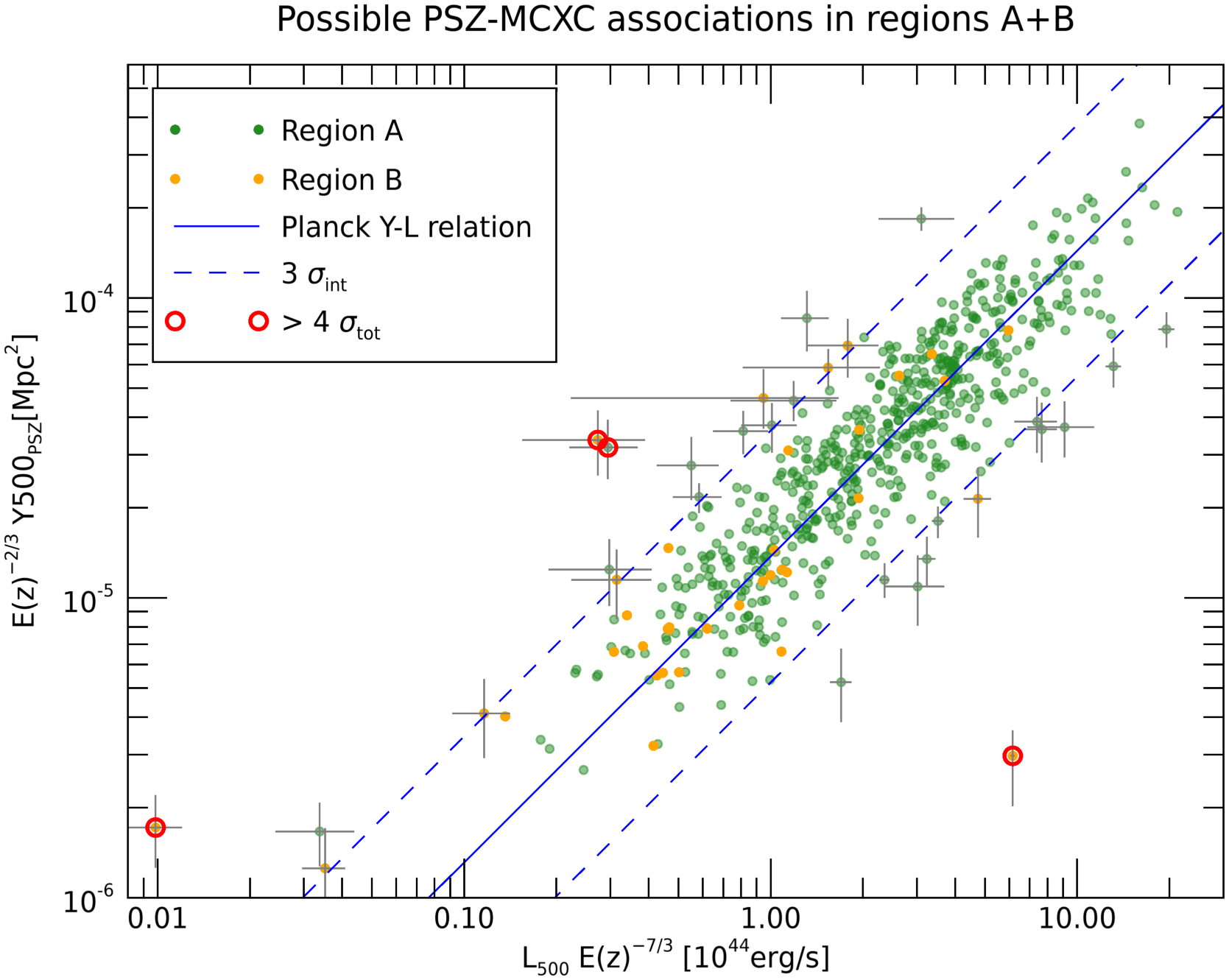}
\caption{\footnotesize Criteria for matching PSZ candidates to MCXC-II clusters. Left: Distance between each PSZ candidate and its possible two-way MCXC-II association versus their relative distance in terms of the MCXC-II cluster size $\theta_{500}$. Regions A, B, and C are shown in green, yellow, and red, respectively. Green/red dots indicate whether the association is finally kept/rejected. Blue pluses and black crosses correspond, respectively, to new and removed  associations with respect to the original PSZ1 and PSZ2 associations. Right: Comparison between PSZ $Y_{500}$ (recomputed at the MCXC-II redshift) and MCXC-II $L_{500}$ for possible PSZ-MCXC-II associations falling in  regions A and B of the left panel, represented by green and yellow dots, respectively. The expected $Y_{500}-L_{500}$ relation from \cite{PlanckEarlyResXI} is shown as a solid blue line, with dashed lines indicating 3 times its intrinsic scatter. Errorbars are shown for associations outside these dashed lines. Red empty circles mark associations that we discard because their position in the $Y_{500}-L_{500}$ plane is at an orthogonal distance from the relation greater than $4 \sigma_{\rm tot}$. } 
\label{fig:PSZ_MCXC_crossmatch}
\end{figure*}

\section{Cross-identification with MCXC metacatalogue}\label{sec:mcxc}

The Meta-Catalogue of X--ray detected Clusters \citep[MCXC;][]{MCXC} is a compilation of 1743 galaxy clusters constructed from publicly available ROSAT All Sky Survey-based and
serendipitous cluster catalogues, as well as the Einstein Medium Sensitivity Survey. The PSZ1 and PSZ2 catalogues include cross-identifications with MCXC clusters; however, the identification criteria differ between the two catalogues, and for PSZ1, the originally published MCXC was extended with some additional clusters. Moreover, the ESZ catalogue does not provide any MCXC cross-identification. To achieve a homogeneous and updated cross-identification across the entire PSZ metacatalogue, we conducted a cross-identification with the MCXC-II metacatalogue \citep{MCXC2024}. MCXC-II is an updated version of MCXC containing 2221 clusters distributed over the whole sky, along with their coordinates, redshifts, and standardised [0.1-2.4] keV band luminosities ($L_{500}$) measured within $R_{500}$. 

We performed a 2-way matching between the PSZ and MCXC-II metacatalogues and defined three regions in the $d$-$d/\theta_{500}$ plane, where $\theta_{500}$ is the MCXC-II cluster size. Region A, defined as $d<5$ and $d<\theta_{500}$, corresponds to a priori good associations. Region B is defined as $d<5$ and $\theta_{500}<d<2\theta_{500}$ (B1), or $5<d<10$ (B2), or $d>10$ and $d<\theta_{500}$ (B3). 
This is an intermediate region which will be checked in an individual basis through visual inspection. 
Finally, region C, defined as $d>10$ and $d>\theta_{500}$, corresponds to bad associations. With this positional matching, we found 
644 potential associations: 604 in region A and 
40 in region B (2 in B1,  
34 in B2, and 4 in B3). 
Left panel of Fig.~\ref{fig:PSZ_MCXC_crossmatch}  shows these three regions.

As a second step, we verified the potential associations by looking at the coherence between the SZE signal and the X--ray luminosity. We use the MCXC-II $L_{500}$, and recompute $Y_{500}$ at the redshift of the MCXC-II cluster for PSZ1 and PSZ2 candidates. In the right panel of Fig.~\ref{fig:PSZ_MCXC_crossmatch}, we compare our results with the expected $Y_{500}-L_{500}$ relation \citep{PlanckEarlyResXI}: we consider as good associations those whose position in the $Y_{500}-L_{500}$ plane are at an orthogonal distance to the relation of less than four times $\sigma_{\rm tot}$, where $\sigma_{\rm tot}$ takes into account quadratically the intrinsic scatter ($\sigma$ = 0.14 \cite{PlanckEarlyResXI}), and the measurement errors on both $Y_{500}$ and $L_{500}$. Based on this criterion we discarded one association in region A (PSZ2 G086.28+74.76), two in region B2 (PSZ2 G355.22-70.03 and PSZ1 G259.04-83.24), and one in region B3 (PSZ2 G150.64-14.21). 
They are highlighted in the right panel of Fig.~\ref{fig:PSZ_MCXC_crossmatch}. For clusters not in PSZ1 or PSZ2 (one in region A, three in region B), we cannot recompute $Y_{500}$, so we rely on visual inspection.
  
The last step is the visual inspection of the potential associations in region B, and those in region A for which $Y_{500}$ could not be recomputed (only one case: PLCKESZ G275.21+43.92). 
Among the 40 clusters in region B, we discarded 
8 of the 34 possible associations in B2 
(PSZ1 G213.95+68.28, PSZ2 G126.61-37.63, \hyperlink{PSZ2 G212.93-54.04}{PSZ2 G212.93-54.04}, \hyperlink{PSZ2 G215.19-49.65}{PSZ2 G215.19-49.65}, \hyperlink{PSZ2 G254.96+55.88}{PSZ2 G254.96+55.88}, \hyperlink{PSZ2 G264.60-51.07}{PSZ2 G264.60-51.07}, discussed in App.~\ref{app:eRASS}, and the two discarded in step 2), 
and 3 of the 4 possible associations in B3 (\hyperlink{PSZ2 G225.18-33.61}{PSZ2 G225.18-33.61}, PSZ2 G319.16+26.63, and PSZ2 G150.64-14.21 discarded in step 2).  In these last 3 cases two distinct and spatially separated X--ray structures are visible, each associated with one of the clusters in the potential match.
This yields 29 final associations in region B. In region A we keep 
602 associations (all except PSZ2 G086.28+74.76, discarded in step 2, and PSZ2 G093.94-38.82=\hyperlink{PSZ1 G093.84-38.80}{PSZ1 G093.84-38.80}, a duplicate discussed in App.~\ref{app:psz12dup}). Left panel of Fig.~\ref{fig:PSZ_MCXC_crossmatch} shows the associations that are finally kept (631 in total) and discarded. 

Among these 631 associations, three correspond to \planck\ candidates without a redshift, for which the MCXC-II cluster provides a new redshift (1 MACS-DR3, 1 new SGP, 1 RXGCC). 
The remaining associations already have \planck\ redshifts, which coincide with that of MCXC-II in most cases, while 71 show redshift differences. We decided to update the redshift of \planck\ clusters that have not been updated before Sect.~\ref{sec:PSZupfu}  
only when the MCXC-II redshift is expected to be more reliable, i.e. in the following cases:
1) the original \planck\ redshift source is MCXC, 
2) the original \planck\ redshift source is not MCXC and its type is less accurate than that of MCXC-II (a photometric redshift updated to spectroscopic, an unknown redshift updated to spectroscopic or photometric), and
3) the original \planck\ redshift is spectroscopic from XMM and MCXC-II provides another spectroscopic redshift. This gives 37 updated redshifts thanks to MCXC-II. 
The remaining 71-37=34 discrepant cases correspond to small redshift differences that can be explained by the different redshift sources and their uncertainties (30 cases with less than 4\% redshift difference, and 4 cases between 8\% and 12\% corresponding to RXGCC clusters with SP or U redshift type). We also updated the {\tt Z\_TYPE} of a few clusters whose redshifts were originally taken from MCXC and for which the {\tt Z\_TYPE} was either missing or incorrectly assigned in the original \planck\ catalogues.

\begin{figure*}[t]
\centering
\includegraphics[width=0.9\columnwidth]{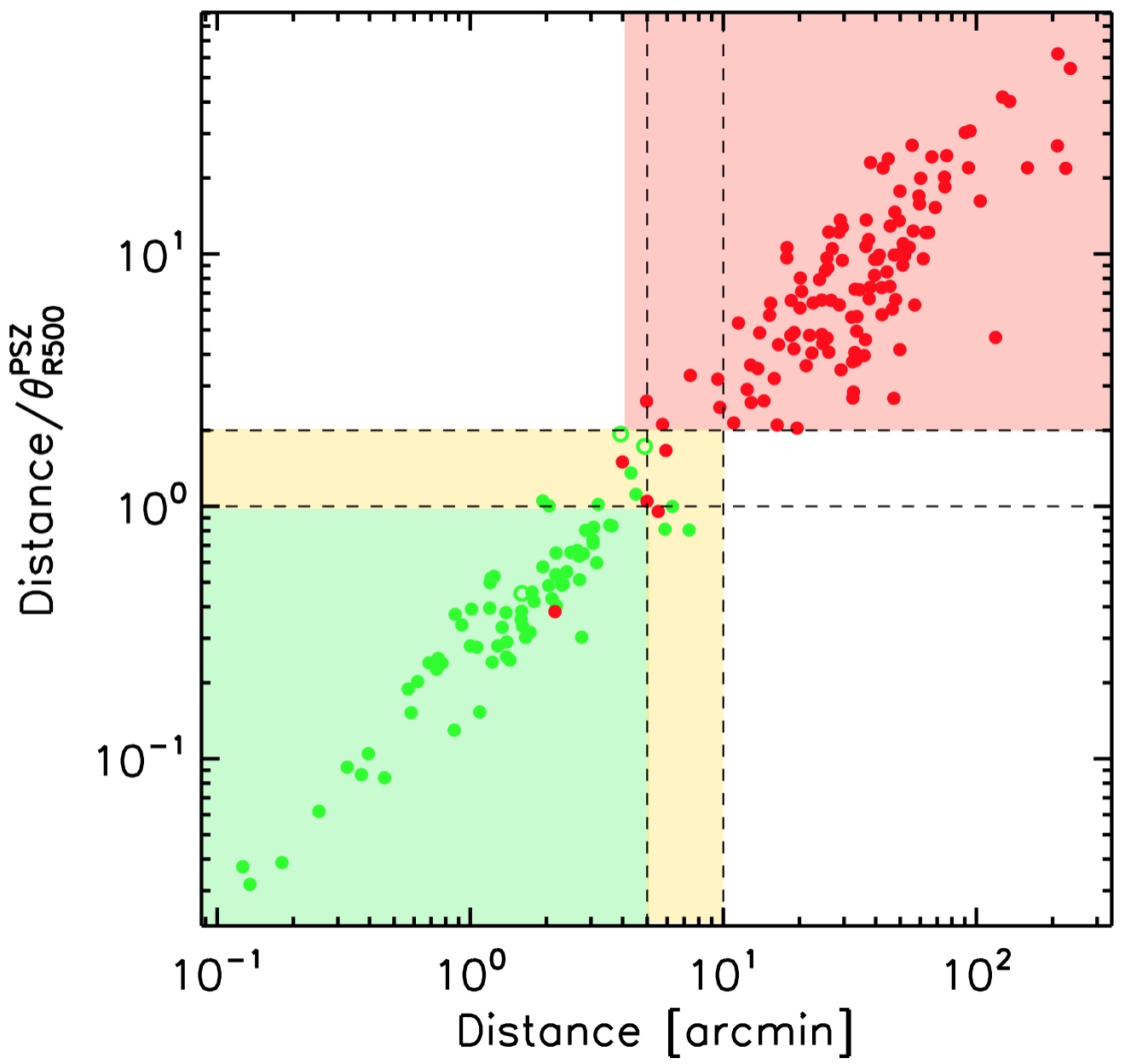}
\hspace{0.5cm}
\includegraphics[width=0.9\columnwidth]{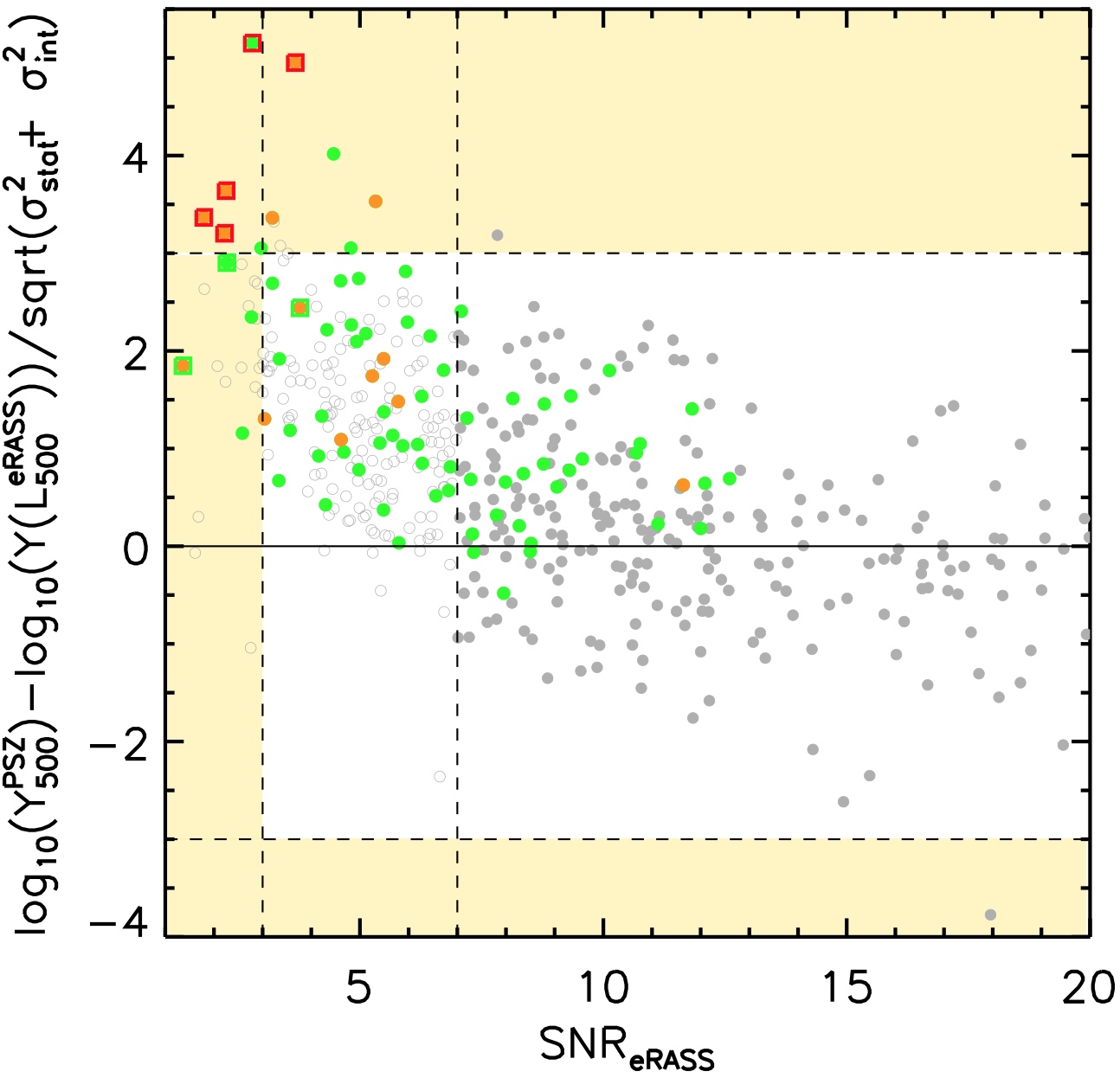}
\caption{\footnotesize Criteria for matching PSZ candidates to eRASS clusters. {\it Left}: Distance between each PSZ candidate and its possible two-way eRASS association versus their relative distance in terms of  $\theta_{500}$ computed at the eRASS redshift. Regions A, B, and C are shown in green, yellow, and red, respectively. Green/red dots indicate whether the association is finally kept/rejected. {\it Right}: Deviations from the $Y_{500}$-$L_{500}$ relation as a function of the S/N ratio of the eRASS detection. Grey points correspond to matching clusters used to calibrate the relation (filled circles) and objects with same distance and redshift constrains at lower  S/N (open circles). Green/orange points are PSZ candidates  in the green/yellow regions of the left panel. Candidates in the yellow region are manually checked. Empty red squares mark the associations that we finally discarded (one confusion case and 4 false candidates). Three candidates, marked with green open circle in the left panel and green squares in the right panel, were confirmed from further cross-match with other catalogues. }
\label{fig:PSZ_eRASS_crossmatch}
\end{figure*}

We evaluated the consistency between our crossmatch with MCXC-II and the original MCXC crossmatch included in PSZ1 and PSZ2, finding a very good agreement. We found 578 \planck\ clusters matched in both crossmatches, with 576 associated to the same MCXC cluster\footnote{PSZ1 G184.23-44.26 is associated with MCXC J0326.8-0043. In PSZ1 and MCXC, the MCXC name incorrectly appears with a + sign.}. The remaining two cases, PSZ2 G107.39-31.48 and PSZ2 G113.02-64.68, are now matched to MCXC J2350.5+2929 and MCXC J0034.2-0204, respectively, since their original PSZ2 associations (MCXC J2350.5+2931 and MCXC J0034.6-0208) were identified as duplicates in \cite{MCXC2024} and removed from the catalogue. Additionally, we identified new associations for 53 \planck\ clusters: 48 due to new MCXC-II clusters, 4 associations for ESZ clusters, and 1  for \hyperlink{PSZ2 G210.01+50.85}{PSZ2 G210.01+50.85} (see Appendix \ref{app:eRASS}). 
13 \planck\ clusters originally associated with MCXC are no longer matched in our crossmatch. 
Three of these, \hyperlink{PSZ2 G212.93-54.04}{PSZ2 G212.93-54.04}, \hyperlink{PSZ2 G254.96+55.88}{PSZ2 G254.96+55.88}, and \hyperlink{PSZ2 G264.60-51.07}{PSZ2 G264.60-51.07}, were incorrectly associated in PSZ1 and PSZ2 (see Appendix \ref{app:eRASS}), another one, PSZ2 G093.94-38.82=\hyperlink{PSZ1 G093.84-38.80}{PSZ1 G093.84-38.80}, is a complex cluster with MCXC being one of its components (see Appendix \ref{app:psz12dup}), and 
the remaining 9 were originally crossmatched to clusters that are not in MCXC-II (8 MACS clusters without redshift and 1 XCS that were added in the extended MCXC catalogue used in PSZ1). Left panel of Fig.~\ref{fig:PSZ_MCXC_crossmatch} highlights the differences between our crossmatch and the one included in PSZ1 and PSZ2.

\section{New validation and redshift update from cross-match with external catalogues}
\label{sec:PSZupcat}

\subsection{eRASS} 
\label{sec:PSZerass}

We first used the eRASS catalogue of clusters detected by SRG/eROSITA  in a 13 116 deg$^2$ region in the western Galactic half of the sky \citep{eRASS}. The search for new redshifts follows the same method\footnote{eRASS catalogue includes  a  cross-identification  with \planck\ catalogues. As also noted by \citet{eRASS}, their simple matching criteria  ($D<2\arcmin$) is too conservative for \planck\ sources.} as described in Sect.~\ref{sec:mcxc}. Figure \ref{fig:PSZ_eRASS_crossmatch} shows the position of  the 198 pairs of PSZ clusters without redshift and the closest eRASS object in the $D$-$D/\Tv$ plane.   
Regions C, B, A correspond to bad associations, 14 intermediate cases to be checked individually  (hereafter Class B candidates),  and a priori good associations.  We also checked the second closest eRASS object in regions A and B and found 3 cases of double X--ray clusters (consistent $z$ and similar mass), resolved by eRASS but not  by \planck. In these cases, we adopted the redshift of the (second) most luminous component.

We  checked the consistency of the SZE and X--ray signal  using  the $Y_{500}$-$L_{500}$ relation.   We could not use the relation from  \citet{PlanckEarlyResXI}. It is derived  for  $\Lv$ defined in the $[0.1$--$2.4]$keV, within  $\Rv$ determined  from $\Lv$--$\Mv$ relation calibrated using  X--ray mass, while the  published  $\Lv^{\rm eRASS}$ is  in the  $0.2-2.3$ keV detection band with a $\Tv$ aperture derived  from $\Lv$--$\Mv$ relation calibrated on lensing masses. We thus recalibrate the relation, using a conservative selection of PSZ-eRASS matching clusters: $D/\theta_{500}<1$, $D<5\arcmin$ and $\vert z_{\rm eRASS}-z_{\rm PSZ}\vert/(1+z_{\rm PSZ})<0.05$. We note that the deviation increases  with the  S/N of the eRASS detection\footnote{This is reminiscent of a Malquist bias} and we empirically further select clusters with S/N$>7$. The best fit parameters are similar, with a slope of $\alpha = 1.04$, a  normalisation of $B=10^{-3.88}$, and an intrinsic scatter of $\sigma_{\rm int}= 0.14$. The deviation of each data point to the best fit relation normalised to the quadratic sum of the intrinsic errors and  intrinsic scatter, $\sigma_{\rm tot}$, is shown in Fig.~\ref{fig:PSZ_eRASS_crossmatch} for the clusters used in the fit (filled circles). The matching clusters at lower S/N$_{\rm eRASS}$ with same distance and redshift constrains  are plotted  as open circles showing the systematic increase of the deviation towards low S/N. We thus also checked individually all the possible cases with S/N$<3$ or deviation greater than $3\sigma_{\rm tot}$ (yellow region, 7 additional Class B candidates). 

Of the 21 class B candidates, 13 objects are confirmed:  8 with {\tt STATUS=C1} and 5 with {\tt STATUS=C2}.
The latter  includes PSZ2 G291.18+21.78, PSZ2 G292.77-23.80, PSZ2 G319.16+26.63,  as well as \hyperlink{PSZ2 G213.73-56.15}{PSZ2 G213.73-56.15} and  \hyperlink{PSZ2 G319.64-65.11}{PSZ2 G319.64-65.11}, further discussed in Appendix \ref{app:eRASS}. Further 3 candidates, with uncertain status from eRASS data alone, were confirmed using other data.  \hyperlink{PSZ1 G245.21-65.29}{PSZ1 G245.21-65.29}, at $D=1.7\Tv$, was confirmed using improved full mission position used in the PSZ-MCMF catalogue (see Sect.~\ref{sec:pszmcmf} and Appendix \ref{app:eRASS}).
\hyperlink{PSZ1 G223.80+58.50}{PSZ1 G223.80+58.50} and PSZ2 G339.74-51.08, with very poor eRASS data (S/N$=1.4$ and  S/N$=2.3$, respectively), proved to be genuine clusters from  ACT cross-identification (see App.~\ref{app:eRASS}) and PSZ-MCMF cross-match (Sect.~\ref{sec:pszmcmf}), respectively. The other 5 candidates are discarded. They are marked with a red square  in the right panel of Fig.~\ref{fig:PSZ_eRASS_crossmatch}. They include 
the 2 objects with  the  largest positive deviations from $Y_{500}$-$L_{500}$ relation, PSZ2 G191.82-26.64 and PSZ2 G225.18-33.61, and the 3 objects  with  the lowest X--ray S/N in the $>3\sigma$ deviation region, PSZ1 G345.44-28.03, PSZ2 G235.96+38.21, and \hyperlink{PSZ2 G236.68-37.71}{PSZ2 G236.68-37.71}. The last one is a complex case of confusion with possible noise contamination, as discussed in App.~\ref{app:eRASS}. Its {\tt STATUS} is set to {\tt Confusion} with no redshift assigned. The other 4 candidates are clearly noise dominated from the SZE extended morphology and/or low $Q_{\rm neural}$ and/or very high $Y_{500}$ as compared to the value expected from $L_{500}$ of the potential counterpart.  Two objects, PSZ2 G191.82-26.64 and PSZ2 G235.96+38.21, are also flagged {\tt Flag=3} by \citet{Bar20} from optical follow-up, the eRASS data thus confirming their {\tt False} status.

In the above search, we only consider eRASS  clusters with estimated $L_{500}$.  We further found 5 PSZ candidates  for which the closest eRASS cluster, without  $L_{500}$ estimate, is at a distance $D<10\arcmin$, i.e. a potential match.   In one case, PSZ2 G289.50-47.49, the second closest eRASS cluster, at a distance of $2.4\arcmin$, is actually the counterpart, with consistent $Y_{500}$ and $L_{500}$ values.  A second case is PSZ2 G285.80-26.46, for which the closest  eRASS cluster has a high detection likelihood of 472, but surprisingly a very large count rate error. The eRASS image shows a bright  cluster coincident with the SZE image  ($D=1.3\arcmin$).  The other 3 cases can be discarded, the eRASS object  having a very low detection significance and being at larger distance.

In summary, this cross-match  confirms and provides new redshifts for  73  candidates.  Most of them are firmly confirmed: 65 objects with  {\tt STATUS = C1} and  the  3 double clusters with  {\tt STATUS\,=\,Complex} described  in the {\tt COMMENT} field.  We set  {\tt STATUS\,=\,C2} for 5 Class B candidates in view of the  offset between the X--ray and SZE position or  deviation from $Y_{500}$-$L_{500}$ relation.  Two  PSZ candidates are flagged with {\tt STATUS\,=\,False}. 

We also used eRASS to find potential redshift updates for already confirmed \planck\ clusters. There are 7 clusters in region A+B  with non-spectroscopic redshift  and a  $\zs$ in eRASS, including the newly confirmed MCXC-II clusters PSZ2 G301.90+20.17 and PSZ2 G264.92+44.70, with a $\zp$ updated with MCXC-II value. The redshifts are consistent in 5 cases ($\dzn < 0.01$ or 4.5$\%$), with two outliers,  PSZ2 G212.80+50.63 (see Note (a) of Table~\ref{tab:BH24}) and \hyperlink{PSZ1 G279.00-24.89}{PSZ1 G279.00-24.89}, discussed in Appendix \ref{app:eRASS}. 
In all cases, we adopted the eRASS value.

We further identified  inconsistency between PSZ 
and eRASS redshifts as they may indicate problematic PSZ identification or confusion cases.  We consider objects outside the  $95\%$ range of $\dzn$, computed for different combinations of eRASS and PSZ redshift types,  separately. We identified 18 outliers:  1 $\zs$-$\zs$ outlier (PSZ2 G254.96+55.88), 7 $\zs$-$\zp$ outliers, 5 $\zp$-$\zp$ outliers, and 5 $z_{\rm NED}-\zp$ outliers. 
As discussed in Appendix \ref{app:eRASS}, we  keep the PSZ redshift for  \hyperlink{PSZ2 G275.64-49.09}{PSZ2 G275.64-49.09}, actually associated with the second closest eRASS cluster at consistent redshift, and for  \hyperlink{PSZ2 G270.78+36.83}{PSZ2 G270.78+36.83} and \hyperlink{PSZ2 G292.74+33.49}{PSZ2 G292.74+33.49}, two cases of line of sight confusion. The PSZ redshift is further kept for 6 cases of  redshift difference likely due to $\zp$ uncertainties (see Appendix \ref{app:eRASS} for \hyperlink{PSZ2 G268.51-28.14}{PSZ2 G268.51-28.14} and  \hyperlink{PSZ2 G269.02+22.27}{PSZ2 G269.02+22.27}, the two cases of unchanged $\zs$-$\zp$ outliers).
The redshift is updated for the 9 other outliers. Seven cases, 
\hyperlink{PSZ2 G181.71-68.65}{PSZ2 G181.71-68.65},  
\hyperlink{PSZ2 G215.19-49.65}{PSZ2 G215.19-49.65}, 
\hyperlink{PSZ2 G224.53-30.27}{PSZ2 G224.53-30.27}, 
\hyperlink{PSZ2 G254.96+55.88}{PSZ2 G254.96+55.88}, 
\hyperlink{PSZ2 G285.87-74.93}{PSZ2 G285.87-74.93}, 
\hyperlink{PSZ2 G307.72-77.87}{PSZ2 G307.72-77.87} and 
\hyperlink{PSZ2 G357.75-41.77}{PSZ2 G357.75-41.77}, correspond to complex cases of redshift mismatch which are clarified in Appendix \ref{app:eRASS}. Finally,  we favour the eRASS $\zp$ for the remaining two $\zp$-$\zp$ outliers, PSZ2 G281.09-42.51 and  PSZ2 G286.68+23.18, with supporting  $\zp$ value from PSZ-MCMF.

\subsection{RASS-MCMF}\label{sec:rassmcmf} 

RASS-MCMF \citep{RASSMCMF2023} is a catalogue of 8449 galaxy clusters identified in the second ROSAT All-Sky Survey (RASS) source catalogue (2RXS; \cite{2RXS2016}). It was constructed by searching across 25000 deg$^2$ of the extragalactic sky covered by the DESI Legacy Survey DR10 \citep{DESI2019} with the Multi-Component Matched Filter (MCMF) algorithm, resulting in a catalogue with 90\% purity.

\begin{figure}[t]
\centering
\includegraphics[width=0.99\columnwidth]{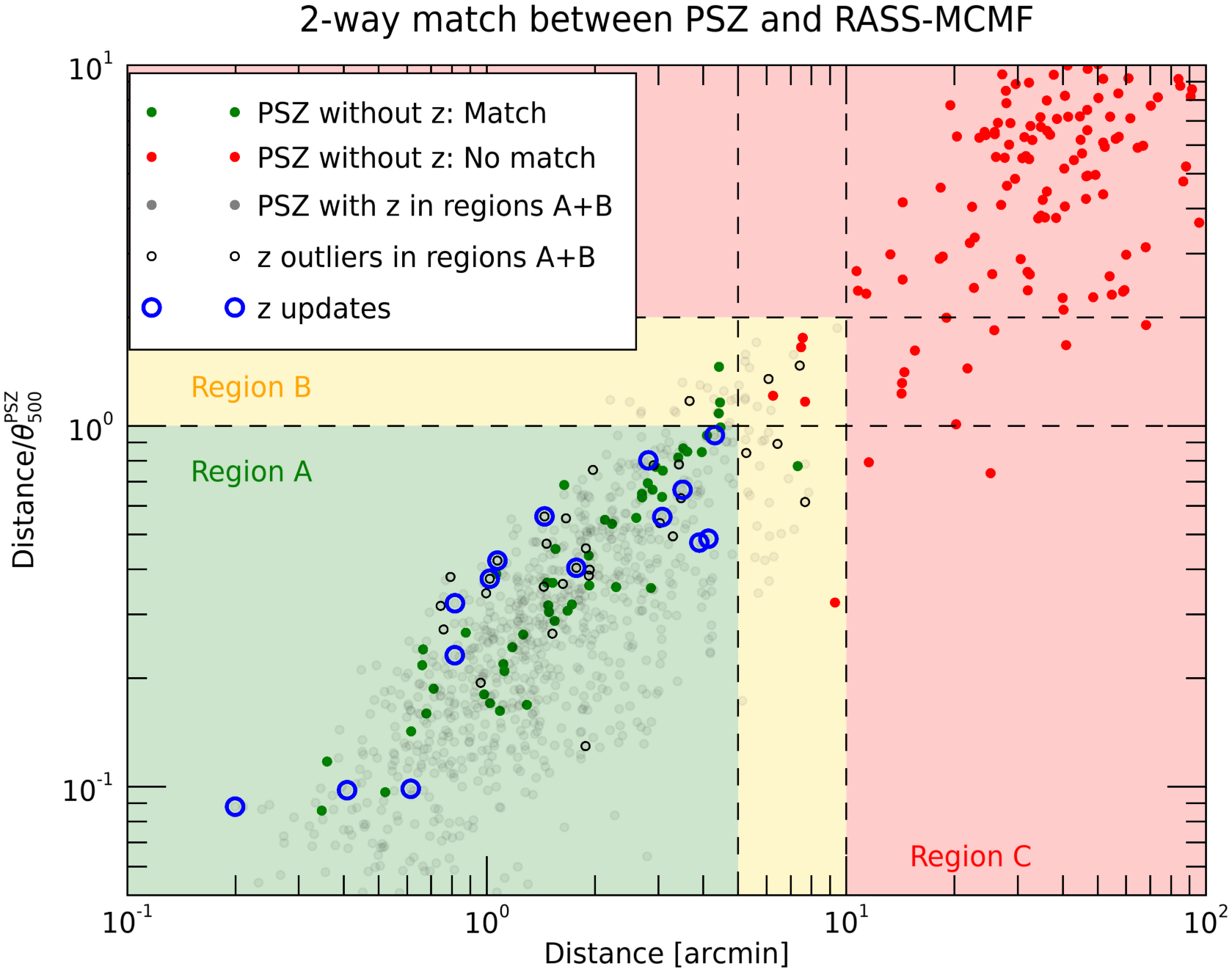}
\caption{\footnotesize Criteria for matching PSZ candidates to RASS-MCMF clusters. The distance between each PSZ candidate and its possible two-way RASS-MCMF association is plotted against their relative distance in terms of the PSZ cluster size $\theta_{500}$ recalculated at the RASS-MCMF redshift. Regions A, B, and C are shown in green, yellow, and red, respectively. Green/red dots indicate, for PSZ candidates (without z), whether the association is finally kept/rejected. The rest of the points (in grey) correspond to PSZ clusters with a redshift (plotted only in regions A and B, for clarity). Black empty circles highlight PSZ clusters for which there is a redshift difference of $\dzn>0.05$. Those coloured in blue are the ones for which we have updated the redshift. }
\label{fig:PSZ_RASSMCMF_xmatch}
\end{figure}

We used RASS-MCMF to validate PSZ candidates that remained unconfirmed after the optical follow-up and MCXC-II cross-identification, and to refine the redshifts of confirmed clusters. Following the method of Sect.~\ref{sec:mcxc},
Fig.~\ref{fig:PSZ_RASSMCMF_xmatch} illustrates the possible 2-way associations in the $d$-$d/\theta_{500}$ plane, where $\theta_{500}$ is the PSZ cluster size recomputed at the redshift of the RASS-MCMF cluster. 
This positional matching results in 943 
possible pairs in regions A and B: 56 PSZ candidates without redshift (47 in region A and 9 in region B) and  887 confirmed PSZ clusters.

Regarding the 56 PSZ candidates, we checked the consistency between the \planck\  SZE mass and the RASS-MCMF X--ray mass to confirm the association. We discarded the potential associations with a difference between the \planck\ mass recomputed at the RASS-MCMF redshift and the X--ray mass of the RASS-MCMF cluster greater than 4 times the total uncertainty $\sigma_{\rm tot}$, which considers, quadratically, the \planck\ mass error and the X--ray mass error (derived from the X--ray count error). With this criterion we discarded 2 associations in region B. The 47 PSZ candidates in region A pass this test, so we consider them as good associations, resulting in 47 new redshifts for the PSZ metacatalogue. We included a comment in the catalogue for one case, where the mass difference is greater than $3\sigma_{\rm tot}$.
The last step is the visual inspection of the 9 possible associations in region B. The 3 pairs at d<5 arcmin (corresponding to PSZ2 G227.44-31.24, \hyperlink{PSZ2 G267.30-46.19}{PSZ2 G267.30-46.19}, and PSZ1 G279.17-80.06) are separated by a relative distance of less than 1.5$\theta_{500}$, the mass difference is small (less than $2\sigma_{\rm tot}$), and the SZE signal appears to partially cover the X--ray emission and the optical counterpart, so we decided to associate the clusters. 
The only valid association above 5 arcmin was found for PSZ2 G285.85-46.36: the SZE signal has an offset relative to the X--ray peak, but appears to partially cover it ($D\sim0.8\Tv$). 
\hyperlink{PSZ2 G236.68-37.71}{PSZ2 G236.68-37.71} is a complex case of confusion described in Appendix \ref{app:eRASS} that we also decided not to associate. In the other cases, the optical RASS-MCMF cluster is outside the SZE emission and/or there is a mismatch between X--ray and SZ masses. 

In summary, the RASS-MCMF cross-identification provides new redshifts for 51 PSZ candidates. 34 of them also have a new redshift from eRASS, which is in agreement with the RASS-MCMF redshift. In these cases, we decided to assign the eRASS value. For the remaining 17, the PSZ cluster is validated and the RASS-MCMF redshift is assigned.

We also used RASS-MCMF to find potential redshift updates for already confirmed PSZ clusters. First, we focused on photometric to spectroscopic improvements. We identified 13 possible pairs in region A (none in region B), and we adopted the RASS-MCMF $\zs$ value in all cases based on the following considerations. For 7 of the 13 pairs, the redshift difference is small ($\dzn<0.01$). For additional 4 pairs, the difference is moderate ($0.01<\dzn<0.05$): \hyperlink{PSZ2 G255.07+54.84}{PSZ2 G255.07+54.84} and \hyperlink{PSZ1 G279.00-24.89}{PSZ1 G279.00-24.89}, are described in Appendix \ref{app:eRASS}, while for PSZ2 G089.39+69.36 ($\zp= 0.68$ from targeted  search in SDSS, $\zs =0.738$), and PSZ2 G176.27+37.54 ($\zp=0.5669$ from cross-identification with RMJ084000.2+442313.3,  $\zs=0.6318$), the RASS-MCMF optical counterpart is a rich cluster ($\lambda=147$ and $\lambda=124$) at small distance ($D=0.2\arcmin$ and $D=0.8\arcmin$), and the redshift difference is likely due to $\zp$ uncertainty at such high $z$. Finally, two pairs show a higher z difference ($\dzn>0.05$): PSZ2 G107.11-39.50 and PSZ2 G140.90-52.52. In both cases the \planck\ $\zp$ comes from redMaPPer ($\zp$=0.53, and $\zp$=0.54, respectively) and is close to the SDSS limit, while RASS-MCMF provides slightly higher but mutually consistent spectroscopic and photometric redshifts: ($\zs$=0.69, $\zp$=0.67, $\lambda$=147, and $\zs$=0.62, $\zp$=0.63, $\lambda$=146, respectively). Given the higher reliability of the RASS-MCMF spectroscopic measurements, we adopt the RASS-MCMF $\zs$ for both clusters.

We then analysed the 7 cluster pairs with large spectroscopic redshift differences ($\dzn>0.05$) in regions A and B. Two of them are not good associations, PSZ2 G139.00+50.92 and PSZ1 G232.76+32.70, since the PSZ and RASS-MCMF detections correspond to different structures. Five cases (PSZ2 G028.63+50.15, PSZ2 G103.40-32.99, PSZ2 G165.95+41.01, PSZ2 G254.96+55.88, and PSZ2 G264.60-51.07) involve the superposition of two clusters. For those, we decided to keep the redshift of the cluster that contributes more to the \planck\ SZE signal, which corresponds in all the cases to the PSZ redshift. 

Finally, we considered 21 possible associations with significant differences ($\dzn>0.05$) between PSZ redshifts and RASS-MCMF photometric estimates.  In general, we decided not to modify the PSZ redshift, as there is not enough evidence to conclude that the RASS-MCMF redshift is better. However, we analysed more in detail three of these 21 clusters, since they are listed in \citet[][Table A1]{RASSMCMF2023} as PSZ2 clusters with possible incorrect redshifts. We adopted the RASS-MCMF redshift for PSZ2 G281.09-42.51, and the eRASS redshift, in perfect agreement with the RASS-MCMF redshift, for \hyperlink{PSZ2 G181.71-68.65}{PSZ2 G181.71-68.65} (see Appendix \ref{app:eRASS}). PSZ2 G287.00-35.24 corresponds to a superposition of two clusters along the line of sight. In this case, we prefer to keep the original PSZ2 redshift.  
 
In summary, the RASS-MCMF cross-identification provides a redshift update for 15 PSZ clusters.  
Four of them, \hyperlink{PSZ2 G181.71-68.65}{PSZ2 G181.71-68.65}, PSZ2 G281.09-42.51, PSZ2 G199.61+53.41, and PSZ1 G279.00-24.89, also have a redshift update from eRASS, which is in agreement with the RASS-MCMF redshift. In these cases, we decided to assign the eRASS value. For the remaining 11, the RASS-MCMF redshift is assigned.

\subsection{PSZ-MCMF}\label{sec:pszmcmf} 
\begin{figure}[t]
\centering
\includegraphics[width=0.99\columnwidth]{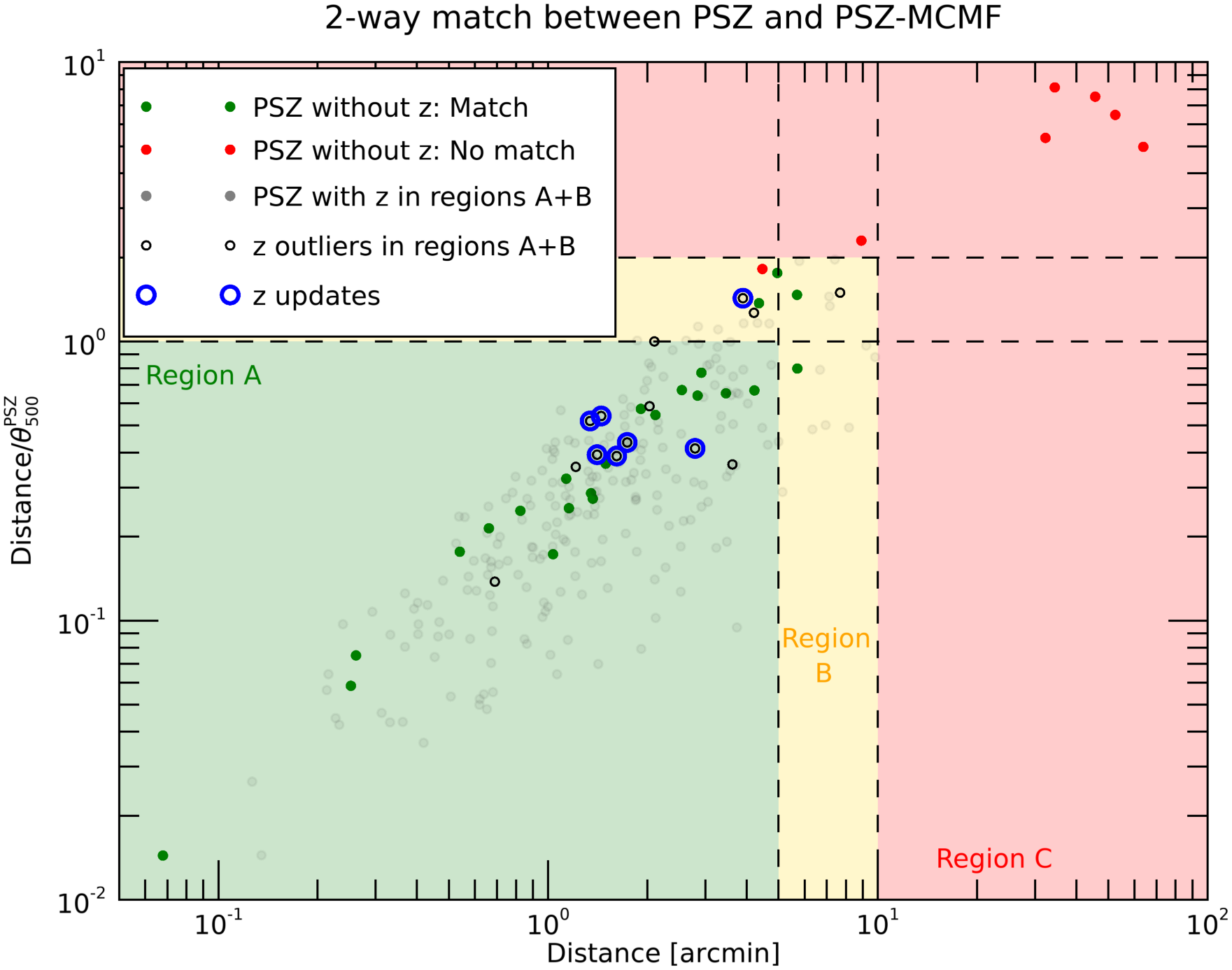}
\caption{\footnotesize Criteria for matching PSZ candidates to PSZ-MCMF clusters. The distance $d$ between each PSZ candidate and possible two-way PSZ-MCMF association is plotted against their relative distance in terms of the PSZ cluster size $\theta_{500}$ recalculated at the PSZ-MCMF redshift. The symbols are the same as in Fig.~\ref{fig:PSZ_RASSMCMF_xmatch}.}
\label{fig:PSZ_PSZMCMF_xmatch}
\end{figure}

PSZ-MCMF \citep{PSZMCMF2023} is a catalogue of 853 galaxy clusters constructed from a systematic follow-up of \planck\ SZE-selected candidates down to S/N = 3  over the 5000 deg$^2$ surveyed by the Dark Energy Survey DR1 \citep{DES_DR1}. The MCMF cluster confirmation algorithm was used to identify optical counterparts, resulting in a catalogue with 90\% purity. 

As with RASS-MCMF, we used PSZ-MCMF to validate PSZ candidates that remained unconfirmed after the optical follow-up and MCXC-II cross-identification, and to refine the redshifts of confirmed clusters, if possible. To do this, we cross-identified the PSZ clusters with the optical counterpart in the PSZ-MCMF catalogue with a two-way crossmatch (see Fig.~\ref{fig:PSZ_PSZMCMF_xmatch}), resulting in 263 possible pairs in regions A and B: 24 PSZ candidates without redshift and 239 confirmed PSZ clusters. 
We do not compare masses to identify bad associations, as both the PSZ and PSZ-MCMF catalogues derive their mass estimates from the same source (\planck\  SZE maps).  
On the other hand, we checked that the possible pairs do correspond to the same SZE detection. The PSZ-SN3 and \planck\  SZE positions are consistent within the errors, except for 2 very nearby and irregular clusters (less than $2\sigma$ difference and less than $\Tv$).

19 of the 24 PSZ candidates (without redshift) with a possible match lie in region A, so we consider them as good associations. The 
5 remaining pairs, in region B, were analysed individually. We decided to match \hyperlink{PSZ2 G213.73-56.15}{PSZ2 G213.73-56.15}, \hyperlink{PSZ2 G227.44-31.24}{PSZ2 G227.44-31.24}, \hyperlink{PSZ1 G245.21-65.29}{PSZ1 G245.21-65.29}, and \hyperlink{PSZ2 G319.64-65.11}{PSZ2 G319.64-65.11} with their closest PSZ-MCMF, and we discarded the association of \hyperlink{PSZ2 G236.68-37.71}{PSZ2 G236.68-37.71}, a case of confusion mentioned in Sect.~\ref{sec:PSZerass}. The description of these cases can be found in Appendix \ref{app:eRASS}.
 
In summary, the PSZ-MCMF cross-identification provides new redshifts for  
23 PSZ candidates. Of these, 
17 also have new redshifts from eRASS, which agree with the PSZ-MCMF redshifts. In these cases, we adopted the eRASS value. Another 2 candidates have new redshifts from RASS-MCMF, which are also consistent with the PSZ-MCMF values; for these, we adopted the RASS-MCMF redshift. For the remaining 4 cases, the PSZ cluster is validated and the PSZ-MCMF redshift is adopted.

We also used PSZ-MCMF to 
identify potential redshift problems in already confirmed PSZ clusters. 
Among the 239 PSZ clusters with a possible PSZ-MCMF counterpart in regions A and B,
14 have a redshift difference of $\dzn>0.05$ and were analyzed individually.
Six of them,  
\hyperlink{PSZ2 G215.19-49.65}{PSZ2 G215.19-49.65}, 
\hyperlink{PSZ2 G224.53-30.27}{PSZ2 G224.53-30.27}, 
PSZ2 G281.09-42.51, 
\hyperlink{PSZ2 G285.87-74.93}{PSZ2 G285.87-74.93}, 
\hyperlink{PSZ2 G307.72-77.87}{PSZ2 G307.72-77.87} and 
\hyperlink{PSZ2 G357.75-41.77}{PSZ2 G357.75-41.77}, were analysed in the eRASS crossmatch (Sect.~\ref{sec:PSZerass}), and their redshift was updated to the eRASS value, in agreement with the PSZ-MCMF redshift. For \hyperlink{PSZ2 G276.09-41.53}{PSZ2 G276.09-41.53} we updated the redshift to the PSZ-MCMF value (see Appendix \ref{app:eRASS}). For the remaining 7 cases, we kept the original catalogue redshift. Among them, 3 present two redshift peaks in PSZ-MCMF, with the secondary peak consistent with the original \planck\ redshift. For two systems, the available information does not allow us to determine which redshift is more reliable, and we therefore kept the original value. Finally, in 2 cases the PSZ-MCMF optical counterpart and the PSZ2 cluster are different clusters.

\subsection{eRASS+PSZ-MCMF}
\label{sec:erass+pszmcmf}

\begin{figure}[t]
\centering
\includegraphics[width=\columnwidth]{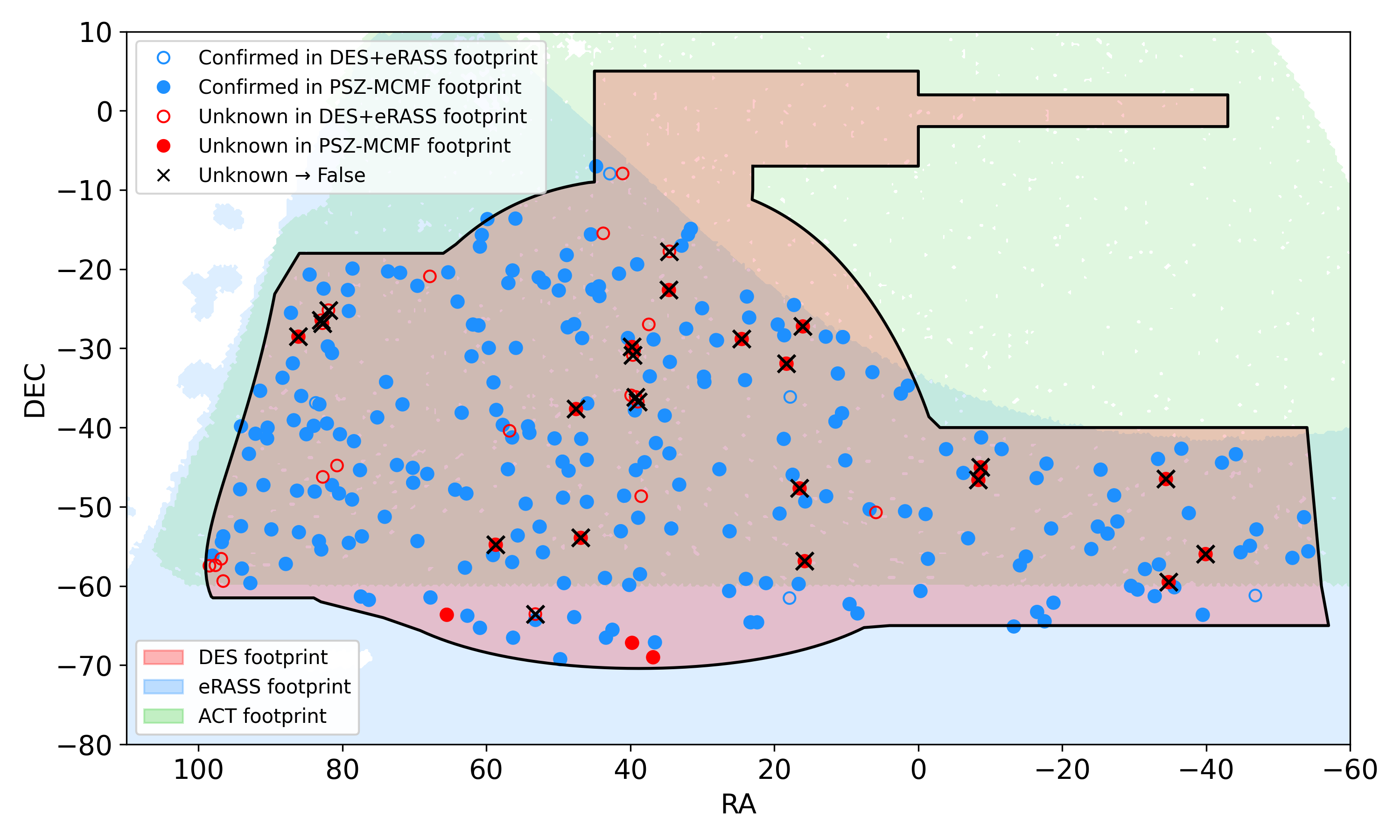}
\caption{\footnotesize \label{fig:DES} Confirmed (blue) and unknown (red) PSZ candidates in the area covered by eRASS and DES. Filled circles represent objects in the PSZ-MCMF footprint (i.e. where PSZ-MCMF searched for optical counterparts). Black crosses mark invalidated candidates. Blue/green/red regions represent the eRASS/ACT/DES footprints, respectively.}
\end{figure}

Cross identification with an external catalogue can confirm candidates. On the other hand, the lack of counterpart is not sufficient to discard the candidate, key issues being catalogue depth at candidate location and cluster multi-wavelength properties. However, we may  identify {\tt STATUS=False} candidates in overlapping regions of catalogue footprints, a bona-fide cluster becoming less-likely to appear in none of the catalogues. Here we consider the overlap of eRASS and DES area (hereafter DES$\cap$eRASS), 
also covered by RASS-MCMF cluster search (see Fig.~\ref{fig:DES}). Due to different \planck\ source selection (see Sect.~\ref{sec:pszmcmf}), the PSZ-MCMF study does not include all PSZ sources in the DES region\footnote{Missing PSZ sources are objects detected by PWS or MMF1 method but not by MMF3 at $S/N>3$ in the last \planck\ maps}. In the following, we define the PSZ-MCMF footprint as the set of sky positions corresponding to the input catalogue used for the PSZ–MCMF counterpart search.

We first assessed confirmed PSZ objects. There are 228 such objects in the DES$\cap$eRASS area, 222 are in the PSZ-MCMF footprint (optical search), and 196 of those at DEC$>-60^{\circ}$ in the ACT footprint (i.e. with independent SZE search). Considering pair positions in the $D$--$D/\Tv$ plane, with manual inspection in B region,  and redshift consistency check (as performed in Sect.~\ref{sec:PSZerass} and Sect.~\ref{sec:pszmcmf}),  we found that 194 of the 196 confirmed clusters have either an eRASS or a PSZ-MCMF counterpart.  
180 have counterparts in both catalogues, 8 only in eRASS, and 6 only in PSZ-MCMF. Only 2 objects, PSZ2 G267.30-46.19 and PSZ2 G208.57-44.31, correspond to neither an eRASS nor a PSZ-MCMF cluster.  However, the former, at $\zp=0.4251$, is confirmed with RASS-MCMF and also coincides with  ACT-CL J0356.0-5607 ($\zp=0.4143$). The latter is a very distant cluster at $z=0.82$ \citep[][optical follow-up]{vdB16}, with a low X-ray mass $6.\,10^{13}\msun$ \citep[][\xmm\ follow-up]{2022A&A...665A..24P}, 10 times lower than the PSZ mass (high Malmquist bias). This may explain why it is not detected by ACT.

We then turn to the 42 candidates with still unknown {\tt STATUS} in the DES$\cap$eRASS area. 19 of those are in the PSZ-MCMF footprint and were manually checked. We set  {\tt STATUS=False} for the 16 objects at DEC$>-60^{\circ}$.  Inspection of \planck\ and eRASS maps,  with the position of known clusters overlayed, confirms they do not match any PSZ-MCMF, eRASS or RASS-MCMF object (consistently with the {\tt STATUS}), no X--ray emission is apparent in eRASS map and no ACT counterpart is found. Furthermore the quality of the detection is generally low: 13 are detected by one method only, with 8 PSZ1 only objects (i.e. not passing the \planck\ S/N threshold in the latest maps). Conservatively,  we keep {\tt STATUS=U} for the other 3 clusters at DEC$<-60^{\circ}$. The 23 candidates not in the PSZ-MCMF footprint are detected at S/N <3 by MMF3 in the last maps by construction, actually by only one method, with 15 PSZ1 only candidates. 4 PSZ2 candidates are invalidated with \xmm\ archival data in Sect.~\ref{sec:xmmval} and we set {\tt STATUS=False} for 4 other candidates of very low $Q_{\rm neural}<0.03$ and no visible emission in eRASS image.

\subsection{Extended Planck catalogue from \citet{bur17}}
\label{sec:bur17}

The extended catalogue from \citet{bur17} is constructed from \planck\ Compton map in the SDSS field. The confirmation and redshift is obtained from redMaPPer catalogue \citep{Rozo2015} and/or  SDSS-DR10 and WISE data. 
The catalogue contains 465 PSZ2 objects, as identified by the author. All objects have a known {\tt STATUS} after the PSZ update described in previous sections. However,  the catalogue  provides a $\zs$  value
for  13 clusters   with $\zp$ (all from \planck\ catalogues).  The $\zp$ and $\zs$ values are consistent, except for  3 outliers with $|\zs-\zp|/(1+\zs)>0.05$. Those outliers,  \hyperlink{PSZ2 G039.34+73.28}{PSZ2 G039.34+73.28}, \hyperlink{PSZ2 G199.75+46.59}{PSZ2 G199.75+46.59} and \hyperlink{PSZ2 G317.52+59.94}{PSZ2 G317.52+59.94}, are discussed in Appendix~\ref{app:bur17}. The $\zp$ are updated to $\zs$ value of \citet{bur17}, except for PSZ2 G317.52+59.94. We also preferred the value $\zs=0.2965$ for PSZ2 G094.00+27.41=H1821+643, based on more than 100 spectroscopic members \citep{2018MNRAS.480.1187B}.

\subsection{Further validation with \xmm}
\label{sec:xmmval}
\xmm\ observation is a unique tool to unambiguously  distinguish between true and false candidates, as shown by  the \xmm\  DDT programme of validation of \planck\  sources  \citep[e.g.][]{xmmfu1}.  For candidates with {\tt STATUS=U},  we thus search for all available observations in  xmmmaster (as of April 2026), the master observation table of \xmm{}, obtaining \xmm\ data for 15 candidates. Four observations cannot be used because of mode (window mode), too large offset angle and/or flares. The total $[0.5$--$4.5]$ keV image\footnote{This is updated information of M2C database performed with the Xamin Command Line tool, provided by HEASARC, using the RA and DEC of each object  and using the default search radius of XMM. The images summing available pointings are produced as described in https://www.galaxyclusterdb.eu/m2c/documentation/images/} clearly confirms 3 objects: PSZ2 G278.79+08.54, PSZ1 G288.27+11.71 and PSZ1 G292.00-43.64. The latter matches XCLASS 2254 at $\zp= 0.55$ \citep{2026arXiv260323195M}. The remaining 8 candidates were set with {\tt STATUS=False}, no extended emission being detected at the PSZ position, with converging evidence from SZE properties. One object, PSZ2 G359.67-07.23, coincides with the \xmm\ target of the very bright X-ray binary V4046 Sgr, a radio source  likely  at the origin of the \planck\ detection. The noise nature of PSZ2 G210.37-37.00 is supported by a very low $Q_{\rm neural}=0.001$, together with a very high $S/N=9.8$. The other 6 candidates are detected by only one method, at low quality for PSZ2 G230.28-28.57 ($Q_{\rm neural}=0.03$),  PSZ2 G243.00-65.94 ($Q_{\rm neural}=0.17$) and  PSZ2 G278.74-45.26  ($Q_{\rm neural}<0.002$).

We further check the consistency of candidate invalidation  and \xmm\ data. We found 15 {\tt STATUS=False} objects with public \xmm\ data. PSZ1 G083.35+76.41 coincides with the target point source, BH CVn, a radio emitter likely at the origin of the \planck\ detection. For 13 objects (11 with low $Q_{\rm neural}<0.1$)  no X--ray diffuse emission is consistently observed. Although the observation is highly flared, a faint diffuse emission is observed at the position of PSZ2 G112.54+59.53, the target of the \xmm\ observation. It was set to {\tt STATUS=False} due to the low richness of the potential counterpart at $\zp=0.83$ \citep[optical mass less than 1/4 of the SZE mass]{Zoh19}. We changed the status to {\tt STATUS=C2}, the SZE detection likely being boosted by noise. 

Conversely, we checked {\tt STATUS=C2} objects with \xmm\ data and we changed the status to {\tt STATUS=C1} for 3 objects: 2 with C2 validation from optical follow-up and PSZ2 G319.16+26.63 from identification with a low S/N eRASS cluster, detected serendipitously in \xmm. On the other hand, no emission is detected at the position of PSZ2 G138.61-10.84 and we changed its status to {\tt STATUS=False}.

\begin{figure}[t]
\centering
\includegraphics[width=0.9\columnwidth]{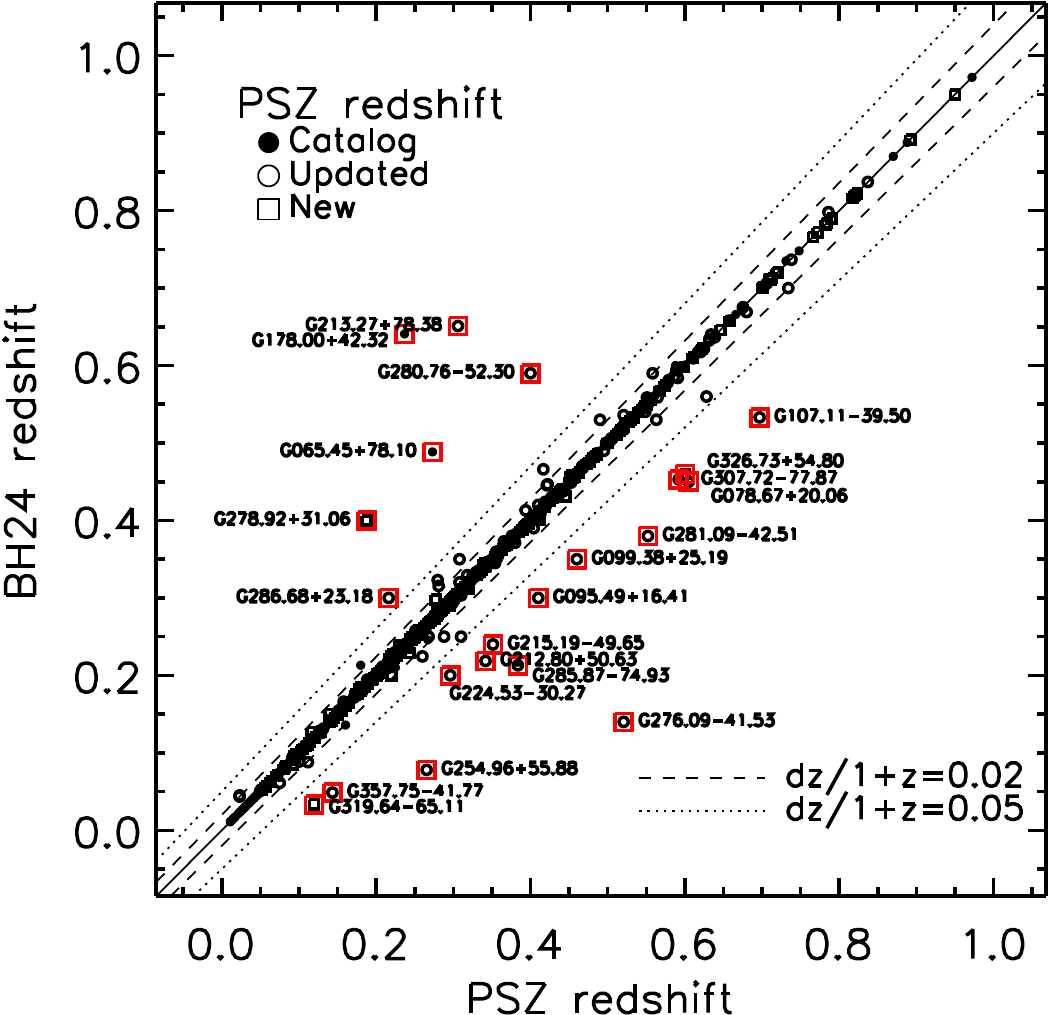}
\caption{\footnotesize \label{fig:BH24} Comparison of  redshifts from the present updated PSZ with that of the updated PZ2 catalogue of \citet{BH24}. Open boxes and circles denote new and updated values, respectively.  Prominent outliers are labelled and discussed Tab.~\ref{tab:BH24} and Sect.~\ref{sec:BH24z}. }
\end{figure}

\begin{table*}[!ht]
\caption{\footnotesize \label{tab:BH24} Discrepant redshift with  update of \citet[][BH24]{BH24}.}
\resizebox{\textwidth}{!} {
\begin{tabular}{lllrlccccrrl}
\toprule
\toprule
     \multicolumn{5}{c}{{PSZ updated catalogue}} &
    \multicolumn{4}{c}{{ BH24 updated PSZ2 catalog}} &  
    \multicolumn{2}{c}{Difference}  & Note\\
    \multicolumn{1}{c}{NAME}  &
    \multicolumn{1}{c}{{\tt z}} &  
   \multicolumn{1}{c}{{\tt Type}} &  
    \multicolumn{1}{c}{{\tt Flag}} &  
    \multicolumn{1}{c}{{\tt Reference}} &  
    \multicolumn{1}{c}{{\tt z} }&
    \multicolumn{1}{l}{{\tt Type}} &  
    \multicolumn{1}{l}{{\tt Val}} &  
    \multicolumn{1}{l}{{\tt Flag} }&
    \multicolumn{1}{l}{$\Delta z/(1+z)$} &
    \multicolumn{1}{l}{$\Delta z/z$} \\  
\midrule 
&&&&& \multicolumn{4}{l}{From PSZ2 catalog} \\
\cline{6-9}
PSZ2 G107.11-39.50& 0.6967 &spec & Updated  &  RASS-MCMF &  0.5330 & spec & 13 & V &  0.097 &  23~$\%$    & see Sect.~\ref{sec:rassmcmf}\\ 
PSZ2 G212.80+50.63& 0.3417 &spec & Updated  &  eRASS     &  0.2187 & spec & 13 & V &  0.092 &  36~$\%$    & (a)\\ 
\hyperlink{PSZ2 G254.96+55.88}{PSZ2 G254.96+55.88}  & 0.2654 &spec & Updated & eRASS      &  0.078 & spec & 20 & V &  0.148 &  71~$\%$    &see Appendix~\ref{app:eRASS}\\ 
PSZ2 G281.09-42.51& 0.5572 &phot & Updated          &  RASS-MCMF &  0.3800 & phot & 20 & V &  0.114&  32~$\%$    & see Sect.~\ref{sec:rassmcmf}\\ 
PSZ2 G286.68+23.18& 0.2163 &phot & Updated          &   eRASS    &  0.3000 & phot & 20 & V & -0.069 &  39~$\%$     &see Sect.~\ref{sec:PSZerass}\\ 
\hyperlink{PSZ2 G285.87-74.93}{PSZ2 G285.87-74.93}& 0.3836 &phot & Updated &    eRASS   &  0.2130 & spec & 20 & V &  0.123 &  44~$\%$    &see Appendix~\ref{app:eRASS}\\ 
\hyperlink{PSZ2 G307.72-77.87}{PSZ2 G307.72-77.87}& 0.5920 &spec & Updated &  \citet{Hilton2021}    &  0.4530 & phot & 20 & V &  0.087 &  23~$\%$ & see Appendix~\ref{app:eRASS}\\\ 
PSZ2 G078.67+20.06& 0.6050 &spec & Updated & NOT Dahle+ priv. com &  0.4500 & spec & 20 & V &  0.097 &  26$~\%$& (b) \\ 
PSZ2 G095.49+16.41& 0.4100 &spec & Updated & NOT Dahle+ priv. com &  0.3000 & spec & 20 & V &  0.078 &  27~$\%$ & (b)\\ 
PSZ2 G099.38+25.19& 0.4600 &spec & Updated & NOT Dahle+ priv. com &  0.3500 & spec & 20 & V &  0.075 &  24~$\%$ & (b)\\ 
\hyperlink{PSZ2 G357.75-41.77}{PSZ2 G357.75-41.77} &0.1434 & phot & Updated & eRASS & 0.0487 & spec & 21 & V & 0.083 & 66~$\%$ & see Appendix \ref{app:eRASS}\\
\hyperlink{PSZ2 G276.09-41.53}{PSZ2 G276.09-41.53}& 0.5207 &phot & Updated &             PSZ-MCMF  &  0.1400 & phot & 22 & V &  0.250 &  73$\%$ & see Appendix \ref{app:eRASS}\\
\hyperlink{PSZ2 G280.76-52.30}{PSZ2 G280.76-52.30}& 0.4000 &spec & Updated &  \citet{Bocquet2019}  &  0.5900 & phot & 22 & V & -0.136 &  47~$\%$ & see Appendix~\ref{app:psz12dup}\\ 
\hyperlink{PSZ2 G224.53-30.27}{PSZ2 G224.53-30.27}& 0.2960 &phot & Updated & eRASS  &  0.2001 & phot & 30 & V &  0.074 &  32~$\%$ & see Appendix~\ref{app:eRASS} \\ 
\hyperlink{PSZ2 G215.19-49.65}{PSZ2 G215.19-49.65}& 0.3513 &phot & Updated & eRASS  &  0.2399 & spec & 30 & V &  0.082 &  32~$\%$ & see Appendix~\ref{app:eRASS}\\ 
\midrule 
&&&&& \multicolumn{4}{l}{From follow-up} \\
\cline{6-9}
PSZ2 G326.73+54.80& 0.6000 &phot &     New &  \citet{Zoh19} &  0.4600 & phot & 50 & V &  0.088 &  23~$\%$ & (c) \\ 
\midrule 
&&&&& \multicolumn{4}{l}{From external catalog} \\
\cline{6-9}
\hyperlink{PSZ2 G065.45+78.10}{PSZ2 G065.45+78.10}& 0.2730 &     & Catalogue &  \citep[][NED]{psz2} &  0.4885 & spec & 60 & V & -0.170 &  79~$\%$ & see Appendix \ref{app:BH24}\\ 
\hyperlink{PSZ2 G178.00+42.32}{PSZ2 G178.00+42.32}& 0.2368 &     & Catalogue &   \citep[][NED]{psz2} &  0.6408 & spec & 60 & V & -0.327 & 171~$\%$ & see Appendix \ref{app:BH24} \\ 
PSZ2 G213.27+78.38& 0.3060 &spec & Updated &  \citet{Str18} &  0.6510 & spec & 60 & V & -0.264 & 113~$\%$ &  (d) \\ 
PSZ2 G278.92+31.06& 0.1875 &spec &     New &  eRASS  &  0.4000 & spec & 61 & V & -0.179 & 113~$\%$ & (e) \\ 
\midrule 
&&&&&\multicolumn{4}{l}{From galaxy catalog} \\
\cline{6-9}
\hyperlink{PSZ2 G319.64-65.11}{PSZ2 G319.64-65.11}& 0.1196 &spec &     New &  eRASS  &  0.0336 & spec & -1 & S &  0.077 &  72~$\%$ & (f) \\ 
\bottomrule
\end{tabular}
}
\tablefoot{ 
Column (8) BH24 validation code: 13: RedMaPPer non blind; 20: PSZ1 legacy; 50: CIO follow-up ; 60: \citet{bur17}; 61: \citet[][SPT]{Bocquet2019};  -1: galaxy catalogue search.

(a): New $z$ from identification with 1eRASS J100048.7+203907=RMJ100050.3+203923.0 at 4.6$\arcmin$ with possible contribution of lower richness RMJ100040.7+204654.0=WHL J100044.5+204728 ($\zs=0.2108$ ) at $3.7\arcmin$ in the north. The latter corresponds to the PSZ2 value from redMaPPer non blind search.

(b) The original PSZ1 redshift from NOT follow-up used in PSZ1 catalogue is updated (see Sect.~\ref{sec:PSZupfu})

(c): The BH24 redshift is from \citet{Bar20} follow-up,  $\zp=0.46\pm0.05$. We prioritize  the  higher precision  $\zp=0.60\pm0.02$ from \citet{Zoh19},  which also corresponds  to a good match between optical and SZE mass.

(d)  The updated value from \citet{Str18} is based on SDSS DR12 data, with an optical counterpart  at $D=0.88\arcmin$ and a  $\zs$  based on three galaxy redshifts. The optical counterpart identified by \citet{bur17} is further away ($D=3.8\arcmin$) and the redshift is based on one galaxy. 

(e) We did not find a close SPT cluster and  the BH24 value is likely due to a misprint; New $z$ from identification with 1eRASS J111857.7-272517 ($\zs=0.1875$) at $D\,=\,2.2\arcmin\,=\,0.4 \, \Tv$,  consistent with that of  RASS-CL J111858-2725.5 ($\zp\,=\,0.192\pm0.005$) at $D\,=\,1.9\arcmin\,=\,0.36 \, \Tv$.

(f)  New $\zs=0.1196$ from identification with  1eRASS J000654.8-503210  at $5.9\arcmin= 0.8 \, \Tv$, coincident with  WHY J000657.3-503147 at the same $\zs$. Visual inspection of the eRASS and SZE images confirms the association. 
}
\end{table*}

\section{Comparison with the PSZ2 update of \citet{BH24} }
\label{sec:BH24}

\subsection{Input data and method}
During the course of this work, \citet[][hereafter BH24]{BH24} published an update of the PSZ2 catalogue, based  on optical follow-up publications and  cross-match with  external  catalogues (see their Table 1). The source of the retained redshift is identified by  the {\tt Val} code, as given in the PSZ2 catalogue ({\tt Val}<50) or in their Table 1 for updated values. They  performed   further  validation,  using  galaxy redshift catalogues  ({\tt Val}=-1). 
	
As in our work,  a higher priority is put on spectroscopic redshift and follow-up work. However, their cross-match with external catalogues  is based on  fixed distance criteria (adapted to each catalogue) rather than the more sophisticated analysis described in Sect.~\ref{sec:matching}. They  also do not consider PSZ1 only objects and do not include very recently published MCXC-II and eRASS catalogues or \xmm\  information, considered in the present work.  
	 
We carefully compared the BH24 updated redshift and validation status for PSZ2 clusters and candidates with that in the updated PSZ  (1649 objects).  We identified   21 prominent redshift discrepancies and different validation status for 78 objects, as discussed in Sect.~\ref{sec:BH24z} and Sect.~\ref{sec:BH24status}   below.

\subsection{Discrepant redshift}
\label{sec:BH24z}
There are 1289 PSZ2 clusters with redshift both in BH24 and in the updated PSZ. The vast majority (1000) being identical as expected.  The histogram of $\dzn$ for the other clusters is bimodal, the first peak below $\dzn<0.00016$ likely simply related to data formatting, with 245 clusters with larger redshift differences. The redshifts are compared in Fig.~\ref{fig:BH24}. 
Most of the differences are small and are linked to input data (e.g., PSZ $\zp$ update to $\zs$ not included in BH24). However, there are 21 prominent outliers. They are listed in Table~\ref{tab:BH24} with relevant information and notes on individual cases. Nineteen  PSZ redshifts are updated or new values (15 of those are original PSZ2 values in BH24 catalogue), which we think are robust as detailed in the footnote of the table or in Appendix (see Note column). The PSZ redshift of the other 2 clusters, PSZ2 G065.45+78.10 and PSZ2 G178.00+42.32,  are  from the PSZ2 catalogue from cross-identification with GMBCG J204.74580+32.97396 and NSC J090659+430556, respectively. They have been updated by BH24 from \citet{bur17}. We discuss these two cases in Appendix~\ref{app:BH24} and retain the BH24 updated value for PSZ2 G065.45+78.10. The other redshift is unchanged.

\subsection{Discrepant validation status }
\label{sec:BH24status}
(1) There are 38 confirmed clusters in the updated PSZ,  still lacking validation information in BH24 catalogue.  All are new validation  ({\tt z\_flag=New}), mostly from input  data not included in BH24. The redshift of 33 clusters  comes  from our cross-match with eRASS catalogue (Sect~\ref{sec:PSZerass}), one cluster PSZ2 G112.54+59.53 was validated with \xmm\ (see Sect.~\ref{sec:xmmval}) and the redshift of one cluster (PSZ2 G156.88+13.48) is  from Dahle  private comm. However, we failed to understand  the remaining three cases not found in BH24: PSZ2 G246.91+24.65  with $z$ from \citet{Zoh19};  PSZ2 G339.74-51.08 and PSZ2 G018.64-83.11 from cross-match with PSZ-MCMF. The distances of the PSZ-MCMF counterpart are 1.13\arcmin and 0.26\arcmin,  respectively, smaller than the 3\arcmin cross-match distance used  by BH24 for this catalogue.

\smallskip
\noindent (2) There are  8 confirmed  clusters in BH24 catalogue, without redshift in the updated PSZ: the complex case of confusion, \hyperlink{PSZ2 G236.68-37.71}{PSZ2 G236.68-37.71}, as discussed in Sect.~\ref{sec:PSZerass} and Appendix \ref{app:eRASS}, and 7 candidates with {\tt STATUS=U}. Those  includes:
\begin{itemize}[noitemsep,topsep=0pt,label=$-$] 
\item PSZ2 G098.38+77.22  with  BH24 $z$ from \citet[][]{boa19} ({\tt Val=54}). The optical counterpart is uncertain as we discussed in Sect.~\ref{sec:PSZupfu}.
\item \hyperlink{PSZ2 G011.36-72.93}{PSZ2 G011.36-72.93}, \hyperlink{PSZ2 G014.72-62.49}{PSZ2 G014.72-62.49} and PSZ2 G278.94-17.62, validated by BH24 from cross-match with ACT/SPT. The first two clusters are further discussed in Appendix \ref{app:BH24status}, PSZ2 G011.36-72.93 is updated with consistent ACT $\zs$ value and 
and we retain the SPT value for PSZ2 G014.72-62.49.
On the other hand, we assume that there is a typing error in the BH24 catalogue for the last object, the closest SPT cluster being at distance $D=9^\circ$.
\item  \hyperlink{PSZ2 G017.25-70.71}{PSZ2 G017.25-70.71}, \hyperlink{PSZ2 G031.37-71.95}{PSZ2 G031.37-71.95} and \hyperlink{PSZ2 G327.27+11.05}{PSZ2 G327.27+11.05}, three newly validated objects from BH14 study ({\tt Val=-1, f\_val=S}), all qualified as 'strong' candidates. We further discussed these objects in Appendix~\ref{app:BH24status}. We retain the BH24 $z$ for PSZ2 G017.25-70.71 and PSZ2 G031.37-71.95 with {\tt STATUS=C2} and C1, respectively and kept the PSZ flag for PSZ2 G327.27+11.05.
\end{itemize} 

\smallskip
\noindent (3) There are 32 objects with redshift in BH24, invalidated in the updated PSZ ($z=-10$, {\tt STATUS=False}). The difference comes mostly  from our different validation criteria. This includes:
\begin{itemize}[noitemsep,topsep=0pt,label=$-$] 
\item 15 objects with BH24 redshift from the follow-up conducted at the CIO\footnote{Some are mis-labelled in BH24 as {\tt Val=-1} (from galaxy surveys)} with {\tt Flag=3} (e.g., too low mass potential counterpart). 
\item  PSZ2 G051.48-30.87 and PSZ2 G110.69-46.25, not detected ({\tt Flag=ND}) by \citet{str19} in the CIO follow-up.  BH24 validated these objects at $z = 0.135$ and $z=0.086$, respectively,  using  galaxy redshift catalogue ({\tt Val=-1}). These cases are discussed in the Sect.~5.2  of their paper, noting the multiple clustering signal. We keep the PSZ flag.
\item PSZ2 G157.07-33.63, PSZ2 G165.41+25.93 and  PSZ2 G191.82-26.64, with contradictory results in various follow-up that we discussed in Sect.~\ref{sec:PSZupfu}. 
\item Two objects, PSZ2 G133.92-42.73 and PSZ2 G235.96+38.21 with BH24 redshift from \citet{bur17}, also studied by  \citet{Agu19} and  \citet{Bar20}, respectively. In both cases, we verified that the potential optical counterpart is the same (close position and redshift). The potential counterpart of PSZ2 G133.92-42.73 was discarded by \citet[][Sect.~5.2]{Agu19}, while the PSZ2 G235.96+38.21 one can be discarded in view of its too low mass ({\tt Flag=3}). The case of \hyperlink{PSZ2 G133.92-42.73}{PSZ2 G133.92-42.73} is revisited in App.~\ref{app:BH24status} where we set a {\tt STATUS=C2} with redshift of a nearby ACT cluster.
\end{itemize}
\noindent The {\tt STATUS} of 4 objects relies on information not available in BH24. This includes:
\begin{itemize}[noitemsep,topsep=0pt,label=$-$] 
\item \hyperlink{PSZ2 G225.18-33.61}{PSZ2 G225.18-33.61}, noise dominated from our combined study of \planck\ and eRASS data (Sect.~\ref{sec:PSZerass}). The BH24 validation of this candidate from galaxy study ({\tt Val=-1, f\_val=S}) is  discussed in Appendix~\ref{app:BH24status}.
\item PSZ2 G215.25-87.14 and PSZ2 G224.86-79.51, invalidated by our combined eRASS-PSZ-MCMF study in the DES region (Sect.~\ref{sec:erass+pszmcmf}). The  redshift, $z=0.1201$ and $z = 0.095$,  from BH24 galaxy study ({\tt Val=-1, f\_val=S}) is likely that of foreground galaxies.
\item PSZ2 G048.09+27.18, validated in the PSZ1 catalogue from SDSS data,  in-validated from later NOT follow-up (Dahle, priv. comm).
\end{itemize}

\noindent Finally, PSZ2 G138.61-10.84 is invalidated from \xmm\ study and five candidates are invalidated from the mass study described below in Sect.~\ref{sec:mass} below.

\section{Mass correction for selection effects} \label{sec:mass}

\subsection{Masses published in the PSZ1 and PSZ2 catalogues}
\label{sec:pubmass}

\begin{figure}[]
\centering
\includegraphics[width=0.99\columnwidth]{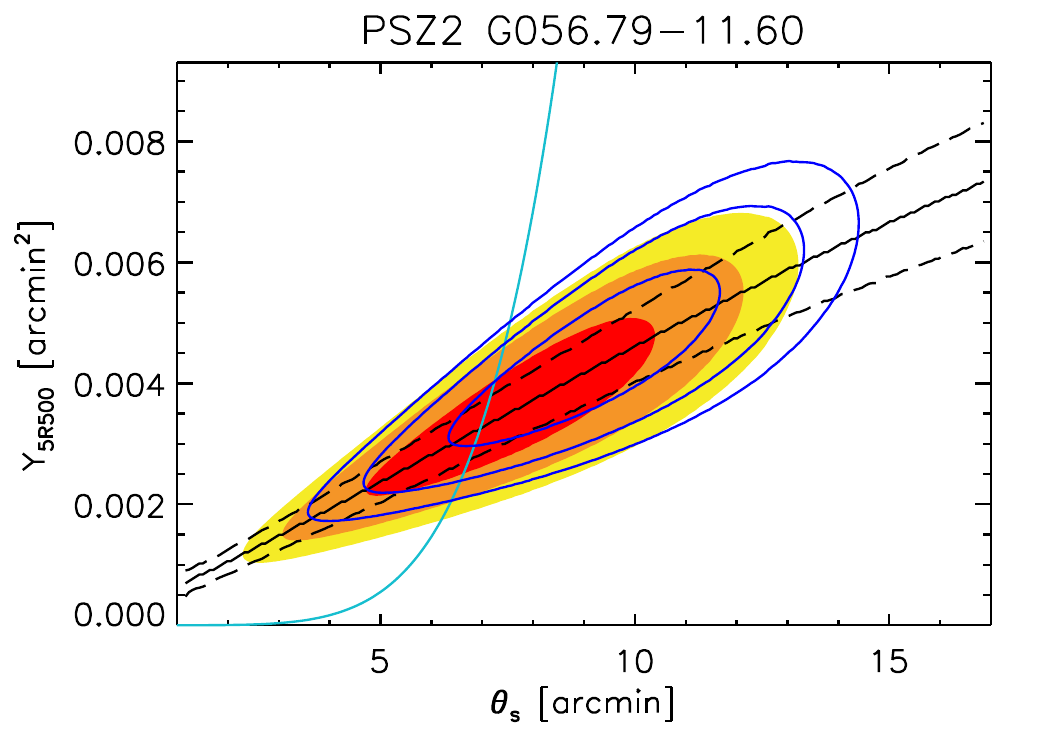}
\caption{\footnotesize Degeneracy contours corrected (filled coloured contours) and uncorrected (solid blue line) for selection effects for \mbox{PSZ2 G056.79-11.60}. The contours show the 68\%, 95\% and 99\% confidence limits from the innermost to the outermost. Solid and dashed black lines are ridge and $\pm$68\% degeneracy lines for the corrected contours. The X--ray prior at the cluster redshift $z=0.12$ is shown as the solid cyan line. Correcting for selection effects pushes the fluxes, sizes and the masses towards lower values.}
\label{fig:degecont}
\end{figure}

The masses published in the PSZ1 and PSZ2 catalogues were estimated from the integrated SZE flux $\Ytot$ measured by the cluster extraction algorithms. In \planck\ data, $\Ytot$ is strongly degenerate with the size of the filter used to extract clusters. Indeed, the filter can only weakly constrain the cluster size $\ts$ because it is of the same order as the \planck\ resolution (arcmin scale). This is known as the size-flux degeneracy \citep{PlanckEarlyResXI}. The \planck\ collaboration therefore provided posterior probability contours in the $\Ytot-\ts$ plane, also named degeneracy contours, for the clusters in the PSZ1 and PSZ2 catalogues. If a cluster was detected by multiple algorithms, a degeneracy contour was provided for each detection algorithm. The masses were obtained in breaking the size-flux degeneracy in the $\Ytot-\ts$ posterior probability contours with a prior on the $\Mv$--$\Yv$ scaling relation, calibrated from X--ray data. Combined with the $\Mv$--$\Rv$ relation and the universal pressure profile, this provides the expected scaling $\Ytot-\ts$ relation if $z$ is known \citep[see  Eq.~7, Table~1 and Eq.~9 in][]{psz1cosmo}. Details of the calculation are given in Sect.~7.2.2 of \cite{psz1} and Sect.~5.3 of \cite{psz2}. The  intersections of the observed and prior relation provides the $\Mv$ and $\Yv$ values, as illustrated in Fig.~\ref{fig:degecont}. Difficulties may arise when the intersection occurs outside the range within $\ts$ is defined; this is further discussed in Sect.~\ref{sec:newmass}.

\begin{figure}[t]
\centering
\includegraphics[width=0.99\columnwidth]{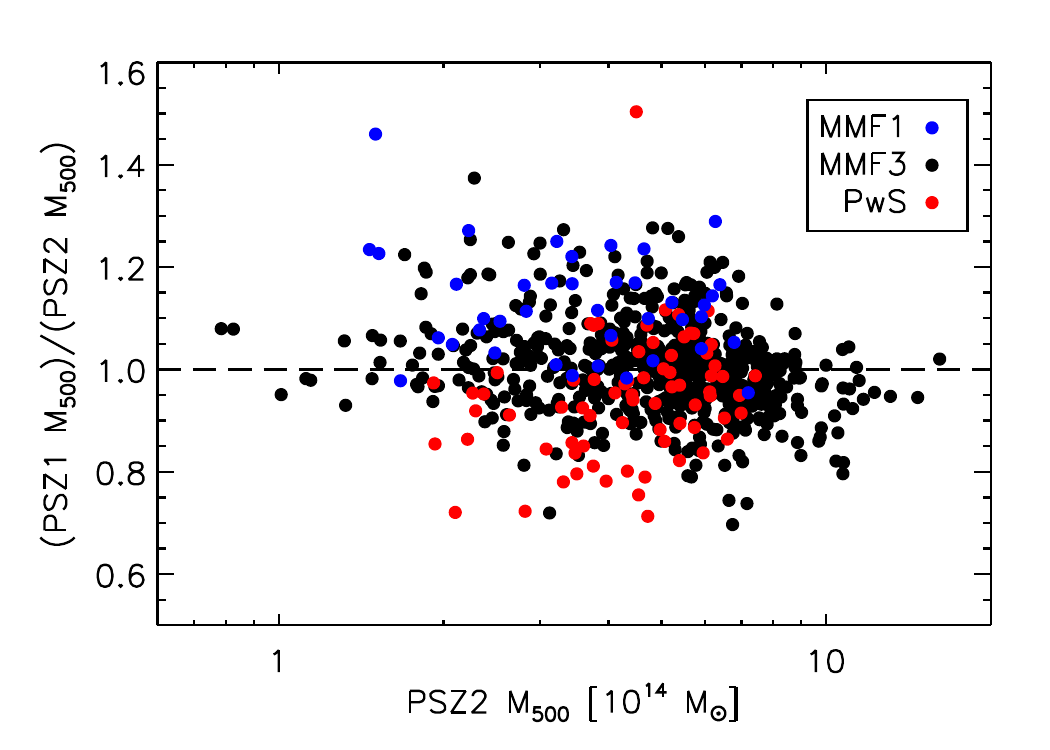}
\caption{\footnotesize Ratio between the PSZ1 and the PSZ2 masses as a function of PSZ2 masses. The published PSZ1 masses are systematically overestimated (resp. underestimated) for the MMF1 (resp. PwS) detection algorithm with respect to the PSZ2 masses. The PSZ1 and PSZ2 masses are consistent for MMF3.}
\label{fig:PSZ12masscomparison}
\end{figure}

In the Planck Legacy Archive (PLA), masses for the PSZ1 are available in the file {\tt COM\_PCCS\_SZ-validation\_R1.13.fits} as {\tt M\_YZ\_500}. For the PSZ2, the masses are given as {\tt MSZ} in {\tt HFI\_PCCS\_SZ-union\_R2.08.fits}.
We compare both quantities for the three detection algorithms in Fig.~\ref{fig:PSZ12masscomparison}, for clusters with a redshift difference $\Delta z < 0.05$ between PSZ1 and PSZ2. The PSZ1 masses provided for the algorithm MMF1 (resp. PwS) are systematically overestimated (resp. underestimated) with respect to the listed PSZ2 masses by about 10\% (resp. 5\%). This is not the case for MMF3, for which the masses are consistent between the two \planck\ catalogues. To investigate this effect, we calculated the PSZ1 masses directly from the published PSZ1 degeneracy contours. We did not find the systematic shifts for MMF1 and PwS when masses were obtained from the contours. In summary, there is an inconsistency between the published PSZ1 masses and the PSZ1 degeneracy contours. The PSZ1 degeneracy contours are in agreement with PSZ2 masses and PSZ2 degeneracy contours, which points towards possible systematic errors in the calculation of the PSZ1 masses for MMF1 and PwS in {\tt COM\_PCCS\_SZ-validation\_R1.13.fits}. To overcome this issue, we therefore propose to undertake a new and consistent estimation of PSZ1 masses from the published PSZ1 degeneracy contours for the three algorithms. 

We provide the masses for a grid of redshifts as was done for the PSZ2 catalogue. This will allow the user to associate an SZE mass to a detection when an optical counterpart is found. The tables providing the masses as a function of redshift were not available in the PSZ1 public release.

\subsection{New estimation of PSZ1 and PSZ2 masses}
\label{sec:newmass}

The limitations on the published PSZ1 masses discussed in Sect.~\ref{sec:pubmass}, and the need for consistency between PSZ1 and PSZ2 masses, led us to undertake a new and consistent estimation of the PSZ1 and PSZ2 masses based on the published degeneracy contours. For each degeneracy contour $P(\ts,\Ytot)$, we extracted the ridge line (the maximum of the probability for each $\ts$) and the 68\% upper and lower degeneracy lines (the 68\% limits of the probability for each $\ts$). We then intersected these three lines with the prior obtained from X--ray observations for a set of 100 redshifts, ranging from $z=0.01$ to~1 in steps of 0.01. The intersections provide values for $\ts$ and $\Ytot$, which are then converted into the mass $M_{500}$($z$) and its upper and lower 68\% confidence limits using Eq.~7 and Table~1 of \cite{psz1cosmo}. Figure~\ref{fig:degecont} shows an illustration of the method for \mbox{PSZ2 G056.79-11.60}. 

\begin{figure}[t]
\centering
\includegraphics[width=0.99\columnwidth]{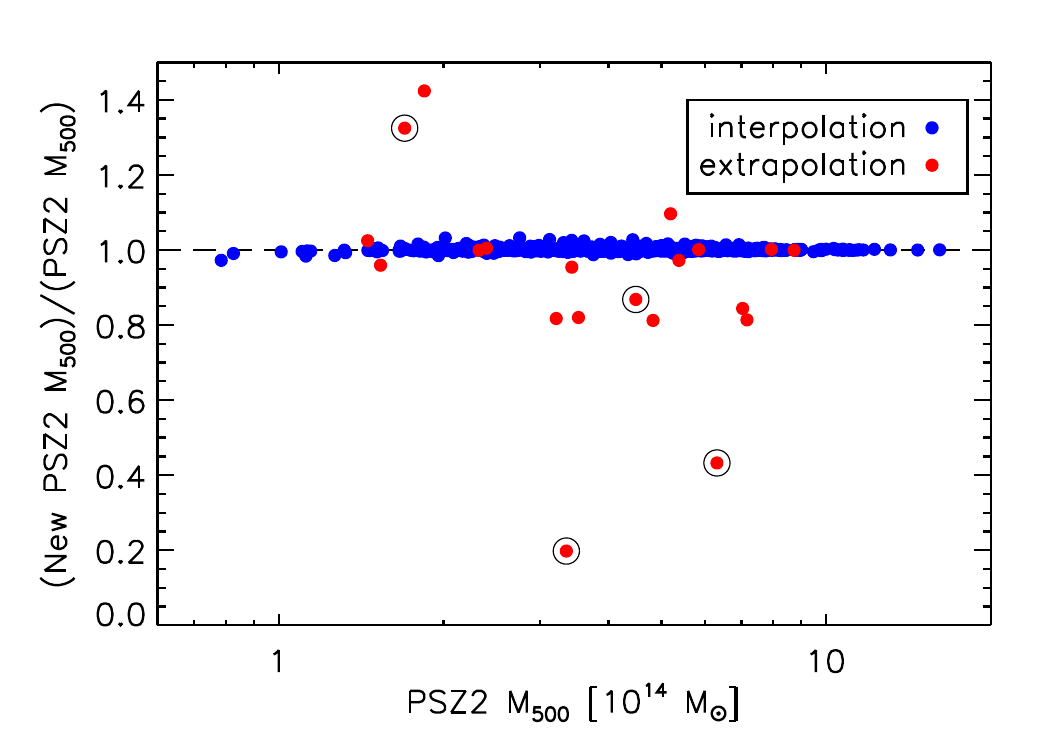}
\caption{\footnotesize Ratio between the new and the original PSZ2 masses as a function of the original PSZ2 masses. Blue (resp. red) points are obtained by interpolation (resp. extrapolation) of the degeneracy lines. Masses are in very good agreement, with only 20 masses obtained by extrapolation differing from the original values. Those are discussed in Sect.~\ref{sec:newmass}. The detections flagged as false are surrounded by a black circle.}
\label{fig:newvsoldPSZ2mass}
\end{figure}

As mentioned in Sect.~\ref{sec:pubmass}, difficulties arise when the intersection occurs outside the range within which $\ts$ is defined, in which case we need to extrapolate the ridge line and the upper and lower confidence limits beyond the definition range. 
We developed a method for extrapolation when needed\footnote{The extrapolation method is not detailed in~\cite{psz2}.}. This is obtained by fitting the $\log(\ts)$-$\log(\Ytot/Y^{\rm prior}_{5R500})$ relation in the range where $\ts$ is defined with a second-degree polynomial, and then finding the closest root located outside the definition range of $\ts$, which corresponds to the expected $\ts$ value for the intersection of the degeneracy lines and the prior. This gives us the MSZ($z$) mass curves and their uncertainties, for each cluster and each detection algorithm. It is then straightforward to compute the mass proxy and associated errors by interpolating the mass curves at the cluster redshift. Figure~\ref{fig:newvsoldPSZ2mass} compares the new masses to the published masses for the PSZ2 catalogue. The masses are in very good agreement. The only significant differences come from the extrapolation method adopted in our work. 

There are 19 cases where the intersection lies well outside the 99\% confidence range of $\ts$, and one case (PSZ2 G302.41+21.60) where the intersection lies outside the 95\% confidence range. For these 20 objects, the model (the expected  $\Ytot-\ts$ scaling relation at the considered redshift) and the data (the observed $\Ytot(\ts)$ with errors) are formally inconsistent at very high significance (>99\% confidence level), casting doubt on the detection. We examined these objects individually: 
\begin{itemize}[noitemsep,topsep=0pt,label=$-$] 
\item Eleven objects are well known local ($z<0.1$) massive MCXC/Abell clusters of size larger than the \planck\ PSF ($10\arcmin<\Tv<50\arcmin$). From the literature and from visual inspection of X--ray maps, they are all disturbed, unrelaxed, clusters, including  two double systems (PSZ2 G263.19-25.19=ACO 3395 and PSZ2 G125.37-08.67= MCXC J0107.7+5408).  We therefore expect deviation from the simple universal pressure profile used in the extraction, typically towards a flatter, more inflated, distribution. An example is PSZ2 G125.68-64.12=ACO 119, a merging cluster \citep{2023ApJ...955..103W} with an inflated  pressure profile, as shown from  combined analysis of spatially resolved \planck\ and \xmm\ data  \citep[][Fig. C1]{pip_pressure}. 
\item Nine clusters are validated objects in the PSZ2 catalogue, using NED or {\tt redMaPPer} searches at the candidate location,  or dedicated \planck\ follow-up. They are discussed in Appendix~\ref{app:mass}. We confirm that four of these are genuine clusters. The inconsistency between model and data may be due to the specific characteristics of \hyperlink{PSZ2 G079.88+14.97}{PSZ2 G079.88+14.97} and \hyperlink{PSZ2 G246.49-35.31}{PSZ2 G246.49-35.31},  and to SZE signal contamination for \hyperlink{PSZ2 G078.67+20.06}{PSZ2 G078.67+20.06} and \hyperlink{PSZ2 G294.89-37.19}{PSZ2 G294.89-37.19}. In all four cases, the  difference between the original and new mass values is small, being respectively 0, 2, 10, and 19\%. One object, \hyperlink{PSZ2 G006.84+50.69}{PSZ2 G006.84+50.69} is the detection of the complex system A2023-A2029, as a whole. On the other hand, we concluded that four sources, 
\hyperlink{PSZ2 G087.25-41.86}{PSZ2 G087.25-41.86}, \hyperlink{PSZ2 G146.82+40.97}{PSZ2 G146.82+40.97}, \hyperlink{PSZ2 G153.29+36.56}{PSZ2 G153.29+36.56} and \hyperlink{PSZ2 G250.17+73.53}{PSZ2 G250.17+73.53}, are noise dominated and their status was set to {\tt STATUS=False}.
\end{itemize}

\subsection{Correction for selection effects}
\label{sec:corrmass}

The published \planck\ mass estimates were not corrected for selection effects \citep[see e.g.,][]{bat16}. The $P(\ts,\Ytot)$ posterior probability contours assume equal probability for all observed $\ts$ and $\Ytot$. This is not the case in reality, as clusters are distributed inhomogeneously in the $(\ts,\Ytot)$ plane.
Counting SZ-detected clusters from the underlying halo mass function, one is more likely to detect a lower mass cluster with an up-scattered SZE flux than a higher mass cluster with a down-scattered SZE flux \citep[e.g.][]{aem11}. 
We implemented a correction of the posterior probability contours for this effect by weighting them with the underlying cluster number counts, such that:
\begin{equation}
 P_{\rm corr.}(\ts,\Ytot)= {{P(\ts,\Ytot) \times {dN \over d\ts d\Ytot}} \over {\int d\ts \int \, d\Ytot \, P(\ts,\Ytot) \times {dN \over d\ts d\Ytot}}}
\end{equation}
where $dN / d\ts\, d\Ytot$ are the cluster number counts expressed in the $(\ts,\Ytot)$ plane. These are obtained from the Tinker halo mass function $dN / dz\, d\Mv$ in the $(z,\Mv)$ plane \citep{Tinker2008}. We use the same scaling relations as before ($\Mv$--$\Yv$ from the X--ray prior, and $\Mv$--$\Rv$ relation) to convert the counts from the $(z,\Mv)$ to the $(\ts,\Ytot)$ plane. For this conversion, we take into account the intrinsic scatter in the $\Mv$--$\Yv$ relation and use $\sigma_{\log Y}=0.075$ given in Table~1 of \cite{psz1cosmo}.
Figure~\ref{fig:degecont} illustrates the typical extent to which  the contours are modified when applying the correction. $P_{\rm corr.}(\ts,\Ytot)$ corresponds to the filled coloured contours while $P(\ts,\Ytot)$ is displayed with solid blue contours. To first order, the correction shifts the contours towards lower fluxes and sizes.  The intersection with the X-ray prior (cyan line) corresponds to a lower flux and, therefore a lower mass, than in the uncorrected case.

\begin{figure}[t]
\centering
\includegraphics[width=0.99\columnwidth]{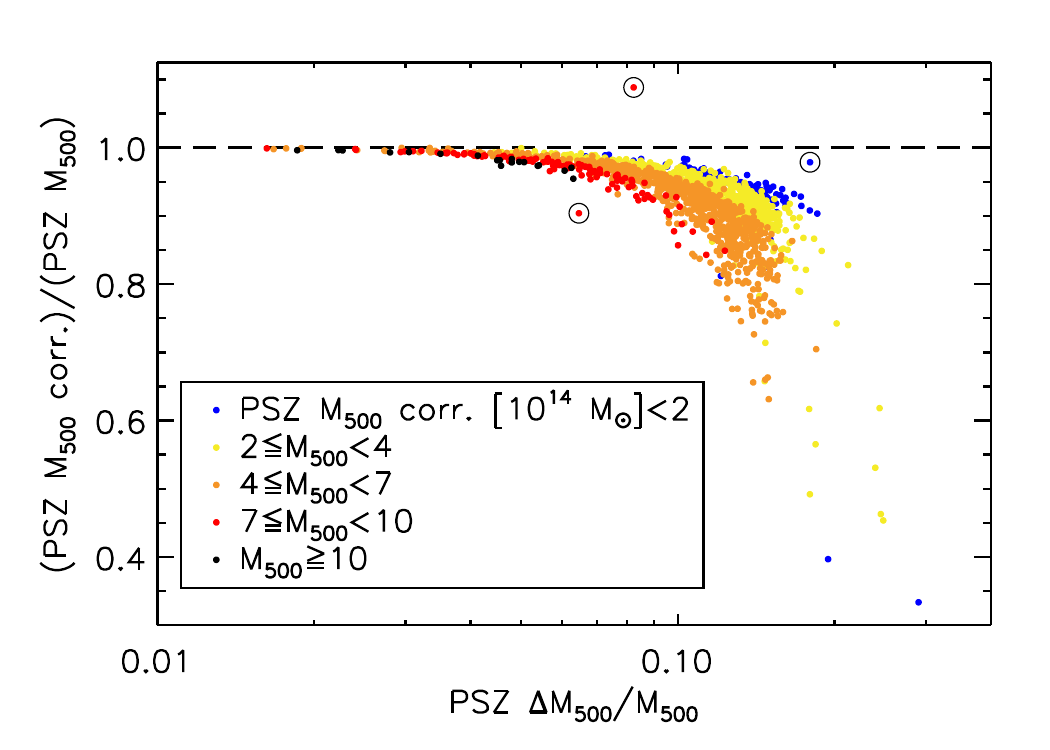}
\caption{\footnotesize Ratio between the corrected and uncorrected PSZ masses as a function of the relative error on the uncorrected mass. The correction increases with the relative error on the uncorrected mass, reaching approximately 20-30\% for a relative error on the uncorrected mass of 10-15\%. The three clear outliers are surrounded by a black circle and discussed in the text.}
\label{fig:corr_vs_snr}
\end{figure}

Figure~\ref{fig:corr_vs_snr} shows the ratio of the corrected to uncorrected masses for the PSZ catalogue as a function of the relative error on the uncorrected mass. The correction increases from 0\% to 20\%-30\% as the relative error on the uncorrected mass increases from a few percent to 10-15\%. At fixed relative error on the uncorrected mass, the correction increases with increasing mass. One can readily spot three outliers (surrounded by a black circle) in the distribution. From left to right in the Figure, these objects are PSZ2 G123.35+25.39, PSZ2 G107.83-45.45, and PSZ2 G269.36-47.20. They were manually inspected. In all three cases, the degeneracy contours from the SZE extraction algorithms are far from the X-ray prior, a configuration discussed in Sect.~\ref{sec:newmass}. 
We found that PSZ2 G123.35+25.39 is actually noise dominated (see App.~\ref{app:mass}) and we set {\tt STATUS=False} for this detection. 
PSZ2 G269.36-47.20 is an unrelaxed PSZ2-MCXC cluster identified in Sect.~\ref{sec:newmass}, and PSZ2 G123.35+25.39 is a newly-validated object affected by a radio source. We added a {\tt COMMENT} in the PSZ catalogue for these two clusters.

We tested the sensitivity of the correction to the assumed value of the intrinsic scatter. Assuming zero intrinsic scatter in the $\Mv$--$\Yv$ relation changes the corrected masses by less than 1\% on average, and thus only has a small impact on the correction. We also tested the sensitivity to the assumed cosmological parameters for the calculation of $dN / dz\, d\Mv$. As a baseline, we adopted the \planck\ cosmological parameters from the primary CMB ($h=0.6736, \Omega_{\rm m}=0.3153, \sigma_8=0.8111, 1-b=0.62$). We computed the correction using the \planck\ cosmological parameters from cluster counts ($h=0.738, \Omega_{\rm m}=0.31, \sigma_8=0.76, 1-b=0.78$), and found that the corrected mass changes by less than 1\% on average with respect to our baseline, showing that our results are unaffected by the assumed cosmology as long as the cluster counts remain the same.

With this work, we release the new degeneracy contours and corrected masses. These are described in Appendix~\ref{app:PSZfield}.

\section{Summary and final quantities} \label{sec:PSZprop}

\begin{figure*}[t]
\centering
\resizebox{0.9\textwidth}{!} {
 \includegraphics[width=0.99\columnwidth]{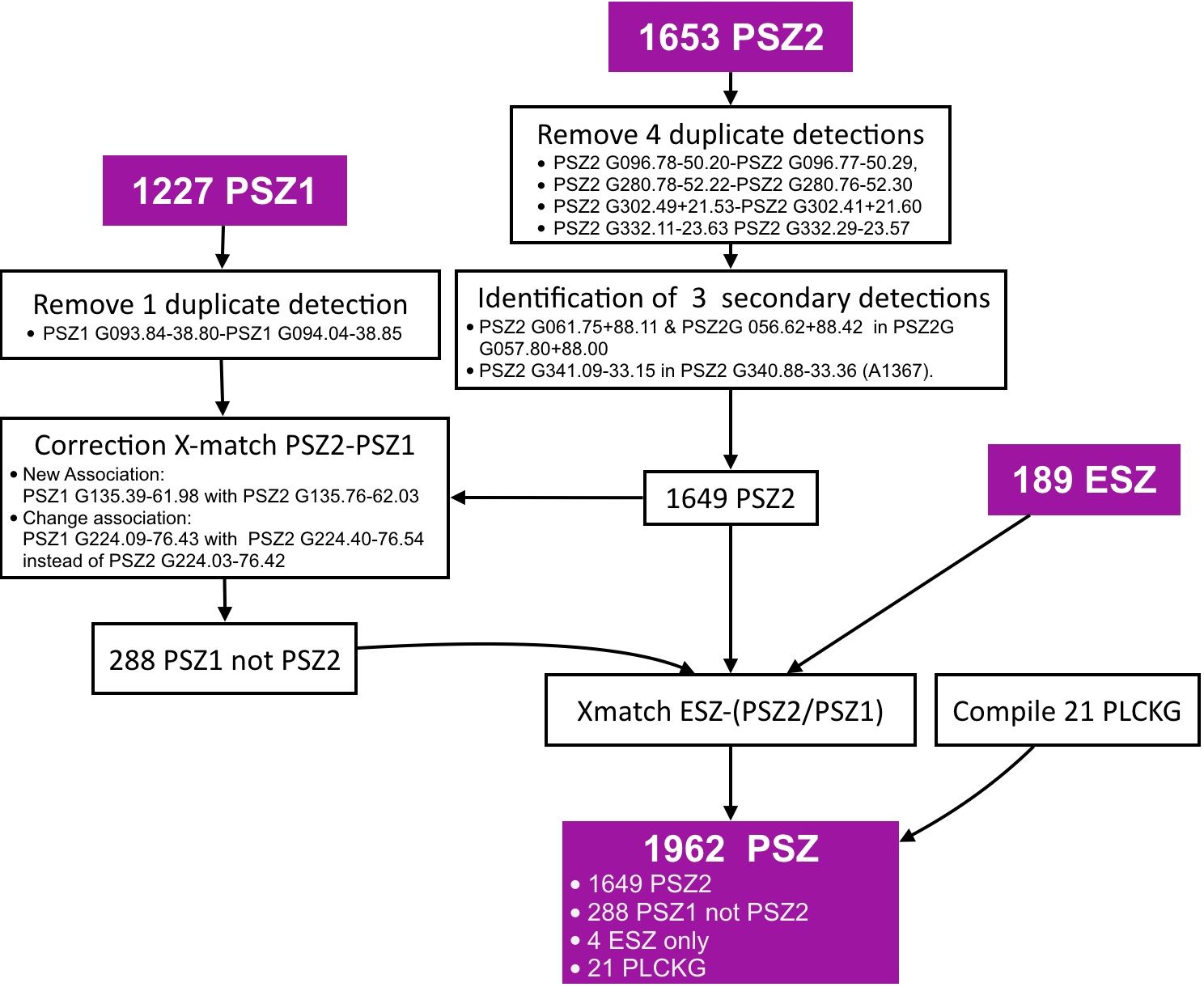}
 \hspace{0.5cm}
\includegraphics[width=0.99\columnwidth]{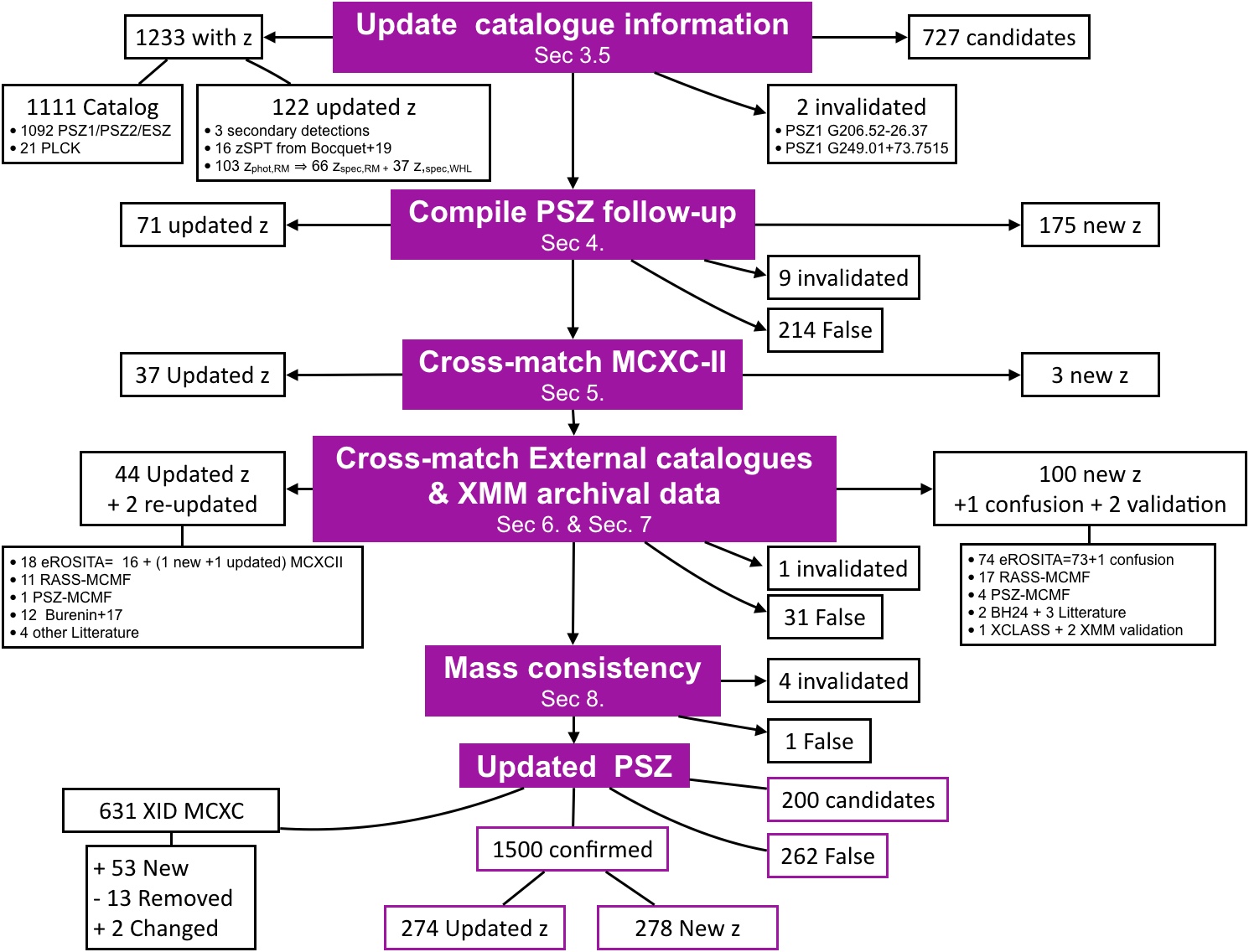}}
\caption{\footnotesize Procedure for the construction of the PSZ. {\it Left:} Construction of the union of \planck\ catalogues. Main modifications of the original catalogues are indicated. 
{\it Right} Summary of the procedure for the update of PSZ candidates status and redshift, with reference to corresponding  sections. The source and number of objects of various status at each step are indicated on the left and right of the tree. The status of 5 objects changed at more than one step. They are only counted once, at the invalidating step for the False candidates and where the $z$ is established for confirmed objects.}
\label{fig:PSZconstruction}
\end{figure*}

\begin{figure}
\centering
\includegraphics[width=0.8\columnwidth, trim={0mm 0mm 0mm 0mm}, clip]{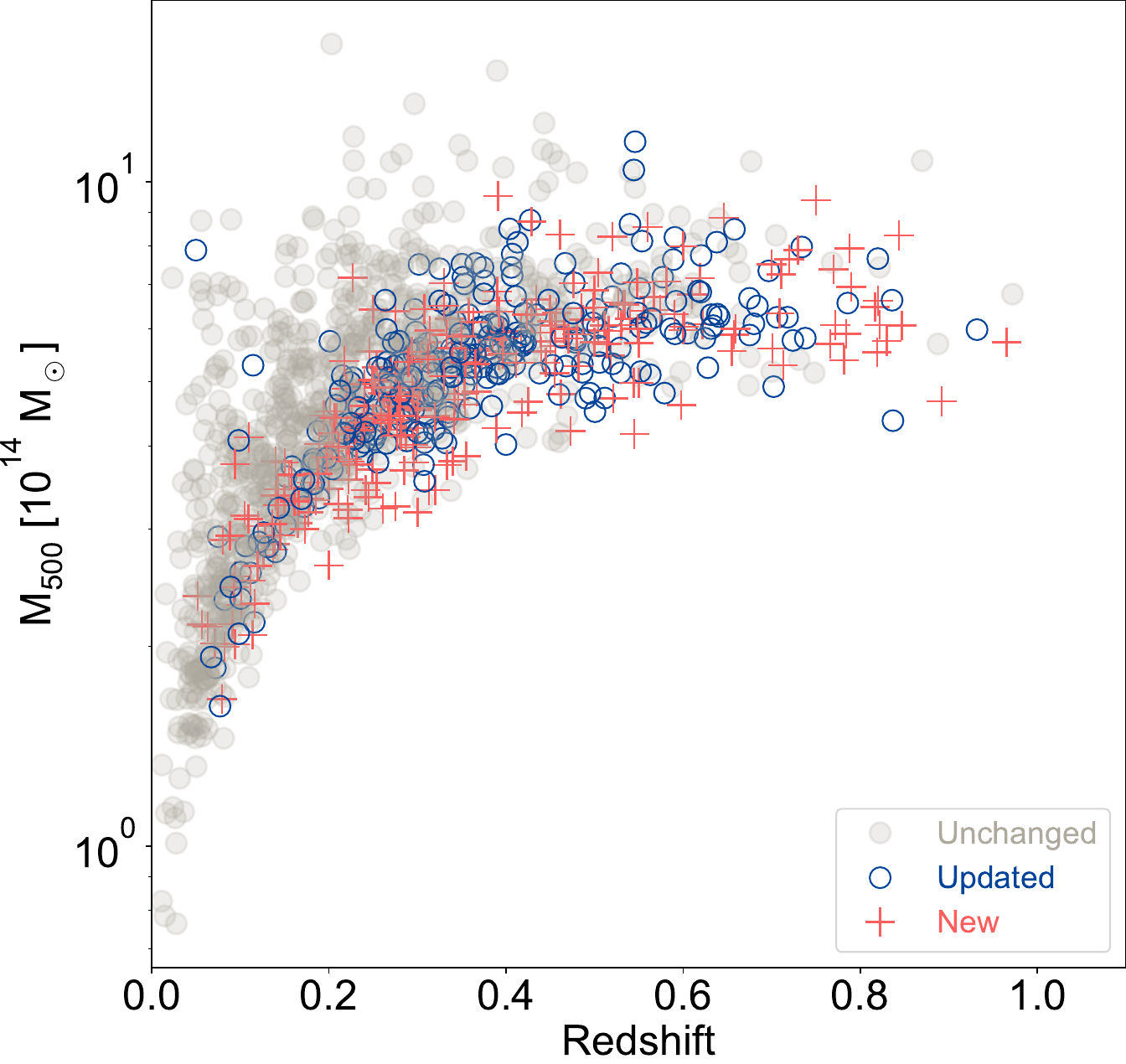}
\caption{\footnotesize Distribution of the confirmed PSZ clusters in the $z$--$\Mv$ plane. Newly confirmed objects and updated redshifts are indicated by red crosses and blue open circles, respectively.}
\label{fig:PSZ_zM}
\end{figure}

\subsection{Catalogue content and properties}
\label{sec:PSZcont}
In summary, we have created a union \planck\ catalogue, PSZ, with updated validation status, redshift,  mass, and MCXC cross-identification. It contains  1962 objects and 33 fields, whose names, units, and descriptions are given in Appendix~\ref{app:PSZfield}. 
The main information on each cluster includes its position, updated redshift, and $\Mv$ corrected for selection effects. We also provide uncorrected mass, integrated Compton parameters and size, the index in the original catalogue(s), and the MCXC-II counterpart, if it exists. We provide additional information beyond the redshift value: the redshift type {\tt Z\_TYPE}, the bibliographic reference {\tt Z\_REF}, given as far as possible in the form of BIBCODE, and a flag, {\tt Z\_FLAG}, that refers to the PSZ revision ({\tt new}, {\tt updated} or from {\tt catalog}). 
The {\tt STATUS} of each object, as described in Sect.~\ref{sec:method}, goes beyond the simple distinction  between confirmed cluster ({\tt STATUS=C1}), noise dominated detection ({\tt STATUS=False}) or candidate ({\tt STATUS=U}). 
More complex situations, {\tt STATUS=Complex, Confusion} or {\tt C2}, as described in Table~\ref{tab:PSZfield}, are  identified as far as possible. The {\tt COMMENT} field summarizes the rationale for invalidating a candidate ({\tt STATUS=False}) or for assigning {\tt STATUS=C2}. This field also provides complementary information for {\tt STATUS=Complex} or  {\tt Confusion} cases. 
The identification of these cases is complete  for manually checked objects mentioned in the construction or update of the PSZ catalogue. For other objects, we used information from follow-up paper tables (see Sect.~\ref{sec:PSZupfu}) or from comments in the input catalogues, so the identification is not exhaustive. In addition to the comments arising from the present work, the {\tt COMMENT} field also includes relevant notes from the PSZ1 and  PSZ2 publications, identified with prefix [PSZ2] and [PSZ1], respectively. For PSZ1, we selected  the physical information on clusters in the {\tt COM\_PCCS\_SZ-union\_comments\_R1.11} document and for PSZ2, we extracted this information from  the catalogue {\tt COMMENT} field. 

\begin{figure}[t]
\centering
\includegraphics[width=1.015\columnwidth]{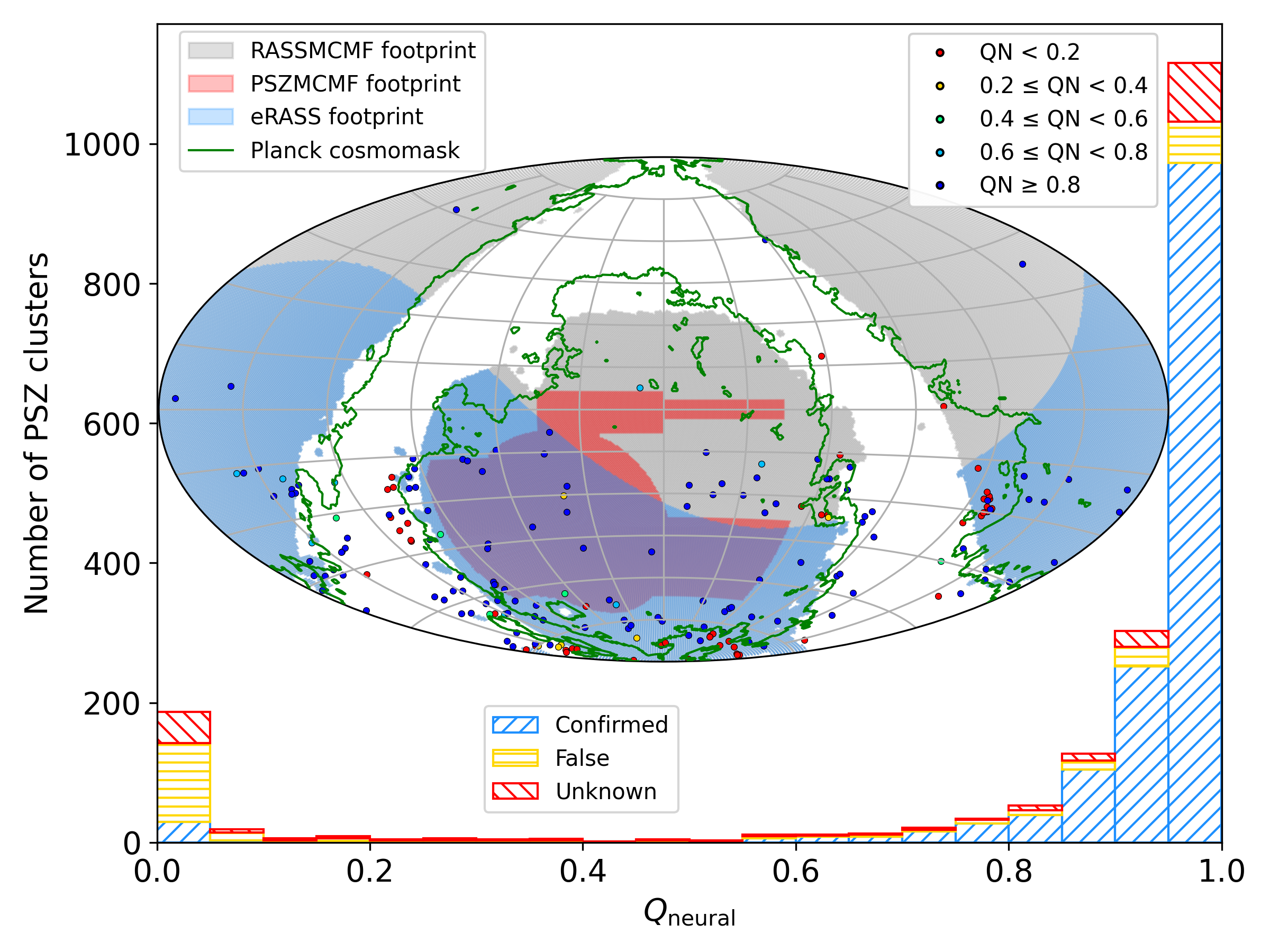}
\captionof{figure}{\footnotesize \label{fig:histQNskymap} Histogram of $Q_{\rm neural}$ for confirmed, false, and unknown PSZ objects with available $Q_{\rm neural}$. $Q_{\rm neural}$ is not available for 25 clusters: 21 PLCKG and 4 ESZ-only. Inset: sky distribution of the PSZ candidates that remain unvalidated colour-coded by their $Q_{\rm neural}$. The grey/blue/red shaded regions represent the RASS-MCMF/eRASS/DES footprints, respectively. The green line represents the \planck\ cosmological mask. The coordinate labels (same as in Fig. \ref{fig:skymap}) were omitted for clarity. }
\end{figure}

Of the 1962 PSZ objects, there are 1500 confirmed clusters, 281 clusters being newly confirmed (278 with redshift), 274 redshift are updated, and 262 candidates are now invalidated ({\tt STATUS = False}). Of the latter, 16 were assigned  a redshift in the original \planck\ catalogues and subsequently invalidated.  
This includes the 2 clusters invalidated in Sect.~\ref{sec:PSZ_physprop}, 9 sources invalidated with the optical follow-up, 1 with \xmm\ archive, and 4 invalidated in the mass analysis (Sect.~\ref{sec:newmass}). Those are objects with low $Q_{\rm neural}$ or non-consolidated SDSS or Pan-STARRS cross-match. The PSZ construction flowchart and differences with the published catalogue content are summarised in the left panel of Fig.\ref{fig:PSZconstruction}, while the right panel summarizes the procedure for status and redshift update.  Figure~\ref{fig:PSZ_zM} shows the distribution of the PSZ clusters in the $z$--$\Mv$ plane. New confirmations extend the catalogues to lower mass and higher redshift. 

Figure \ref{fig:histQNskymap} shows the histogram of $Q_{\rm neural}$ for PSZ1+PSZ2 objects (the $Q_{\rm neural}$    value is not available for 4 ESZ and 21 PLCKG objects).
This extends the histogram presented in \citet{Agu19} and \citet{str19}, which included only the PSZ2-North sample. The histogram is bimodal with two well separated peaks. Most confirmed clusters have good quality; nevertheless, 39 of the 1475 PSZ1+PSZ2
confirmed clusters have $Q_{\rm neural}<0.4$. In contrast, while
most of the 262 invalidated candidates have poor quality, 87  have $Q_{\rm neural}>0.9$ and 124 $Q_{\rm neural}\!>\!0.4$. 
The probability that a confirmed cluster has $Q_{\rm neural}\!>\!0.4$ is thus $P(Q\!> \!0.4\vert \texttt{TRUE})=0.97$, and the probability that a false detection has $Q_{\rm neural}\!<\!0.4$ is $P(Q\!<\!0.4\vert \texttt{FALSE})=0.53$. 
Therefore, the $Q_{\rm neural}$
parameter provides a useful discriminator\footnote{Its performance is quite insensitive to the choice of the cut in between the two peaks, as they are well separated.} between bona fide clusters and false detections.
These probabilities are intrinsic properties of the application of the $Q_{\rm neural}$ to \planck\ detections and do not depend on the false fraction in the considered sub-sample.
However the probabilities $P(\texttt{FALSE}|Q\!<\!0.4)$$\ {\rm and}\ $P(\texttt{TRUE}|$Q\!>\!0.4$) do depend on the sub-sample under consideration. 
Assuming a fraction of false candidates of $\sim\!19\%$
(Sect.~\ref{sec:val}), and applying Bayes' theorem, we obtain  a probability that a
candidate with $Q_{\rm neural}\!<\!0.4$ is false of $P(\texttt{FALSE}|Q\!<\!0.4) = 0.82$.
Conversely, the probability that a candidate with $Q_{\rm neural}\!>\!0.4$ is a
bona-fide cluster is $P(\texttt{TRUE}|$Q > 0.4$) = 0.9$. 

Finally we note that the 39 confirmed candidates with low $Q_{\rm neural}<0.4$ are optically validated, the majority ($80\%$) with {\tt STATUS=C2}.  On the other hand, no cluster confirmed from X-ray (MCXC, eRASS, RASS-MCMF validation) or SZE data has a low $Q_{\rm neural}$ value. This reinforces the robustness of the X-ray/SZE validation, as expected from the tight relation between SZE and X-ray signals, rendering the mass proxy consolidation unambiguous.

\subsection{Validation status}
\label{sec:val}

\begin{figure}[t]
\centering
\includegraphics[width=\columnwidth,trim=1.7cm 0.2cm 2.5cm 0.5cm, clip]{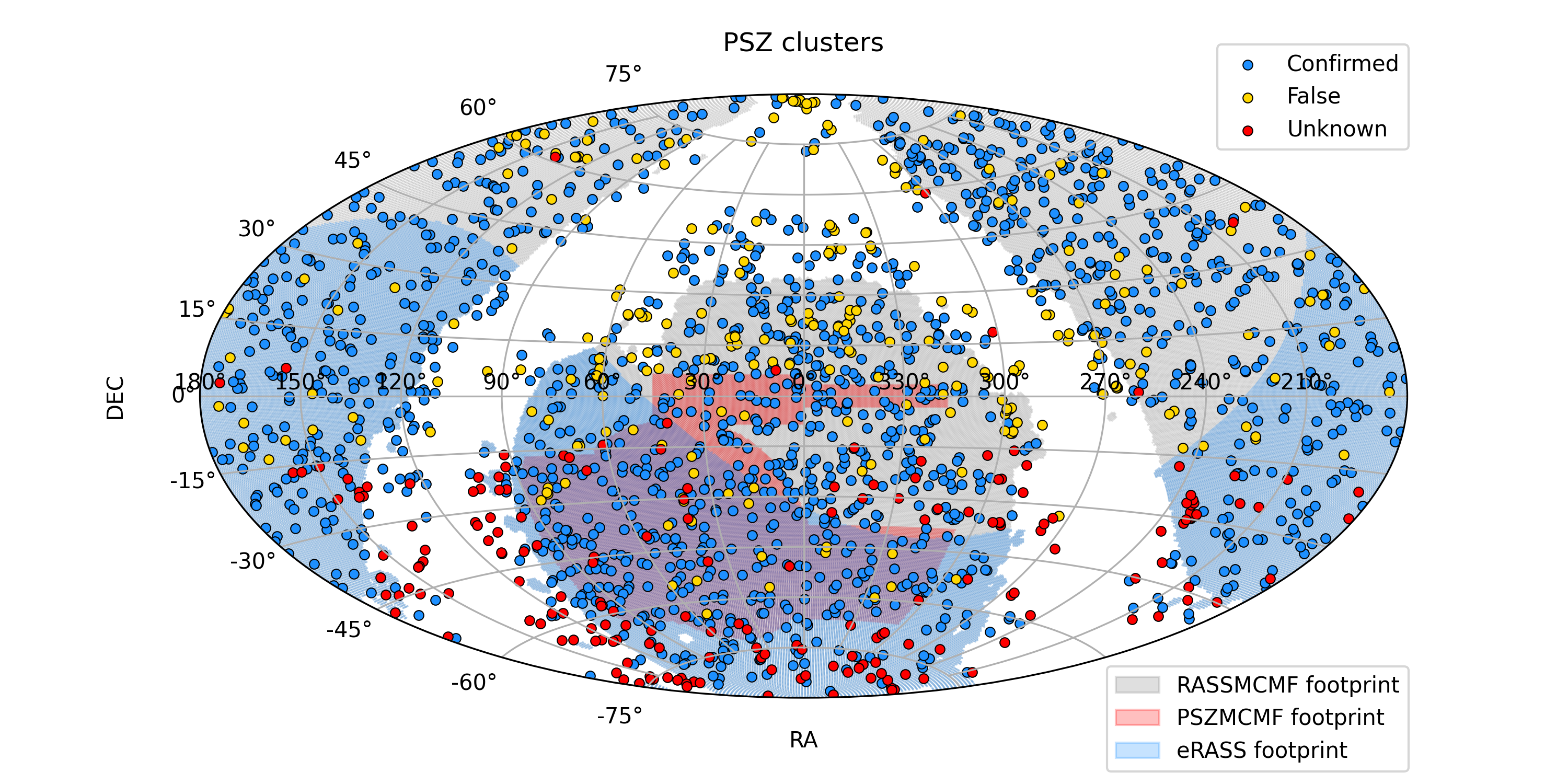}
\caption{\footnotesize \label{fig:skymap} Sky distribution of the PSZ objects colour-coded by their final status. The grey/blue/red shaded regions represent the RASS-MCMF/eRASS/DES footprints, respectively. }
\end{figure}

Figure \ref{fig:skymap} shows the sky distribution of the PSZ catalogue, color-coded by the object {\tt STATUS}. 
In the northern sky (DEC$>-15^{\circ}$), only ten candidates remain unvalidated; the rest have been either validated (960) or invalidated (232) by the intensive follow-up effort. This corresponds to at least 79.9\% purity and 19.3\% of false detections in this part of the sky. In the southern sky (DEC$<-15^{\circ}$), there are 540 confirmed clusters and 30 false detections, but 190 candidates still remain to be validated. There are 63 objects with $Q_{\rm neural}\!<\!0.4$ among the 200 unvalidated objects, a larger fraction, $F\!=\!0.315$, than in the general population. We thus expect\footnote{p= (F-(1-P1))/(P2 + P1-1) with $P1\!=\!P(Q\!>\!0.4\vert\texttt{TRUE})$ and $P2\!=\!P(Q\!<\!0.4\vert\texttt{FALSE})$} a larger proportion of {\tt FALSE} objects, namely 57\% from the probabilities given in Sect.~\ref{sec:PSZcont}, or equivalently, confirm less than  $\sim\!86$ new clusters. This is not surprising. In the south, the current validation essentially relies on X-ray data (eRASS, RASS or MCXC), i.e. the confirmation of bona-fide clusters, while the identification of false objects is restricted to the DES region (Sect.~\ref{sec:erass+pszmcmf}) and in the small \xmm\ footprint (Sect.~\ref{sec:erass+pszmcmf}). We thus have yet a more complete confirmation of true clusters than the identification of noise dominated objects. 

If we consider the region used for the \planck\ cosmological analysis, which masks the regions close to the galactic plane (see Fig. \ref{fig:histQNskymap}), there are 87 candidates that remain unvalidated in the unmasked cosmological region, 83 of them with good $Q_{\rm neural}\!>\!0.4$. These are the most interesting objects to follow-up, since they are probably true clusters and would complete the validation of the catalogue in this region. Most of them lie within the eRASS or RASS-MCMF footprints and could therefore be assessed using existing data, as discussed next.

Considering the subset of 200 unvalidated candidates (see Fig. \ref{fig:histQNskymap}), their distribution is as follows. 103 lie within the eRASS footprint and could be validated or invalidated through a targeted eRASS search at the SZE position. An additional 23 candidates lie within the RASS–MCMF footprint and could be assessed using the IKI eRASS data. Together, these 103 + 23 candidates could also be investigated with the MCMF framework, searching for optical counterparts in the DESI Legacy Imaging Surveys \citep{DESI2019}, as in \citet{ACTMCMF2024}. Finally, the remaining 74 candidates lie outside these footprints and would require alternative validation, for example through optical follow-up observations. The implementation of these validation strategies is beyond the scope of this paper.

\section{Conclusion} \label{sec:conclusion}
Comprising both confirmed objects and cluster candidates, and subject to ongoing follow-up work, the \planck\ catalogues represent a subject of continuing interest in the community. In the present paper, we have presented the construction and properties of the PSZ meta-catalogue, derived from the ESZ, PSZ1,
and PSZ2 \planck\ sub-catalogues. 

After a thorough examination of the detections within each individual \planck\ sub-catalogue and a merging of the resulting lists (Sect.~\ref{sec:PSZconstruction}), we undertook a  rigorous 
validation and redshift update procedure (see Fig.\ref{fig:PSZconstruction}, Sects.~\ref{sec:PSZupfu}, \ref{sec:mcxc}, \ref{sec:PSZupcat}, \ref{sec:BH24}). 
Internal consistency, both between various follow-up efforts, and in the cross-matching with external catalogues has been systematically checked, ensuring a particularly robust update.  
As a result of this process, significant numbers of candidates have been newly-confirmed; conversely, equally significant numbers of candidates have been invalidated, 
and the number of candidates for which the validation status is not known at the present time has been rigorously quantified.
This is summarised for all PSZ detections in the field {\tt STATUS}. In particular, the {\tt STATUS=False} flag, corresponding to noise dominated detections, constitutes new information which is important for the exploitation of the  PSZ catalogue.

A new, homogeneously-derived, mass estimate 
was added to the PSZ catalogue (Sect.~\ref{sec:mass}). This estimate includes a correction for selection effects due to intrinsic scatter and the properties of the underlying mass function, but does not correct for any hydrostatic bias. The selection effect correction is  essential for future cross-comparison with other survey-selected cluster samples. 

In the course of our work and building on cross-match methods introduced in the original \planck\ papers, we demonstrate that cross-identification between catalogues using simple fixed-distance matching is insufficient.
We confirm the importance of additional consistency checks for validation, object matching, identification of complex objects, and erroneous data. 
Comparison between mass proxies, redshifts, and distance versus angular physical  size are critical tests, and should all be taken into account, as explained in detail in Sect.~\ref{sec:matching}.
Where inconsistencies are identified, additional ancillary data should be used, if available. A very extensive number of manual checks were undertaken in the course of the construction of the PSZ catalogue, some of which are detailed in Appendix~B of the present paper.

The present work has yielded the PSZ, which is both a homogeneous \planck\ cluster catalogue and a list of detections which await validation. We will extend this work to the full panoply of SZE cluster catalogues in the upcoming Meta-Catalogue of SZ-detected Clusters (MCSZ). At the same time, the ever-growing number and scale of cluster  catalogues and the need for subsequent follow-up is demonstrating the need for new approaches to the problem of cross-correlation and cross-matching between catalogues. In the future, it is likely that machine learning techniques will play an increasingly prominent role in this effort.

\begin{acknowledgements} 
The results reported in this article are based on observations obtained with \planck\ (http://www.esa.int/Planck), an ESA science mission with instruments and contributions directly funded by ESA Member States, NASA, and Canada. This research has made use of data and products from the Planck Legacy Archive (PLA).
This research has made use of observations obtained with the Dark Energy Spectroscopic Instrument (DESI); with \xmm, an ESA science mission with instruments and contributions directly funded by ESA Member States and NASA; and with eROSITA, the soft X-ray instrument aboard SRG, a joint Russian-German science mission supported by the Russian Space Agency (Roskosmos), in the interests of the Russian Academy of Sciences represented by its Space Research Institute (IKI), and the Deutsches Zentrum für Luft- und Raumfahrt (DLR). The SRG spacecraft was built by Lavochkin Association (NPOL) and its subcontractors, and is operated by NPOL with support from the Max Planck Institute for Extraterrestrial Physics (MPE). The development and construction of the eROSITA X-ray instrument was led by MPE, with contributions from the Dr. Karl Remeis Observatory Bamberg \& ECAP (FAU Erlangen-Nuernberg), the University of Hamburg Observatory, the Leibniz Institute for Astrophysics Potsdam (AIP), and the Institute for Astronomy and Astrophysics of the University of Tübingen, with the support of DLR and the Max Planck Society. The Argelander Institute for Astronomy of the University of Bonn and the Ludwig Maximilians Universität Munich also participated in the science preparation for eROSITA. This research has made use of data obtained from the XMM-Newton Science Archive (XSA), provided by the European Space Agency (ESA); from the Chandra Data Archive provided by the Chandra X-ray Center (CXC); from the High Energy Astrophysics Science Archive Research Center (HEASARC), which is a service of the Astrophysics Science Division at NASA/GSFC; from the Aladin sky atlas developed at CDS, Strasbourg Observatory, France; and from the SZ-Cluster Database operated by the Integrated Data and Operation Center (IDOC) at the Institut d'Astrophysique Spatiale (IAS) under contract with CNES and CNRS.
The research leading to these results has received funding from the European Research  Council  under  the  European  Union’s  Seventh  Framework Programme (FP72007-2013) ERC grant agreement no 340519. 
P.T. acknowledges support from the Spanish grant PID2022-138560NB-I00, funded by MCIN/AEI/10.13039/501100011033/FEDER, EU.
\end{acknowledgements}

\bibliographystyle{aa} 
\bibliography{mcsz}

\begin{appendix}
\nolinenumbers
\section{Catalogue content}

\label{app:PSZfield}
\begin{table*}[b]
 \centering
\resizebox{0.95\textwidth}{!}{
 \begin{threeparttable}[h]
  \caption{\footnotesize Summary overview of catalogue fields.   }
\begin{tabular}{llll}
\toprule
\toprule
     \multicolumn{1}{c}{{\bf Field Name}} &
    \multicolumn{1}{l}{{\bf FORMAT }} &  
    \multicolumn{1}{l}{{\bf UNIT}} &
    \multicolumn{1}{l}{{\bf DESCRIPTION}} \\
\midrule 
{\tt INDEX}  			& INT  &                  &	Cluster index  \\
{\tt NAME\_PLANCK}	 	& STRING &                  &	Name in \planck\ catalogues  \\
{\tt GLON}	 		    & DOUBLE  & 	deg 			&	Galactic longitude  \\
{\tt GLAT} 			    & DOUBLE	 & 	deg 			&	Galactic latitude  \\
{\tt RA}	 		    & DOUBLE   & 	deg 			&	Right Ascension (J2000)   \\
{\tt DEC}	 		    & DOUBLE  & 	deg 			&	Declination (J2000)   \\
{\tt Z} 				& FLOAT  &                  & Redshift value \\
{\tt Z\_TYPE} 			& STRING &                  & Redshift type (1) \\
{\tt Z\_FLAG} 			& STRING &                  & Redshift flag (2)   \\
{\tt Z\_REF} 			& STRING &                  & Bibliographical reference for the redshift  \\
{\tt M500} 			    & FLOAT & $10^{14}$ solar mass 	& Mass corresponding to a density contrast of 500   \\
{\tt ERRMM500}     & FLOAT & $10^{14}$ solar mass 	& Lower error on {\tt M500}   \\
{\tt ERRPM500}	    & FLOAT & $10^{14}$ solar mass 	& Upper error on {\tt M500}   \\
{\tt THETA500} 			& FLOAT & arcmin 	& Cluster size corresponding to a density contrast of 500   \\
{\tt YSZ500\_MPC2}      & FLOAT  & $10^{-4}$ Mpc$^2$  &  Integrated Compton parameter within R500 (spherical) \\
{\tt ERRMYSZ500\_MPC2}  & FLOAT  & $10^{-4}$ Mpc$^2$  & Error inf. on {\tt YSZ500\_MPC2} \\
{\tt ERRPYSZ500\_MPC2}  & FLOAT  & $10^{-4}$ Mpc$^2$  & Error sup. on {\tt YSZ500\_MPC2} \\
{\tt YSZ500\_ARCMIN2}   &FLOAT   & $10^{-4}$ arcmin$^2$ & SZE flux corresponding to {\tt YSZ500\_MPC2}    \\
{\tt ERRMYSZ500\_ARCMIN2} &FLOAT & $10^{-4}$ arcmin$^2$ & Error inf. on {\tt YSZ500\_ARCMIN2}  \\
{\tt ERRPYSZ500\_ARCMIN2} &FLOAT & $10^{-4}$ arcmin$^2$ &   Error sup. on {\tt YSZ500\_ARCMIN2}     \\
{\tt M500UNCORR}			    & FLOAT & $10^{14}$ solar mass 	& Mass corresponding to a density contrast of 500, uncorrected for selection effects (3)\\
{\tt ERRMM500UNCORR}    & FLOAT & $10^{14}$ solar mass 	& Lower error on {\tt M500UNCORR}   \\
{\tt ERRPM500UNCORR}	    & FLOAT & $10^{14}$ solar mass 	& Upper error on {\tt M500UNCORR}   \\
{\tt SNR\_PLANCK} &FLOAT &   &   SNR in \planck\ catalogues     \\
{\tt Q\_NEURAL} & FLOAT &   &   $Q_{\rm neural}$ in \planck\ catalogues     \\
{\tt STATUS}        & STRING &                  & Validation status (4) \\
{\tt COMMENT}           & STRING &                  & Individual comment on the cluster\\
{\tt SUB\_CATALOGUE} 	& STRING & 	&	Sub-catalogue name (survey)  \\
{\tt NAME\_MCXC}        & STRING &                  & Name of the MCXC-II counterpart  \\              
{\tt IND\_MCXC}         & LONG   &                  & Index in the MCXC-II catalogue \citep{MCXC2024}\\
{\tt IND\_PSZ2 }        & LONG   &                  & Index in the PSZ2 catalogue \citep{psz2}\\
{\tt IND\_PSZ1}         & LONG   &                  & Index in the PSZ1 catalogue \citep{psz1rev} \\
{\tt IND\_ESZ}          & LONG   &                  & Index in the ESZ catalogue \citep{esz}\\
\midrule 
\bottomrule
\end{tabular}

\begin{tablenotes}
\item[] (1) redshift type: spectroscopic (spec), photometric (phot), a combination of photometric and spectroscopic values (SP), estimated (E) or unknown (empty string).
\item[] (2) Redshift flag:  {\tt Catalog}: original value from catalogue;
 {\tt New}: No redshift was available in the source catalogue; the new  redshift origin is given in {\tt Z\_REF}; {\tt Revised}: redshift in the catalogue has been updated from {\tt Z\_REF}.
 \item[] (3) The corresponding cluster size and SZE flux can be computed as $\theta_{500}^{\rm uncorr} = \theta_{500}^{\rm corr}*(M_{500}^{\rm uncorr}/M_{500}^{\rm corr})^{1/3}$ and $Y_{500}^{\rm uncorr} = Y_{500}^{\rm corr}*(M_{500}^{\rm uncorr}/M_{500}^{\rm corr})^{1.79}$.
\item[] (4) Validation status:
 {\tt False} False candidate; {\tt U} Unconfirmed candidate (status unknown); {\tt C1} Confirmed cluster;
 {\tt C2} There is an optical or X--ray counterpart in the \planck\ error box but the association is uncertain (low mass proxy for SZE mass or not available and/or likely high Malmquist Bias);
{\tt Confusion} More than one cluster contributes to the SZE \planck\ signal with SZE peak not centered on any of the clusters; 
{\tt Complex} The counterpart have several components at the same redshift; 
{\tt Secondary} Secondary \planck\ detection in a massive large cluster.
\end{tablenotes}
\label{tab:PSZfield} 
 \end{threeparttable}
 }
\end{table*}

We provide the meta-catalogue of \planck\ cluster candidates (PSZ) in the format given in Table~\ref{tab:PSZfield}. It is available at the Centre de Donn{\'e}es astronomiques de Strasbourg (CDS) via anonymous ftp to cdsarc.u-strasbg.fr (\url{ftp://130.79.128.5}) or via \url{http://cdsarc.u-strasbg.fr/} {\bf to be updated when available}.

At the CDS, we also provide the uncorrected and corrected $M_{500}(z)$ curves, for the PSZ1 and the PSZ2, and for each of the three detection algorithms: MMF1, MMF3 and PwS. We adopted the same format as used for the official products in the Planck Legacy Archive. The masses are provided as an array with dimensions $100 \times 4 \times N_{\rm det}$ as in the extension 3 of the official individual catalogues. The first dimension (length 100) corresponds to the assumed redshift, which is linearly spaced from $z=0.01$ to 1. The second dimension (length 4) corresponds to redshift, $M_{500}(z)$, and the 68\% lower and upper limits on $M_{500}(z)$. The third dimension (length $N_{\rm det}$) corresponds to the cluster. $N_{\rm det}$ is the number of detections in the considered catalogue. Details of this format are provided in \url{https://wiki.cosmos.esa.int/planck-legacy-archive/index.php?title=Catalogues#SZ_Catalogue}. 

For completeness, we provide the degeneracy contours corrected for selection effects, for PSZ1 and PSZ2, and for each of the three detection algorithms. They are provided at \url{https://zenodo.org/}{\bf to be updated when available}. We also adopted the same format as used for the official products in the Planck Legacy Archive. The contours are provided as an array with dimensions $256 \times 256 \times N_{\rm det}$ as in extension 2 of the official individual catalogues. The first two dimensions ($256 \times 256$) correspond to a grid in ($\theta_s$,$Y_{5R500}$) with limits given in extension 1 of the official catalogues. The third dimension corresponds to the cluster, with $N_{\rm det}$ being the number of detections in the considered catalogue.

\section{Special cases in the PSZ construction}\label{app:psz}

\subsection{Duplicates in PSZ1 and PSZ2 catalogues}\label{app:psz12dup}

\begin{figure*}[t]
\centering
\includegraphics[width=0.99\textwidth]{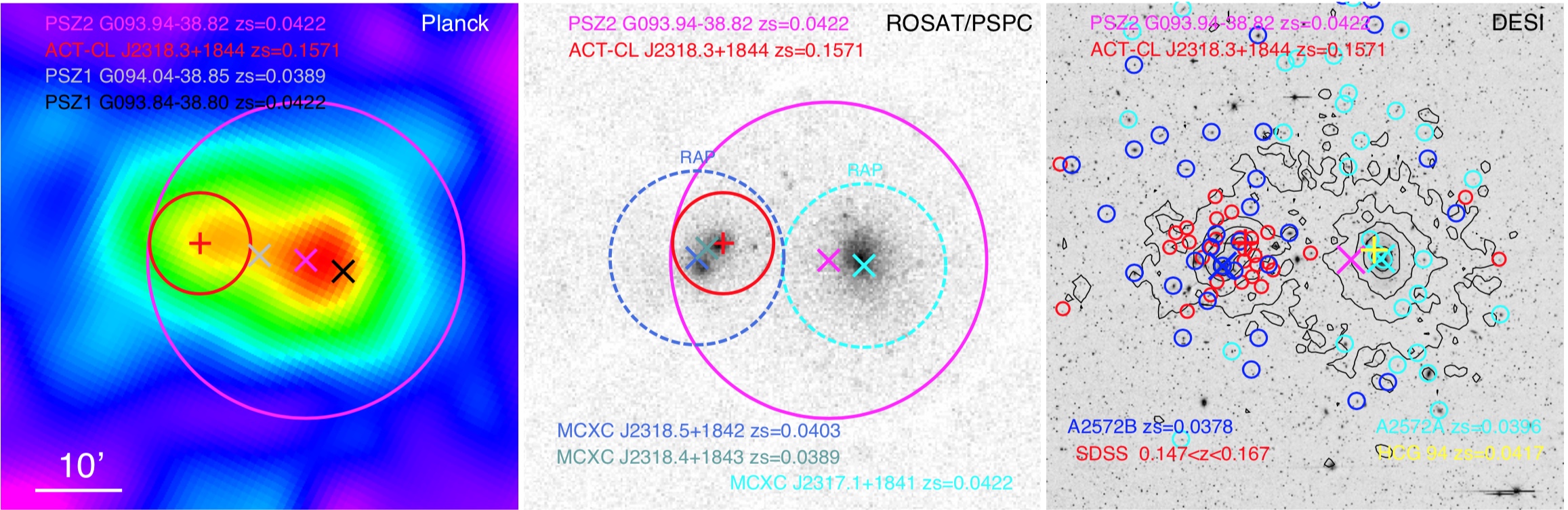} 
\caption{\footnotesize \planck\ filtered map (left), ROSAT pointed 
observation (middle) and DESI image (right) of the A2572 region
including \protect\hyperlink{PSZ1 G094.04-38.85}{PSZ1 G094.04-38.85}, \protect\hyperlink{PSZ1 G093.84-38.80}{PSZ1 G093.84-38.80}, \protect\hyperlink{PSZ2 G093.94-38.82}{PSZ2 G093.94-38.82} and 
ACT-CL J2318.3+1844. 
SZE, MCXC and optical clusters are listed and color-coded with their position  marked with cross or plus.  Plain circles for SZE clusters have a size of $\Tv$ in radius and  dotted circles correspond to the X--ray flux integration region for the MCXC clusters. In the right panel,  blue and cyan circles indicate spectroscopic galaxies of the A2572A and A2572B components published by \citet{smi04} with ROSAT contours. Red circles  indicate galaxies found in NED with redshifts close to the ACT redshift ($\zs=0.157$).}
\label{fig:A2572}
\end{figure*}

\noindent{\bf \hypertarget{PSZ1 G093.84-38.80}{PSZ1\,G093.84-38.80}, and \hypertarget{PSZ1 G094.04-38.85}{PSZ1\,G094.04-38.85}:} This region of the sky is dominated in X--ray by a triple system around A2572
with one component  to the west and  two close-by components to the east, at $d\sim18\arcmin$ (Fig.~\ref{fig:A2572}). The west and southeast components
correspond to A2572A=HCG94 ($z=0.0396$) and A2572B ($z=0.0378$),  respectively \citep{Ebe95, smi04}. The northeast component is a  background cluster, as first suggested by  \citet{Ebe95}, and corresponds to ACT-CL J2318.3+1844 ($\zs=0.157$).  
The PSZ1 catalogue includes two detections in this region, PSZ1 G093.84-38.80 centered on the western component, and PSZ1 G094.04-38.85  on the middle of the complex. They were not merged due to their distance ($9.8\arcmin$) and  associated with A2572A and A2572B, respectively. However their estimated size ($\Tv=18\arcmin$) covers the whole system, and they are detected by different methods (MMF1/PWS and MMF3, respectively)  with consistent SZE signal. Furthermore only one source, \hypertarget{PSZ2 G093.94-38.82}{PSZ2 G093.94-38.82},  appears in the PSZ2 catalogue,  peaking at   A2572A with extension towards the eastern component (Fig.~\ref{fig:A2572} left). We thus consider  PSZ1 G093.84-38.80, and PSZ1 G094.04-38.85 as duplicates, and keep only one PSZ source,  PSZ2 G093.94-38.82,  cross-identified with PSZ1 G093.84-38.80.  PSZ2 G093.94-38.82 corresponds to the complex system as a whole, thus the catalogue association with MCXC J2317.1+1841=A2572A is removed.  \\

\noindent{\bf \hypertarget{PSZ2 G096.78-50.20}{PSZ2\,G096.78-50.20}  and \hypertarget{PSZ2 G096.77-50.29}{PSZ2\,G096.77-50.29}:}
They are detected at S/N=4.82 by PWS, and  at S/N=5.43, respectively. In the PSZ2 catalogue, PSZ2\,G096.77-50.29 is identified with MCXC cluster RXC J2344.9+0911 (A2567) at $z=0.04$. From redMaPPer analysis of SDSS data at the \planck\ position, PSZ2\,G096.78-50.20 was validated as a cluster at $\zp=0.07$. The estimated redshifts of the two sources are consistent, as well as their S/N and their  angular separation, $5.6\arcmin$, is only  one third of the $\theta_{500}$ value estimated for  PSZ2\,G096.77-50.29.  A visual inspection of the \planck\ and \xmm\ images  clearly confirms that PSZ2\,G096.78-50.20   and PSZ2\,G096.77-50.29 are the same cluster, detected independently by MMF3 and PWS. In the PSZ2 catalogue,  the parameters (position, mass, etc.) of clusters detected  by more than one method are taken from the detection with the highest S/N, considered as the prime detection.  We keep this philosophy, keeping PSZ2\,G096.77-50.29  in the PSZ and  removing the second detection, PSZ2\,G096.78-50.20. \\

\begin{figure}
\includegraphics[width=\columnwidth]{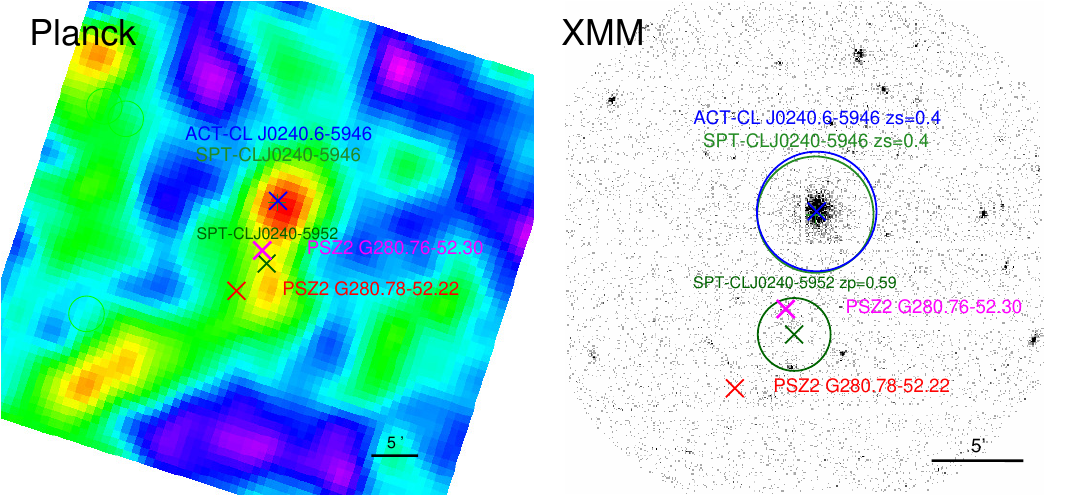}
\caption{\footnotesize \planck\ filtered map (left) and \xmm\ (right) images of the region including \protect\hyperlink{PSZ2 G280.76-52.30}{PSZ2 G280.76-52.30} and \protect\hyperlink{PSZ2 G280.78-52.22}{PSZ2 G280.78-52.22} with overlays  as in Fig.~\ref{fig:A2572}.}
\label{fig:PSZ2G280.76}
\end{figure}

{\noindent\bf  \hypertarget{PSZ2 G280.76-52.30}{PSZ2\,G280.76-52.30} and \hypertarget{PSZ2 G280.78-52.22}{PSZ2 G280.78-52.22}:} They were detected  at S/N=4.52  by MMF1 and  S/N=4.82 by MMF3, respectively. Their  separation of $5.1\arcmin$ is close to the adopted  threshold for merging sources and  their S/N are consistent.
Moreover, the  filtered \planck\ map actually shows a very extended bimodal north-south structure, with a size of ($\sim 15\arcmin$) three times larger than the source separation (see Fig.~\ref{fig:PSZ2G280.76}). 
This indicates that PSZ2\,G280.76-52.30 and PSZ2\,G280.78-52.22 are the same object, and we therefore removed  PSZ2\,G280.78-52.22,  further away from the brightest peak.
PSZ2\,G280.76-52.30  was originally cross-identified with SPT-CLJ0240-5952 ($z=0.59$, $D=1.46\arcmin$) and its redshift taken from this source. However, the brightest PSZ2 peak, $5.4\arcmin$ north of the nominal (centroid) position, is coincident with another SZE cluster, SPT-CLJ0240-5946=ACT CLJ0240-5946,  from catalogues  not available at the time of PSZ2 construction. Twice more massive and at lower redshift  ($z=0.4$) than SPT-CLJ0240-5952, SPT-CLJ0240-5946 is clearly visible in the \xmm\ image and dominates the \planck\ signal. We thus changed the  cross-identification of PSZ2\,G280.76-52.30 from SPT-CLJ0240-5952 to SPT-CLJ0240-5946, and changed its redshift to  $\zs=0.4$,  accordingly.\\

\noindent{\bf  \hypertarget{PSZ2 G302.49+21.53}{PSZ2\,G302.49+21.53} and \hypertarget{PSZ2 G302.41+21.60}{PSZ2\,G302.41+21.60}:}
They are detected  at S/N=8.4 by MMF1 and  S/N=9.5 by MMF3, respectively.  They  were identified with the same X--ray cluster  MCXC J1248.7-4118 at $z=0.0114$ in the PSZ2 catalogue and their angular separation  is $5.9\arcmin$ or  $\sim 0.1 \theta_{500}$.  The \planck\ and \xmm\ images  confirm that this is the same cluster. We kept the highest S/N detection, PSZ2\,G302.41+21.60,   and removed  PSZ2\,G302.49+21.53 from the PSZ catalogue. PSZ2\,G302.41+21.60 is identified with the PSZ1 cluster,  PSZ1 G302.47+21.60 in the PSZ2 catalogue. \\

\noindent{\bf \hypertarget{PSZ2 G332.11-23.63}{PSZ2\,G332.11-23.63} and \hypertarget{PSZ2 G332.29-23.57}{PSZ2\,G332.29-23.57}:}
They are detected  at S/N=4.81 with MMF1 and S/N=5.23 with MMF3, respectively. PSZ2\,G332.29-23.57 was identified as the very large REFLEX cluster  RXC J1847.3-6320 (S0805) at $z=0.0146$,  while  PSZ2\,G332.11-23.63 is a cluster candidate. The distance between the two sources, $10.7\arcmin$, is less than 1/3 of $\theta_{500}$. A visual inspection of the \planck\ and \xmm\ images show that PSZ2\,G332.29-23.57 is well centered on the X--ray peak while PSZ2\,G332.11-23.63 is offset to the south. PSZ2\,G332.29-23.57  is clearly the prime detection and we removed PSZ2\,G332.11-23.63  from  PSZ.

\subsection{Secondary detections in PSZ2 catalogue}\label{app:psz2sec}
\noindent{\bf \hypertarget{PSZ2 G057.80+88.00}{PSZ2\,G057.80+88.00} (Coma), PSZ2\,G061.75+88.11 and PSZ2\,G056.62+88.42:}
Two sources are detected within $\theta_{500}$  of the centre of PSZ2\,G057.80+88.00 (Coma): PSZ2\,G061.75+88.11 with PWS  at  $10.6\arcmin\ (0.2\theta_{500}$) and  PSZ2\,G056.62+88.42 with MMF3 at  $25.3\arcmin\ (0.5\theta_{500}$).  Their photometric redshifts, published in the PSZ2 catalogue,  were  estimated from redMaPPer analysis of SDSS data at the \planck\ position.  Both redshfits, $z=0.044$ and $z=0.045$, are consistent with the Coma redshift.   PSZ2\,G061.75+88.11 and  PSZ2\,G056.62+88.42  are actually sub-structures of Coma. PSZ2\,G056.62+88.42, in the southwest, is at the position of the well-know infalling  NGC4639 group, while PSZ2\,G061.75+88.11, in the northwest, is likely associated with the western shock front discussed by \citet{coma13}. \\

\noindent{\bf \hypertarget{PSZ2 G340.88-33.36}{PSZ2\,G340.88-33.36} (A3667) and \hypertarget{PSZ2\,G341.09-33.15}{PSZ2\,G341.09-33.15}:} Similarly, PSZ2\,G341.09-33.15, detected by  PWS, is a secondary detection  of PSZ2\,G340.88-33.36 (PSZ1 G340.86-33.36 or A3667). Its  center is 16.8\arcmin\ away in the northwest, at $0.8\theta_{500}$  of PSZ2\,G340.88-33.36 center.  It is possibly associated with the dynamics around the northwest relics. 

\begin{figure}
\centering
\includegraphics[width=\columnwidth]{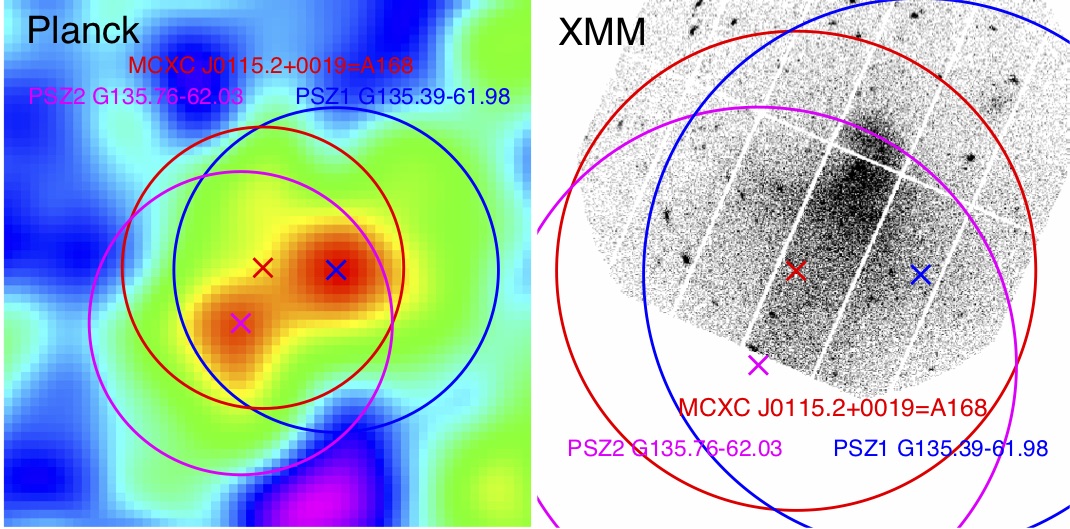}
\caption{\footnotesize \planck\ (left) and \xmm\ (right) images of region around \protect\hyperlink{PSZ1 G135.39-61.98}{PSZ1 G135.39-61.98} and  \protect\hyperlink{PSZ2 G135.76-62.03}{PSZ2 G135.76-62.03}. Those sources correspond to  the same cluster MCXC J0115.2+0019. The clusters are listed on the image and color-coded. The corresponding crosses indicate their center and the big circles represent their size $\Tv$ in radius.}
\label{fig:PSZ2G135.76-62.03}
\end{figure}

\begin{figure}[t]
\centering
\includegraphics[width=0.49\columnwidth]{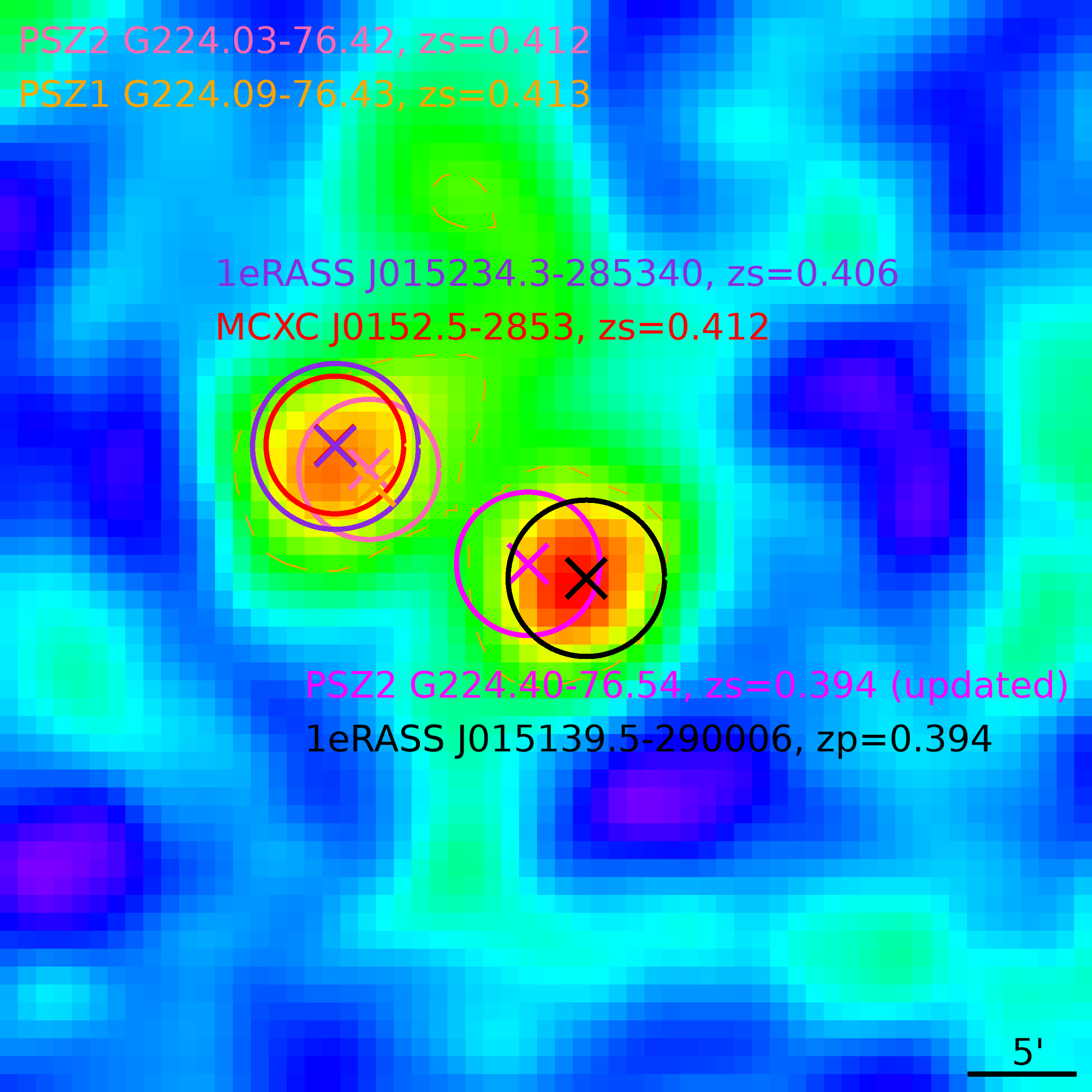}
\includegraphics[width=0.49\columnwidth]{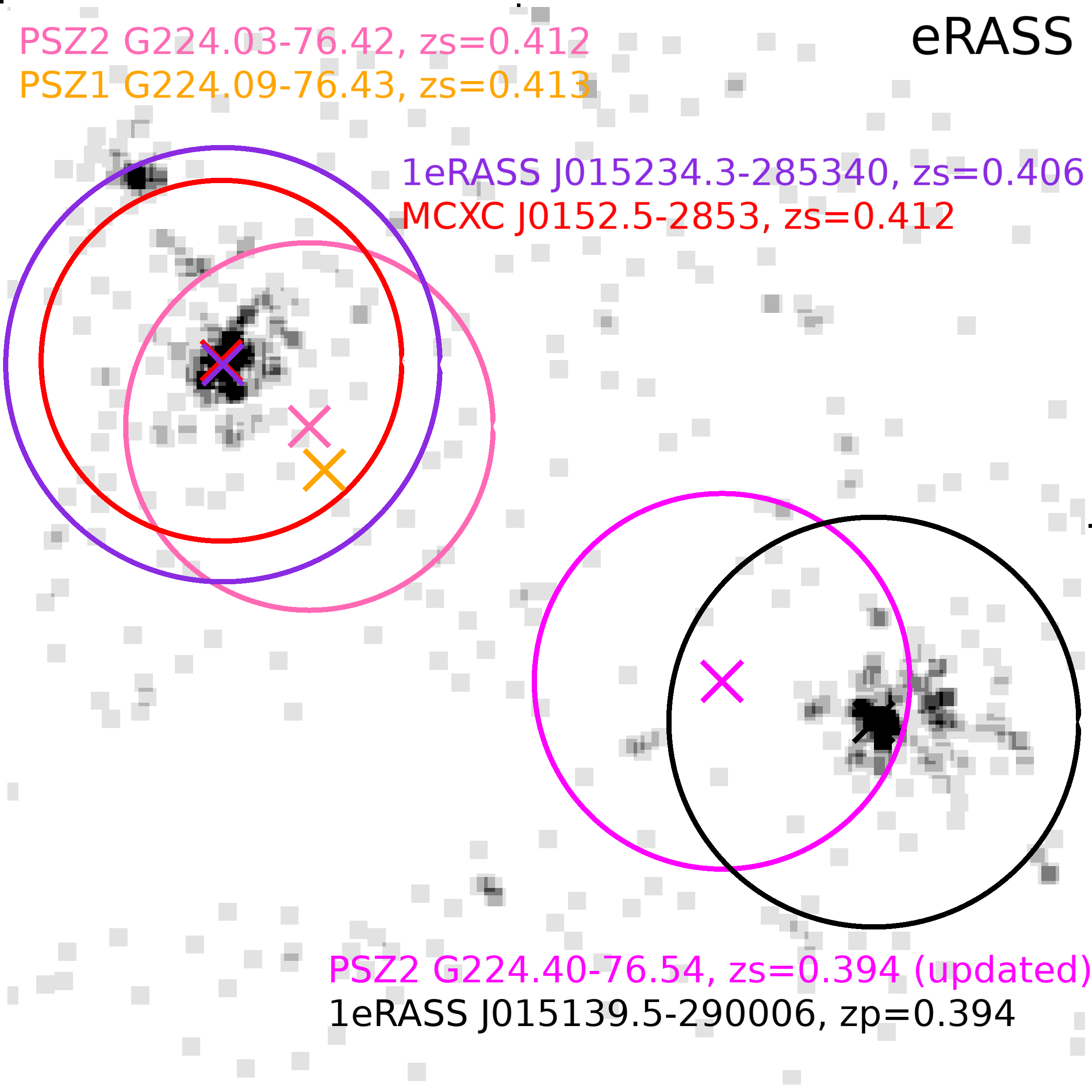}
\caption{\footnotesize  \planck\ (left) and eRASS (right) images of \protect\hyperlink{PSZ2 G224.03-76.42}{PSZ2 G224.03-76.42},  \protect\hyperlink{PSZ1 G224.09-76.43}{PSZ1 G224.09-76.43}, and PSZ2 G224.40-76.54. Overlays as in Fig.~\ref{fig:PSZ2G135.76-62.03}.} 
\label{fig:PSZ2G224.40-76.54}
\end{figure}

 \subsection{Changed PSZ1-PSZ2 association} \label{app:psz1psz2}
\noindent{\bf \hypertarget{PSZ1 G135.39-61.98}{PSZ1 G135.39-61.98} and \hypertarget{PSZ2 G135.76-62.03}{PSZ2 G135.76-62.03}} are identified with the same MCXC cluster, MCXC J0115.2+0019 ($z=0.045$), in the PSZ1 and PSZ2 catalogues, respectively.  They were not cross-matched in the PSZ2 catalogue, as their \planck\ separation of $\theta=10.7\arcmin$ is larger than  $10\arcmin$. However the separation  is less than $0.7\,\Rv$.  Furthermore, MCXC J0115.2+0019 is  a  nearby merger cluster \citep{yan04b,hal04}. The \planck\ image shows a bimodal morphology with the PSZ2\,G135.76-62.03 position centered in the brightest peak in the north and PSZ1~G135.39-61.98 position at the second peak in the south. This roughly corresponds  to the position  of the two merging sub-clusters \citep{yan04a}.  We thus cross-identified the PSZ1 and PSZ2 objects.  Fig~\ref{fig:PSZ2G135.76-62.03} shows the \xmm\ and \planck\ images.\\

\noindent{\bf  \hypertarget{PSZ2 G224.03-76.42}{PSZ2 G224.03-76.42} and \hypertarget{PSZ1 G224.09-76.43}{PSZ1 G224.09-76.43}} are identified with the same MCXC cluster, MCXC J0152.5-2853 ($z=0.413$), in the PSZ1 and PSZ2 catalogues. However, PSZ1 G224.09-76.43 was cross-matched with another PSZ2 cluster, PSZ2\,G224.40-76.54 at a distance  of $8.2\arcmin$ by \citet{psz2}. This cross-match is incorrect.  At a distance of $0.8\arcmin$, PSZ2\,G224.03-76.42 is closer to PSZ1 G224.09-76.43, consistent with their respective  cross-match with MCXC J0152.5-2853.  The position of the two clusters  are shown on the eRASS and \planck\ images on Fig~\ref{fig:PSZ2G224.40-76.54}: PSZ2 G224.03-76.42=PSZ1 G224.09-76.43, associated with MCXC J0152.5-2853 ($z_{\rm spec}=0.41$) in the north, and PSZ2 G224.40-76.54 in the south cross-identified with 1eRASS J015139.5-290006 ($z_{\rm phot}=0.39$).  We thus de-associated PSZ1 G224.09-76.43 and PSZ2\,G224.40-76.54, and associated it with PSZ2\,G224.03-76.42. Furthermore, we changed the catalogue redshift of PSZ2\,G224.40-76.54 ($z_{\rm spec}=0.41$, inherited from PSZ1 G224.09-76.43) to the eRASS  value.

\section{Special cases in the PSZ status and redshift update}\label{app:PSZupz}

\subsection{Cross-match with external catalogues: MCXC-II, eRASS, RASS-MCMF or PSZ-MCMF}
\label{app:eRASS}
\subsubsection{New Validation }
 \begin{figure*}[]
\centering
    \includegraphics[width=0.66\columnwidth]{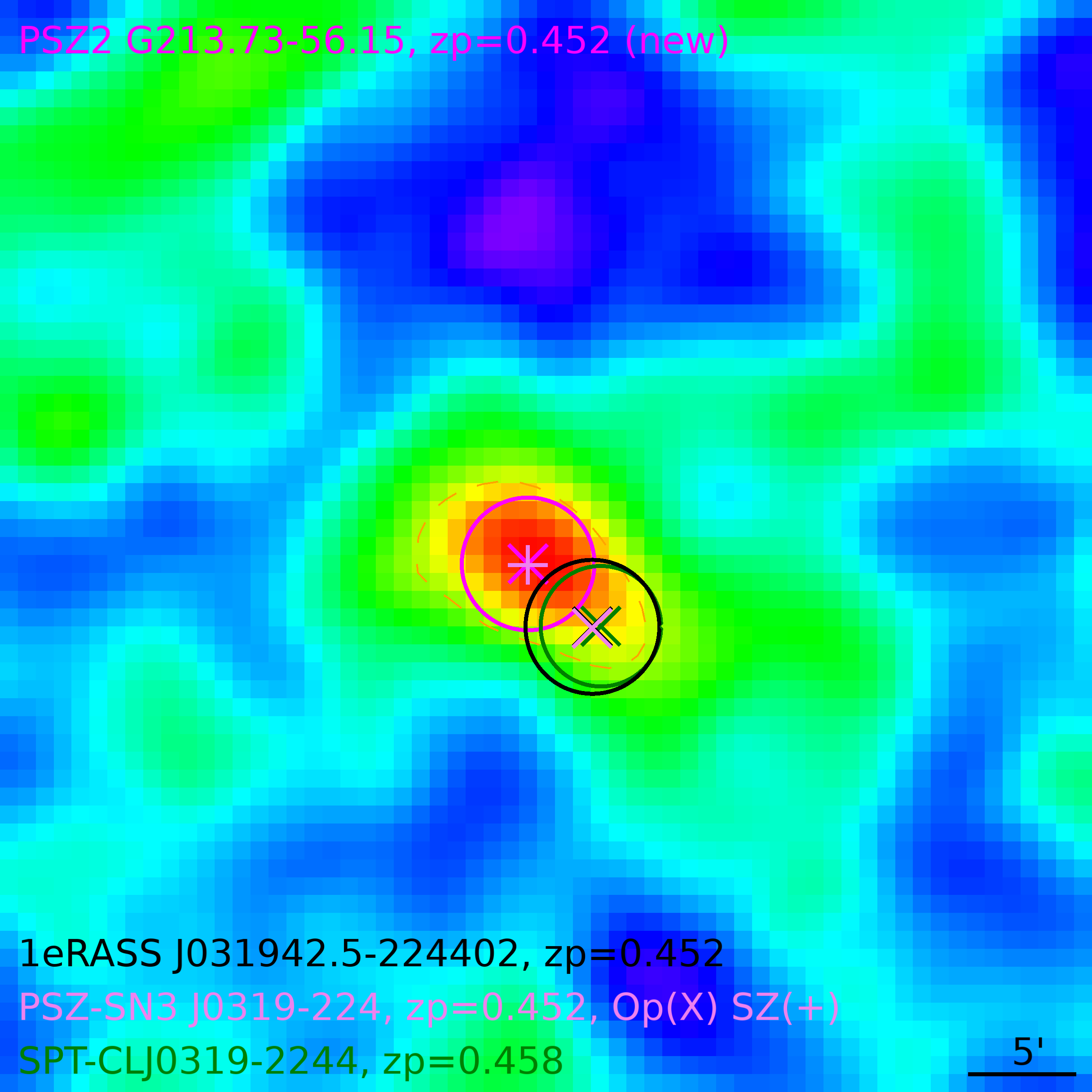}
    \includegraphics[width=0.66\columnwidth]{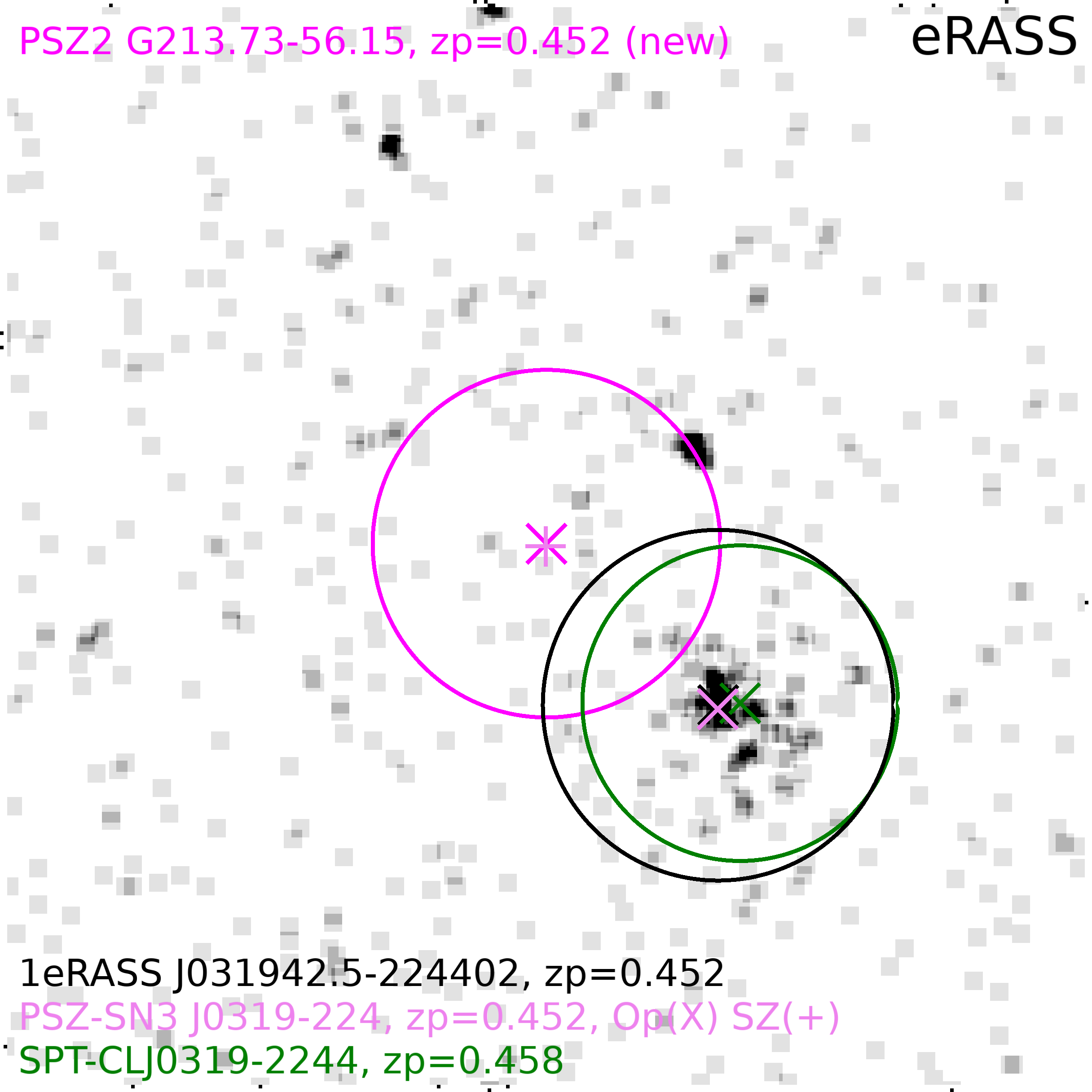}
    \includegraphics[width=0.66\columnwidth]{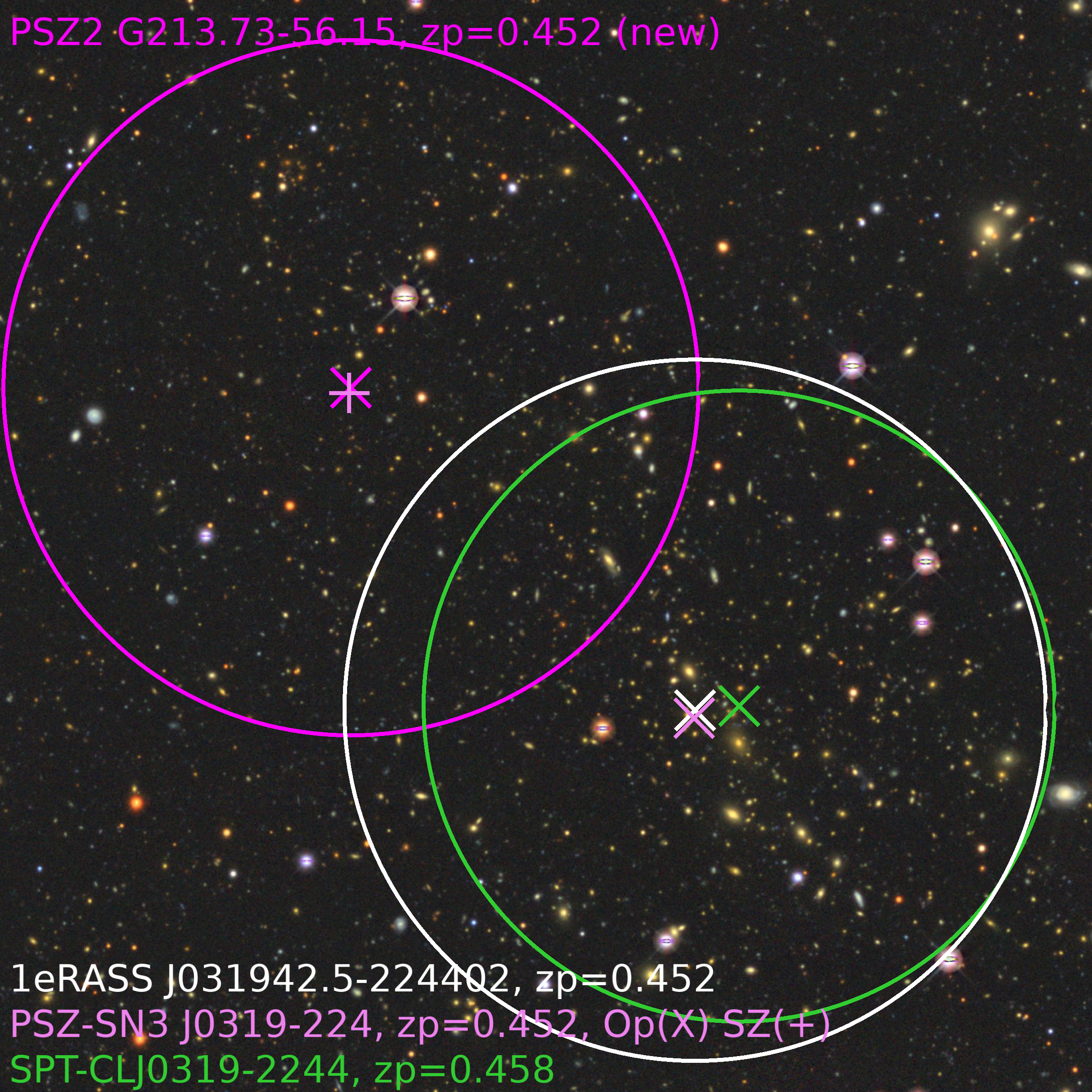}
    \includegraphics[width=0.66\columnwidth]{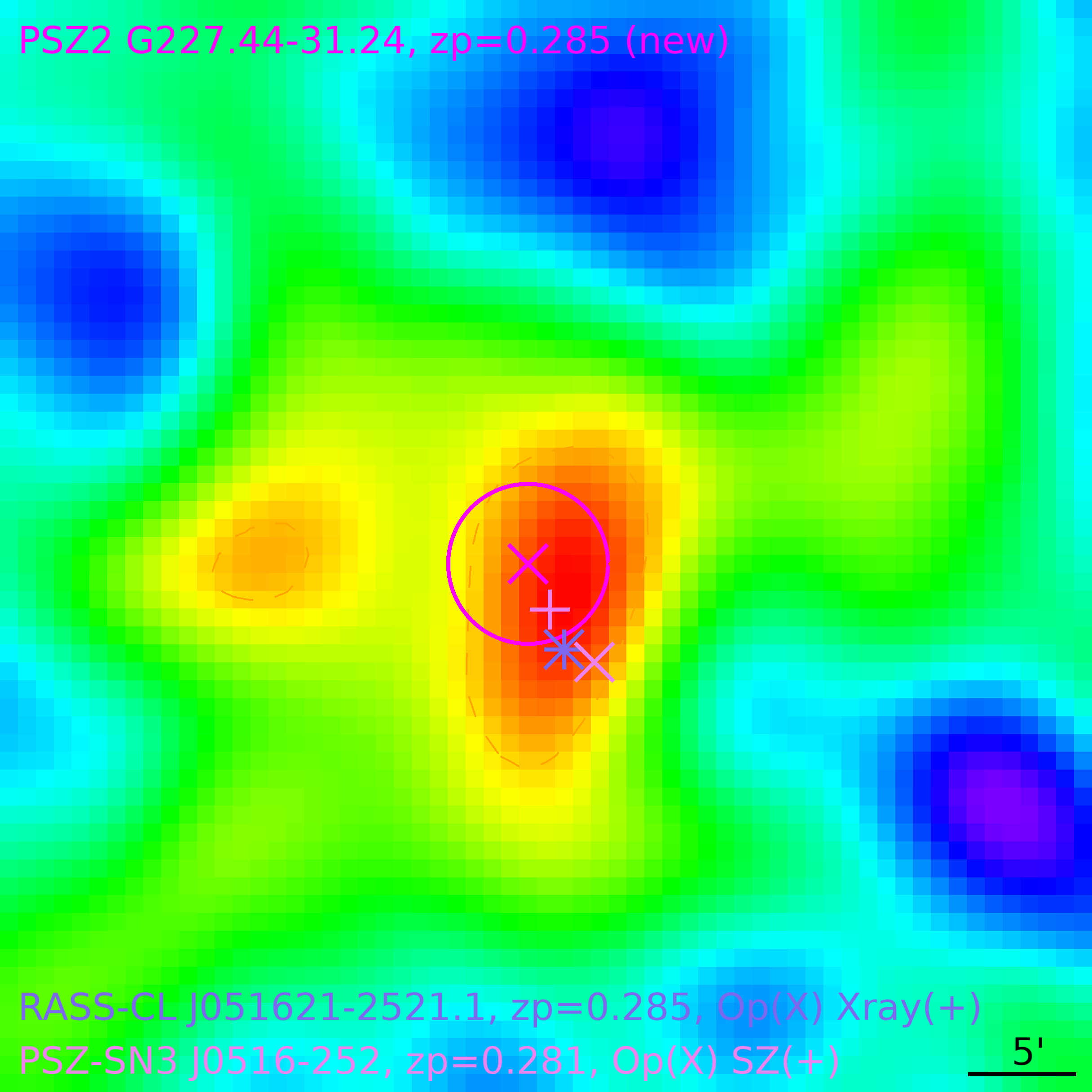}
    \includegraphics[width=0.66\columnwidth]{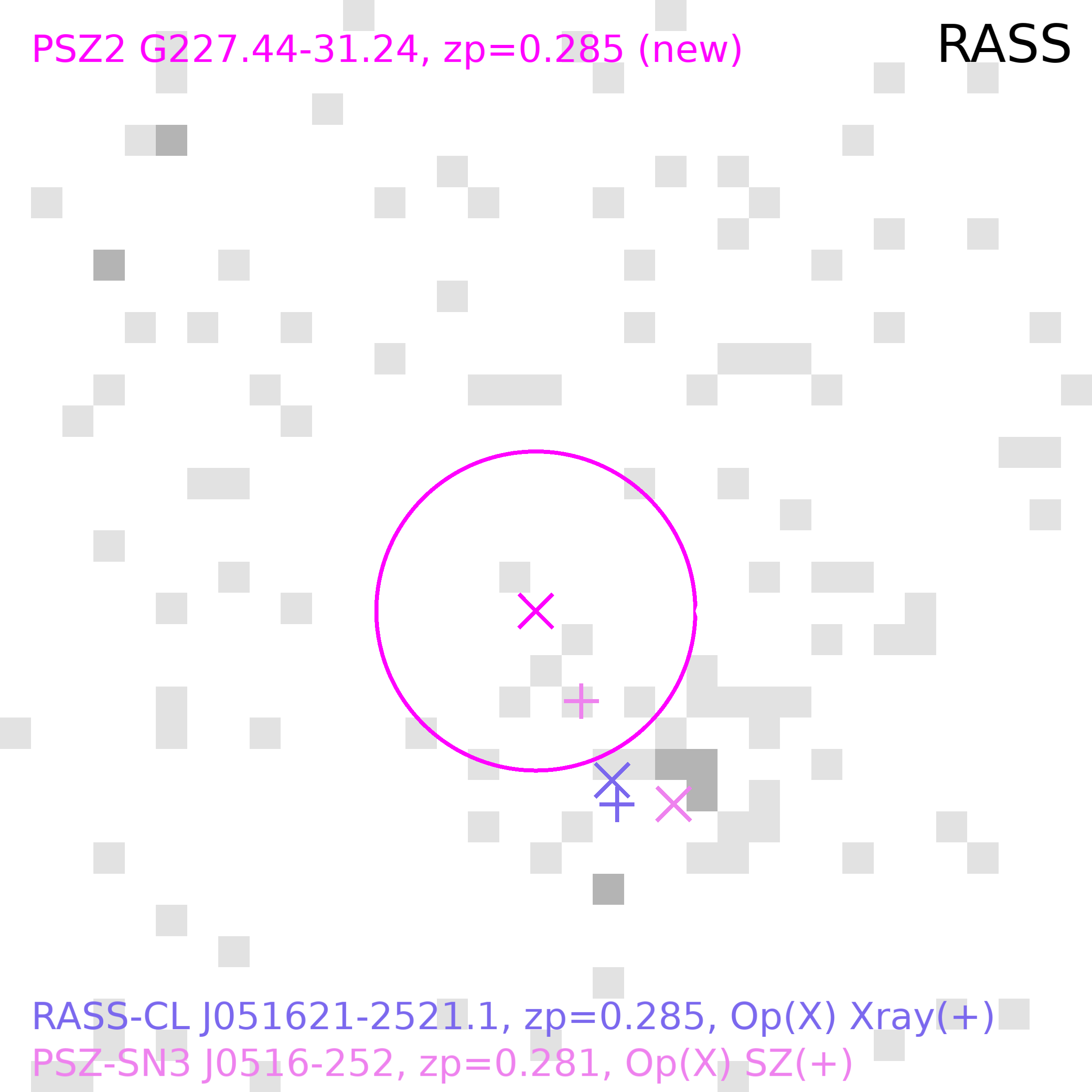}
    \includegraphics[width=0.66\columnwidth]{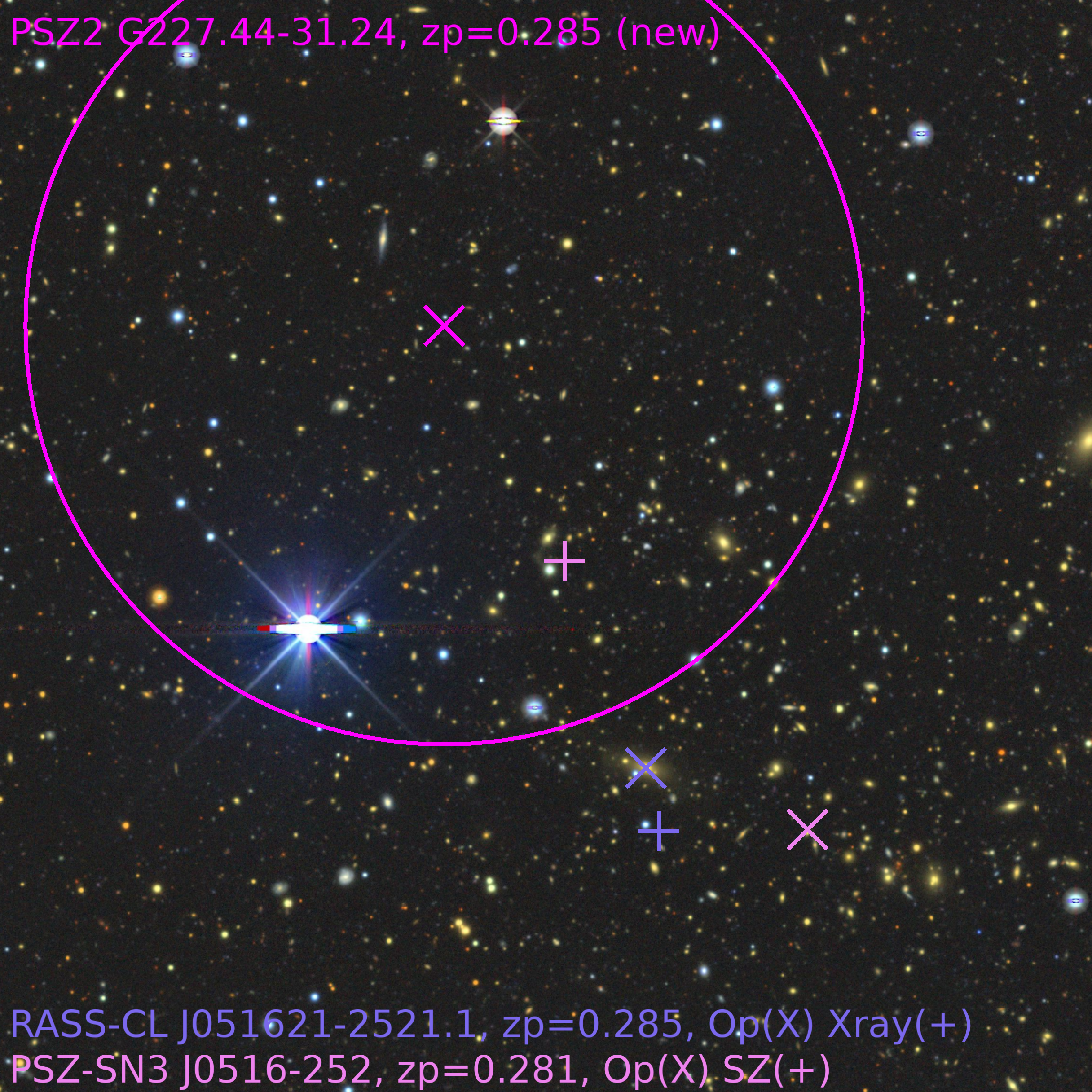}
\caption{\footnotesize \planck\ filtered map (left column), X--ray image (middle column),  and optical DESI image (right column) for \protect\hyperlink{PSZ2 G213.73-56.15}{PSZ2 G213.73-56.15} and \protect\hyperlink{PSZ2 G227.44-31.24}{PSZ2 G227.44-31.24}. The origin of the X--ray image is given in the figure. Clusters from other X--ray/SZE or optical catalogues discussed in the text are colour-coded and listed in each panel with their redshift. The corresponding crosses indicate their centers (optical position for RASS-MCMF and PSZ-MCMF clusters). The X--ray and SZE center of  RASS-MCMF and PSZ-MCMF clusters are indicated by a plus sign. For PSZ, ACT, SPT, and eRASS or MCXC-II clusters, the big circles represent their sizes $\theta_{500}$, as given in the corresponding catalogue. }
\label{fig:B1_3images_a}
\end{figure*}

\begin{figure*}[] 
 \centering
 \includegraphics[width=0.9\textwidth]{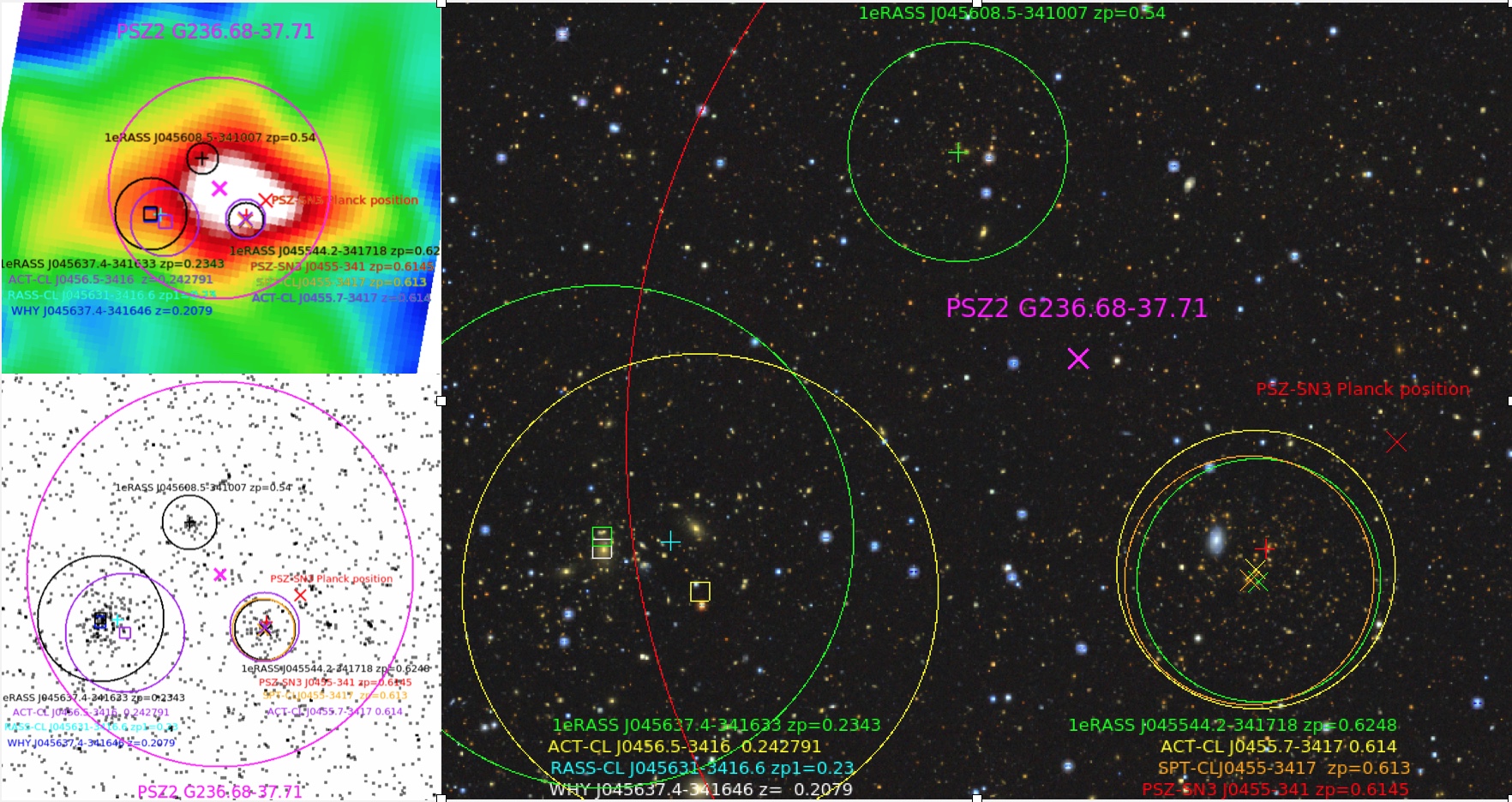}
 \caption{\footnotesize \planck\ MMF3 filtered image of \protect\hyperlink{PSZ2 G236.68-37.71}{PSZ2 G236.68-37.71} (left-top panel) with corresponding eRASS  (bottom-left) and DESI (right panel) images.   
 Clusters from ACT, SPT, eRASS, RASS-MCMF, PSZ-MCMF and SDSS catalogues  are listed in each panel, with their redshift. They are color-coded,  as well as their marked center position. Big circles have  a radius of $\Rv$ computed from corresponding eRASS, ACT or SPT survey data. The large magenta circle is the scale size of the MMF3 detection. The \planck\ source is a complex confusion of 3 clusters.}
 \label{fig:PSZ2G236p68}
\end{figure*}

\begin{figure*}[t]
\centering
    \includegraphics[width=0.66\columnwidth]{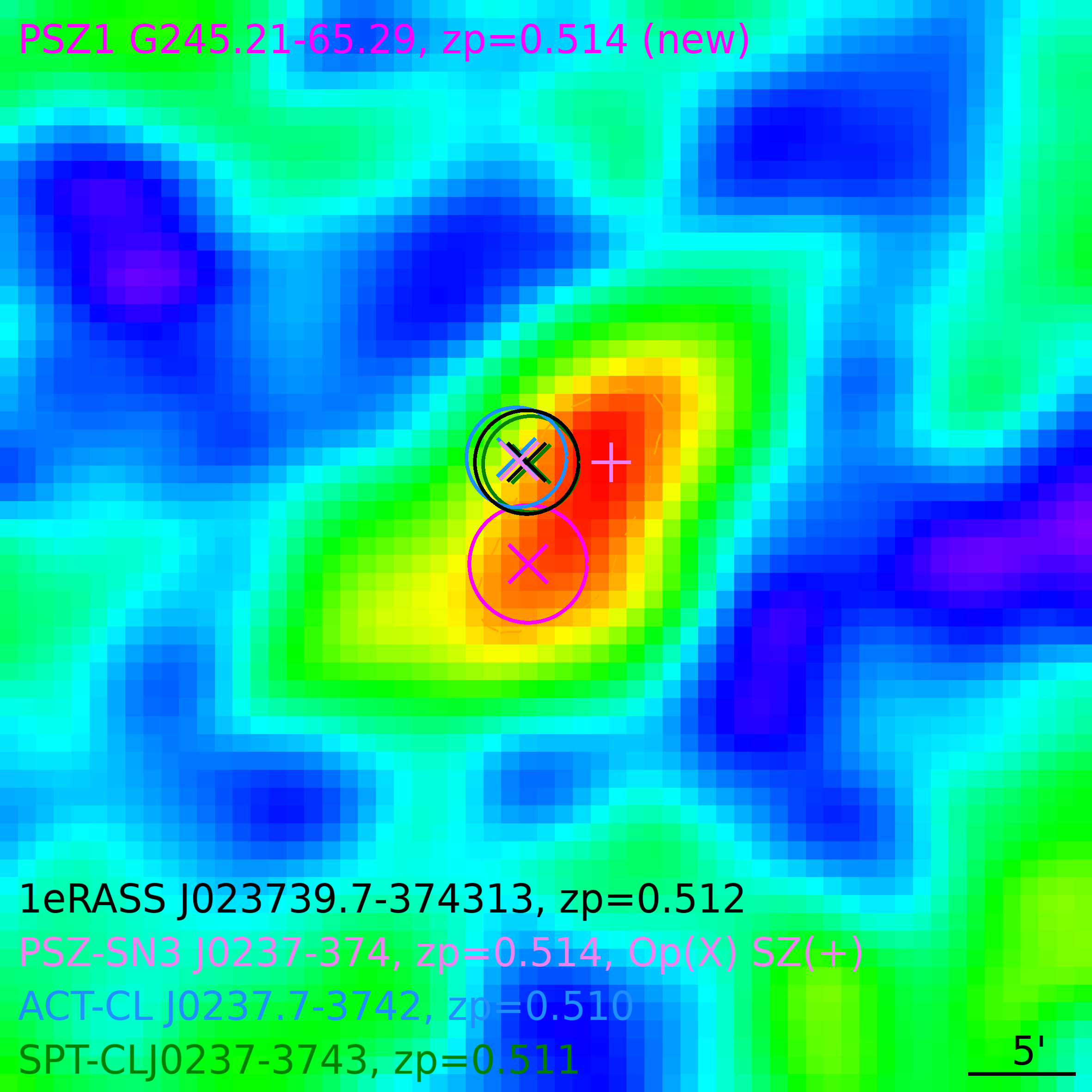}
    \includegraphics[width=0.66\columnwidth]{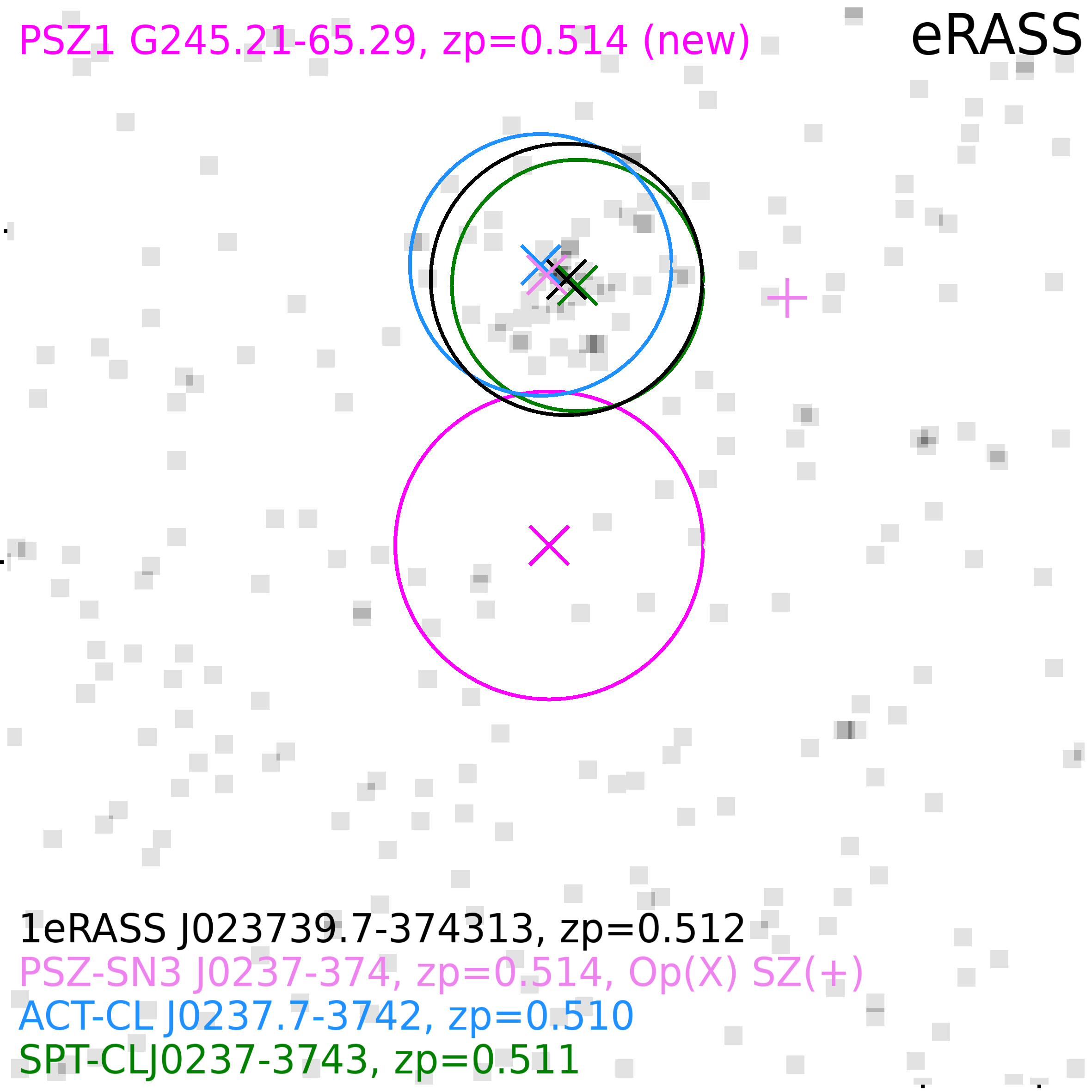}
    \includegraphics[width=0.66\columnwidth]{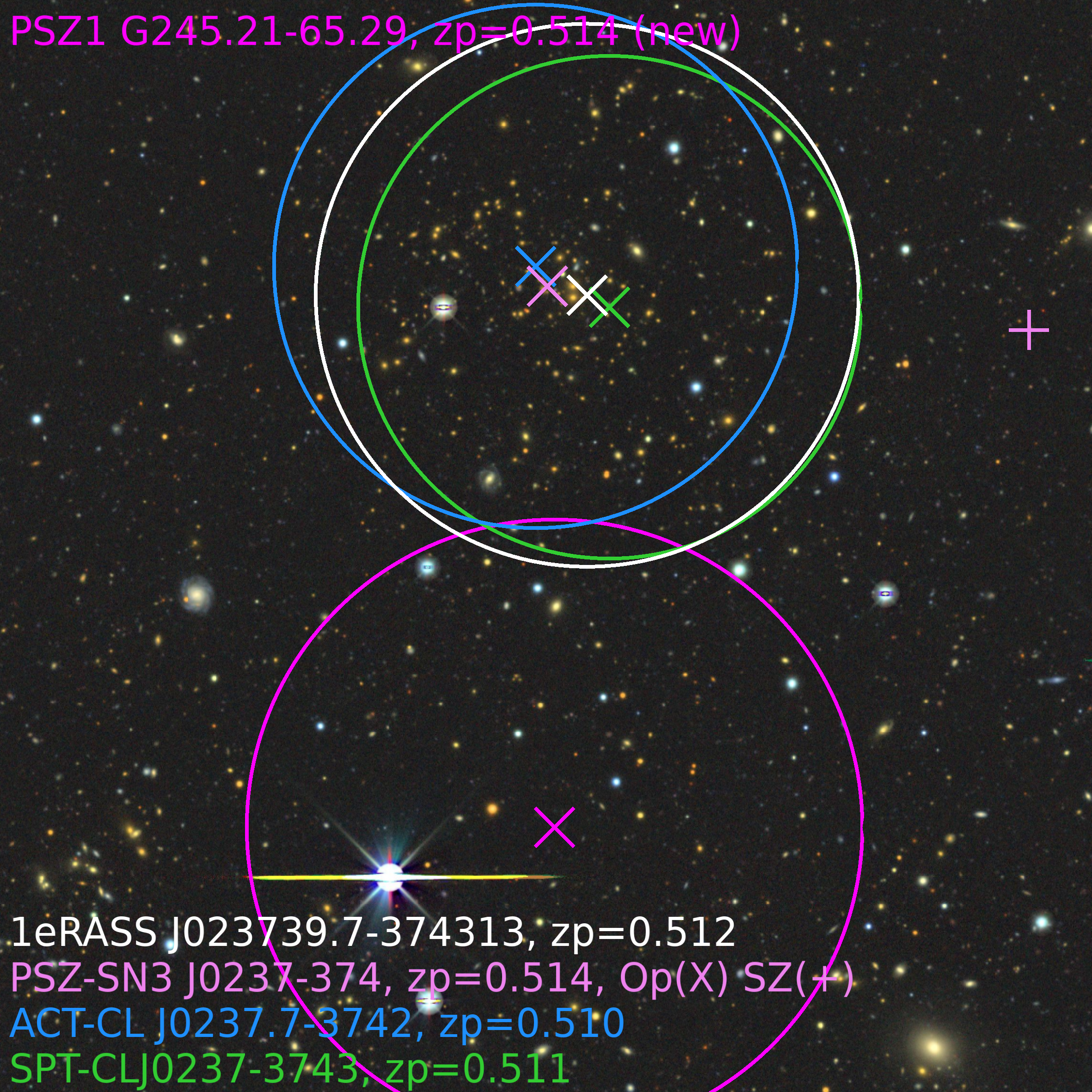}
 	\caption{\footnotesize Same as Fig.~\ref{fig:B1_3images_a} for \protect\hyperlink{PSZ1 G245.21-65.29}{PSZ1 G245.21-65.29}}  
	\label{fig:B1_3images_b}
\end{figure*}

\noindent{\bf \hypertarget{PSZ2 G213.73-56.15}{PSZ2 G213.73-56.15:} }
This PSZ2 candidate  corresponds to the PSZ-MCMF cluster PSZ-SN3 J0319-224. The identified  optical counterpart is a rich cluster ($\lambda=96$, $\zp=0.45$), well visible in the DESI image (Fig.~\ref{fig:B1_3images_a}) and matching  1eRASS J031942.5-224402 ($\zp=0.45$) and  SPT-CLJ0319-2244 ($\zp=0.46$) at the same position and redshift.   The offset with the \planck\ position, $D=4.4'=1.4\theta_{500}$, is significant, about twice the PSZ2 error. However,  
the \planck\ signal extends toward the $z=0.45$ cluster, which lies inside its S/N=3 contour. Furthermore, a single cluster is detected in eRASS and DESI maps and  the PSZ2 mass computed at the eRASS redshift is in excellent agreement with the eRASS mass, and compatible with the SPT mass within $2\sigma_{\rm tot}$.  We  have thus associated the PSZ2 candidate with both eRASS and SPT clusters and assigned the eRASS redshift. We set {\tt STATUS=C2} in view of the offset between the PSZ and X--ray position.\\

\noindent{\bf \hypertarget{PSZ2 G227.44-31.24}{PSZ2 G227.44-31.24:}}
An optical cluster  is visible in the DESI image (see Fig.~\ref{fig:B1_3images_a}) coinciding with the X--ray peak in the RASS image at the location of RASS-MCMF cluster 
 RASS-CL J051621-2521.1 ($\zp=0.2849$) and the identified  counterpart of PSZ-SN3 J0516-252 ($\zp=0.2808$). The PSZ2 position is located at $D=4.45'=\,1.16\theta_{500}$ 
and $D=5.7'\,=\,1.4\theta_{500}$ from the X--ray and optical positions, respectively. Despite the distance, the \planck\ SZE signal is extended and covers the position of the counterpart within the S/N=3 contour. 
Furthermore the position error is $6.6\arcmin$ and the RASS-MCMF and PSZ2 masses are consistent within the PSZ2 errors. Therefore, we have associated these clusters and adopted the RASS-MCMF redshift.\\ 

\begin{figure*}[t]
\centering
     \includegraphics[width=0.66\columnwidth]{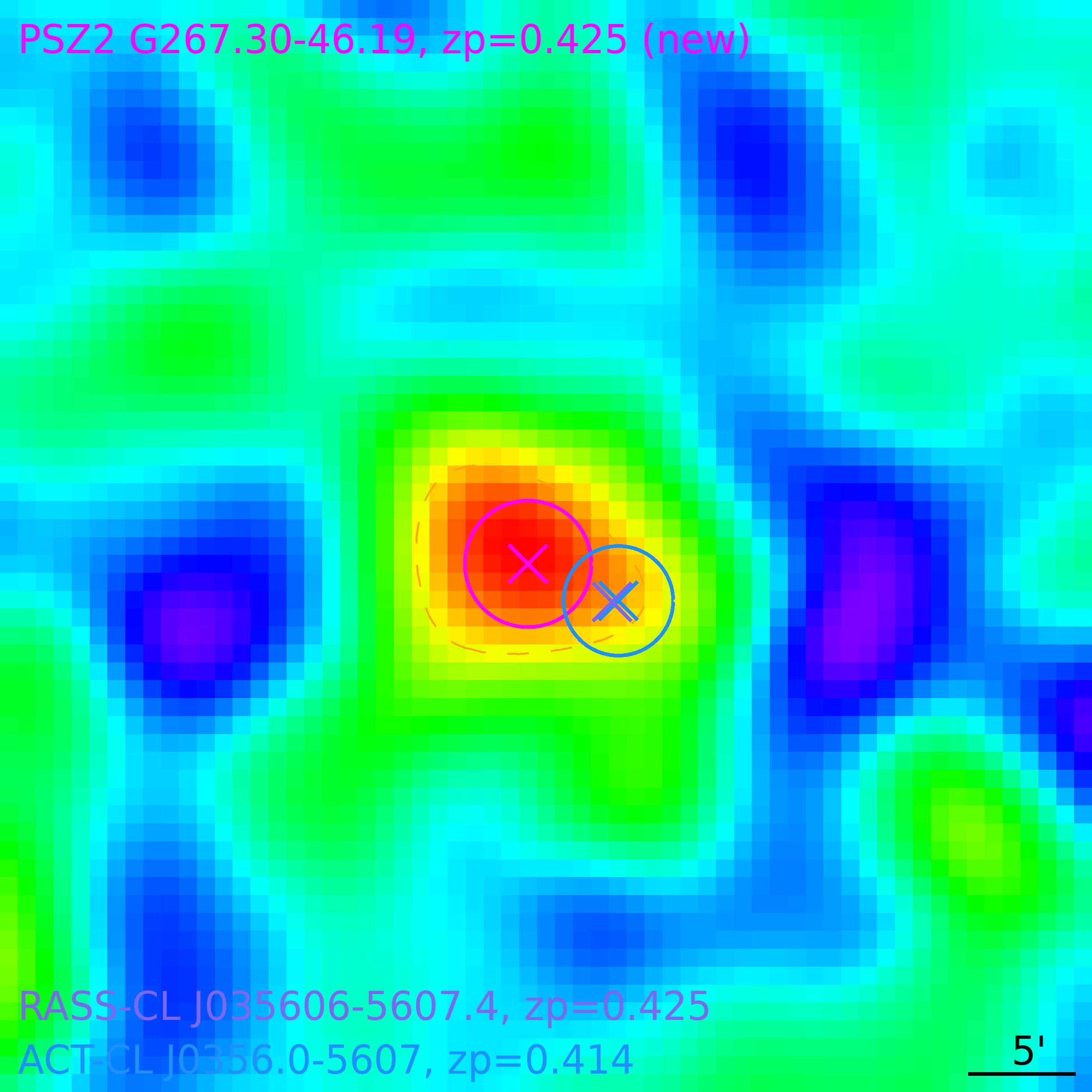}
     \includegraphics[width=0.66\columnwidth]{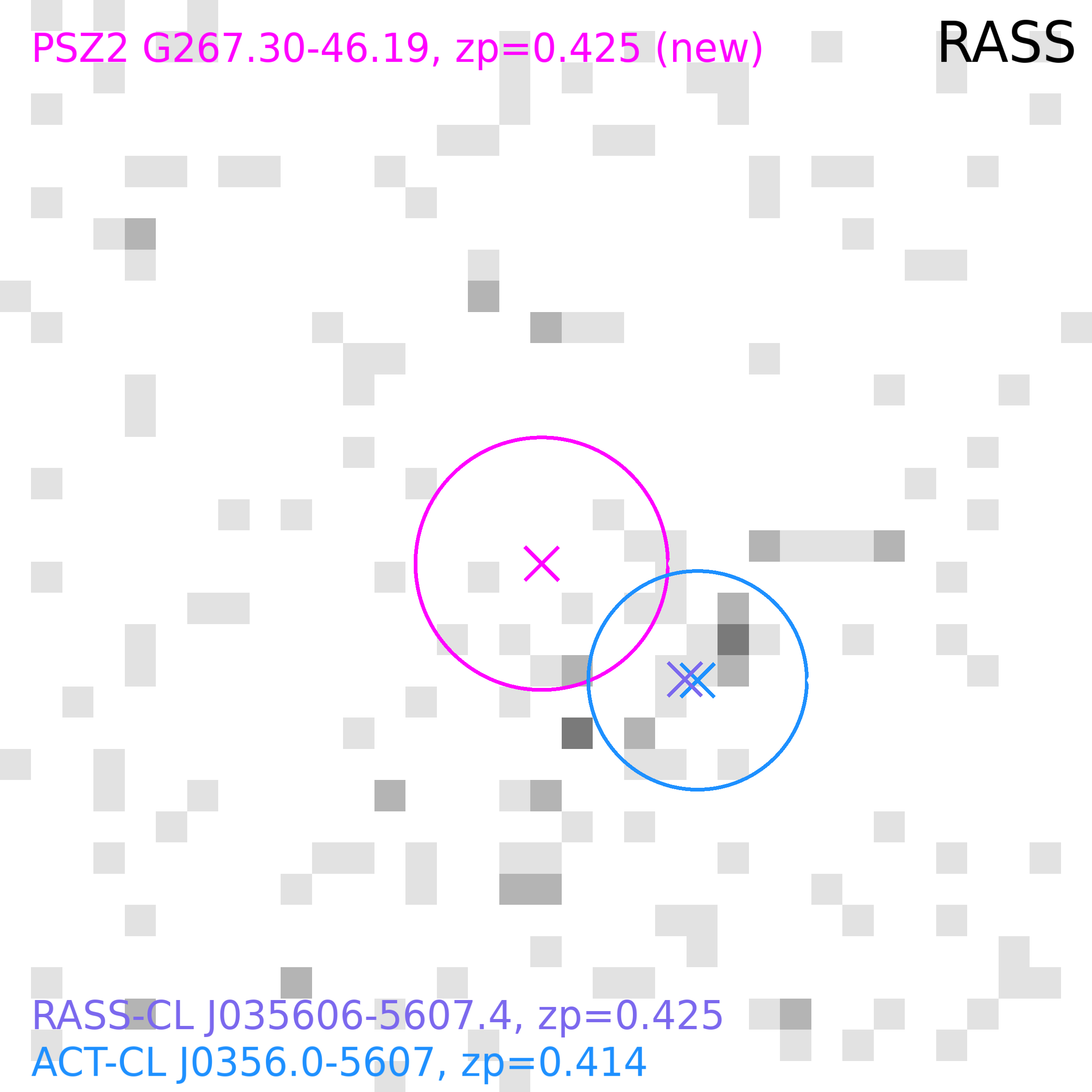}
    \includegraphics[width=0.66\columnwidth]{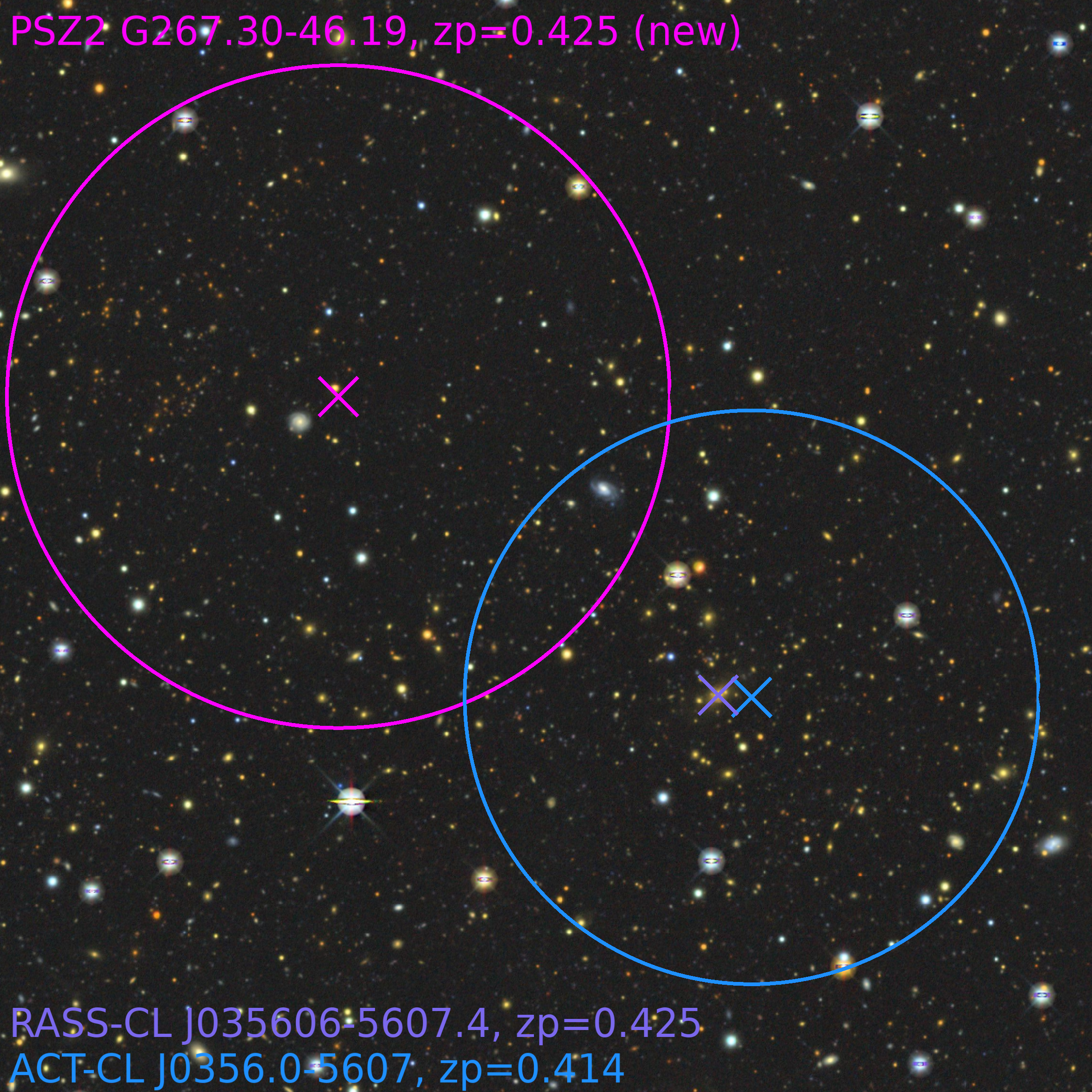}
    \includegraphics[width=0.66\columnwidth]{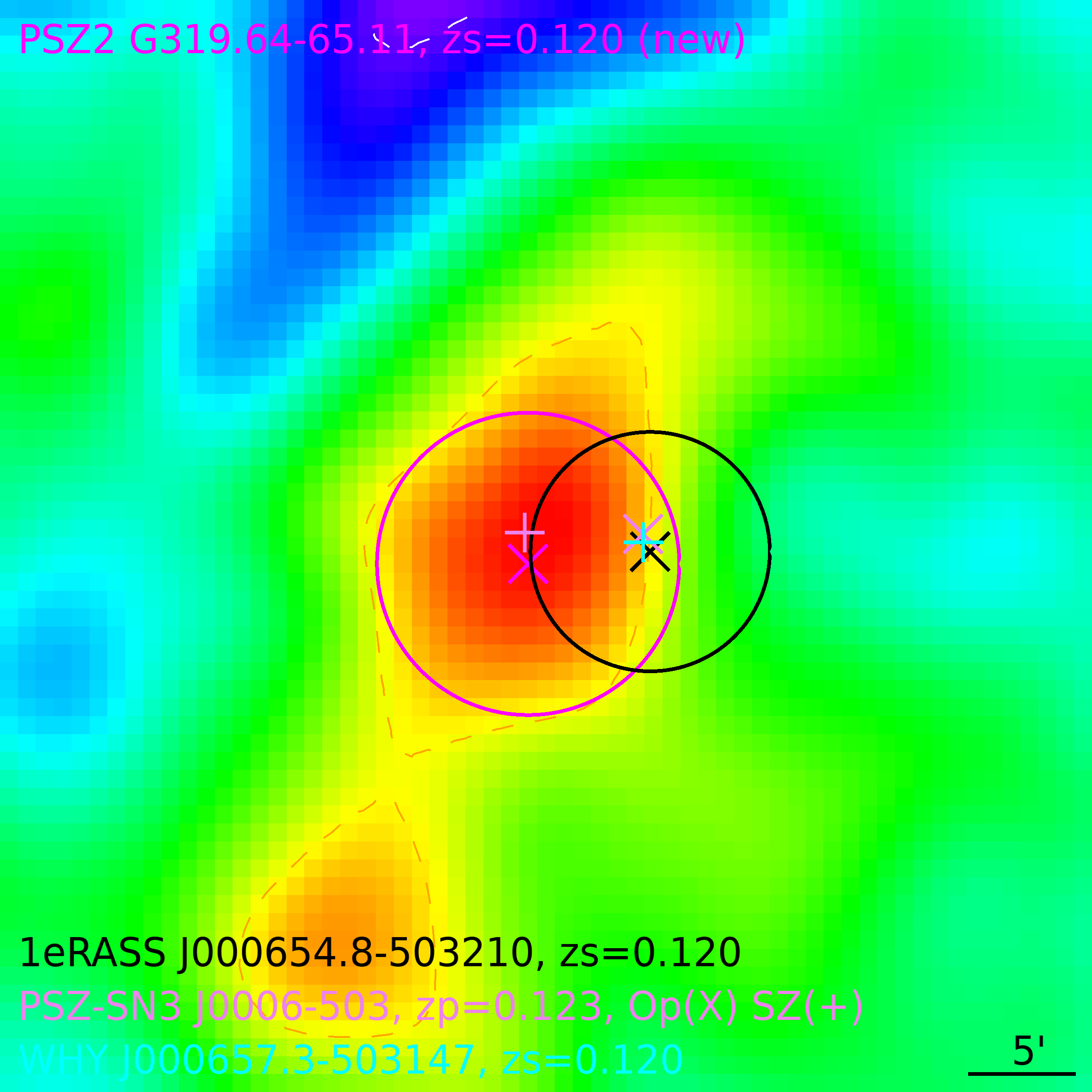}
   \includegraphics[width=0.66\columnwidth]{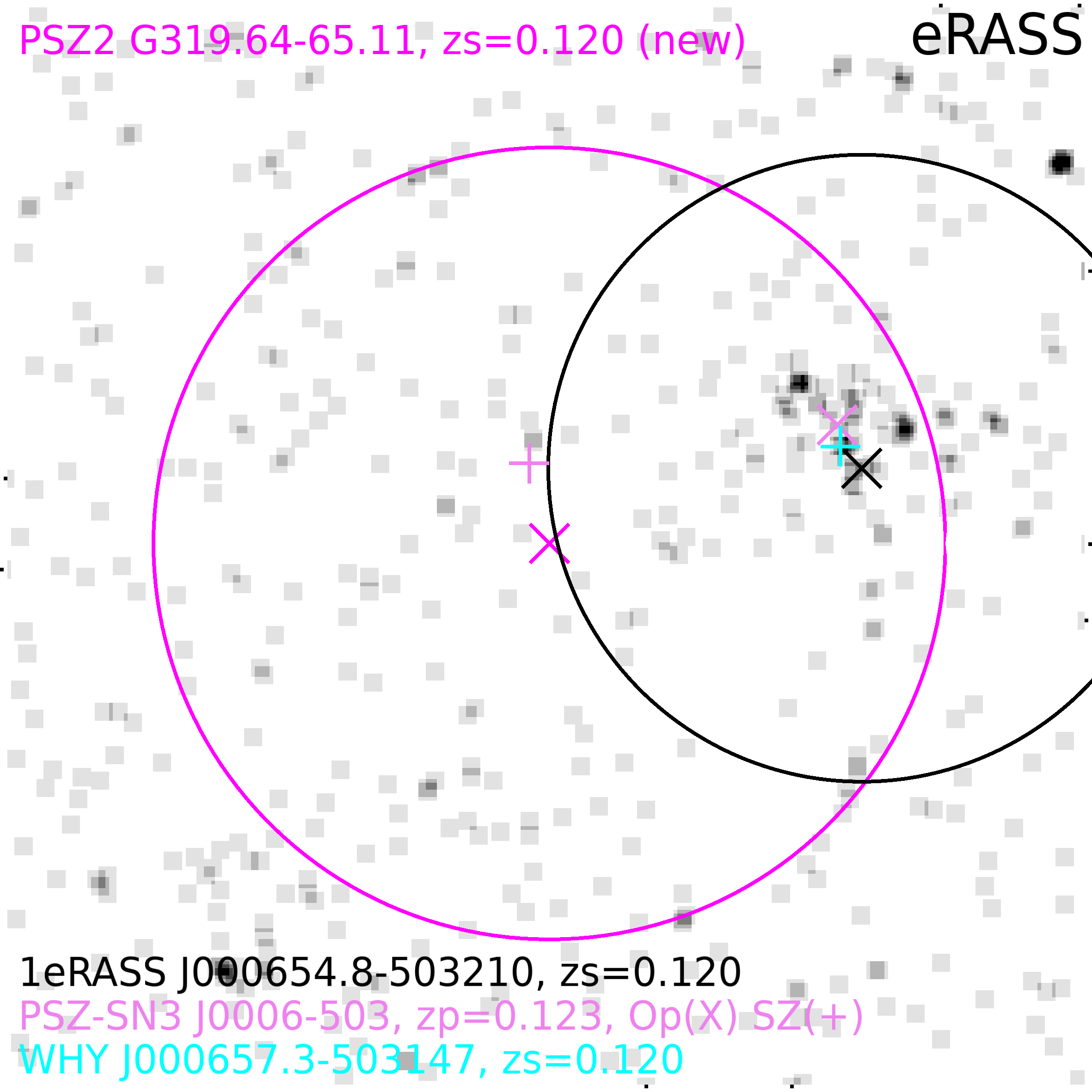}
  \includegraphics[width=0.66\columnwidth]{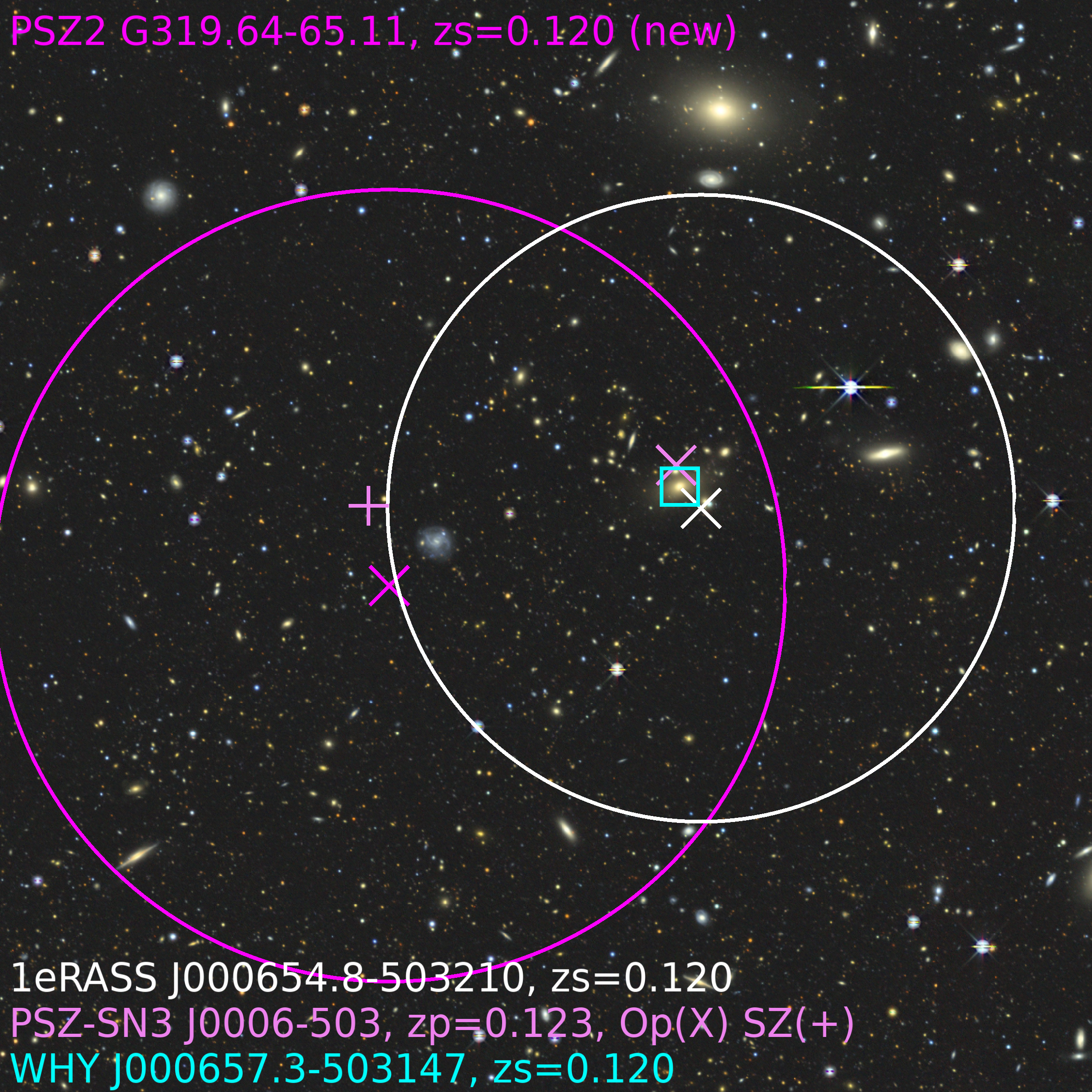}
 	\caption{\footnotesize Same as Fig.~\ref{fig:B1_3images_a} for  
    \protect\hyperlink{PSZ2 G267.30-46.19}{PSZ2 G267.30-46.19}  (top)
    and  \protect\hyperlink{PSZ2 G319.64-65.11}{PSZ2 G319.64-65.11} (bottom)}
	\label{fig:B1_3images_c}
\end{figure*}

\noindent{\bf \hypertarget{PSZ2 G236.68-37.71}{PSZ2 G236.68-37.71:}}
It presents a very extended morphology (see Fig.~\ref{fig:PSZ2G236p68}), with only the MMF1 detection passing the catalogue S/N threshold.  It is detected at S/N= 4.1 with the MMF3 algorithm and corresponds to PSZ-SN3 J0455-341 in the PSZ-MCMF catalogue. The  identified  optical counterpart is a rich cluster ($\lambda=139$), well visible in the DESI image. This distant cluster is also detected in ACT, SPT and eRASS surveys, as ACT-CL J0455.7-3417 ($\zp=0.6138$),  SPT-CLJ0455-3417 ($\zp =0.6131$) and  1eRASS J045544.2-341718 ($\zp=0.625$), respectively. The PSZ-SN3 MMF3 center is  $2.9\arcmin$ from the optical position, while the PSZ2/MMF1 position is somewhat farther away at $4.5\arcmin$. The published  $\Mv$ are  formally consistent, but  within the large uncertainties: $\Mv^{\rm ACT}= [4.71\,\pm\,0.82]\, 10^{14}\msun$, $\Mv^{\rm SPT}=[3.37\pm0.6]\, 10^{14}\msun$, $\Mv^{\rm eRASS}= [3.27\pm0.62]\, 10^{14}\msun$.  The PSZ2 mass computed at that redshift, $\Mv= [5.55\pm0.79]\, 10^{14}\msun$,  is  consistent with the ACT value ($0.7\sigma$ difference) and $2.3\sigma$ higher than the SPT value. However the large extent of the \planck\ signal, with a MMF3 scale radius of  $12.9\arcmin$,   is inconsistent with that of a $z=0.6$ cluster.  Actually,  the \planck\ signal encompasses 2 other eRASS clusters: 1eRASS J045637.4-341633 ($\zp=0.54$) in the north and 1eRASS J045637.4-341633  ($\zp=0.234$) in the southeast (left-bottom panel). The latter  coincides with ACT-CL J0456.5-3416 ($\zp=0.2428, \Mv= [2.80\pm 0.70]\, 10^{14}\msun$), RASS-CL J045631-3416.6 ($\zp=0.2321 $) and  WHY J045637.4-341646  ($\zp=0.2079 $).  Those certainly contribute to the PSZ signal but the relative contribution of the 3 clusters is  uncertain.  From the ACT masses, we expect an equal SZE flux from the $z=0.6$ and $z=0.2$ clusters. On the other hand,  the SZE flux estimated from eRASS X--ray luminosity  (Sect.~\ref{sec:PSZerass}) of the $z=0.2$ cluster is twice higher than that of the  $z=0.6$ cluster.  Furthermore, in view of its morphology, the \planck\ signal may also be contaminated by noise.  We set {\tt  STATUS=Confusion} for this cluster, with no assign redshift ($z=-1$) in view of the uncertainty on the main component. \\

\noindent{\bf \hypertarget{PSZ1 G245.21-65.29}{PSZ1 G245.21-65.29:}}   
It  is  located $D=4.9 \arcmin$ south of 1eRASS J023739.7-374313  ($\zp=0.512$), a large relative distance ($D=1.7\,\Tv$). This eRASS cluster coincides with ACT-CL J0237.7-3742, SPT-CLJ0237-3743 and the optical counterpart of PSZ-SN3 J0237-374 at the same redshift (see eRASS and DESI images in Fig.~\ref{fig:B1_3images_b}).  PSZ-SN3 J0237-374, from  the PSZ-MCMF catalogue, corresponds to a detection  of S/N=3.96 in the latest \planck\ map (Fig.~\ref{fig:B1_3images_b}, left panel),  below the PSZ2 catalogue threshold. The SZE position is consistent with the PSZ1 position within the large error ($6\arcmin$), but  closer  to ACT/SPT/eRASS position ($D=4\arcmin$). We thus  decided to match  the above clusters and  assign the PSZ-MCMF redshift to the PSZ1 G245.21-65.29 candidate. However, the  PSZ emission extends largely beyond  the $\Tv$ region of the  ACT/SPT/eRASS cluster,  with a very flat morphology. The estimated  PSZ mass,  $[5.7\pm0.9]\,10^{14}\msun$,  is also significantly higher than the consistent eRASS, ACT and SPT values, $[3.88\,\pm\,0.74], [3.44\,\pm\,0.62], [3.06\,\pm\,0.60]\, 10^{14}\msun$, respectively. The signal of the genuine $z=0.5$  cluster is likely boosted by noise  in the \planck\ map and  we set  {\tt STATUS=C2} to PSZ1 G245.21-65.29. \\

\noindent{\bf \hypertarget{PSZ2 G267.30-46.19}{PSZ2 G267.30-46.19}:} Fig.~\ref{fig:B1_3images_c} shows a single X--ray diffuse emission coinciding with the RASS-MCMF cluster RASS-CL J035606-5607.4 ($\zp= 0.425$) and ACT-CL J0356.0-5607 ($\zp= 0.414$). The cluster is also visible in the DESI image. The \planck\ SZE peak has a 
offset with respect to the X--ray emission ($D=4.68' = 1.78\theta_{\rm ACT}$), but covers it. The PSZ2 mass recomputed at the RASS-MCMF cluster redshift is in excellent agreement with the RASS-MCMF mass and agrees with the ACT mass within $2\sigma_{\rm tot}$, so we have associated these clusters and assigned the
RASS-MCMF redshift to PSZ2 G267.30-46.19. \\

\noindent{\bf \hypertarget{PSZ2 G319.64-65.11}{PSZ2 G319.64-65.11:}}
It is located at  $D=5.9\arcmin=0.8\theta_{500}$ west of 1eRASS J000654.8-503210 ($\zs = 0.1196$).  The latter can be matched with WHY J000657.3-503147 and the identified optical counterpart of  PSZ-SN3 J0319-224, well visible in the DESI image at the X--ray position (see Fig.~\ref{fig:B1_3images_c}). The  SZE position of the PSZ2 (based MMF1) and  PSZ-SN3 (based on MMF3) objects are consistent ($D=1.5\arcmin$).  We  have thus associated the above clusters and assigned the eRASS redshift to  PSZ2 G319.64-65.11. However the X--ray and SZE signals deviate by $3.5\sigma$ from the \YL\ relation (Sect.~\ref{sec:PSZerass}) and the X--ray--SZE offset  is 2.4 times the PSZ2 position error. We thus set a {\tt STATUS=C2}.

\subsubsection{Redshift comparison and update}

\noindent{\bf \hypertarget{PSZ2 G181.71-68.65}{PSZ2\,G181.71-68.65:}} The PSZ2 catalogue redshift, $\zs=0.1529$, comes from a cross-identification with MCXC J0206.4-1453 at $D=1.6\arcmin$, whose redshift was obtained from REFLEX follow-up. We cross-identified PSZ2 G181.71-68.65 with 1eRASS J020628.4-145358 ($D=2.1\arcmin$), RASS-CL J020629-1453.8  ($D=1.77\arcmin$), and WHY J020641.1-145351. The $\zp$ value for all those counterparts are consistent: $\zp=0.2983\pm0.0057, \zp=0.2926\pm0.0056, z=0.2471$, respectively. The REFLEX redshift corresponds to a foreground galaxy, as noted by \citet[][Table A1]{RASSMCMF2023}, and we therefore updated the PSZ2 redshift to the eRASS value.\\

\noindent{\bf \hypertarget{PSZ2 G210.01+50.85}{PSZ2 G210.01+50.85:}} The PSZ2 catalogue redshift $z=0.319$ comes from a cross-identification with GMBCG J149.91715+22.43516 using NED. At $D=7.2\arcmin$ from the PSZ2 position, we found MCXC J0959.7+2223 = 1eRASS J095947.2+222415 =  RASS-CL J095946+2224.1, all of them at $\zs=0.24$. The \planck\ SZE map presents an extended double-peaked morphology, with  one peak at the MCXC position and a second peak not coincident with any known cluster. We have associated these clusters, despite their distance, due to the extension of the SZE signal, which covers the X--ray cluster. We have therefore updated the PSZ2 redshift to the MCXC value. \\

\begin{figure}[t]
\centering
\includegraphics[width=0.49\columnwidth]{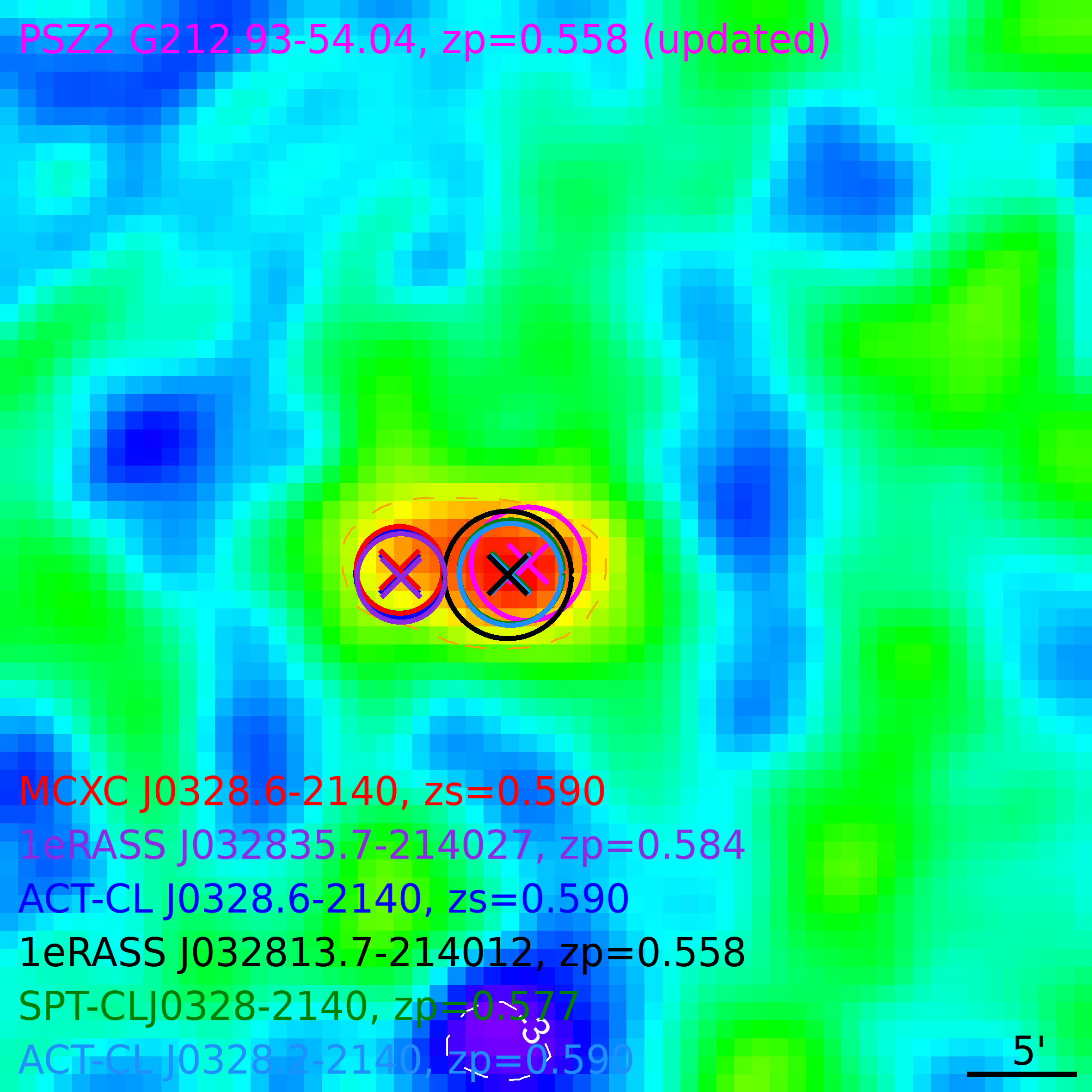}
\includegraphics[width=0.49\columnwidth]{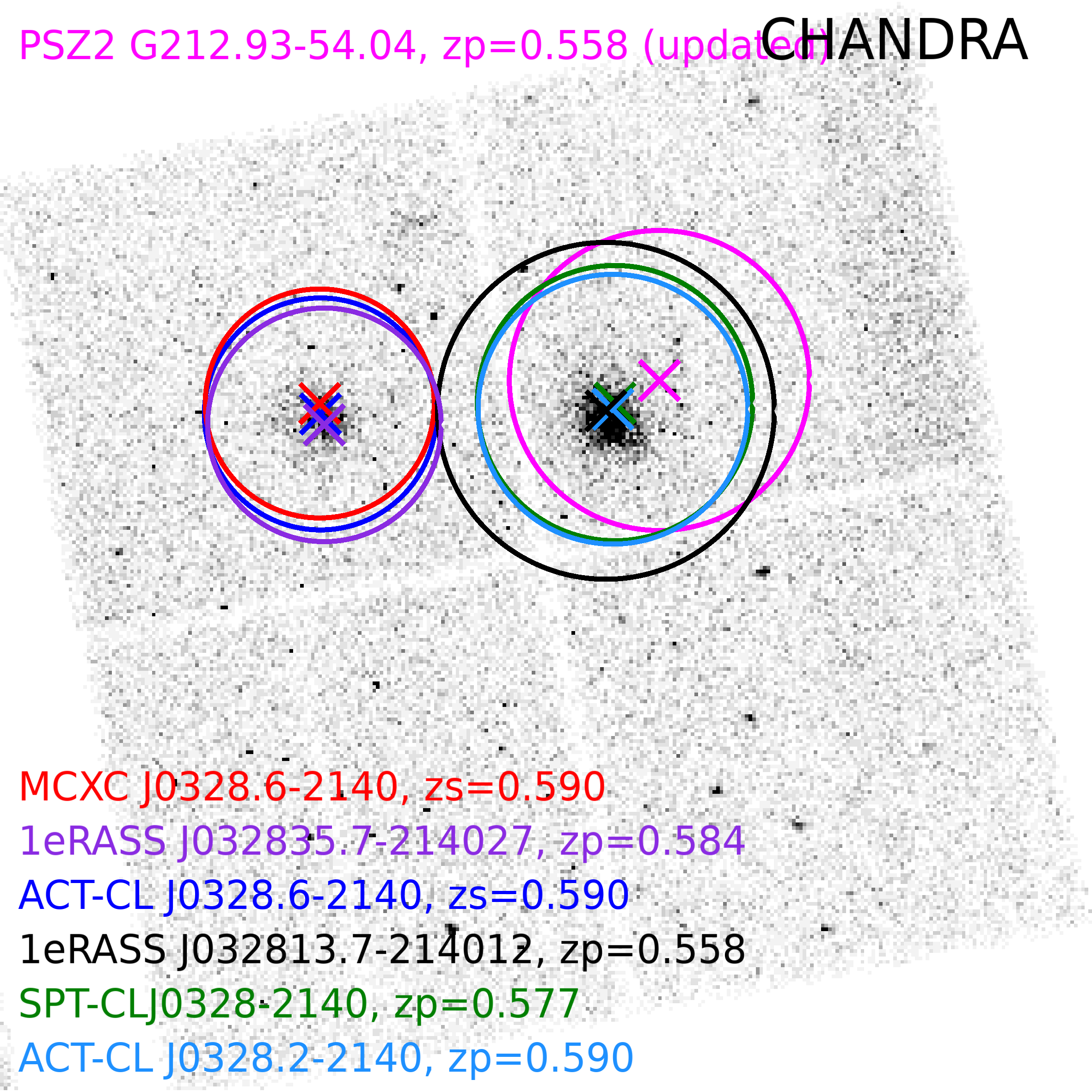}
\caption{\footnotesize \planck\ filtered map (left) and \chandra\ images (right) of  \protect\hyperlink{PSZ2 G212.93-54.04}{PSZ2 G212.93-54.04}. Overlays  as in Fig.~\ref{fig:B1_3images_a}. Catalogue redshift $\zs=0.590$ updated to $\zs=0.558$.}
\label{fig:G212.93-54.04}
\end{figure}

\noindent{\bf \hypertarget{PSZ2 G212.93-54.04}{PSZ2 G212.93-54.04}:} 
This cluster was associated with MCXC J0328.6-2140 ($z_{\rm spec}=$0.59) in the PSZ2 catalogue, but there is a large  distance between the two objects ($D=6.2\arcmin=2.3\Tv$).
The eRASS and Chandra (Fig.~\ref{fig:G212.93-54.04}) images show two different clusters in the vicinity of this PSZ2 detection, at almost identical redshifts and separated by $5.3\arcmin$: 
MCXC J0328.6-2140 on the east, coinciding with  1eRASS J032835.7-214027 ($\zp=0.58$) = ACT-CL J0328.6-2140 ($z_{\rm spec}=$0.59), and 1eRASS J032813.7-214012 $\zp=0.56$ on the west, coinciding with ACT-CL J0328.2-2140 ($\zp=0.59$) = SPT-CLJ0328-2140 ($\zp=0.58$). The position of PSZ2 G212.93-54.04 is closer to this western cluster ($D=0.99\arcmin$), although its SZE signal may be contaminated by the eastern one. We thus discard the association of PSZ2 G212.93-54 with the MCXC cluster. Since the original PSZ2 redshift was taken from the MCXC value, we updated the PSZ redshift to the eRASS value, and added a {\tt COMMENT} describing this update and the potential contamination.\\

\noindent{\bf \hypertarget{PSZ2 G215.19-49.65}{PSZ2 G215.19-49.65}:} 
In the PSZ2 catalogue,  PSZ2 G215.19-49.65 is associated with Abell cluster ACO3168 with its redshift, $\zs= 0.24$ taken from NED. The NED value refers to \cite{Chon2012}, who reported 5 galaxies with spectroscopic redshift for REFLEXII cluster MCXC J0347.4-2149. However, these galaxies are far from ACO3168 center and probably do not belong to it. On the other hand,  PSZ2 G215.19-49.65  matches 1eRASS J034802.4-214515  ($D=1.3\arcmin, \zp=0.351$), coinciding with ACT-CL J0347.9-2144=SPT-CLJ0348-2144 ($D=0.9\arcmin$ from the ACT/SPT position at consistent $\zp$).  Furthermore, RASS-MCMF, PSZ-MCMF, and WHY18 catalogues also find a similar photometric redshift. We thus  update the PSZ redshift to $\zp=0.35$  and we do not match this cluster to MCXC J0347.4-2149. \\

\begin{figure}[t]
\centering
\includegraphics[width=0.49\columnwidth]{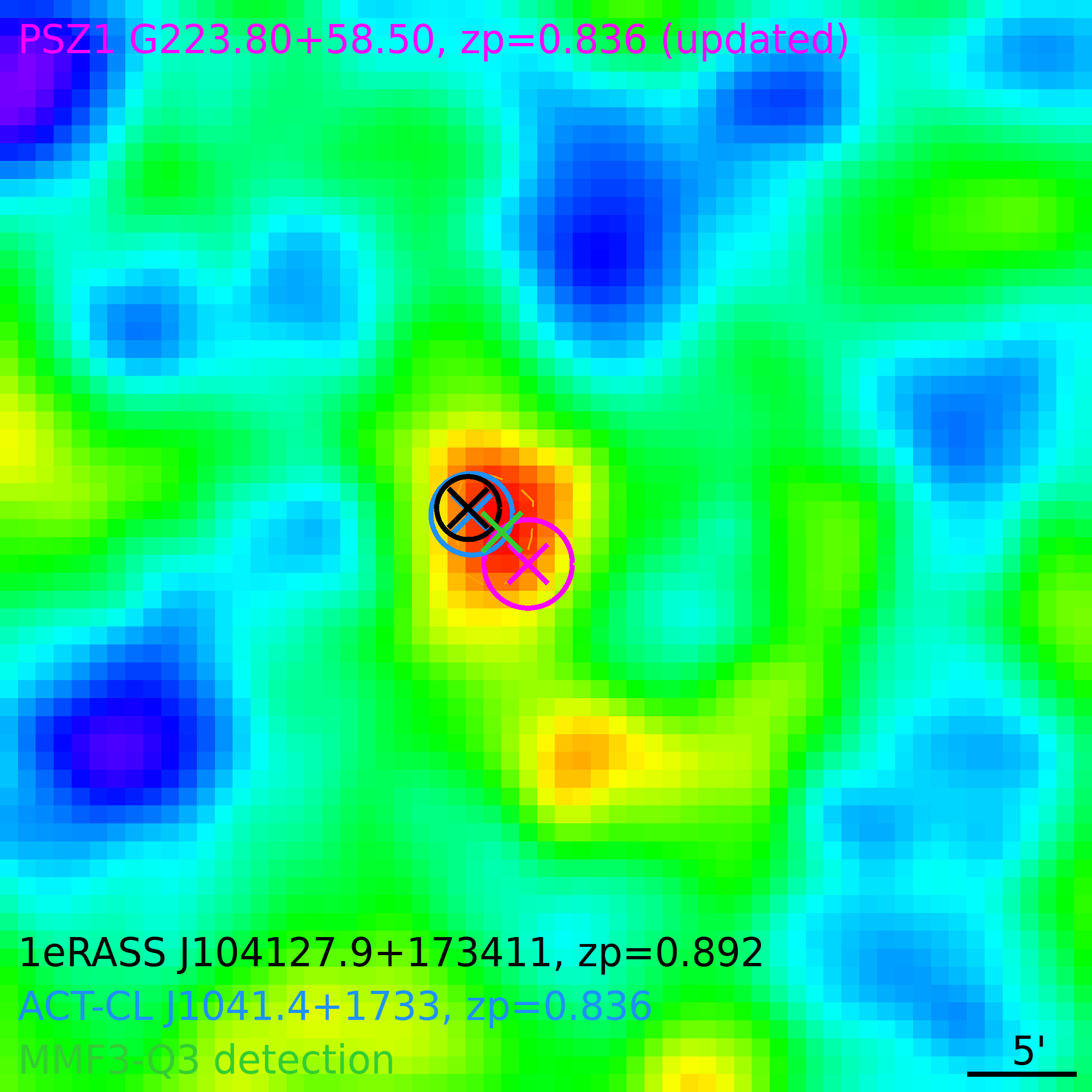}
\includegraphics[width=0.49\columnwidth]{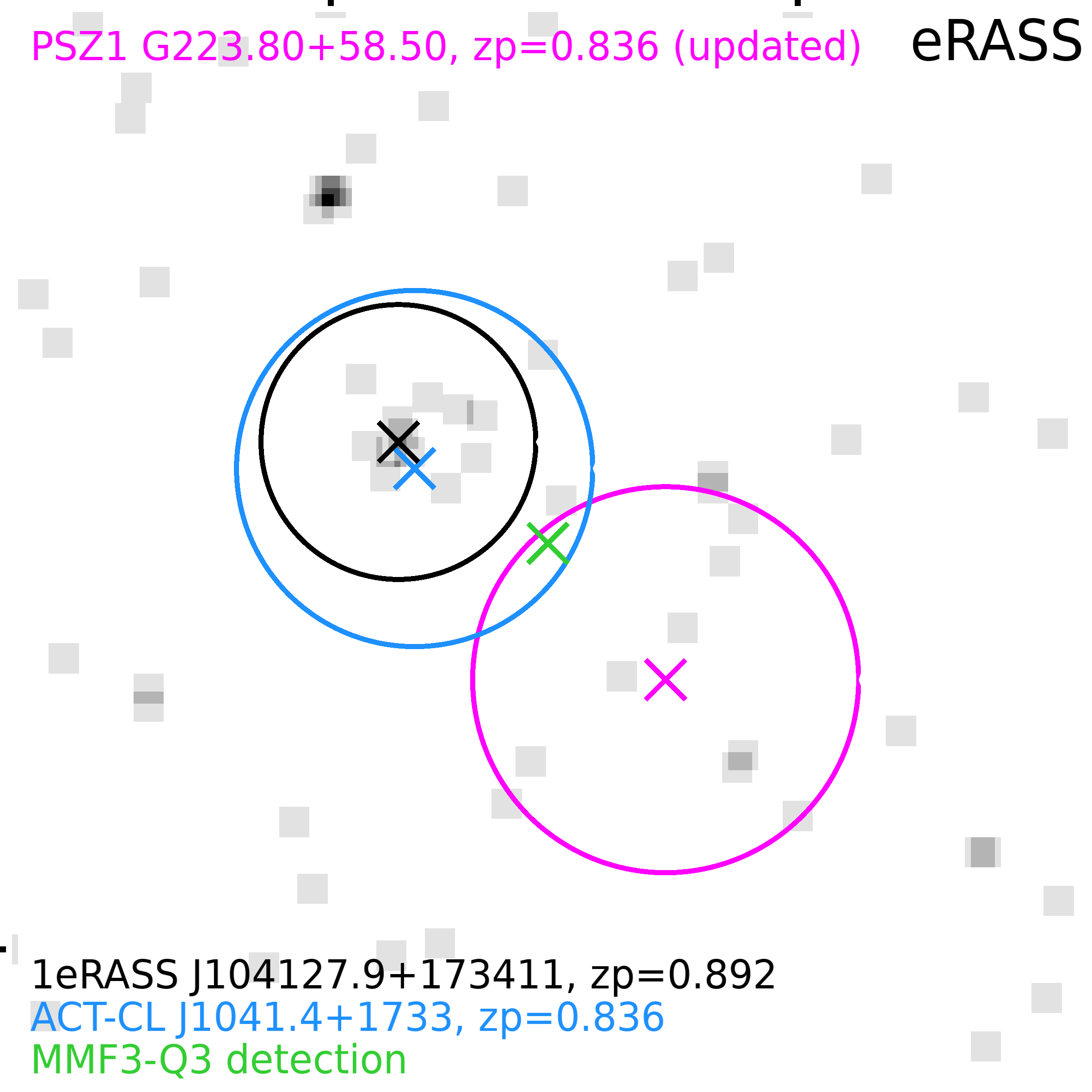}
\caption{\footnotesize 
\planck\ filtered map (left) and eRASS image (right) covering cluster \protect\hyperlink{PSZ1 G223.80+58.50}{PSZ1 G223.80+58.50}.
Overlays as in Fig.~\ref{fig:B1_3images_a}. The green cross is the center of the detection in the latest \planck\ maps. Catalogue redshift $\zp=0.381$ updated to $\zp=0.836$.}
\label{fig:G223.80+58.50}
\end{figure}

\noindent{\bf \hypertarget{PSZ1 G223.80+58.50}{PSZ1 G223.80+58.50}:}
 PSZ1 catalogue redshift  $\zp=0.38$ comes from the PanSTARRS study of  \cite{Liu15}. However, in the optical follow-up of \cite{Bar20}, PSZ1 G223.80+58.5 was  labelled 'ND' (non-detection = no galaxy overdensity) so it was flagged as a false candidate during the first step of the construction of the PSZ metacatalogue (see Sect.~\ref{sec:PSZupfu}). Our further cross-match with external catalogues shows that PSZ1 G223.80+58.50 is close to the distant object 1eRASS J104127.9+173411 ($\zp=0.89, D=3.6\arcmin$), detected in the eRASS survey at low S/N, cross-identified with  ACT-CL J1041.4+1733  ($\zp=0.84$, see Fig~\ref{fig:G223.80+58.50}).   The redshift  is at the detection limit of the optical follow up ($z=0.85$ according to \cite{Bar20}), the PSZ1 detection could indeed correspond to this distant cluster. There is also a faint detection in the \planck\ maps of PSZ2 (S/N=3.6, below the S/N threshold of the PSZ2 catalogue), with its peak coinciding better with the ACT or eRASS position ($D=1.7\arcmin$). The PSZ1 mass recomputed at the ACT redshift and the ACT mass are compatible within the error bars ($M_{500}^{\rm PSZ1}=6.6\pm1.1\cdot10^{14} M_{\odot}$, $M_{500}^{\rm ACT}=5.2\pm0.8\cdot10^{14} M_{\odot}$). We thus concluded that PSZ1 G223.80+58.50 is a true cluster at the ACT redshift value. \\

\noindent{\bf \hypertarget{PSZ2 G224.53-30.27}{PSZ2 G224.53-30.27}:}
In the PSZ2 catalogue, PSZ2 G224.53-30.27 is associated with Abell S0519 with its  redshift, $z= 0.20$ taken from NED. The NED value is an  estimated redshift from \cite{Coziol2009}. 
PSZ2 G224.53-30.27 can be cross-matched with  1eRASS J051657.9-22370 ($D=1.47\arcmin$) at a larger redshift, $\zp=0.2960\pm0.0063$. The latter coincides with ACT-J0516.9-2237,  SPT-CLJ0516-2236, RASS-CL J051657-2237.0 and  PSZ-SN3 J0516-223, at consistent  redshift ($\zp=0.3020\,,0.3023\,,0.292,\,0.2949$, respectively).
eRASS image shows a single extended emission at the position of these 5 counterparts which lie within less than $0.6\arcmin$ of each other. Thus we have updated the PSZ redshift to $\zp=0.2960$. \\

\begin{figure}[t]
\centering
\includegraphics[width=0.49\columnwidth]{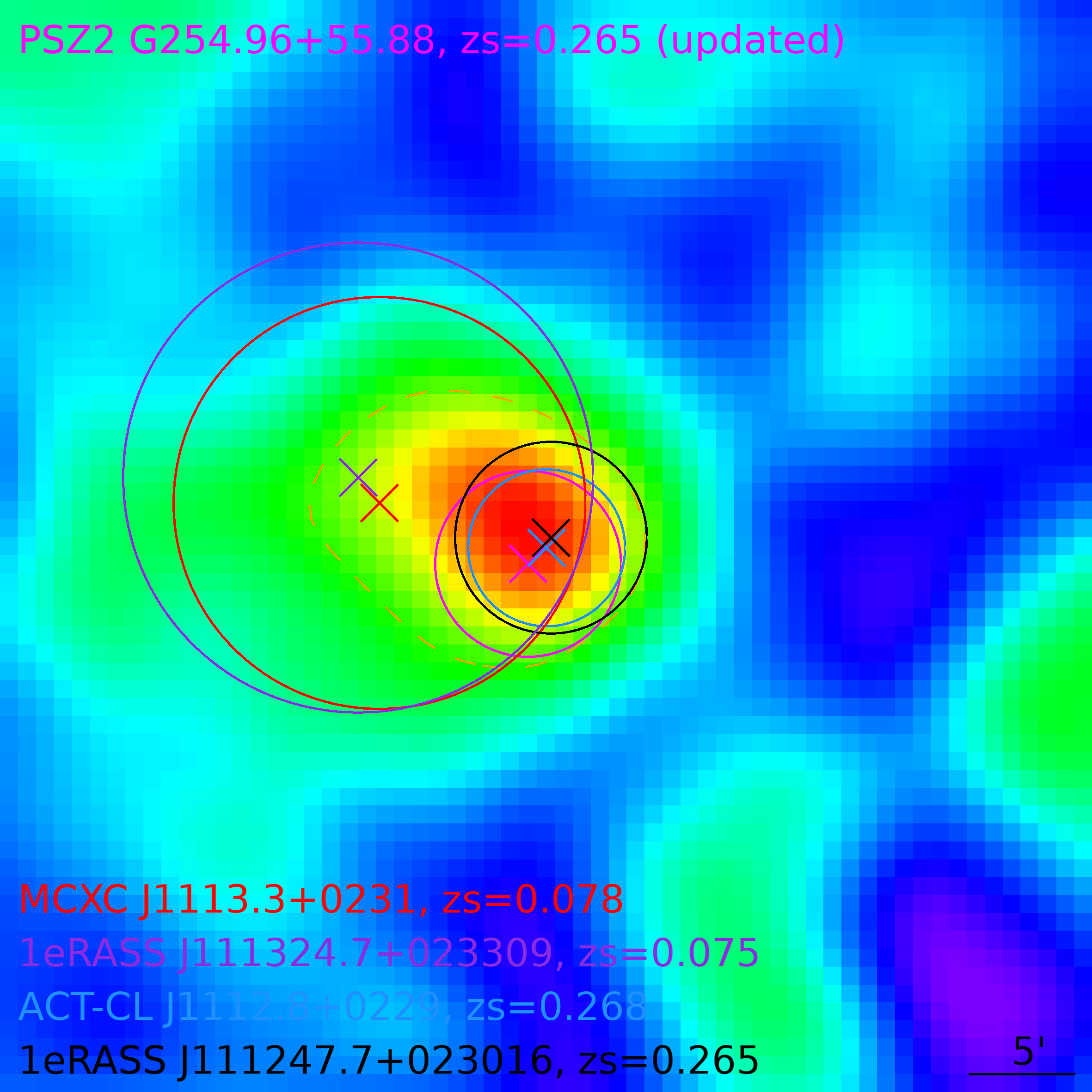}
\includegraphics[width=0.49\columnwidth]{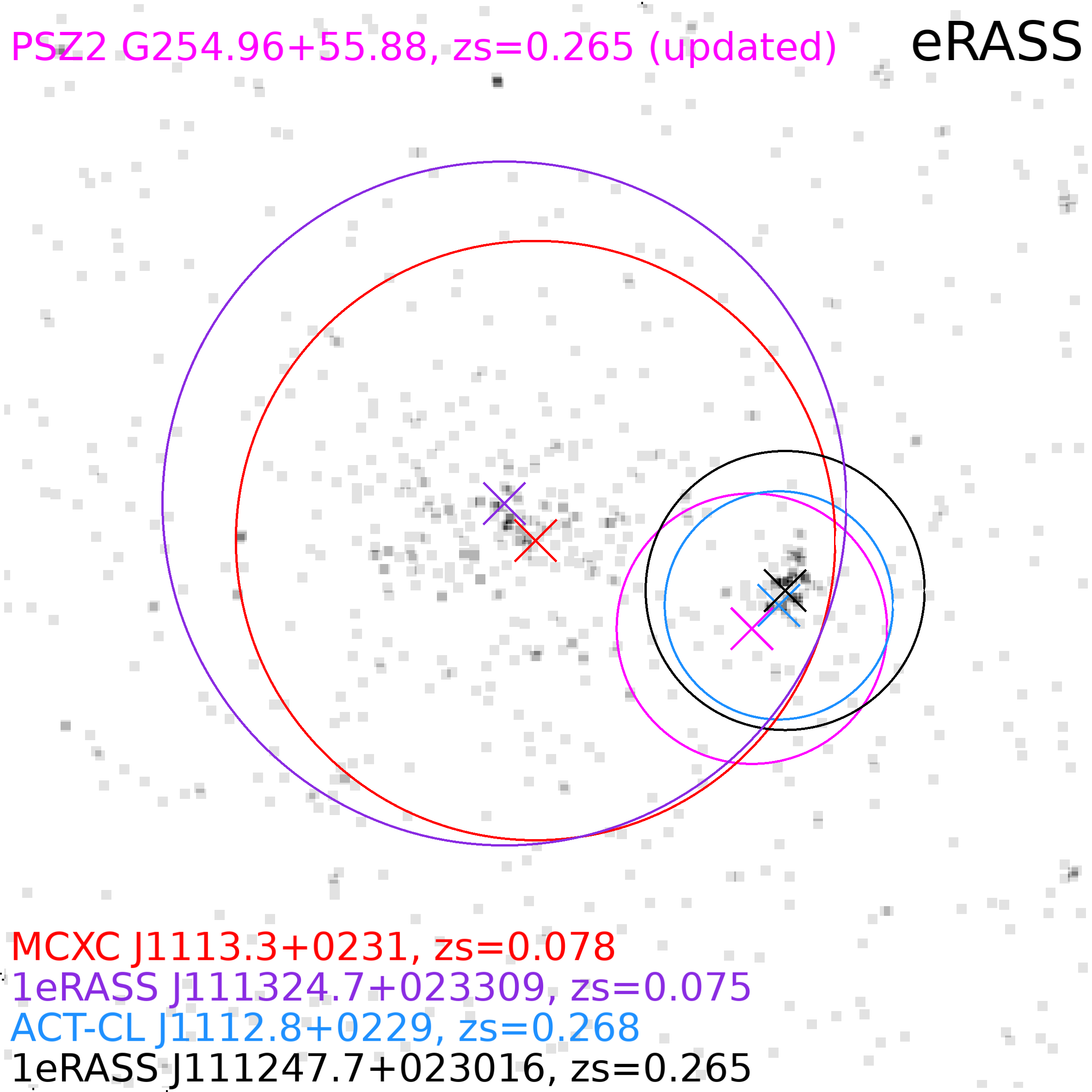}
\caption{\footnotesize 
 \planck\ filtered map (left) and eRASS image (right) covering
cluster \protect\hyperlink{PSZ2 G254.96+55.88}{PSZ2 G254.96+55.88}. Overlays as in Fig.~\ref{fig:B1_3images_a}. Catalogue redshift $\zs=0.078$ updated to $\zs=0.265$. }
\label{fig:PSZ2 G254.96+55.88}
\end{figure}

\noindent{\bf \hypertarget{PSZ2 G254.96+55.88}{PSZ2 G254.96+55.88}:}
Fig.~\ref{fig:PSZ2 G254.96+55.88}
shows the \planck\ and  eRASS 
image around PSZ2 G254.96+55.88.
The  catalogue redshift, $\zs=0.078$, comes from PSZ1 identification with the MCXC-NORAS cluster,  RXC J1113.3+0231. The closest eRASS cluster is a higher redshift object, 1eRASS J111324.7+023309 ($\zs=0.2654$, $\Mv=(2.7\pm 0.3)\,10^{14}\msun$) at $D=  1.67\arcmin$. It coincides with  ACT-CL J1112.8+0229 at same redshift and mass ($\zs=0.2680$, $\Mv=(3.0\pm0.6)\,10^{14}\msun$). The second closest cluster is 1eRASS J111324.7+023309 ($\zs=0.0751$, $\Mv=(5.3\pm 0.8)\,10^{14}\msun$) at $D=9.14\arcmin$ in the east, at the redshift and position of  RXC J1113.3+0231.  The \planck\ SZE signal extends over the two components but clearly  peaks at the position of the ACT cluster, with diffuse extension towards the east.
We thus update the redshift to the eRASS value, $\zs=0.2654$, remove the association with MCXC J1113.3+0231, and note that it  contributes  to the SZE signal  in the {\tt COMMENT} field.  
\\

\begin{figure*}[t]
\centering
\includegraphics[width=0.66\columnwidth]{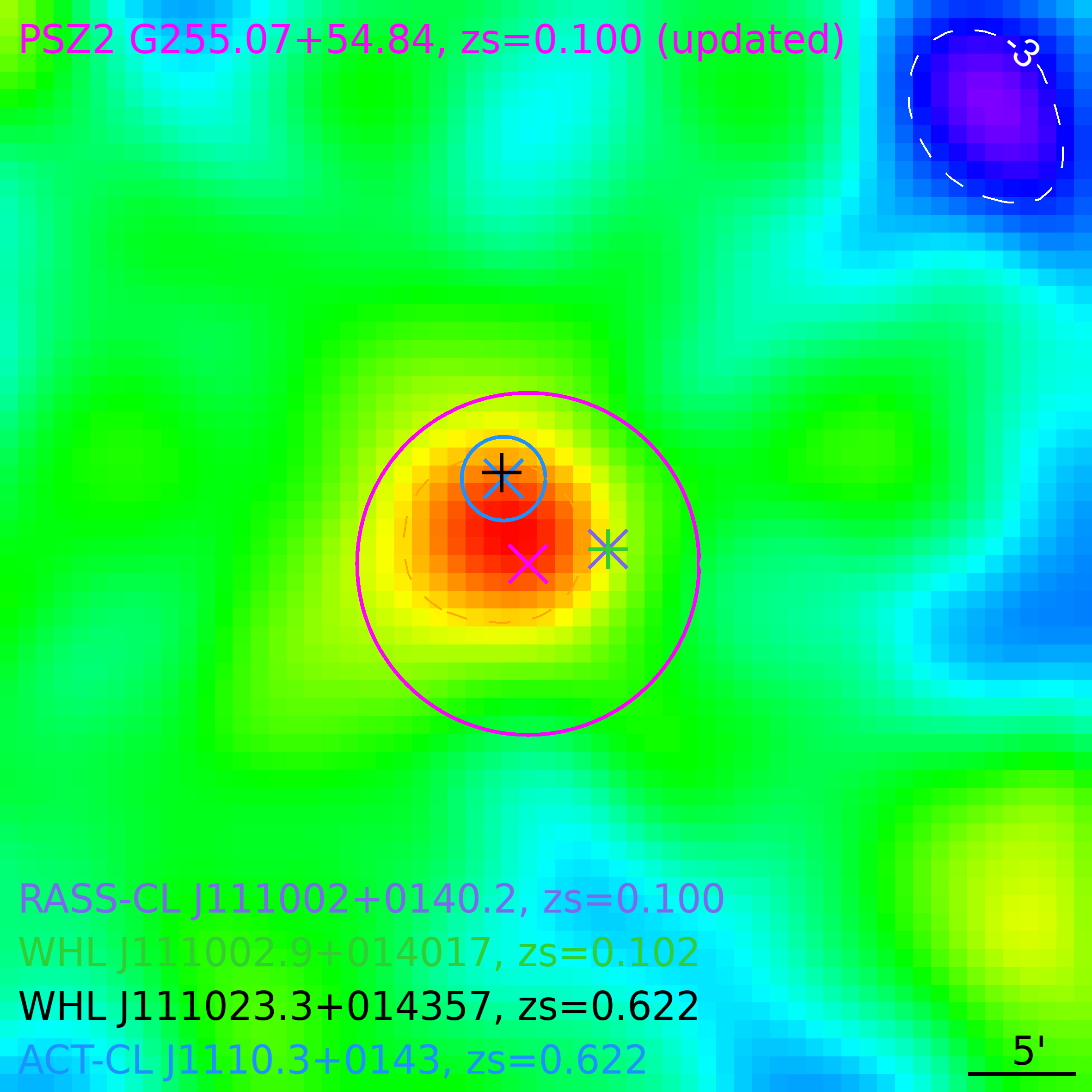}
\includegraphics[width=0.66\columnwidth]{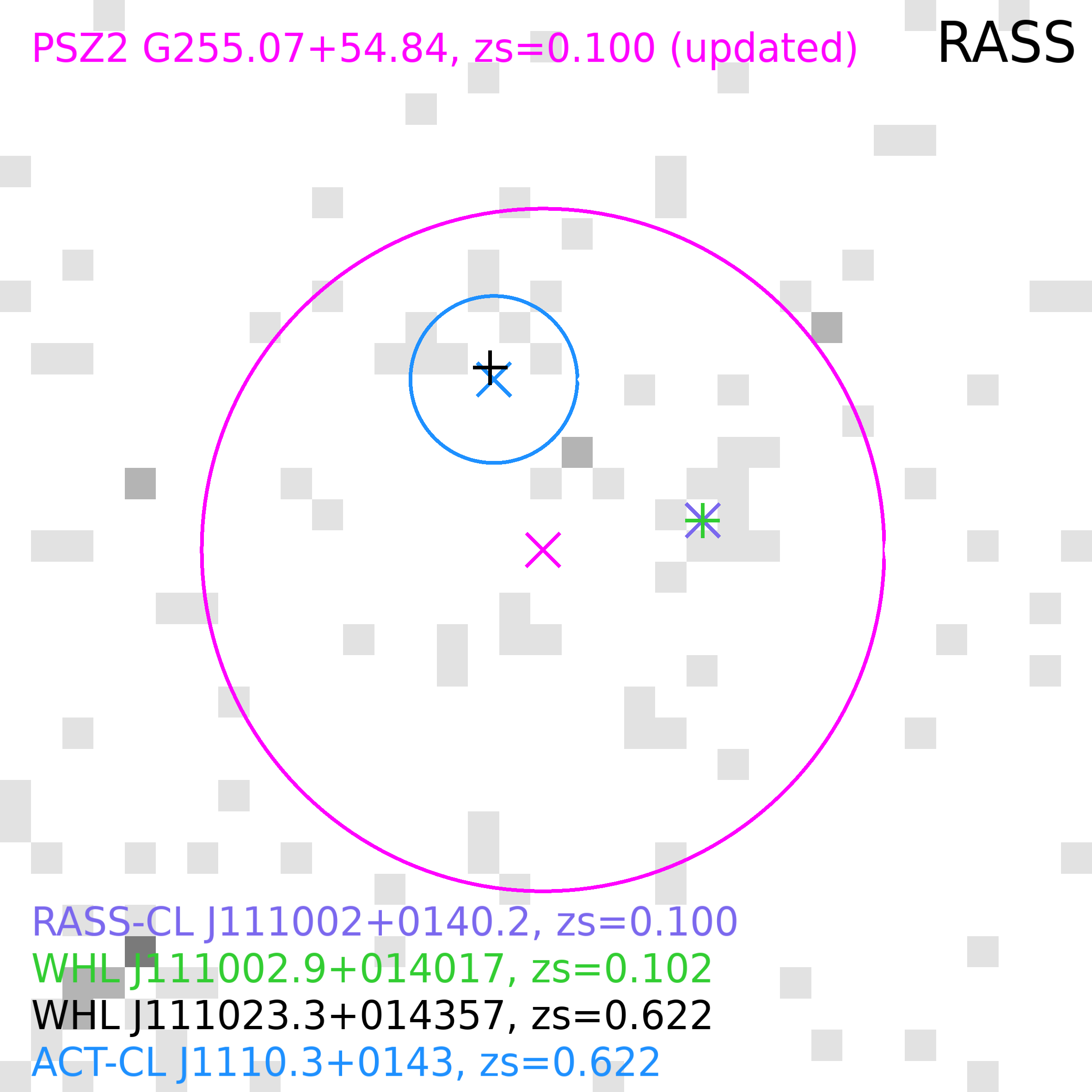}
\includegraphics[width=0.66\columnwidth]{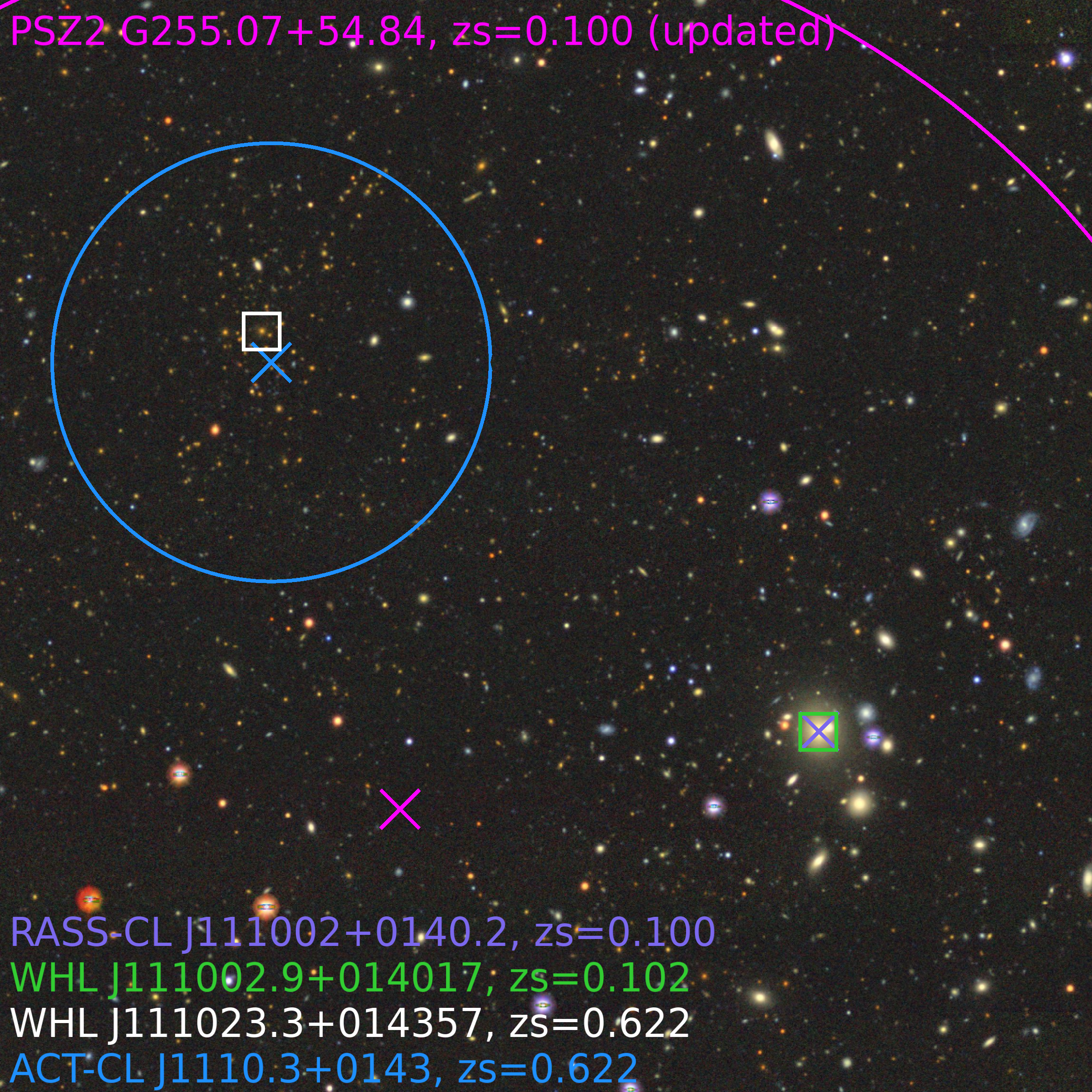}
\caption{\footnotesize Same as Fig.~\ref{fig:B1_3images_a} for \protect\hyperlink{PSZ2 G255.07+54.84}{PSZ2 G255.07+54.84}. Catalogue redshift $\zp=0.089$ updated to $\zs=0.100$.}
\label{fig:F1_3images_b}
\end{figure*}

\noindent{\bf \hypertarget{PSZ2 G255.07+54.84}{PSZ2 G255.07+54.84}:} RASS and DESI images (see Fig.~\ref{fig:F1_3images_b}) show two different clusters at different redshifts separated $D=6.0'$. The cluster on the southwest, at lower redshift, corresponds to RASS-MCMF cluster RASS-CL J111002+0140.2 ($z_\mathrm{spec}=0.10$) and WHL J111002.9+014017 ($z_\mathrm{spec}=0.1018$). The cluster on the northeast is ACT-CL J1110.3+0143 ($z_\mathrm{spec}= 0.62$) = WHL J111023.3+014357 ($z_\mathrm{spec}= 0.6218$). The PSZ2 detection is located at $D=3.9'$ from RASS-CL J111002+0140.2 and $D=4.3'$ from ACT-CL J1110.3+0143. The original PSZ2 redshift ($z_\mathrm{phot}= 0.089$), from redMaPPer, is compatible with the RASS-CL J111002+0140.2 $z_\mathrm{spec}=0.10$, and their masses are compatible within $3\sigma_{\rm tot}$. Thus, we have associated these two clusters and updated the PSZ2 redshift to the spectroscopic value. The ACT cluster, incompatible in terms of redshift and mass,
is a different object, although it may contaminate the
PSZ2 SZE signal. We have thus set {\tt STATUS=Confusion}. \\

\begin{figure}
\centering
\includegraphics[width=0.49\columnwidth]{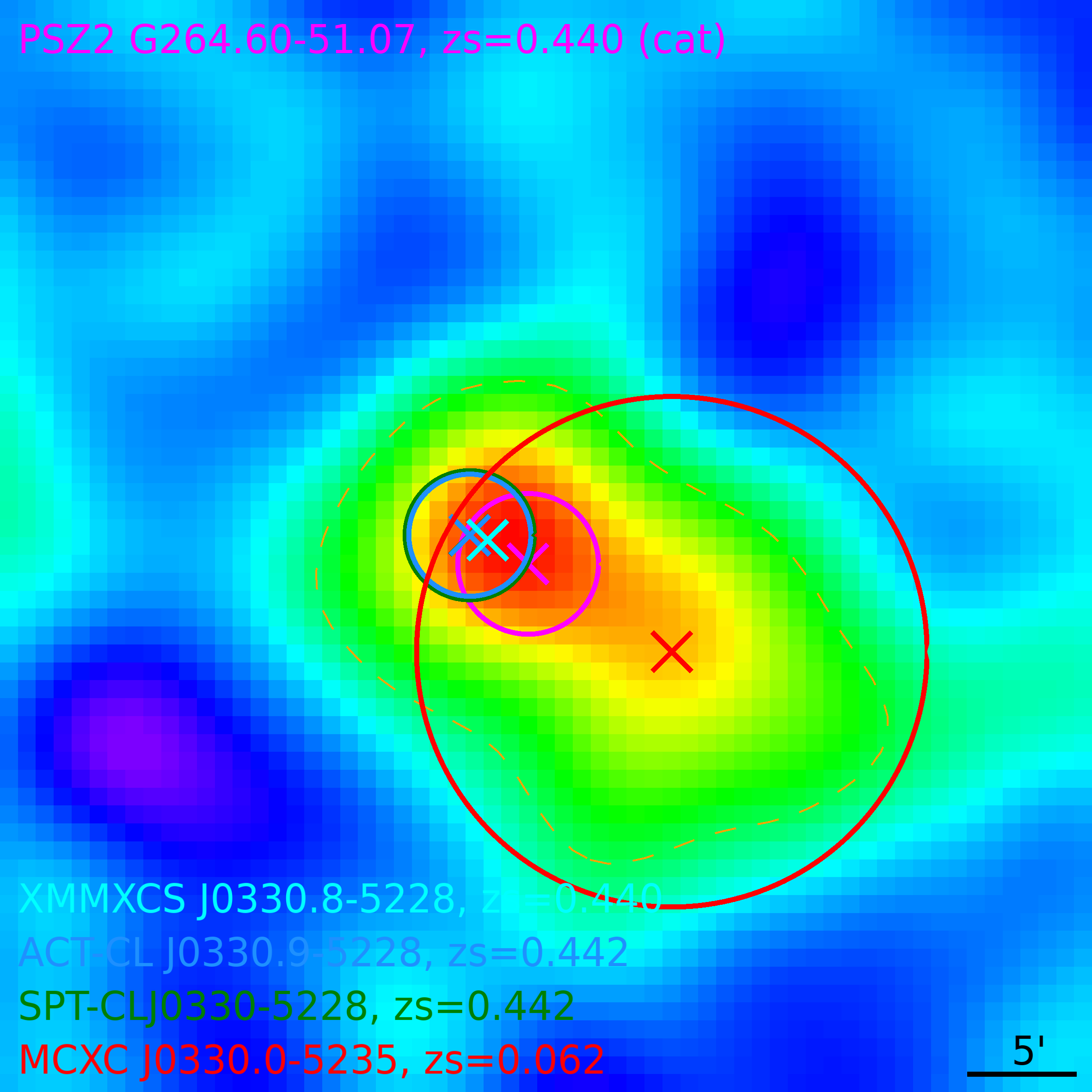}
\includegraphics[width=0.49\columnwidth]{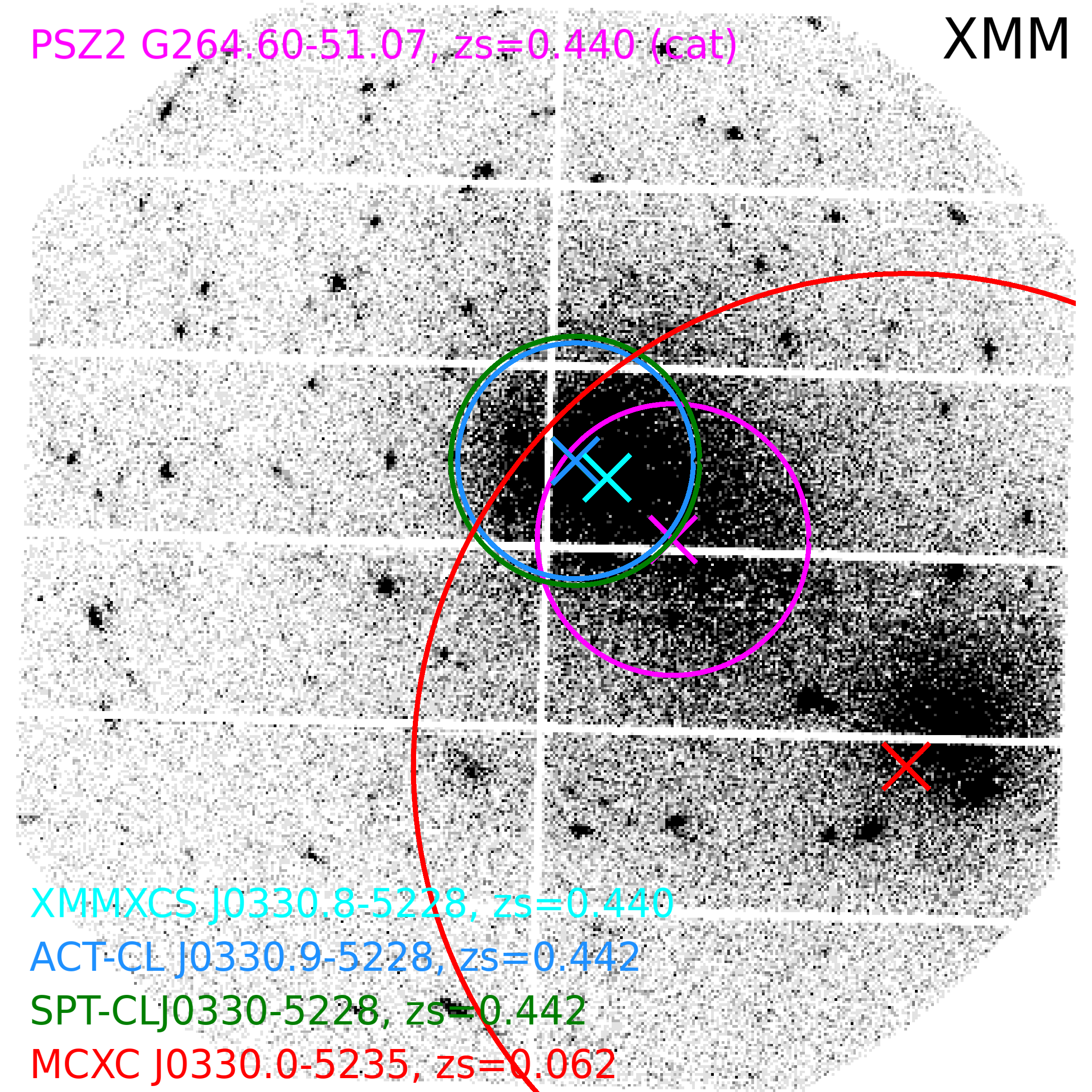}
\caption{\footnotesize \planck\ filtered map (left) and \xmm\ image (right) for \protect\hyperlink{PSZ2 G264.60-51.07}{PSZ2 G264.60-51.07}.   Overlays  as in Fig.~\ref{fig:B1_3images_a}.}
\label{fig:PSZ2 G264.60-51.07}
\end{figure}

\noindent{\bf \hypertarget{PSZ2 G264.60-51.07}{PSZ2 G264.60-51.07}:}
In the original PSZ2 catalogue, PSZ2 G264.60-51.07 was associated with MCXC J0330.0-5235 ($\zs=0.0624$) and ACT-CL-J0330-5227=SPT-CLJ0330-5228 ($\zs=0.442$), noting the SZE projection of the low and high z clusters. The redshift, $\zs=0.44$, was inherited from PSZ1 redshift derived from  cross identification with XMMXCS J0330.8-5228=A3128-NE, incorrectly named as a ROSAT cluster (RXC J0330.8-5228). There is a large distance between the MCXC and PSZ clusters ($D=8.1\arcmin=2.4\Tv$). 
The \xmm\ image (Fig.~\ref{fig:PSZ2 G264.60-51.07}) confirms  the presence of two clusters in this region: the low redshift cluster MCXC J0330.0-5235 in the southwest and the higher redshift cluster in the northeast,  ACT-CL J0330.9-5228=SPT-CLJ0330-5228 coinciding with XMMXCS J0330.8-5228. 
The PSZ2 detection is between the two X--ray peaks, but closer to the ACT/SPT cluster, and  the filtered \planck\ map shows that the \planck\ signal mainly comes from this cluster. We thus  decided to dissociate  PSZ2 G264.60-51.07 and  MCXC J0330.0-5235. 
However, the \planck\ signal is also influenced by this low redshift cluster, so its mass may be overestimated.\\

\begin{figure}[t!]
\centering
\includegraphics[width=0.7\columnwidth]{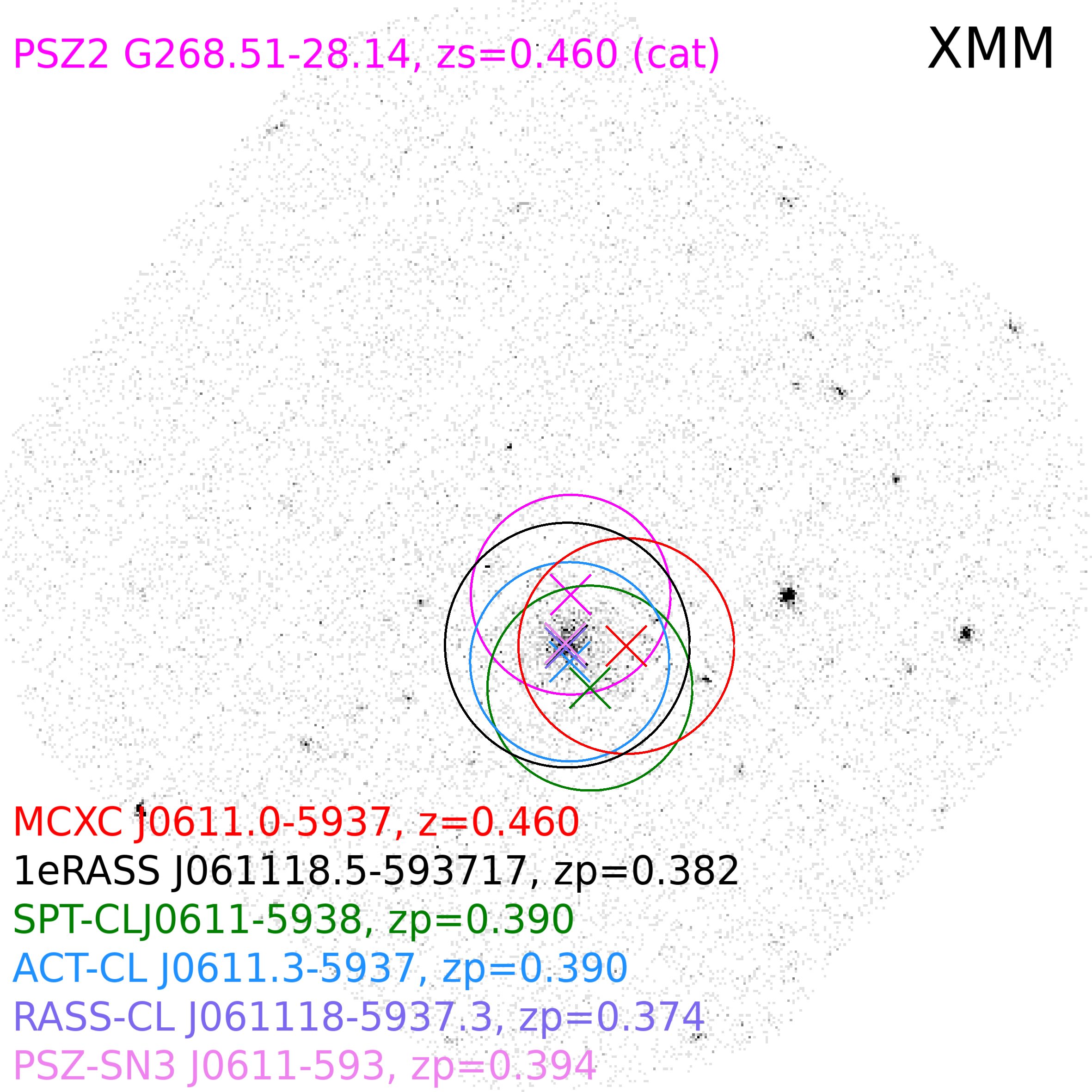}
\caption{\footnotesize \xmm\ image covering cluster \protect\hyperlink{PSZ2 G268.51-28.14}{PSZ2 G268.51-28.14}.
Overlays  as in Fig.~\ref{fig:B1_3images_a}.}
\label{fig:PSZ2 G268.51-28.14}
\end{figure}

\noindent{\bf \hypertarget{PSZ2 G268.51-28.14}{PSZ2 G268.51-28.14}:}
The PSZ2 catalogue redshift, $\zs=0.46$, was obtained from \xmm\ spectroscopy (see Fig.~10 of \cite{xmmfu_pip1}). PSZ2 G268.51-28.14 is cross-matched with MCXC J0611.0-5937 (RXGCC  237) at $D=2.07\arcmin$, whose redshift is taken from the PSZ2 redshift, thus not independent. 
PSZ2 G268.51-28.14 is cross-identified with 1eRASS J061118.5-593717 ($D=1.38\arcmin$, $\zp=0.3817$), and RASS-CL J061118-5937.3 ($D= 1.47\arcmin$, $\zp=  0.3742$). It also close to   ACT-CL J0611.3-5937 and SPT-CLJ0611-5938 at $\zp=0.390$ and $D=1.83\arcmin$ and $D=2.6\arcmin$, respectively. All the $\zp$ values are consistent, as well as with that estimated  for the MMF3 detection in PSZ-MCMF (PSZ-SN3  J0611-593 at  $D=1.33\arcmin$, $\zp= 0.394$). The XMM image (Fig.~\ref{fig:PSZ2 G268.51-28.14}) shows a single extended X--ray emission, confirming the presence of only one cluster and the $\zp$ values lie within the $90\%$ error of the XMM value. We thus associated all these clusters, kept the catalogue value, and included a comment in the catalogue to signal the redshift difference.
\\

\noindent{\bf \hypertarget{PSZ2 G269.02+22.27}{PSZ2 G269.02+22.27}:}
The PSZ2 catalogue redshift,  $\zs=0.3575$, was obtained from NTT follow-up  of PSZ1 cluster from the identified BCG and at least one other galaxy  \citep[][Sect.~5.3.2]{psz1}.
PSZ2 G269.02+22.27 is cross-matched with 1eRASS J102343.8-304106 at $D=1.75\arcmin$ and $\zp=0.4634\pm0.0088$. Although the difference with PSZ2 $\zs$ is statistically significant, it is only slightly outside the  $95\%$ deviation range  between eRASS $\zp$ and PSZ2 $\zs$ ($\dzn=[-0.09,+0.05]$). 
 Taking also the good agreement with  $\zp\,=\,0.3157$ of  WHY J102402.0-303816  at $D=4.9\arcmin$, we conclude that the difference is due to $\zp$ uncertainties and keep the PSZ2 catalogue value.\\ 

\noindent{\bf \hypertarget{PSZ2 G270.78+36.83}{PSZ2 G270.78+36.83}:}   The candidate has been validated by the SDSS study of \citet{Str18} with two components at $\zp=0.52$ and $\zp=0.22$, unresolved by \planck\ (their Fig.~6, right panel).   \citet{Agu19} spectroscopically confirmed  the high $z$ component as a  massive  cluster ($\sigma_{v}=900$km/s) with 25 cluster members  at $\zs=0.516$.  The eRASS counterpart, 1eRASS J110407.2-191350, at $D=1.89\arcmin$, is also likely a case of confusion along the line of sight. The published redshift, $\zp=0.2185$, is that of the low redshift component. Consistently, we found two WHL counterparts, WHL J110407.0-191348 ($D=1.94\arcmin$, $\zp=0.2214$) and WHL J110417.9-191409 ($D=2.02\arcmin$, $\zp=0.4802$), the latter being richer (RL\_500= 49.5) than the former (RL\_500= 33.9).  PSZ2 G270.78+36.83 is also cross-matched with RASS-CL J110407-1913.7 ($D=1.93\arcmin$), with again 2 components at $\zp=0.223$ and $\zp=0.491$. We therefore flag this object as a clear case of confusion along the line-of-sight ({\tt STATUS=Confusion}) and adopt $\zs=0.516$, the redshift of a priori the main component.\\

\begin{figure}[]
\centering
\includegraphics[width=0.49\columnwidth]{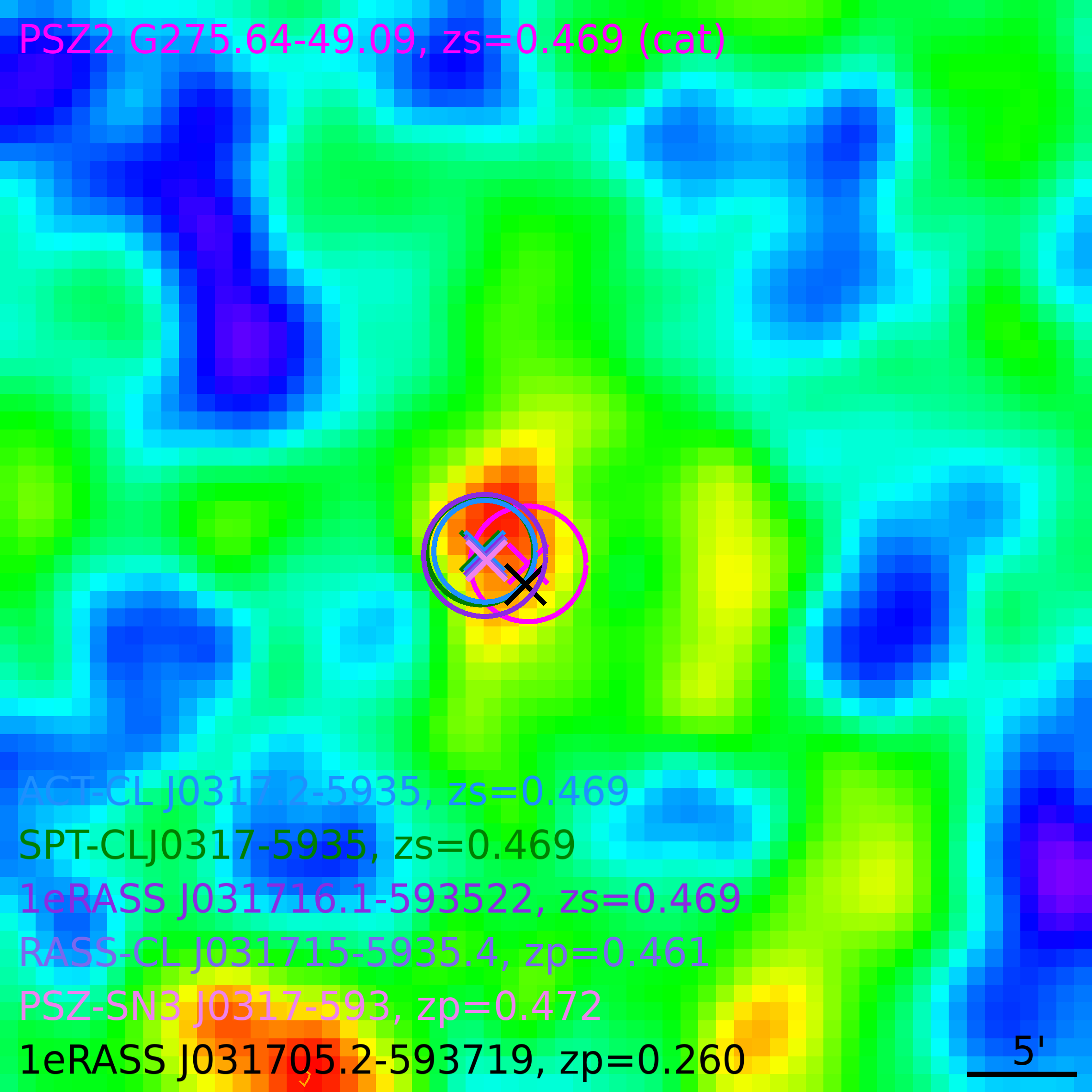}
\includegraphics[width=0.49\columnwidth]{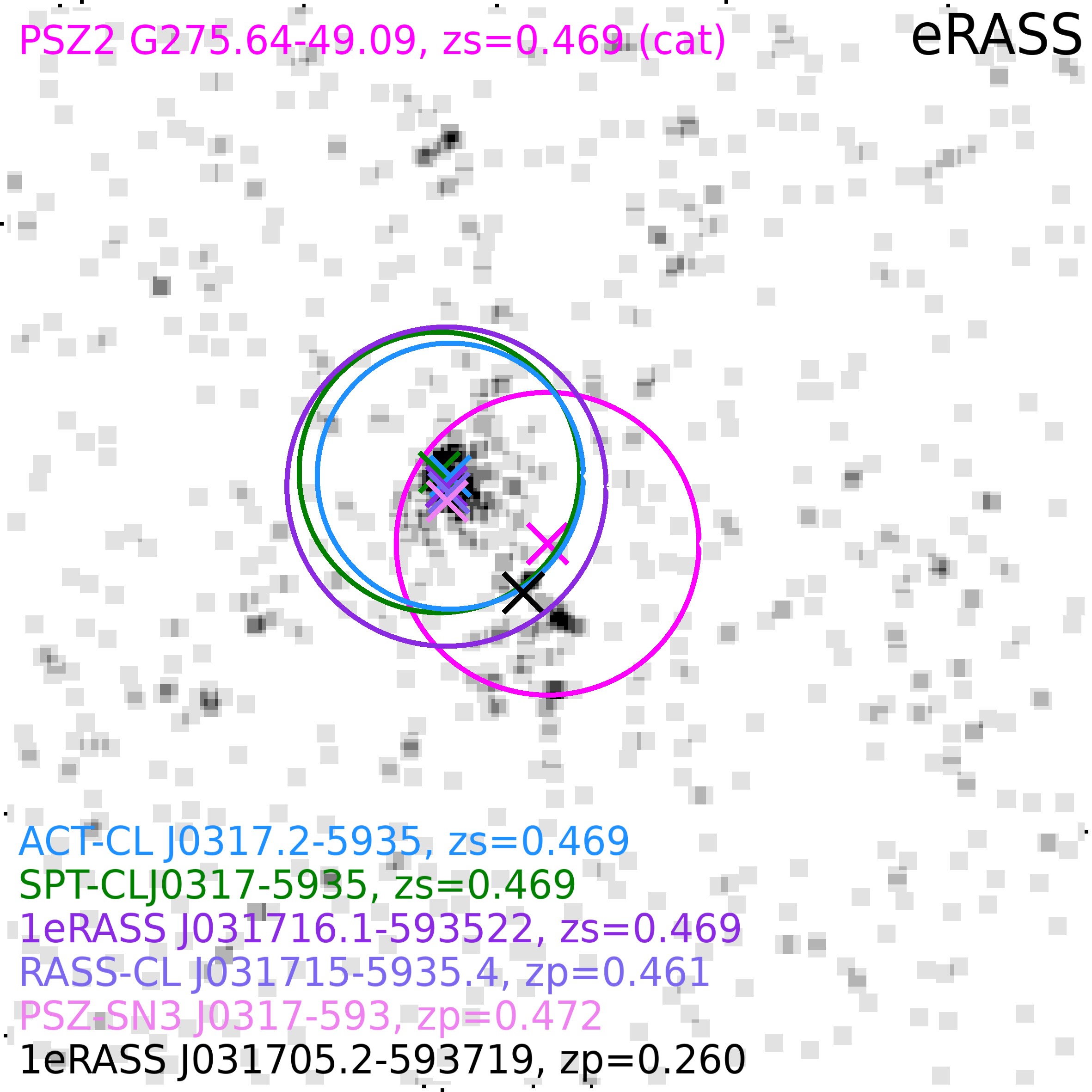}
\caption{\footnotesize 
\planck\ filtered map (left) and eRASS image (right) covering clusters \protect\hyperlink{PSZ2 G275.64-49.09}{PSZ2 G275.64-49.09}.
Overlays as in Fig.~\ref{fig:B1_3images_a}.}
\label{fig:PSZ2 G275.64-49.09}
\end{figure}

\noindent{\bf \hypertarget{PSZ2 G275.64-49.09}{PSZ2 G275.64-49.09}:} The PSZ catalogue redshift, $\zs=0.469$,  is based on cross-identification with SPT-CLJ0317-5935.  PSZ2 G275.64-49.09 is associated with the second closest eRASS cluster (see Fig.~\ref{fig:PSZ2 G275.64-49.09}),  1eRASS J031716.1-593522 at $D\,=\,2.13\arcmin$, at same redshift,  $\zs\,=\,0.4691\pm0.0005$, and consistent $Y_{500}$ and $L_{500}$ values. The closest eRASS cluster, 1eRASS J031705.2-593719, at $D=1.\arcmin$ and $\zp=0.2596\pm0.0066$, is a foreground group (low S/N X--ray emission and no estimated mass). PSZ2 G275.64-49.09 is also cross-identified with ACT-CL J0317.2-5935, RASS-CL J031715-5935.4 at $\zp=0.4607$ and PSZ-SN3 J0317-593 at $\zp=0.4717$. \\

\begin{figure}
\centering
\includegraphics[width=0.49\columnwidth]{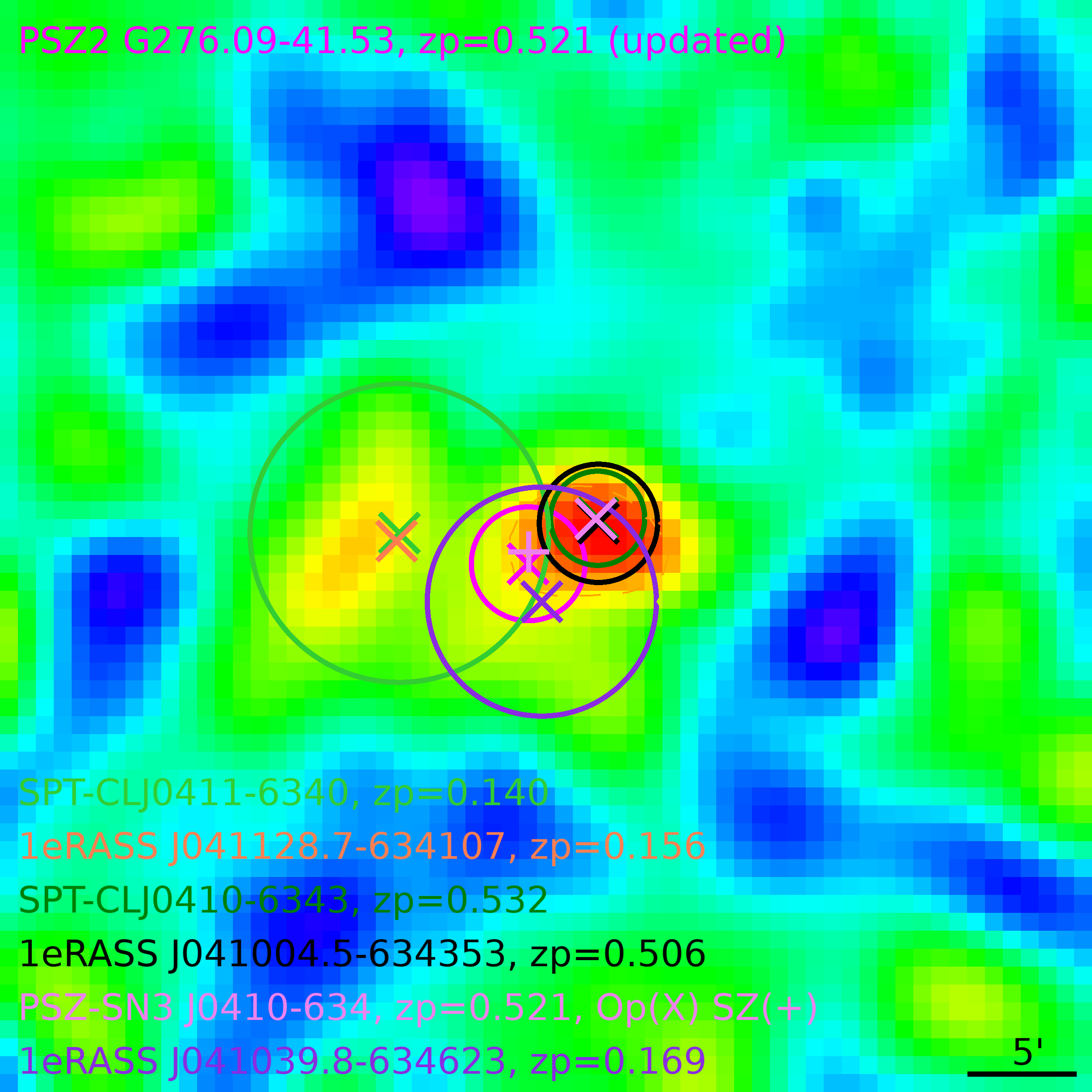}
\includegraphics[width=0.49\columnwidth]{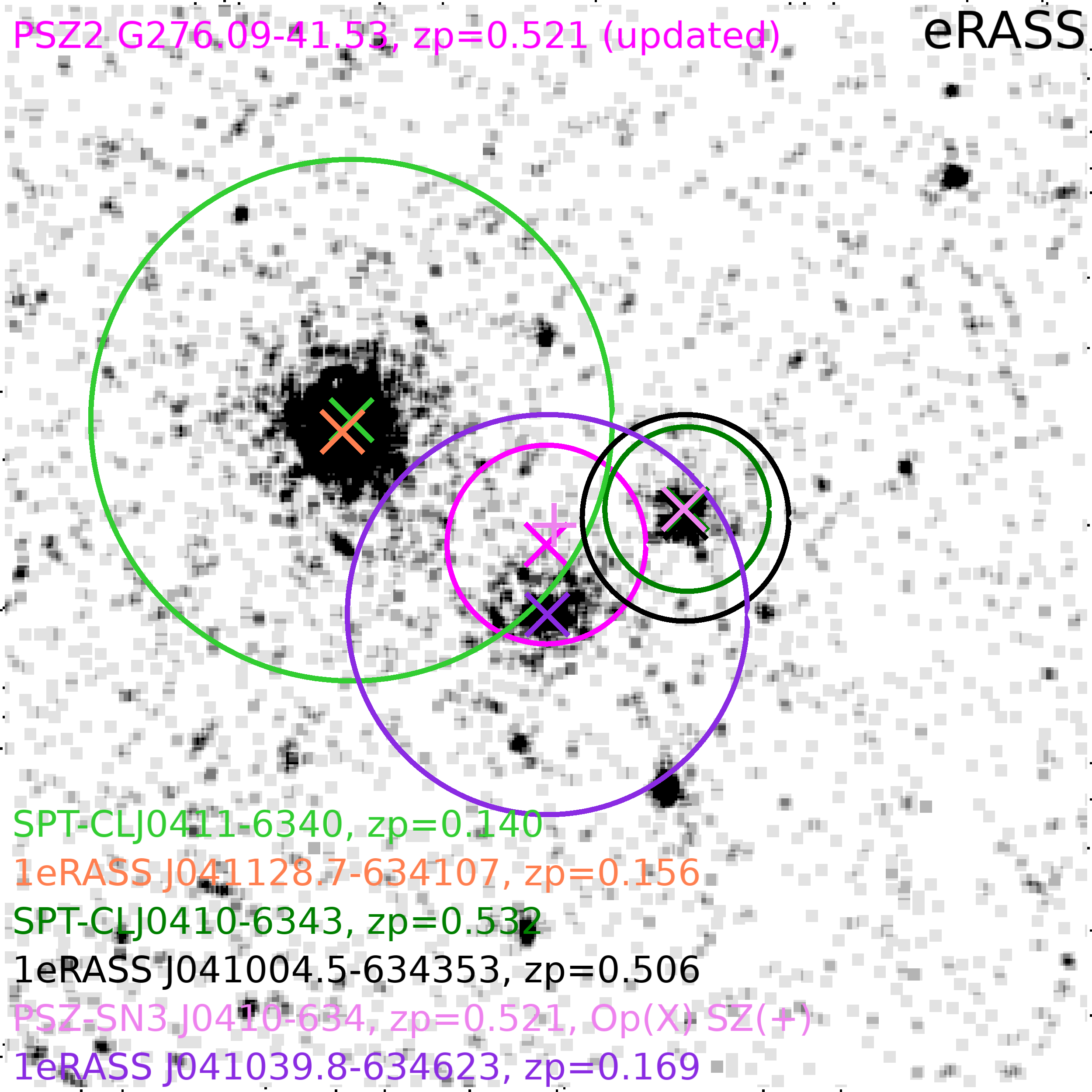}
\caption{\footnotesize Same as Fig.~\ref{fig:PSZ2 G275.64-49.09} for \protect\hyperlink{PSZ2 G276.09-41.53}{PSZ2 G276.09-41.53}. Catalogue redshift $\zp=0.140$ updated to $\zp=0.521$.}
\label{fig:PSZ2 G276.09-41.53}
\end{figure}

\noindent{\bf \hypertarget{PSZ2 G276.09-41.53}{PSZ2 G276.09-41.53:} }
The PSZ2 catalogue redshift ($z_\mathrm{phot}= 0.14$) comes from the association of this cluster with SPT-CLJ0411-6340 in the original PSZ2 catalogue. These clusters are in a complex X--ray emission region (see Fig.~\ref{fig:PSZ2 G276.09-41.53}), where we can distinguish 3 extended emissions. SPT-CLJ0411-6340 ($z_\mathrm{phot}= 0.14$) = 1eRASS J041128.7-634107 ($z_\mathrm{phot}= 0.1556$) is centered on the most extended emission, whereas SPT-CLJ0410-6343 ($z_\mathrm{phot}= 0.532$) = 1eRASS J041004.5-634353 ($z_\mathrm{phot}= 0.5056$) = PSZ-SN3 J0410-634 ($z_\mathrm{phot}= 0.5207$) is centered on the less extended emission on the west. The third emission on the south coincides with 1eRASS J041039.8-634623 ($z_\mathrm{phot}= 0.1688$). The \planck\ detection is between the 3 X--ray peaks, and closer to the one on the south. The \planck\ filtered image shows that the SZE emission mainly comes from the higher redshift cluster on the west (the position of the PSZ2 detection was taken from the PwS algorithm of PSZ2, which can explain the miscentering with respect to the MMF3 peak seen in the figure), and not from SPT-CLJ0411-6340. Therefore, we have decided to update the redshift of the cluster to $z_\mathrm{phot}= 0.5207$. \\ 

\noindent{\bf \hypertarget{PSZ1 G279.00-24.89}{PSZ1 G279.00-24.89}:} The PSZ catalogue redshift  $\zp\,=\,0.14$ was obtained by the optical follow-up of PSZ1 clusters with the Wide-Field Imager (WFI) at MPG/ESO 2.2-m telescope.
PSZ1 G279.00-24.89 is 
identified with two eRASS clusters at the same redshift ($\zs\,=\,0.0982$),  1eRASS J065405.9-682524 at $D\,=\,2.7\arcmin$ and 1eRASS J065258.7-6827 at $D\,=\,4.1\arcmin$. The latter is $>20$ times more massive than the former, and is likely a merging group.  The total eRASS mass is consistent with the PSZ mass computed at $\zs\,=\,0.0982$ and we adopted this redshift value.  We note that the \planck\ SZE signal peaks at the lowest mass system with extension towards the more massive component. PSZ1 G279.00-24.89 is also coincident with RASS-CL J065258-6827.2=WHY J065258.7-682713 ($D=4.1\arcmin$) at consistent $\zs=0.0995$. \\

\noindent{\bf \hypertarget{PSZ2 G292.74+33.49}{PSZ2 G292.74+33.49}:} In the PSZ2 catalogue, it is associated  with A3506, with $z=0.08157$ taken from NED. The eRASS image shows two close-by clusters 1eRASS J121251.4-283956=A3506 ($\zs=0.0816$) and 1eRASS J121255.6-283520 ($\zp=0.2451$), at $D=3.5\arcmin$  and $D=2.7\arcmin$ from the SZE position, respectively. A slightly higher SZE flux is  expected from the former than the latter, in view of their respective X--ray luminosity. We thus set {\tt STATUS=Confusion} for this source.\\

\noindent{\bf \hypertarget{PSZ2 G285.87-74.93}{PSZ2 G285.87-74.93}:} The PSZ2 redshift, $\zs=0.213$, was obtained from NTT follow-up of PSZ1 cluster from BCG identified from pre-imaging and at least one other galaxy \citep[][Sect.~5.3.2]{psz1}.  The cluster can be identified with clusters from several surveys:  1eRASS J011444.3-412337 ($D=1.28\arcmin$, $\zp=0.3836$),   WHY J011443.3-412357  ($D=1.10\arcmin,\zp=0.2957 $), 
ACT-CL J0114.7-4123  ($D=1.39\arcmin,\zp=0.3791$)  and  SPT-CLJ0114-4123 ($D=1.19\arcmin, \zp=0.3910 $). The PSZ-MCMF catalogue provides a similar redshift, $\zp=0.3845$, for PSZ-SN3 J0114-412 ($D=1.41\arcmin$). All counterparts are at distance less than $1.5\arcmin$, with a single cluster  visible in the \xmm\ image (same case as in Fig.~\ref{fig:PSZ2 G268.51-28.14}), and have consistent photometric redshifts,  higher than NTT $\zs$. Fitting the available spectrum from the source in the 4XMM-DR14 catalogue, we estimated a X--ray $\zs=0.404\pm0.04$, confirming the photometric values. We thus conclude that the NTT $\zs$ is that of foreground galaxies and we adopted the eRASS value. \\

\noindent {\bf \hypertarget{PSZ2 G307.72-77.87}{PSZ2 G307.72-77.87}:}
The PSZ2 redshift, $\zp= 0.45$, is a photometric estimation obtained by the optical follow-up of PSZ1 clusters with the Wide-Field Imager (WFI) at MPG/ESO 2.2-m telescope.  The cluster can be matched with 1eRASS J004621.6-391157,  ACT-CL J0046.4-3911, SPT-CLJ0046-3911, and  the optical counterpart of PSZ-SN3 J0046-39 at $D=1.4, 1.9,2.1$ and $1.45\arcmin$, respectively, and  similar, higher,  redshifts:  $\zp=0.588\pm0.006$,  $\zs=0.5920$,  $\zp=0.593$ and  $\zp=  0.5907$, respectively. We have  thus updated the redshift to the ACT spectroscopic value, $\zs=0.5920$.
\\

\noindent{\bf \hypertarget{PSZ2 G357.75-41.77}{PSZ2 G357.75-41.77:}}  The PSZ catalogue redshift comes from  cross-identification with RXC J2103.4-4319, a REFLEX cluster identified with A3736 by \citet{reflex} at $\zs=0.0487$ from \citet{1994ApJ...425..418L}. This redshift is that of the BCG identified by \citet{1995ApJ...440...28P}, defined as the brightest galaxy within a large aperture of  $5h^{-1}$Mpc in radius. They did not publish redshifts of other cluster members. From their table 1, this BCG is located at $D=20.3\arcmin=1.55$Mpc (at $z=0.0487$) from A3736 optical center \citep{1989ApJS...70....1A}. A search in NED shows that this BCG is actually the central galaxy of another cluster, A S0919, with the same $z=0.0487$ in  the Abell catalogue. The redshift of RXC J2103.4-4319=A3567 is thus erroneous.  We cross-identified PSZ2 G357.75-41.77 with 1eRASS J210326.7-431934 at $D= 2.96\arcmin$ and $\zp=0.1434\pm0.0043$  and adopted that redshift. PSZ2 G357.75-41.77 also matches with clusters from other catalogues at consistent redshift: RASS-CL J210326-4320.7 ($\zp=0.1411$), PSZ-SN3 J2103-432 ($\zp=0.1385$), WHY J210328.6-432036 ($z\,=\,0.151$), and ACT-CL J2103.4-4321 ($z_{\rm phot}=0.140$). We associated these PSZ2 and ACT clusters with MCXC J2103.4-4319 ($z_{\rm spec}=0.049$), despite the wrong MCXC redshift. We included a note in the PSZ catalogue to explain the redshift update.

\subsection{Comparison with redshift from \citet{bur17}} \label{app:bur17}
\noindent{\bf \hypertarget{PSZ2 G039.34+73.28}{PSZ2\,G039.34+73.28}:}
The PSZ2 redshift is from cross-identification  with RMJ140649.4+274556.8  at $\zp= 0.5664$, not listed in the published redMaPPer catalogue. However, the optical counterpart found by \citet{bur17} at $\zs=0.6752$ coincides  ($D= 0.2\arcmin$) with WHL J140637.2+274351 at the same redshift ($\zs=0.6752$).  WHL J140637.2+274351 is a rich cluster ($\lambda= 72$) located at $2.02\arcmin$ from the PSZ2 position. We thus update the PSZ value. RMJ140649.4+274556.8 and WHL J140637.2+274351 are likely two different clusters ($D= 3.4\arcmin$). RMJ140649.4+274556.8 may contribute to the SZE signal, depending on its richness. \\
 
\noindent{\bf \hypertarget{PSZ2 G199.75+46.59}{PSZ2\,G199.75+46.59}:}  The PSZ catalogue redshift  $\zp=0.554$ comes from PSZ1 search in SDSS-DR9 data, with no further information on the counterpart.  The  center of the optical counterpart found by \citet{bur17} is $3.8\arcmin$ away from the PSZ2 position, a distance  smaller than the SZE position large  uncertainty ($4.8\arcmin$). The spectroscopic redshift, $\zs=0.706$,  is  based on one galaxy, likely  SDSS J093352.36+280338.4 at that redshift ($\zs =0.7061$) located $3\arcsec$ from the optical center. We thus update the PSZ value. \\

\noindent{\bf \hypertarget{PSZ2 G317.52+59.94}{PSZ2 G317.52+59.94}:}
The PSZ2 catalogue redshift, $\zp=0.3156$ is  from redMaPPer non blind search. Two SDSS clusters are within $5\arcmin$ of the SZE position: WHL J132035.4-020723 ($\zs=0.2809$) at $D=2.5\arcmin$ and  RMJ132026.6-021038.1 ($\zs=0.2264$) at $D=4.7\arcmin$ farther away in the south. The  optical counterpart found by \citet{bur17} coincides with the latter ($D=5\arcsec$  and same $\zs$). 
PSZ2 G317.52+59.94 can be identified with 1eRASS J132026.1-020800 at $\zp=0.2798$  at 2.0$\arcmin$ from the SZE position. WHL J132035.4-020723 is between the X--ray and SZE position, with a consistent redshift. We thus updated the catalogue redshift to the WHL cluster value, noting that RMJ132026.6-021038.1 may contribute to the SZE signal.

\subsection{Comparison with catalogue of  \citet{BH24}} \label{app:BH24}

\subsubsection{Redshift mismatch}
The following clusters  are  two strong outliers when comparing PSZ redshift and redshift from \citet{BH24}.\\

\noindent{\bf \hypertarget{PSZ2 G065.45+78.10}{PSZ2 G065.45+78.10}:} The PSZ catalogue redshift, $\zp=0.2730$, is based on cross-identification with GMBCG J204.74580+32.97396 (at $D=4.2\arcmin$) from NED search.   The optical counterpart from \citet{bur17} adopted by BH24, is at $\zs=0.4885$ (from 3 galaxies), and $3.5\arcmin$ from the SZE position. It is close ($D=1.2\arcmin$) to WHL J090655.0+430249 at the same redshift, $\zs=0.4863$ from 5 galaxies \citep{2015ApJ...807..178W}, and may be the same object. Furthermore, a search in NED within $6\arcmin$ of the PSZ position reveals only one spectroscopic galaxy around $z=0.27$, exactly  at the GMBCG  location and redshift. The latter is likely based on a foreground galaxy. The redshift is thus updated to the WHL value.\\

\begin{figure}[t]
\centering
\includegraphics[width=0.70\columnwidth]{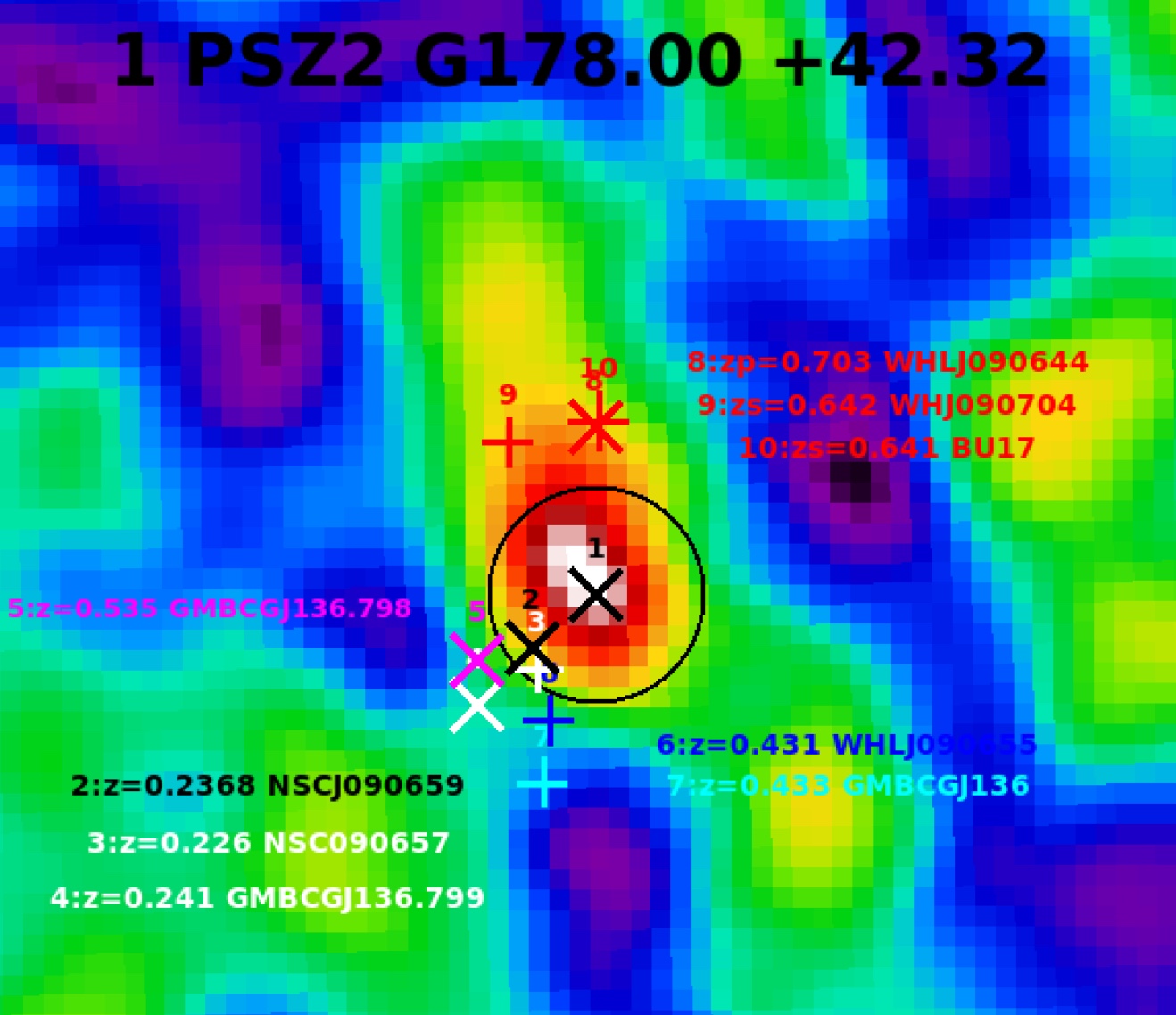}
\caption{\footnotesize \planck\ image of \protect\hyperlink{PSZ2 G178.00+42.32}{PSZ2 G178.00+42.32}. The position of the  clusters from various optical surveys are indicated in the figure and labelled with their redshift. }
\label{fig:PSZ2G178}
\end{figure}

\noindent{\bf \hypertarget{PSZ2 G178.00+42.32}{PSZ2 G178.00+42.32}: }
The PSZ catalogue redshift, $\zp=0.2368$, is based on cross-identification   with  NSC J090659+430556, the closest optical cluster at $D=3.5\arcmin$ in the southeast. The optical counterpart from \citet{bur17} adopted by BH24,   at $\zs=0.6408$ (from 3 galaxies), is further away in the north at  $7.36\arcmin$ from the SZE position. 
Notably, several clusters are detected in optical surveys within $8\arcmin$ from the SZE position (see Fig.~\ref{fig:PSZ2G178}): 
\begin{itemize}[noitemsep,topsep=0pt,label=$-$]
\item A high-z structure in the north:  WH J090704.6+431438   at the same redshift than the  \citet{bur17} counterpart ($\zs=0.6420$) but $D=3.75\arcmin$ away, and WHL J090644.3+431516 at a different redshift ($\zp=0.7028$)  but at the same position.  It is unclear whether this is a single object and/or merging clusters.  \item  3 clusters at  $z\sim0.24$, the original PSZ2 redshift, in the southeast: NSC J090659+430556 ($\zp=0.2368$, $D=3.5\arcmin$),  NSC J090657+430502 ($\zp=0.226$, $D=3.97\arcmin$) and GMBCG J136.79957+43.05787 ($\zp=0.241$, $D=6.9\arcmin$), where $D$ is the distance to the SZE center. They may belong to the same structure. 
\item   WHL J090655.0+430249 ($\zs=0.4306$, $D=7.3\arcmin$) and GMBCG J136.73454+43.00281 ($\zp=0.433$, $D=8.3\arcmin$), likely the same object. 
\end{itemize}
It is likely that all clusters contribute to the SZE emission. The high-z component appears rich in a combined  PanSTARRS WISE image and the \planck\ image shows an elongation towards that direction. We  flag the PSZ2 object as a case of {\tt STATUS=Confusion}, with an assumed main contribution from the high-z component. 

\subsubsection{Validation status mismatch }\label{app:BH24status}

\begin{figure}[t]
\centering
\includegraphics[width=0.49\columnwidth]{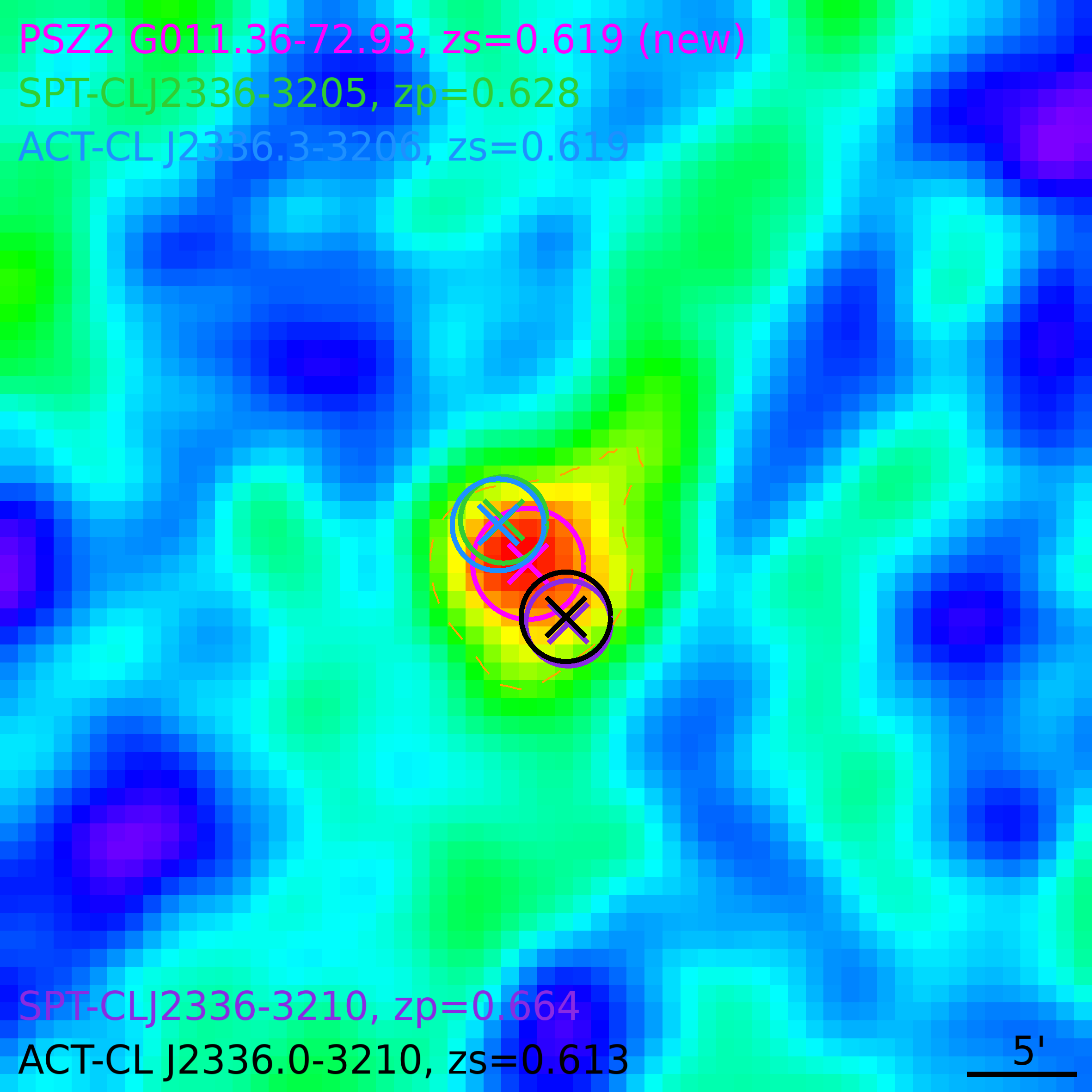}
\includegraphics[width=0.49\columnwidth]{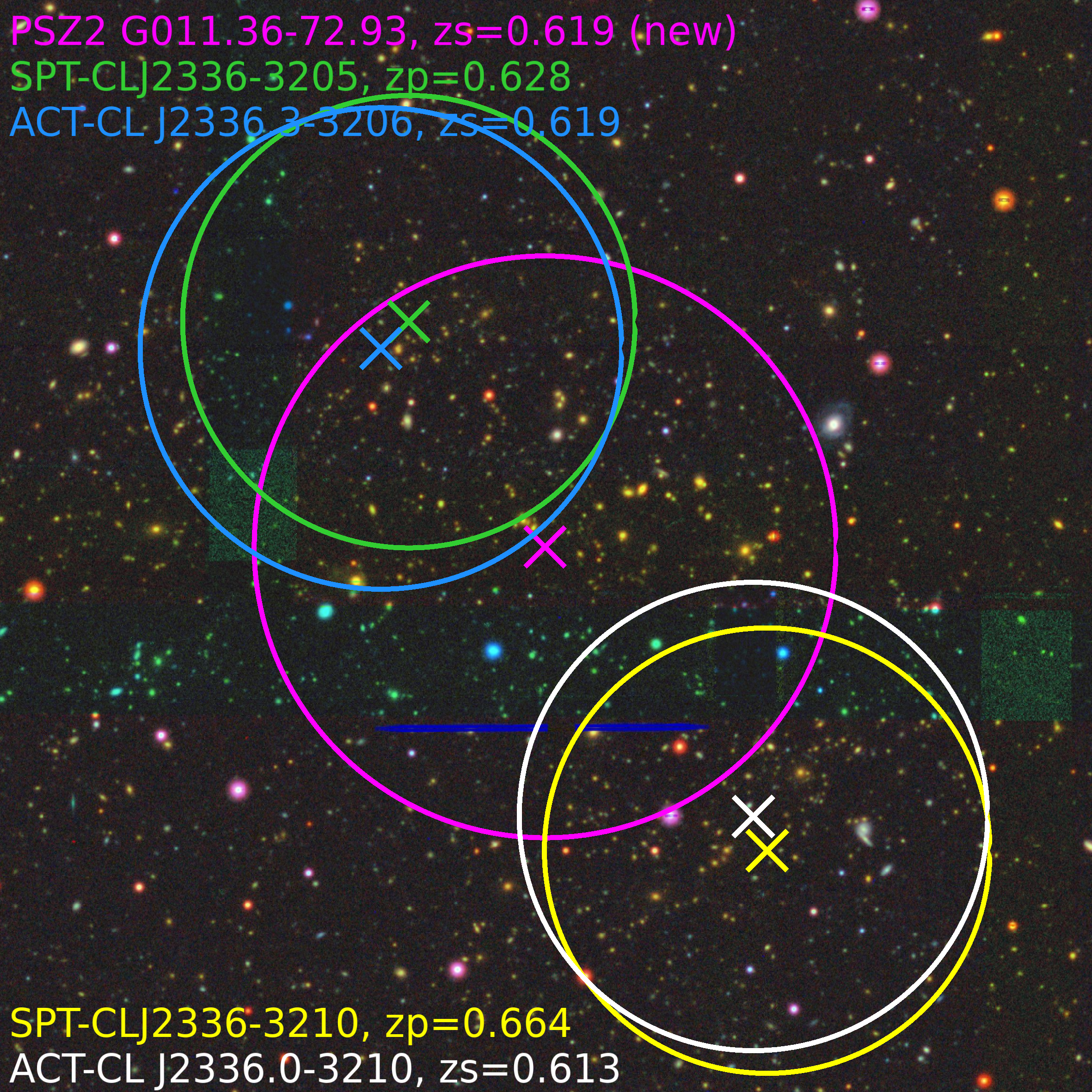}
\caption{\footnotesize \planck\ fileterd map (left) and DESI image (right) covering clusters \protect\hyperlink{PSZ2 G011.36-72.93}{PSZ2 G011.36-72.93}, ACT-CL J2336.3-3206 ($z_{\rm spec}=0.6192$) = SPT-CLJ2336-3205, and ACT-CL J2336.0-3210 ($z_{\rm spec}=0.6133$) = SPT-CLJ2336-3210. } 
\label{fig:PSZ2 G011.36-72.93}
\end{figure}

\noindent{\bf \hypertarget{PSZ2 G011.36-72.93}{PSZ2 G011.36-72.93}:} This cluster is validated by BH24 from cross-match with SPT. Fig.~\ref{fig:PSZ2 G011.36-72.93} shows that this candidate  corresponds to the superposition of two ACT/SPT clusters at the same redshift: ACT-CL J2336.3-3206 ($z_{\rm spec}=0.6192$) = SPT-CLJ2336-3205 on the northeast, and ACT-CL J2336.0-3210 ($z_{\rm spec}=0.6133$) = SPT-CLJ2336-3210 on the southwest. The PSZ2 detection is located between the two clusters, at $D=2.36 \arcmin$ from ACT-CL J2336.3-3206, and $D=3.12\arcmin$ from ACT-CL J2336.0-3210. Both clusters have similar masses (4.06 and 3.66, respectively), and the \planck\ SZE signal covers both of them, so there is no clear main contribution. We have assigned the redshift of the closest and more massive cluster (ACT-CL J2336.3-3206) to the PSZ2 candidate, and set {\tt STATUS = Complex}.\\

\noindent{\bf \hypertarget{PSZ2 G014.72-62.49}{PSZ2 G014.72-62.49}:}
This cluster,  without redshift in PSZ2 catalogue, is validated by BH24 from cross-match with SPT.  The PSZ2 detection is located towards the southwest of  SPT-CLJ2246-3210=ACT-CL J2246.7-3210  ($\zp=0.50$) at $D=3.78'=1.59\theta_{500}^{\rm ACT}$, 1.5 times the PSZ2 position error ($\epsilon_{\rm pos}=2.43'$). However, the \planck\ SZE map shows an extension of the SZE signal towards the ACT/SPT cluster, which lies inside the S/N=3 contour. The PSZ2 mass computed at the ACT redshift is compatible with the ACT and SPT masses within $3\sigma_{\rm tot}$. Thus, we confirm the SPT/ACT association and assign the redshift to $\zp=0.50$.  \\

\noindent {\bf \hypertarget{PSZ2 G133.92-42.73}{PSZ2 G133.92-42.73}.} We originally classified PSZ2 G133.92-42.73 as a false candidate (see Sect.~\ref{sec:PSZupfu}), its potential optical counterpart being  discarded by \citet{Agu19}. However, the presence of ACT-CL J0125.5+1923, at $D=1.53'$ from PSZ2 G133.92-42.73, with a redshift that is compatible with the redshift of the optical counterpart of \citet{Agu19} and that was obtained independently from photometric data from the Dark Energy Camera Legacy Survey, confirms the cluster. Nevertheless, the mass of the PSZ2 cluster recomputed at $z=0.581$ is three times higher than the ACT mass ($M_{500}^{\rm PSZ2}=7.08\cdot10^{14} M_{\odot}$, $M_{500}^{\rm ACT}=2.55\cdot10^{14} \msun$), so we have set {\tt STATUS=C2}. \\

The following four clusters are the clusters newly validated by \citet{BH24} as "Strong candidates", from their study of galaxy redshift catalogues (their Sect.~3.3), and that have a {\tt STATUS=U} in the updated PSZ catalogue (Sect.~\ref{sec:BH24status}). \\

\noindent{\bf \hypertarget{PSZ2 G031.37-71.95}{PSZ2 G031.37-71.95}:}  From galaxy redshift catalogue search,  BH24 estimated a redshift of $\zs=0.137$. The $\zs$ histogram of galaxies within $15\arcmin$ shows  a strong peak at that value (their Fig.~3.). The optical counterpart is likely WHY J233054.0-260920, centered   $2.54\arcmin$ from the SZE position and at the same  redshift, $\zs=  0.1373$ (from one galaxy).  We note that WHY J233021.1-260801, a slightly less  rich cluster ($\lambda=32$ versus $\lambda=47$) at similar redshift ($\zs=  0.1429$) and  located at 5.5\arcmin from the SZE position,  may also contribute to the SZE signal. PSZ2 G031.37-71.95 is also detected in ComPRASS as PSZRX G031.49-72.00 and the SWIFT image shows a faint diffuse emission $\sim2\arcmin$ from the SZE center. 
We thus confirm the cluster with {\tt STATUS = C1} and  $\zs=0.137$ from \citet{BH24}.\\

\noindent{\bf \hypertarget{PSZ2 G327.27+11.05}{PSZ2 G327.27+11.05}: } From galaxy redshift catalogue search, BH24 estimated a redshift of $\zs=0.034$, from  10 galaxies within a search radius of $15\arcmin$. The cluster is outside the SDSS footprint. PSZ2 G327.27+11.05 is also detected in ComPRASS as PSZRX G327.26+11.05 and the SWIFT image shows a faint diffuse emission $\sim2\arcmin$ from the SZE center. Thus, PSZ2 G327.27+11.05 is  likely a genuine cluster. However, the SZE  and X--ray morphology appear very compact for such a low redshift, with a size $\ts\sim1\arcmin$ from degeneracy contours, far away from the estimated $\Tv=15\arcmin$ at $z=0.034$. This redshift, which cannot be consolidated from SZE and mass proxy  comparison, is likely that of a foreground structure. We thus kept the status to {\tt STATUS=U}. \\

\noindent{\bf \hypertarget{PSZ2 G017.25-70.71}{PSZ2 G017.25-70.71}: } From galaxy redshift catalogue search, BH24 estimated a redshift of $\zs=0.139$. The cluster is outside the SDSS footprint.  The SZE morphology appears very compact. We  changed the status to confirmed  with {\tt STATUS=C2} by lack of mass consolidation. \\

\noindent{\bf \hypertarget{PSZ2 G225.18-33.61}{PSZ2 G225.18-33.61}:} It is newly validated by BH24 at $\zs=0.042$. The eRASS cluster 1eRASS J050408.8-241051 ($\zs=0.1640$) is only $2.8\arcmin$ from the SZE position. However, we discarded it as a possible counterpart (Sect.~\ref{sec:PSZerass}) in view of its low S/N<3 and too low luminosity as compared to the SZE signal. Furthermore, the quality flag of PSZ2 G225.18-33.61 is bad $Q_{\rm neural}=0.55$ and we concluded that it is likely a false candidate. We thus keep its status as {\tt STATUS=U}. 
We also note that the second closest eRASS cluster is 
1eRASS J050401.3-235947 at $11.5\arcmin$. It coincides with the MCXC-II cluster MCXC J0504.0-2400 from the RXGCC catalogue, at $z=0.0426$, which was not retained as a possible counterpart in view of its distance (Sect.~\ref{sec:mcxc}). 
The galaxies selected by BH24, at similar redshift,  are likely from the periphery of that cluster  in view of their large search radius.

\section{Special cases in Mass determination}
\label{app:mass}
We detail specific  cases where the degeneracy $Y_{5R500}(\ts)$ contours from the SZE extraction algorithms are far from the X-ray prior, the $Y_{5R500}$--$\ts$ scaling relation (Sect.~\ref{sec:newmass} and Sect.~\ref{sec:corrmass}).

The following objects are genuine clusters, confirmed from independent evidence. Their characteristics explain this mismatch between data and model: \\

\noindent{\bf \hypertarget{PSZ2 G006.84+50.69}{PSZ2 G006.84+50.69}:} It was validated in the PSZ2 catalogue with $\zp=0.0757$ from a {\tt redMaPPer} search at the candidate location. It is located in between A2029=PSZ2 G006.49+50.56 ($D=15.3\arcmin$ in the south) and A2023 ($D=21.4\arcmin$ in the north) at similar redshift,  $\zs=0.0767$ and  $\zs=0.0822$, respectively. The \planck\ signal gives a very large size $\ts>30\arcmin$ at the 90\% confidence level, and likely corresponds to the detection of the complex as a whole. We thus set {\tt STATUS=Complex} for this object, with no mass estimate ($\Mv=-1$). \\

\noindent{\bf \hypertarget{PSZ2 G078.67+20.06}{PSZ2 G078.67+20.06}:} This cluster, at $\zs=0.605$ from NOT optical follow-up, coincides with RASS-CL J185202+4901.2 at consistent redshift ($\zp=0.581$) and mass. However, 
the SZE map of PSZ2 G078.67+20.06 is affected by a point source.\\

\noindent{\bf \hypertarget{PSZ2 G079.88+14.97}{PSZ2 G079.88+14.97}:} This cluster, at $\zs=0.0998$ from PSZ1 optical follow-up, coincides with WHY J192320.7+480957 at consistent  $\zp=0.1202$. It  was noted as X-ray under-luminous in the PSZ2 catalogue. \\

\noindent{\bf \hypertarget{PSZ2 G294.89-37.19}{PSZ2 G294.89-37.19}:} This cluster, at $\zp=0.47$ from PSZ1 optical follow-up, coincides with  1eRASS J025923.0-775206  at consistent redshift ($\zp=0.5346$)  and mass. Its SZE detection may be contaminated by the large cluster PSZ2 G294.68-37.01=MCXC J0303.7-7752 ($\zs=0.274$), at $D=14.7\arcmin$ to the east, biasing the $\Ytot(\ts)$ contours to high size values. \\

\noindent{\bf \hypertarget{PSZ2 G246.49-35.31}{PSZ2 G246.49-35.31}:} It was identified with APMCC 583 at $z=0.0810$ in the PSZ2 catalogue, and coincides with 1eRASS J051336.8-413841  ($\zs=0.0850$) and  PSZ-SN3 J0513-414 ($\zp=0.073$), at consistent z. It was noted as X-ray under-luminous in the PSZ2 catalogue. \\

We consider the following 4 sources as spurious detections:\\

\noindent{\bf \hypertarget{PSZ2 G087.25-41.86}{PSZ2 G087.25-41.86}:}  From their ENO follow-up, \citet{Planck_PSZ1CanaryFollowup} identified PSZ2 G087.25-41.86 with a fossil group at $z=0.048$.  However, the source is detected by only one method, with an extremely low $Q_{\rm neural}$ value of less than $10^{-3}$, the identification was not consolidated with a  mass proxy estimate, and the fossil group is not visible in the RASS. We set {\tt STATUS=False} for this  object.\\

\noindent{\bf \hypertarget{PSZ2 G123.35+25.39}{PSZ2 G123.35+25.39}:} \citet{Bar18} found a optical counterpart at $\zp=0.95$ of Class=2 ({\tt STATUS=C2}). Despite this high redshift, the source is detected at very high S/N (10.9) and large SZE extension ($\sim 10~{\rm arcmin}$).  Furthermore, the $Q_{\rm neural}$ value is very low (0.025) and there is a strong CO cloud at the source location, likely to be the origin of the \planck\ signal.  We thus set {\tt STATUS=False} for this detection. \\

\noindent{\bf \hypertarget{PSZ2 G146.82+40.97}{PSZ2 G146.82+40.97}:} It is identified with GMBCG J144.26171+65.9060 ($z=0.259$) in the PSZ2 catalogue, is only detected with PWS at $Q_{\rm neural}=0.0015$, and at a large  distance of 4.6\arcmin\ from the proposed counterpart. We set {\tt STATUS=False} for this object.\\

\noindent{\bf \hypertarget{PSZ2 G153.29+36.56}{PSZ2 G153.29+36.56}:} From their ENO follow-up, \citet{Planck_PSZ1CanaryFollowup} identified the source with the confusion of two distant clusters at $\zs=0.65$ and $z=0.825$.  However, the source is detected by only one method, with an extremely low $Q_{\rm neural}$ value of $2\,10^{-4}$ and  the identification was not consolidated with a  mass proxy estimate.  We set {\tt STATUS=False} for this object.\\

\noindent{\bf \hypertarget{PSZ2 G250.17+73.53}{PSZ2 G250.17+73.53}:} It was validated from a {\tt redMaPPer} search at the candidate location.  The PSZ2 redshift, $\zp=0.2751$, coincides with that of RMJ115706.0+162935.9=WHL J115706.0+162936 at $D=3\arcmin$. However, the detection is highly spurious with a low-quality SZE signal ($Q_{\rm neural}=0.24$) and very large $\ts$ contour value ($\ts>40\arcmin$), requiring extreme extrapolation. The new mass estimate is only 20\% of the original catalogue value. The PSZ2 detection is likely dominated by noise and we  set {\tt STATUS=False} for this object.

\section{OBSID of \xmm\ observations}\label{app:xmmobsid}

Table~\ref{tab:OBSID} gives the \xmm\ OBSIDs that were used in catalogue validation.

\begin{table}[!ht]
\caption{ \label{tab:OBSID}  List of \xmm\ observations used in this work.}
\centering
\tiny
\begin{tabular}{llll}
\toprule
\planck\ name  &  OBSID  & \planck\ name  &  OBSID \\
\midrule
PSZ2 G028.63+50.15 & 0821810401 & PSZ2 G210.37-37.00 & 0822590701 \\
PSZ2 G029.80-17.40 & 0822591701 & PSZ2 G228.35-66.31 & 0822591601 \\
PSZ2 G032.12-14.96 & 0822591501 & PSZ2 G230.28-28.57 & 0822591101 \\
PSZ2 G035.89-61.39 & 0821871601 & PSZ2 G243.00-65.94 & 0822591401 \\
PSZ2 G044.83+10.02 & 0822591301 & PSZ2 G254.52+62.52 & 0943530501 \\
PSZ2 G058.31+41.96 & 0741033301 & PSZ2 G264.60-51.07 & 0400130101 \\
PSZ2 G075.08+19.83 & 0600830501 & PSZ2 G268.51-28.14 & 0658201101 \\
PSZ2 G078.67+20.06 & 0900510301 & PSZ2 G278.74-45.26 & 0084960201 \\
PSZ2 G096.77-50.29 & 0300210601 & PSZ2 G278.79+08.54 & 0762440101 \\
PSZ2 G096.88+24.18 & 0723160401 & PSZ2 G280.76-52.30 & 0674490101 \\
PSZ2 G100.38+16.73 & 0822590301 & PSZ2 G281.48-08.34 & 0940960701 \\
PSZ2 G100.45+16.79 & 0822590301 & PSZ2 G285.87-74.93 & 0724770901 \\
PSZ2 G107.41-09.57 & 0822590601 & PSZ2 G302.41+21.60 & 0406200101 \\
PSZ2 G112.54+59.53 & 0881960101 & PSZ2 G318.46+83.79 & 0822590901 \\
PSZ2 G125.68-64.12 & 0012440101 & PSZ2 G319.16+26.63 & 0781890601 \\
PSZ2 G135.76-62.03 & 0404410201 & PSZ2 G332.29-23.57 & 0694610101 \\
PSZ2 G138.61-10.84 & 0783881701 & PSZ2 G356.88-11.33 & 0784820101 \\
PSZ2 G145.25+50.84 & 0882720901 & PSZ2 G359.67-07.23 & 0604860401 \\ 
PSZ2 G153.56+36.82 & 0822590101 & PSZ1 G083.35+76.41 & 0785130901 \\
PSZ2 G153.68+36.96 & 0822590101 & PSZ1 G224.73+33.65 & 0083240201 \\ 
PSZ2 G157.43+30.34 & 0723160501 & PSZ1 G252.36-32.75 & 0109460701 \\
PSZ2 G167.43-53.67 & 0652400701 & PSZ1 G288.27+11.71 & 0762440201 \\
PSZ2 G189.79-37.25 & 0723161701 & PSZ1 G292.00-43.64 & 0510181701 \\
PSZ2 G196.65-45.51 & 0679180401 & & \\
\bottomrule 
\end{tabular}
\end{table}
 
\end{appendix}

\end{document}